\PassOptionsToPackage{super, numbers,sort&compress}{natbib}
\documentclass[final,3p,times,authoryear]{elsarticle}

\usepackage{amsthm}
\usepackage{booktabs}
\usepackage{amssymb}
\usepackage{graphicx}
\usepackage{dcolumn}
\usepackage{bm}
\usepackage{mathptmx}
\usepackage{etoolbox}
\usepackage{subcaption}
\usepackage[utf8]{inputenc}  
\usepackage[T1]{fontenc}     
\usepackage{textcomp}        
\usepackage{mathrsfs}
\usepackage{natbib}
\usepackage{amsmath}
\usepackage{bm}
\usepackage[ruled,linesnumbered]{algorithm2e}
\setcitestyle{numbers,square}
\usepackage{hyperref}
\usepackage{mathtools}
\usepackage{multirow}
\usepackage{makecell}
\usepackage{rotating}

\newtheorem{theorem}{Theorem}
\newtheorem{proposition}{Proposition}
\newtheorem{lemma}{Lemma}
\newtheorem{remark}{Remark}
\newtheorem{assumption}{Assumption}

\journal{Journal of Computational Physics}

\begin{document}
\begin{frontmatter}

\title{A Third-Order Weighted Essentially Non-Oscillatory Compact Least-Squares Scheme for Hyperbolic Conservation Laws on Non-Uniform Grids}

\author[fn1,fn2]{Jianhua Pan\fnref{equal}\corref{cor1}}
\author[fn1,fn2]{Luxin Li\fnref{equal}}
\author[fn3]{Wei-Gang Zeng}
\address[fn1]{Zhejiang Provincial Engineering Research Center for the Safety of Pressure Vessel and Pipeline, Ningbo University, Ningbo, 315211, China }
\address[fn2]{Key Laboratory of Impact and Safety Engineering, Ministry of Education, Ningbo University, Ningbo, 315211, China}
\address[fn3]{Academy for Advanced Interdisciplinary Studies, Northeast Normal University, Changchun, 130024, China}
\fntext[equal]{These authors contribute equally to this work.}
\cortext[cor1]{Corresponding author.  \textit{E-mail address}: panjianhua@nbu.edu.cn}


\begin{abstract}
  A third-order weighted essentially non-oscillatory compact least-squares scheme is developed for the finite volume method on structured curvilinear non-uniform grids. The proposed scheme features compact least-squares reconstruction with optimal linear weights for broad-spectrum accuracy and non-linear weights for essentially non-oscillatory property. Through explicit second-order polynomials given for each control volume, the scheme maintains high accuracy on structured meshes with non-uniform grids. By integrating an improved shock detector developed in this work, coefficients with adaptive levels of dissipation are applied to achieve both the high resolution in smooth regions and high robustness in discontinuous regions. Furthermore, the proposed scheme is extended to the Euler equations through a characteristic decomposition technique. Numerical examples including both linear convection equation and nonlinear Euler/Navier-Stokes equations demonstrate the robustness and high-resolution of the proposed method.
\end{abstract}
\end{frontmatter}

\section{Introduction}
Robust discontinuity capturing and multiscale fluid-structure resolution are two key objectives in designing solvers for hyperbolic conservation laws, particularly for supersonic flows. On curvilinear structured grids, weighted essentially non-oscillatory (WENO) schemes \cite{liu_weighted_1994,jiang1996efficient,shu2006essentially,levy2000compact, qiu2002construction,capdeville2008central, feng2012new,baeza2019central,zeng2025high} are undoubtedly the most widely adopted and effective methods since their introduction. This paper proposes a third-order weighted essentially non-oscillatory compact least-squares (WCLS) scheme within the finite volume framework for curvilinear structured grids. Compared with conventional third-order WENO schemes, the proposed scheme demonstrates superior resolution and enhanced robustness in capturing discontinuities.

WENO schemes can be broadly classified into two categories. The first category \cite{liu_weighted_1994,jiang1996efficient,shu2006essentially,qiu2002construction,feng2012new,zeng2025high} employs point values, while the second category \cite{levy2000compact,capdeville2008central,baeza2019central} utilizes polynomial approximations. For point-value-based schemes, multiple lower-order point values are blended nonlinearly through convex combinations. This approach achieves higher-order accuracy in smooth regions while automatically selecting the least oscillatory stencil near discontinuities. However, no explicit polynomials are given in point-value-based WENO schemes. Such explicit polynomials could prove valuable in applications like adaptive $h$-refinement. 
Levy et al. \cite{levy2000compact} introduced the central WENO (CWENO) scheme, which combines a central high-order polynomial with multiple lower-order polynomials using weighting procedures analogous to point-value-based WENO methods. Although WENO strategies excel at discontinuity capturing, they may suffer from order degradation near smooth extrema when nonlinear weights are improperly designed. To mitigate this limitation, enhanced variants such as WENO-Z \cite{jacobs2009high,castro2011high,acker2016improved} and WENO-M \cite{henrick2005mapped,hong2020mapping,li2021efficient} have been developed.

Extensive numerical studies demonstrate that broadening the spectral resolution bandwidth of numerical schemes enhances multiscale fluid structure simulations \cite{tam1993dispersion,cheong2001grid,sun2011class,sun2014sixth,zeng2025high,li2023scale,sengupta2009new,popescu2005finite,wang2013low,lele1992compact}. 
Pioneering this field, Lele \cite{lele1992compact} introduced compact finite difference schemes with spectral-like resolution. Despite their spectral advantages, compact schemes require specialized discontinuity treatments to capture shocks and contact discontinuities.  To synergize the superior spectral properties of compact schemes with the shock-capturing robustness of WENO schemes, researchers have developed hybrid compact-WENO methodologies. These hybrid approaches are categorized into two distinct frameworks. In the first framework \cite{cockburn1994nonlinearly,adams1996high,pirozzoli2002conservative}, a WENO scheme activates upon being detected as a troubled cell and a switch function is integrated to guarantee the smooth transition from the compact scheme to WENO scheme.
In the second framework \cite{deng1997compact,guo2016fifth,guo2025weighted}, one-sided coefficients for the compact scheme are derived in discontinuous regions. The one-sided lower-order coefficients automatically select the least-oscillating stencil, keeping the overall scheme oscillation-free and are recovered to the high-order coefficients in smooth regions. The second framework, denoted as the weighted compact scheme, constitutes a natural extension of WENO methodology to compact formulations.

However, all the aforementioned hybrid compact-WENO schemes are specifically designed for finite difference/volume methods on uniform grids. A notable limitation of extending these methods by coordinate transformation to curvilinear structured grids is their stringent requirement for grid smoothness \cite{gamet1999compact}. In contrast, the finite volume method proposed by Wang and Ren \cite{wang2015accurate} implemented dimension-by-dimension on structured grids imposes less restrictive smoothness requirements. Wang and Ren \cite{wang2015accurate} developed a compact cubic spline reconstruction technique for finite volume methods on curvilinear structured grids, which Huang et al. \cite{huang2018high} subsequently extended to quintic polynomial reconstruction. These spline reconstruction schemes maintain continuity of polynomials at cell interfaces. For shock and contact discontinuity capturing, the schemes incorporate WENO method blending through shock detectors, while artificial dissipation is introduced in smooth regions to ensure numerical stability. Besides, Ren and his coworkers \cite{wang2016compact1, wang2016compact2,wang2017compact,jianhua2018high} also proposed another compact least-squares finite volume method which naturally fits the nonuniform curvilinear structured grids.

This paper presents a novel third-order WCLS (WCLS3) scheme for curvilinear structured grids, featuring several key innovations. First, the scheme provides explicit second-order polynomial, which facilitates adaptive mesh refinement during the mapping process. Unlike the hybrid spline-WENO approaches by Wang $\&$ Ren \cite{wang2015accurate} and Huang et al. \cite{huang2018high,huang2022adaptive}, the state on the cell interface is discontinuous and no artificial dissipation is required in smooth regions. Second, the weighting mechanism fundamentally differs from both point-value-based WENO \cite{liu_weighted_1994,jiang1996efficient,shu2006essentially,qiu2002construction,feng2012new,zeng2025high, guo2025weighted} and polynomial-based CWENO schemes \cite{levy2000compact,capdeville2008central,baeza2019central}. Rather than using convex combinations of lower-order components, we achieve the essentially non-oscillatory property through weighted least-squares reconstruction, which can be treated as a convex combination of same-order components. Third, the WCLS method with the proposed shock detector demonstrates no order degradation near local extrema. The scheme exhibits superior spectral properties compared to conventional third-order upwind methods.  The proposed approach solves block tridiagonal systems to directly obtain non-oscillatory second-order polynomials, which is different from the work of CLS method \cite{wang2016compact1}, where a gradient limiter is essential after the solution of block tridiagonal system. For Euler equations, we implement characteristic-wise reconstruction. While current results focus on third-order accuracy, the framework can be readily extended to higher-order polynomials.  Additionally, higher efficiency can be expected if the proposed WCLS method are coupled into an implicit solver since the tridiagonal matrix do not need to be solved explicitly \cite{wang2016compact2,wang2017compact,jianhua2018high}.

The structure of the paper is organized as follows. In Sec. \ref{sec:wcls}, the basic procedure of the CLS method is reviewed and the idea of WCLS is introduced. In Sec. \ref{sec:linear_and_nonlinear_weights}, the linear and nonlinear parameters of the WCLS3 scheme are carefully designed and analyzed. In Sec. \ref{sec:euler}, the proposed WCLS method is extended to the Euler equations. In Sec. \ref{sec:numerical_example}, numerical examples including linear convection, Euler and Navier-Stokes equations are given to demonstrate the property of the WCLS method. Conclusions are made in Sec. \ref{sec:concl}.

\section{Weighted Compact Least-Squares Scheme \label{sec:wcls}}
Consider the following one-dimensional (1D) linear convection equation,
\begin{equation}
  \label{eq:1dconvection}
  \frac{\partial u}{\partial t} + \frac{\partial f}{\partial x} = 0,
\end{equation}
where $f = au$ with a constant wave speed $a$. The computational domain is $\Omega = \left[x_{1/2}, x_{N+1/2}\right]$ and is divided into $N$ non-overlapping control volumes, each of which is denoted as $\Omega_i = \left[x_{i-1/2}, x_{i+1/2}\right]$. The integration of Eq. (\ref{eq:1dconvection}) over control volume $\Omega_i$ results in the following semi-discretized form as
\begin{equation}
  \label{eq:1dconvectionsemidiscrete}
  \Delta x_i \frac{\partial \overline{u}_i}{\partial t} = -\left(f_{i+1/2} - f_{i-1/2}\right),
\end{equation}
where $\Delta x_i$ is the length of control volume $\Omega_i$ and $\Delta x_i = x_{i+1/2} - x_{i-1/2}$.

Point values of $u$ at cell boundaries $x_{i-1/2},\,i=1,2,\cdots, N+1$  are demanded to forward Eq. (\ref{eq:1dconvectionsemidiscrete}) in temporal direction using time integrators such as Runge-Kutta (RK) method. In this work, an explicit second-order polynomial is assumed on each control volume $\Omega_i$ as
\begin{equation}
  P_i(x) = \overline{u}_i + a_i^{(1)}\frac{x-x_i}{\Delta x_i} + a_i^{(2)} \left(\left(\frac{x-x_i}{\Delta x_i}\right)^2-\frac{1}{12}\right),
\end{equation}
where $a_i^{(j)},\,j=1,2$ are unknowns to be determined and $x_i = \frac{1}{2}\left(x_{i-1/2} + x_{i+1/2}\right)$. Zero-mean basis functions $\psi_i^{(j)} = \left(\frac{x-x_i}{\Delta x_i}\right)^j-\frac{1}{\Delta x_i}\int_{\Omega_i}{\left(\frac{x-x_i}{\Delta x_i}\right)^j \mathrm{d}x}$ are utilized to ensure the conservation of the reconstructed polynomials.

First, the fundamental principles of compact least squares reconstruction \cite{wang2016compact1,wang2016compact2,wang2017compact,jianhua2018high,jianhua2025Hybrid} is reviewed.
In smooth regions, $u$ and its derivatives should be kept continuous across cell interfaces. For the second-order polynomials, the following six equations are included to construct the least-squares problem in a compact stencil for $\Omega_i$,
\begin{equation}
  \begin{aligned}
    \left.P_i\right|_{x_{i-1/2}} & = \left.P_{i-1}\right|_{x_{i-1/2}},\\
    w_{i,1}\left(\left. \frac{\mathrm{d}P_i}{\mathrm{d}x} \right|_{x_{i-1/2}}\right. & = \left. \left.\frac{\mathrm{d}P_{i-1}}{\mathrm{d}x} \right|_{x_{i-1/2}}\right), \\
    w_{i,2}\left(\left. \frac{\mathrm{d}^2 P_i}{\mathrm{d}x^2} \right|_{x_{i-1/2}}\right. & = \left.\left.\frac{\mathrm{d^2}P_{i-1}}{\mathrm{d}x^2} \right|_{x_{i-1/2}}\right), \\
  \end{aligned}
  \,\,\,\,
  \begin{aligned}
    \left.P_i\right|_{x_{i+1/2}} & = \left.P_{i+1}\right|_{x_{i+1/2}},\\
    w_{i,1}\left(\left. \frac{\mathrm{d}P_i}{\mathrm{d}x} \right|_{x_{i+1/2}}\right. & = \left.\left.\frac{\mathrm{d}P_{i+1}}{\mathrm{d}x} \right|_{x_{i+1/2}}\right), \\
    w_{i,2}\left(\left. \frac{\mathrm{d}^2 P_i}{\mathrm{d}x^2} \right|_{x_{i+1/2}}\right. & = \left.\left.\frac{\mathrm{d^2}P_{i+1}}{\mathrm{d}x^2} \right|_{x_{i+1/2}}\right), \\
  \end{aligned}
  \label{eq:linearCLS}
\end{equation}
where $w_{i,1} = w_1 \Delta x_i$ and $w_{i,2} = w_2 \left(\Delta x_i\right)^2$ are the weights keeping all the six equations dimensional consistent. $w_1$ \& $w_2$ are nonzero values and denoted as the linear weights of the scheme. The values of $w_1$ and $w_2$ do not affect the accuracy of the third-order CLS scheme and have been optimized in another work of the authors \cite{jianhua2025Hybrid} to broaden the resolved bandwidth.

Equation (\ref{eq:linearCLS}) constitutes an over-determined linear system for the unknown vector $\vec{a}_i = \left(a_i^{(1)}, a_i^{(2)}\right)^T$ and can be rearranged in the form as follows,
\begin{equation}
\bm{E}_i^{(-1)} \vec{a}_{i-1}  + \bm{E}_i^{(0)} \vec{a}_{i}  + \bm{E}_i^{(1)} \vec{a}_{i+1} = \vec{r}_i,
\label{eq:linearCLSMatrixForm}
\end{equation}
where
\begin{equation}
  \bm{E}_i^{(-1)} = -\left(
\begin{array}{lr}
  \frac{1}{2} & \frac{1}{6}\\
  \frac{w_1}{h_{i-1}}  & \frac{w_1}{h_{i-1}} \\
  0 & \frac{2 w_2}{h^2_{i-1}}\\ 
  0 & 0\\
  0 & 0 \\
  0 & 0
\end{array}
\right),
  \bm{E}_i^{(0)} = \left(
\begin{array}{lr}
  -\frac{1}{2} & \frac{1}{6}\\
  w_1 & -w_1 \\
  0 & 2 w_2\\
  \frac{1}{2} & \frac{1}{6}\\
  w_1 & w_1 \\
  0 & 2 w_2
\end{array}
\right),
  \bm{E}_i^{(1)} = -\left(
\begin{array}{lr}
  0 & 0\\
  0 & 0 \\
  0 & 0 \\
  -\frac{1}{2} & \frac{1}{6}\\
  \frac{w_1}{h_{i+1}} & -\frac{w_1}{h_{i+1}} \\
  0 & \frac{2 w_2}{h^2_{i+1}}
\end{array}
\right),
  \vec{r}_i = \left(
\begin{array}{c}
  \overline{u}_{i-1}-\overline{u}_i \\
  0 \\
  0 \\
  \overline{u}_{i+1}-\overline{u}_i \\
  0 \\
  0 
\end{array}
\right).
\label{eq:linearCLSMatrixDetailed}
\end{equation}
In Eq. (\ref{eq:linearCLSMatrixDetailed}), $h_{i-1} = \frac{\Delta x_{i-1}}{\Delta x_{i}}$ and $h_{i+1} = \frac{\Delta x_{i+1}}{\Delta x_{i}}$.
Multiplying Eq. (\ref{eq:linearCLSMatrixForm}) by $\left(\bm{E}_i^{(0)}\right)^T$ results in Eq. (\ref{eq:linearCLSMatrixFormNorm}) as
\begin{equation}
\bm{M}_i^{(-1)} \vec{a}_{i-1}
+\bm{M}_i^{(0)} \vec{a}_{i}
+\bm{M}_i^{(1)} \vec{a}_{i+1} = \vec{b}_i, \label{eq:linearCLSMatrixFormNorm}
\end{equation}
where
\begin{equation}
  \begin{array}{cc}

 \bm{M}_i^{(-1)} =
\begin{pmatrix}
 -\frac{4 W_1-h_{i-1} }{4 h_{i-1} } & -\frac{12 W_1-h_{i-1} }{12 h_{i-1} } \\
 \frac{12 W_1-h_{i-1} }{12 h_{i-1} } & -\frac{-36 h_{i-1} W_1+144  W_2+h_{i-1}^2 }{36 h_{i-1}^2 } \\
\end{pmatrix}
 , & 
 \bm{M}_i^{(0)} =
 \begin{pmatrix}
  \frac{1}{2} (1+4 W_1) & 0\\
  0& \frac{1}{18} (1+36 W_1+144 W_2)
 \end{pmatrix},\\
 \bm{M}_i^{(1)} =
\begin{pmatrix}
 -\frac{4  W_1- h_{i+1}}{4 h_{i+1}} & \frac{12  W_1- h_{i+1}}{12 h_{i+1}} \\
 -\frac{12  W_1- h_{i+1}}{12  h_{i+1}} & -\frac{144  W_2-36 h_{i+1}  W_1+h_{i+1}^2 }{36 h_{i+1}^2} \\
\end{pmatrix}
 ,&
\vec{b}_i = 
  \left(
\begin{array}{c}
 \frac{1}{2} \left(\overline{u}_{i+1}-\overline{u}_{i-1}\right) \\
 \frac{1}{6} \left(\overline{u}_{i-1}-2 u_{i}+\overline{u}_{i+1}\right) \\
\end{array}
\right),
\end{array}
 \end{equation}
and $W_1 = w_1^2, W_2 = w_2^2$. Eventually, Eq. (\ref{eq:linearCLSMatrixFormNorm}) from all control volumes $\Omega_i,\,i=1,2,\cdots,N$ constitutes a block tridiagonal system as follows which can be inverted efficiently,
\begin{equation}
\bm{M} \vec{\alpha} =\vec{\chi}, \label{eq:clsglobal}
\end{equation}
where 
$\vec{\alpha} = \left(\vec{a}_1^T,\vec{a}_2^T,\cdots,\vec{a}_N^T\right)^T$ and $\vec{\chi} = \left(\vec{b}_1^T,\vec{b}_2^T,\cdots,\vec{b}_N^T\right)^T$.
If a periodic boundary condition is applied, $\bm{M}$ is a cyclic block tridiagonal matrix as
\begin{equation}
  \begin{pmatrix}
 \bm{M}_1^{(0)} &  \bm{M}_1^{(1)}& \bm{0} & \cdots &\cdots& \bm{0}& \bm{M}_1^{(-1)}  \\
 \bm{M}_2^{(-1)} & \bm{M}_2^{(0)}& \bm{M}_2^{(1)}& \bm{0} &\cdots& \bm{0}& \bm{0} \\
 \bm{0} & \bm{M}_3^{(-1)} &  \bm{M}_3^{(0)}& \bm{M}_3^{(1)} & \bm{0} & \cdots&  \bm{0} \\
 \vdots & \vdots & \vdots & \vdots &\vdots & \ddots & \vdots \\
 \bm{M}_{N}^{(1)} & \bm{0} &\cdots&\cdots&\bm{0}&\bm{M}_{N}^{(-1)} & \bm{M}_{N}^{(0)}
  \end{pmatrix}.
\end{equation}
If boundary conditions are applied at the two ends,  $\bm{M}$ becomes a block tridiagonal matrix.

However, Eq. (\ref{eq:linearCLS}) only applies to situation where the physical variables are continuous across cell boundaries. If $\Omega_i$ locates near discontinuity, the weights of the reconstruction stencil should be biased to the smooth side. Keeping this in mind, Eq. (\ref{eq:linearCLS}) is modified as follows,
\begin{equation}
  \begin{array}{cc}
     w_{i,L} \left(
  \begin{aligned}
    \left.P_i\right|_{x_{i-1/2}} & = \left.P_{i-1}\right|_{x_{i-1/2}},\\
    w_{i,1}\left(\left. \frac{\mathrm{d}P_i}{\mathrm{d}x} \right|_{x_{i-1/2}}\right. & = \left. \left.\frac{\mathrm{d}P_{i-1}}{\mathrm{d}x} \right|_{x_{i-1/2}}\right), \\
    w_{i,2}\left(\left. \frac{\mathrm{d}^2 P_i}{\mathrm{d}x^2} \right|_{x_{i-1/2}}\right. & = \left.\left.\frac{\mathrm{d^2}P_{i-1}}{\mathrm{d}x^2} \right|_{x_{i-1/2}}\right), \\
  \end{aligned}
  \right),&
  w_{i,R} \left(
  \begin{aligned}
    \left.P_i\right|_{x_{i+1/2}} & = \left.P_{i+1}\right|_{x_{i+1/2}},\\
    w_{i,1}\left(\left. \frac{\mathrm{d}P_i}{\mathrm{d}x} \right|_{x_{i+1/2}}\right. & = \left.\left.\frac{\mathrm{d}P_{i+1}}{\mathrm{d}x} \right|_{x_{i+1/2}}\right), \\
    w_{i,2}\left(\left. \frac{\mathrm{d}^2 P_i}{\mathrm{d}x^2} \right|_{x_{i+1/2}}\right. & = \left.\left.\frac{\mathrm{d^2}P_{i+1}}{\mathrm{d}x^2} \right|_{x_{i+1/2}}\right), \\
  \end{aligned}
  \right), \\
    l_{i,L} \left(
  \begin{aligned}
    w_{i,1}\left(\left. \frac{\mathrm{d}P_i}{\mathrm{d}x} \right|_{x_{i-1/2}} =  0 \right), \\
    w_{i,2}\left(\left. \frac{\mathrm{d}^2 P_i}{\mathrm{d}x^2} \right|_{x_{i-1/2}} =  0\right), \\
  \end{aligned}
  \right),&
  l_{i,R} \left(
  \begin{aligned}
    w_{i,1}\left(\left. \frac{\mathrm{d}P_i}{\mathrm{d}x} \right|_{x_{i+1/2}} =  0 \right), \\
    w_{i,2}\left(\left. \frac{\mathrm{d}^2 P_i}{\mathrm{d}x^2} \right|_{x_{i+1/2}} =  0 \right), \\
  \end{aligned}
  \right),
\end{array}
  \label{eq:nonlinearCLS}
\end{equation}
where $w_{i,L}$ and $w_{i,R}$ are the weights for the left and right face of $\Omega_i$. $w_{i,L}$ and $w_{i,R}$ are denoted as the nonlinear weights of the WCLS scheme. If there is a discontinuity across $\Omega_i$ and $\Omega_{i-1}$, $w_{i,L} \rightarrow 0$; if $\Omega_i$ and $\Omega_{i-1}$ locate inside smooth regions, $w_{i,L} \rightarrow 1$. Same principle is applied for the right face of $\Omega_i$. $l_{i,L}$ and $l_{i,R}$ are weights for the dissipation terms, which only take in effect when $w_{i,L}$ or $w_{i,R}$ are extremely small. Numerical experiments show that the dissipation terms are necessary to ensure the existence and stability of the solution for block tridiagonal system, even though resulting in a slight order degradation.

Following the same approach for Eq. (\ref{eq:linearCLS}), Eq. (\ref{eq:nonlinearCLS}) are firstly rewritten in the matrix form as,
\begin{equation}
\widetilde{\bm{E}}_i^{(-1)} \vec{a}_{i-1}  + \widetilde{\bm{E}}_i^{(0)} \vec{a}_{i}  + \widetilde{\bm{E}}_i^{(1)} \vec{a}_{i+1} = \vec{\widetilde{r}}_i.
\label{eq:nonlinearCLSMatrixForm}
\end{equation}
The multiplication of $\left(\widetilde{\bm{E}}_i^{(0)}\right)^T$ on both sides of Eq. (\ref{eq:nonlinearCLSMatrixForm}) results in
\begin{equation}
\widetilde{\bm{M}}_i^{(-1)} \vec{a}_{i-1}
+\widetilde{\bm{M}}_i^{(0)} \vec{a}_{i}
+\widetilde{\bm{M}}_i^{(1)} \vec{a}_{i+1} = \vec{\widetilde{b}}_i, \label{eq:nonlinearCLSMatrixFormNorm}
\end{equation}
where
\begin{equation}
 \widetilde{\bm{M}}_i^{(-1)} = 
 W_{i,L}
\begin{pmatrix}
 -\frac{4 W_1-h_{i-1} }{4 h_{i-1} } & -\frac{12 W_1-h_{i-1} }{12 h_{i-1} } \\
 \frac{12 W_1-h_{i-1} }{12 h_{i-1} } & -\frac{-36 h_{i-1} W_1+144  W_2+h_{i-1}^2 }{36 h_{i-1}^2 } \\
\end{pmatrix} ,
\end{equation}
\begin{equation}
 \widetilde{\bm{M}}_i^{(0)} = 
 \begin{pmatrix}
  (\varphi_{i,L}+\varphi_{i,R}) W_1 + (W_{i,L}+W_{i,R}) (\frac{1}{4}+W_1) & (-\varphi_{i,L}+\varphi_{i,R}) W_1 + (W_{i,R}-W_{i,L})(\frac{1}{12}+W_1)\\
  (-\varphi_{i,L}+\varphi_{i,R}) W_1 + (W_{i,R}-W_{i,L})(\frac{1}{12}+W_1)& (\varphi_{i,L}+\varphi_{i,R}) (W_1+4W_2) + (W_{i,L}+W_{i,R}) (\frac{1}{36}+W_1+4 W_2)
 \end{pmatrix},
\end{equation}
\begin{equation}
 \widetilde{\bm{M}}_i^{(1)} = 
 W_{i,R}
\begin{pmatrix}
 -\frac{4  W_1- h_{i+1}}{4 h_{i+1}} & \frac{12  W_1- h_{i+1}}{12 h_{i+1}} \\
 -\frac{12  W_1- h_{i+1}}{12  h_{i+1}} & -\frac{144  W_2-36 h_{i+1}  W_1+h_{i+1}^2 }{36 h_{i+1}^2} \\
\end{pmatrix} ,
\end{equation}
\begin{equation}
\vec{\widetilde{b}}_i = 
  \left(
\begin{array}{c}
 \frac{1}{2} \left(
   W_{i,L}\left(\overline{u}_{i}-\overline{u}_{i-1}\right) 
  +W_{i,R}\left(\overline{u}_{i+1}-\overline{u}_{i}\right) 
 \right) \\
 -\frac{1}{6} 
 \left(
 W_{i,L} \left(\overline{u}_{i}- u_{i-1}\right) 
 +W_{i,R} \left(\overline{u}_{i}- u_{i+1}\right) 
 \right)
\end{array}
\right),
\end{equation}
and $W_{i,L} = w^2_{i,L}$, $W_{i,R} = w^2_{i,R}$, $\varphi_{i,L} = l^2_{i,L}$, $\varphi_{i,R} = l^2_{i,R}$. In this work
\[
\varphi_{i,L} = \left\{
\begin{array}{lr}
  0, & \text{if} \,\,\,\,W_{i,L} \geq 0.01,\\
  1-W_{i,L}, & \text{if} \,\,\,\,W_{i,L} < 0.01,\\
\end{array}\right.,\,\,\,\,
\varphi_{i,R} =\left\{ 
\begin{array}{lr}
  0, & \text{if}\,\,\,\,W_{i,R} \geq 0.01,\\
  1-W_{i,R}, & \text{if} \,\,\,\,W_{i,R} < 0.01.\\
\end{array}
\right.
\]

If the cell is located at boundaries, appropriate boundary conditions should be proposed for the least-squares problem.  
Readers are referred to the work of Wang \cite{wang2016compact1, wang2016compact2} and Pan \cite{jianhua2025Hybrid} for the detailed definition of least-squares problem at boundary cells if we want to introduce the Dirichlet or Neumann boundary condition into the CLS reconstruction.
In this work, another simple but effective boundary treatment is utilized. Taking $\Omega_1$ as an example, we just set $W_{i,L} = 0$ and $\varphi_{i,L} = 1$ in Eq. \ref{eq:nonlinearCLSMatrixFormNorm}. This treatment, which is similar to the natural boundary condition, helps keep the boundness of the solution near boundaries.

Coupling Eq. (\ref{eq:nonlinearCLSMatrixFormNorm}) for all the control volumes and through the solution of either a cyclic block tridiagonal or a block tridiagonal linear system, essentially non-oscillatory second-order polynomials are obtained for each control volume.
Once the second-order polynomials are obtained for each control volume, 
the flux $f_{i+1/2}$ in Eq. (\ref{eq:1dconvectionsemidiscrete}) is calculated by
\begin{equation}
  f_{i+1/2} = \frac{a}{2}\left(u_{i+1/2}^L + u_{i+1/2}^R\right) - \frac{|a|}{2}\left(u_{i+1/2}^R - u_{i+1/2}^L\right),\label{eq:fluxscalar}
\end{equation}
where $u_{i+1/2}^L = P_{i}(x_{i+1/2})$ and $u_{i+1/2}^R = P_{i+1}(x_{i+1/2})$.
At last, a multi-stage Runge-Kutta method, e.g., the third-order strong stability preserving Runge-Kutta (SSP-RK) method \cite{gottlieb2011strong,PAN202324} can be utilized to advance Eq. (\ref{eq:1dconvectionsemidiscrete}) from time $t^n$ to $t^{n+1}$.

\section{Design of Linear and Nonlinear Weights \label{sec:linear_and_nonlinear_weights}}
\subsection{Linear Weights}
A series of optimized linear weights ($W_1$ and $W_2$) have been obtained in the hybrid CLS-CWENO work by the authors \cite{jianhua2025Hybrid}. For the sake of completeness, the optimization process is briefly reviewed here.

Assuming a harmonic distribution of $u(x,t) = A e^{Ikx}$, where $I = \sqrt{-1}$, the modified non-dimensional wavenumber of the CLS scheme in Eq. \ref{eq:linearCLSMatrixForm} can be derived as
\begin{equation}
  \begin{aligned}
Re(\kappa ') & = \frac{-6 \sin(\kappa)\left(
  n_1 + n_2 \cos(\kappa) + n_3 \cos(2 \kappa)
\right)}{
  d_1 + d_2 \cos(\kappa) + d_3 \cos(2\kappa)
},\\
Im(\kappa ') & = \frac{576 W_1 W_2 \left(\sin^6(\kappa /2)\right)}{ d_1 + d_2 \cos(\kappa) + d_3 \cos(2\kappa)},
  \end{aligned}
  \label{eq:dispdissmywork3}
\end{equation}
where
\begin{equation}
\begin{array}{lll}
n_1 = 2 \left(W_1 + 3W_2 + 9 W_1 W_2\right),& n_2 = W_1 -6\left(1+4W_1\right) W_2, & n_3  = 6 W_1 W_2,\\
d_1 = -9\left(W_1 + W_2 + 12 W_1 W_2\right),& d_2 = 8 W_1 \left(-1+18 W_2\right), & d_3 =-\left(W_1 + 9\left(-1+4W_1\right)W_2\right).
\end{array}
\end{equation}

\begin{figure}[!htbp]
  \centering
    \begin{subfigure}[b]{0.3\textwidth}
    \includegraphics[width=\textwidth]{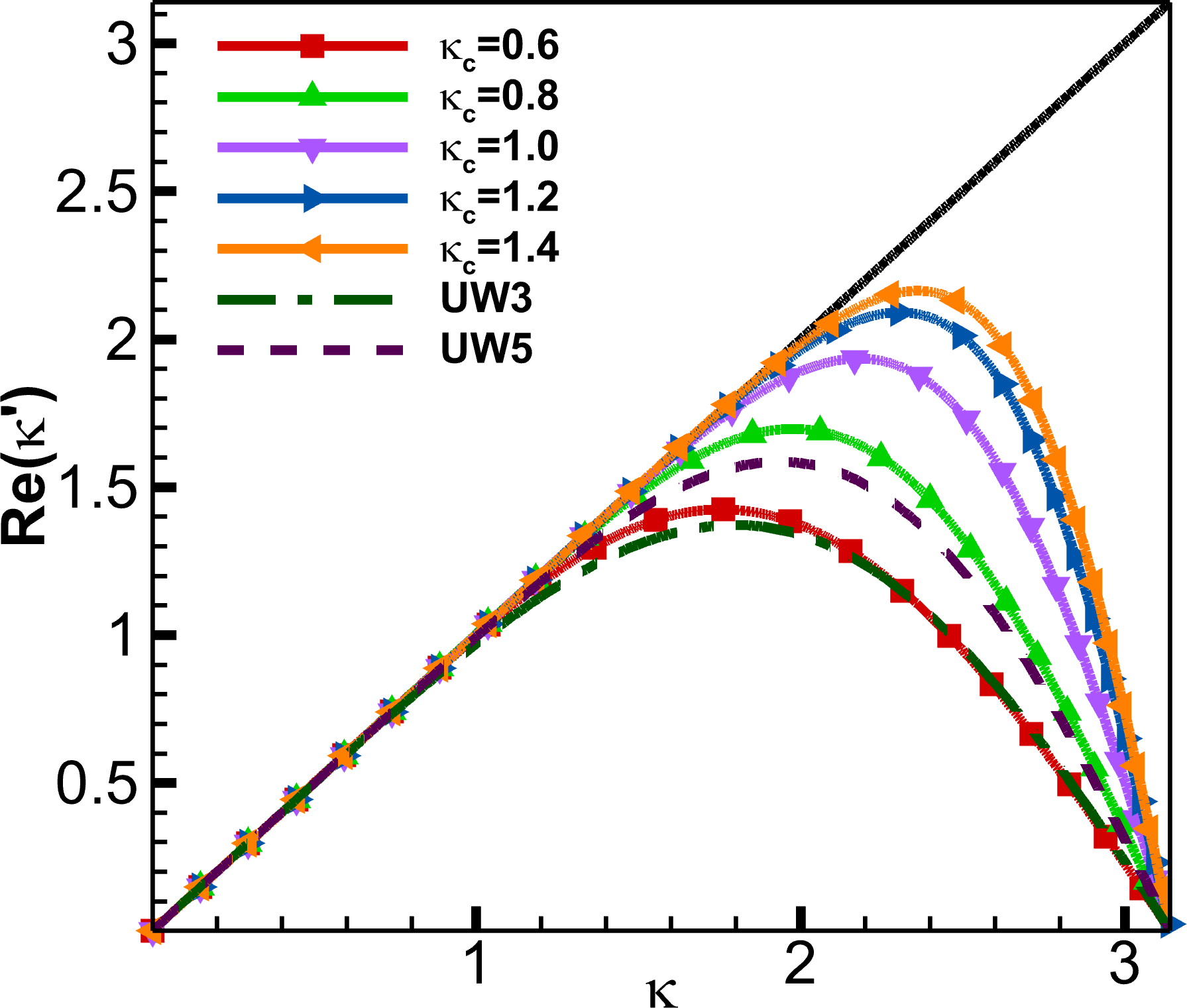}
    \caption{Dispersion.}
    \end{subfigure}
    \quad 
    \begin{subfigure}[b]{0.3\textwidth}
    \includegraphics[width=\textwidth]{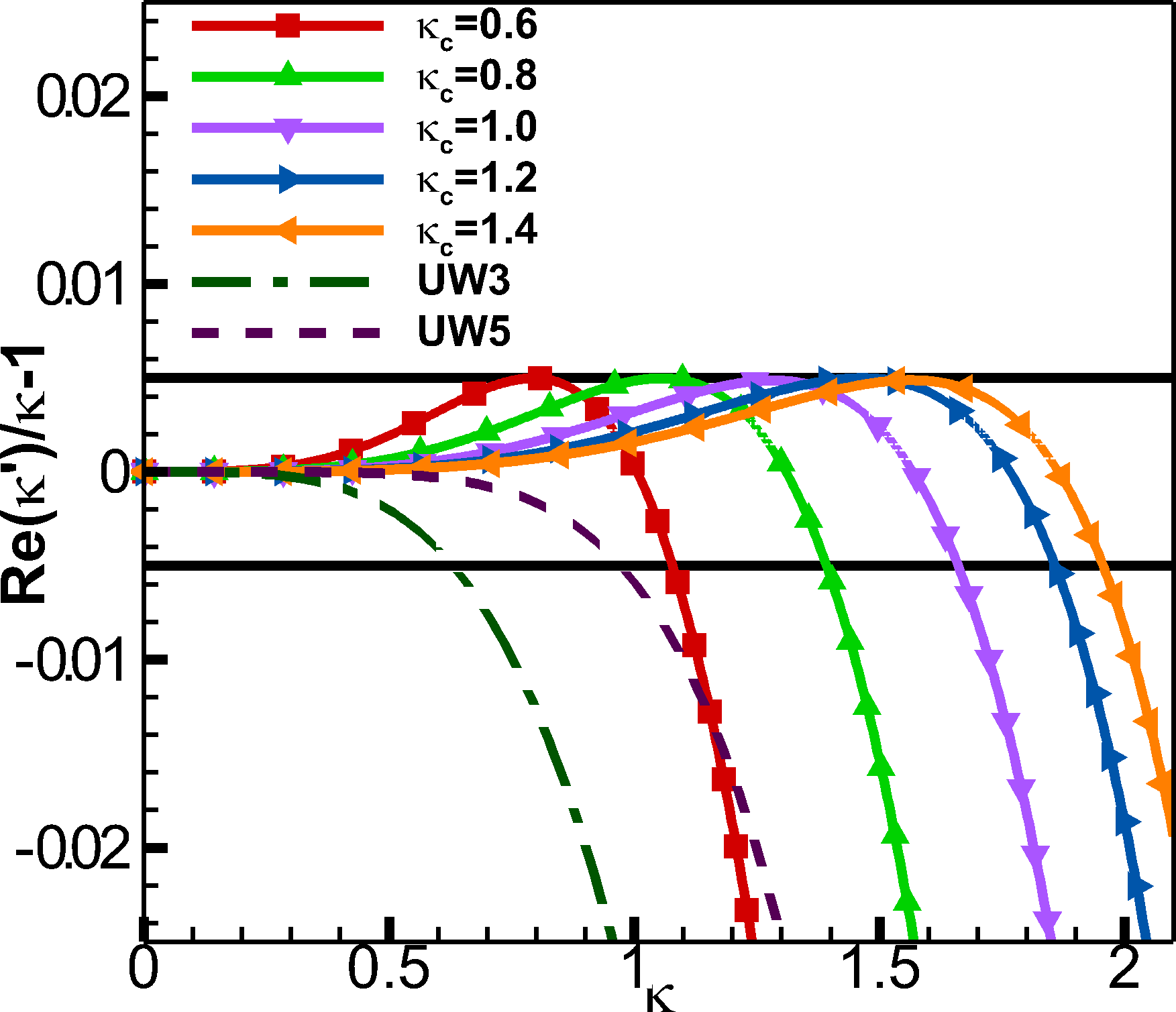}
      \caption{Relative dispersion error.}
    \end{subfigure}\\
    \begin{subfigure}[b]{0.3\textwidth}
    \includegraphics[width=\textwidth]{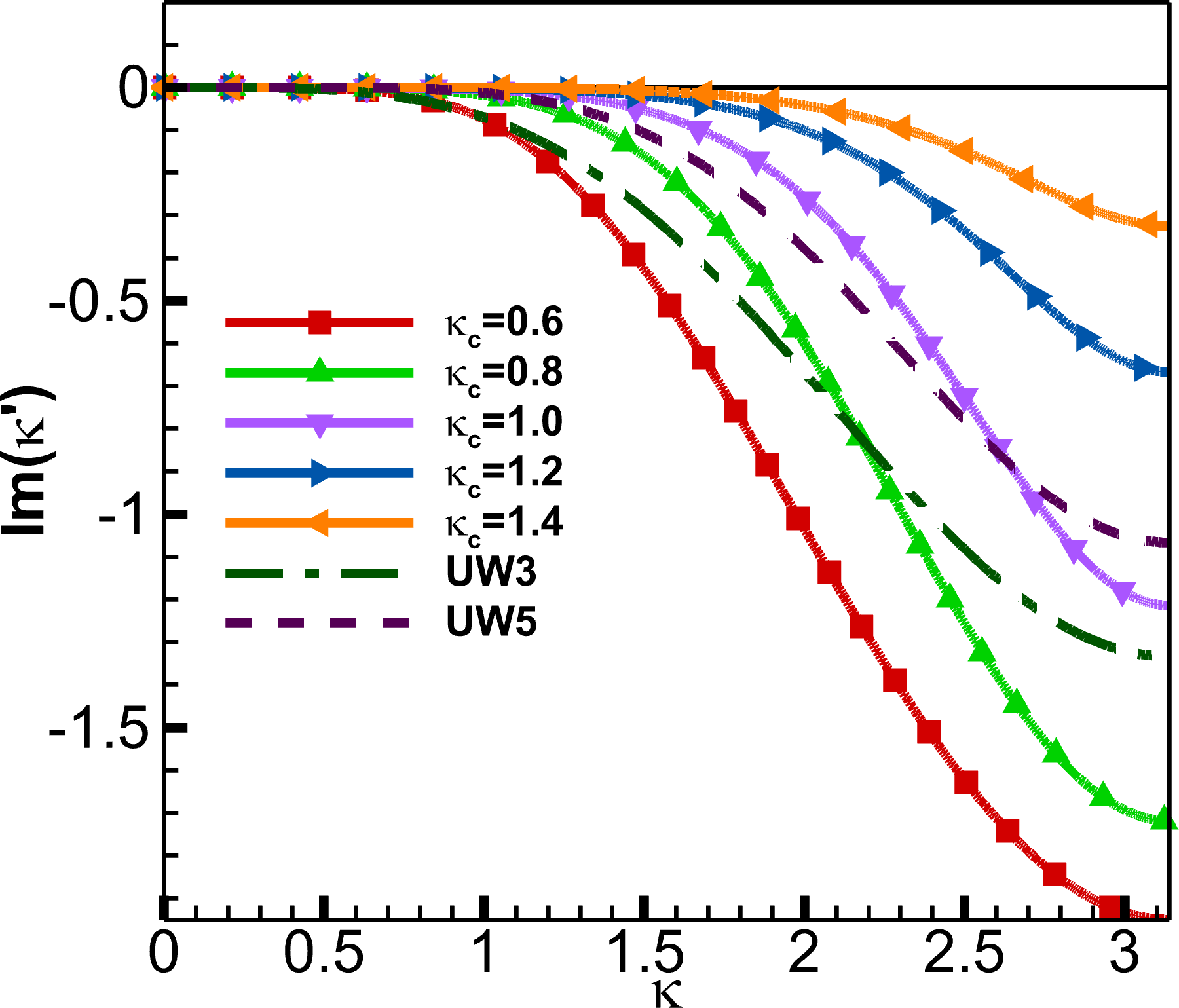}
      \caption{Dissipation.}
    \end{subfigure}
    \quad\quad
    \begin{subfigure}[b]{0.3\textwidth}
    \includegraphics[width=0.95\textwidth]{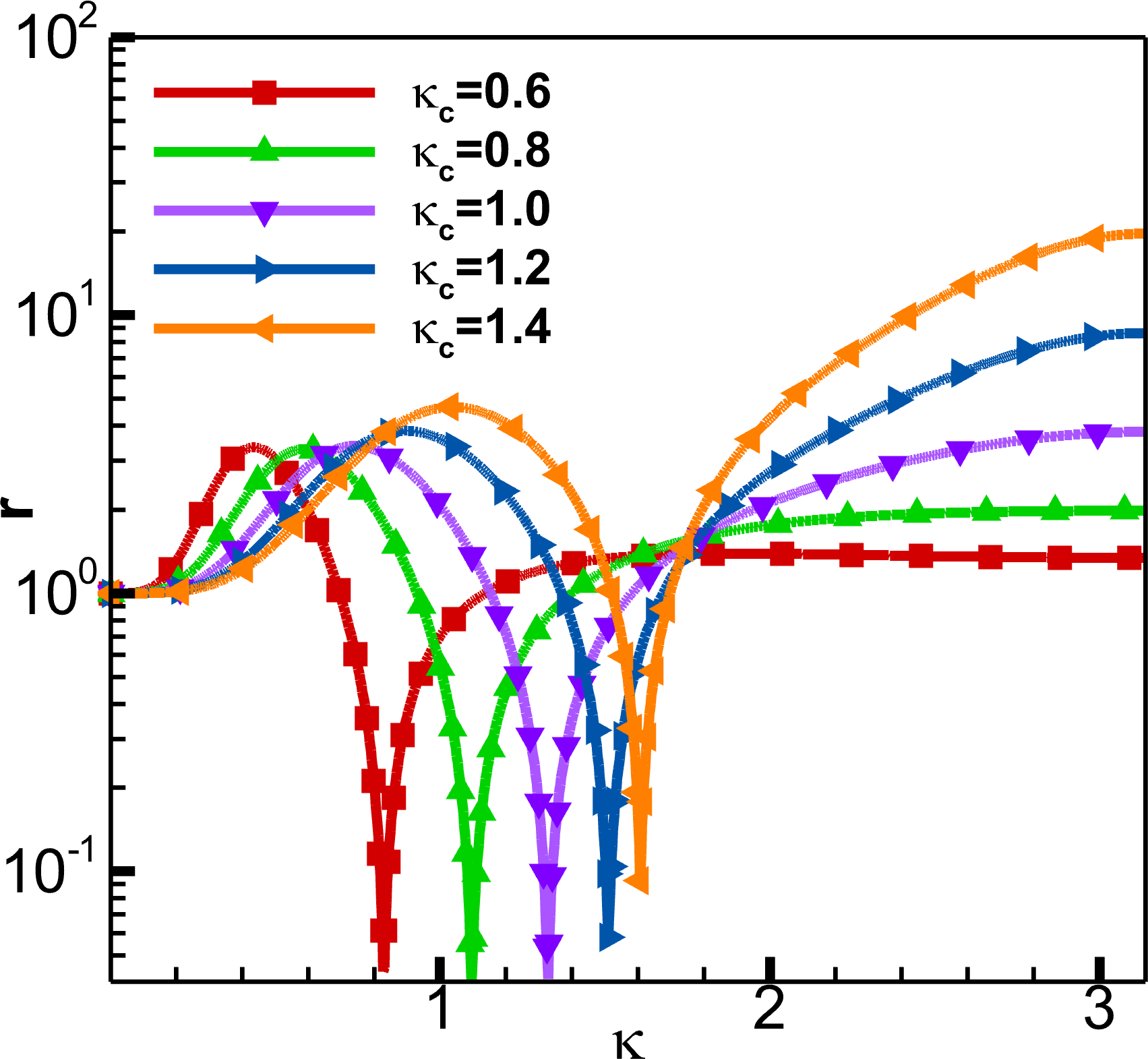}
      \caption{$r$ index.\label{fig:rindex}}
    \end{subfigure}
    \caption{\label{fig:dispdisscls3} Spectral properties for the third-order CLS scheme with $W_1$ and $W_2$ optimized by different $\kappa_c$. UW3: the third-order upwind scheme; UW5: the fifth-order upwind scheme.}
\end{figure}

Then we define the maximum resolved non-dimensional wavenumber as $\kappa_m$ such that
\begin{equation}
  \begin{aligned}
  \forall \kappa \leq \kappa_m,&  \left|\frac{Re(\kappa')}{\kappa}-1\right| \leq 0.5\% \,\,\mathrm{and}\,\, -Im(\kappa') \leq 0.5\%,\\
  \exists \kappa > \kappa_m, &  \left|\frac{Re(\kappa')}{\kappa}-1\right| > 0.5\% \,\,\mathrm{or}\,\, -Im(\kappa') > 0.5\%.
  \end{aligned}
\end{equation}
Given a critical non-dimensional wavenumber $\kappa_c$, the optimized $W_1$ and $W_2$ are obtained such that $W_2$ has the maximum value in the set defined by 
\begin{equation}
\left\{\left.(W_1, W_2)\right| \kappa_m \geq \kappa_c, W_1 \geq W_2, \frac{Re(\kappa')}{\kappa}-1 \leq 0.5\% \right\}.
\label{eq:constrained3rd}
\end{equation}
The optimized linear weights $W_1$ and $W_2$ are listed in Tab. \ref{tab:w1w2} under various critical non-dimensional wavenumber $\kappa_c$ and their spectral properties are shown as in Fig. \ref{fig:dispdisscls3}. In Fig. \ref{fig:rindex}, the $r$-index defined as 
\[r = \frac{\left|\frac{\partial Re(\kappa')}{\partial \kappa} -1\right|+0.001}{\left|-Im(\kappa')\right|+0.001}\] 
is also presented.
The $r$-index measures the ratio between the dispersion and dissipation errors \cite{hu2012dispersion} and is recommended to be smaller than 10.
According to this criterion, the coefficients optimized by $\kappa_c = 1.4$ is too small to have enough dissipation.

\begin{figure}[!htbp]
  \centering
    \begin{subfigure}[b]{0.3\textwidth}
    \includegraphics[width=\textwidth]{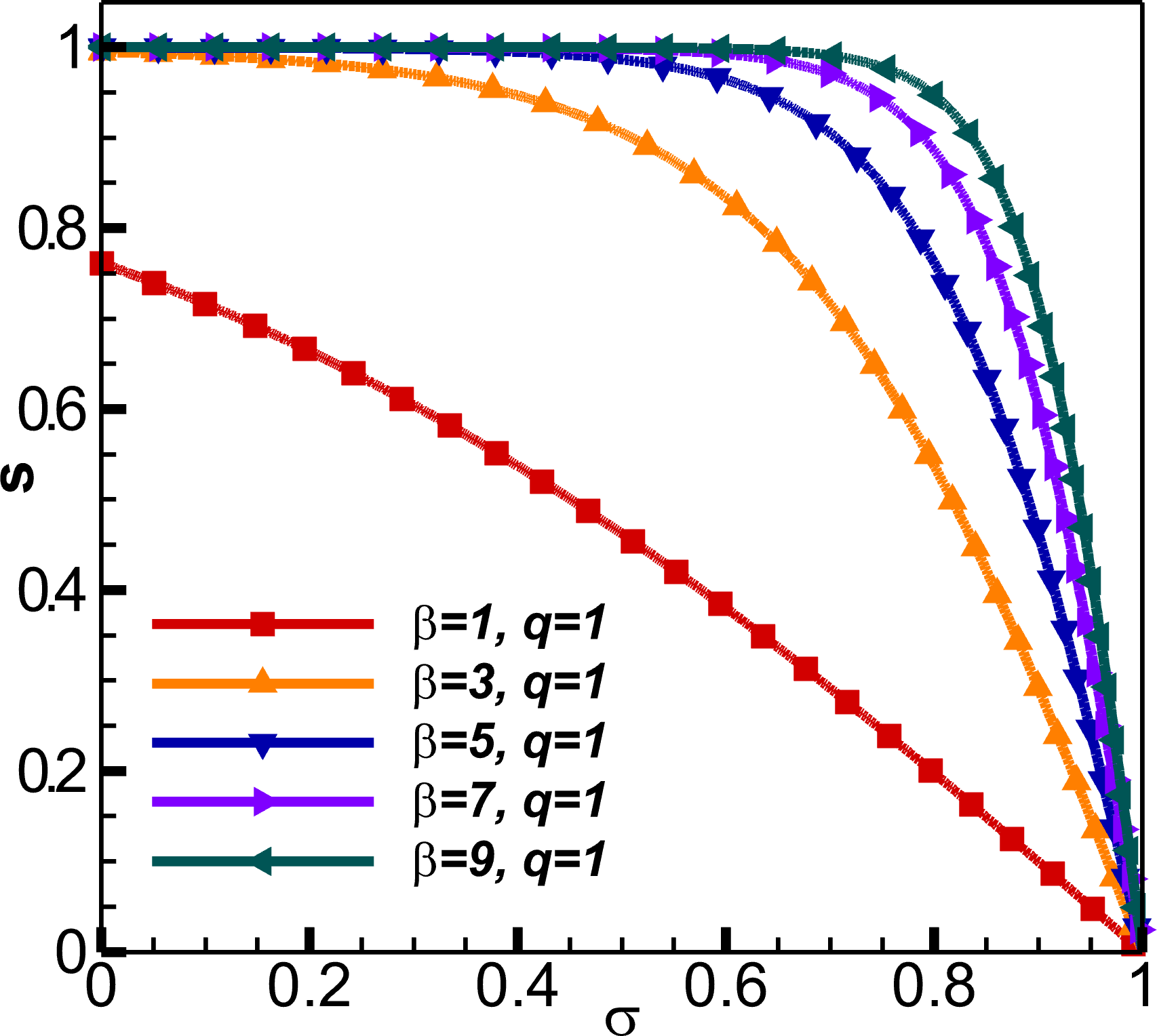}
    \caption{$s-\sigma$ relation with varying $\beta$ and $q=1$.}
    \end{subfigure}
    \quad 
    \begin{subfigure}[b]{0.3\textwidth}
    \includegraphics[width=\textwidth]{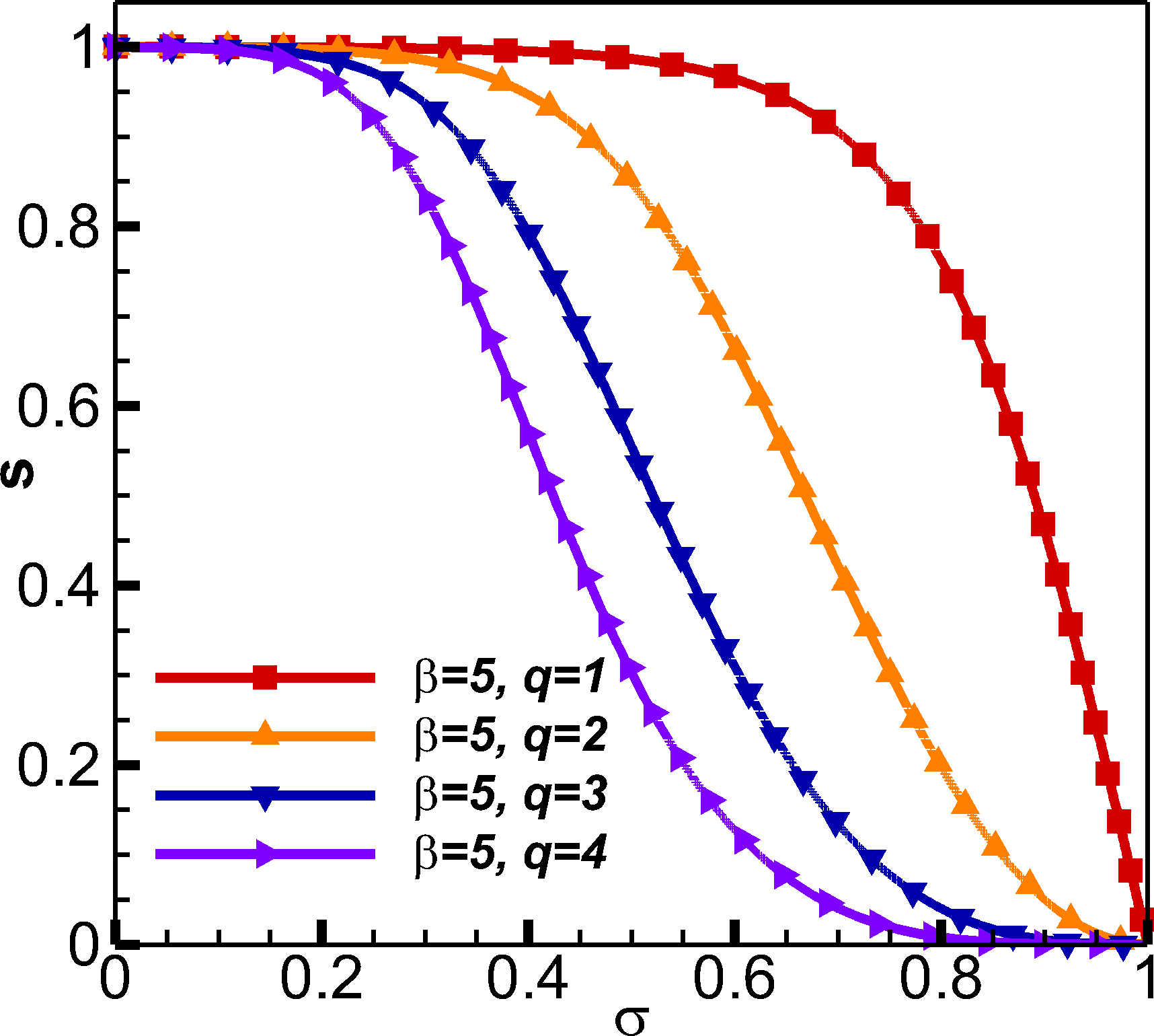}
    \caption{$s-\sigma$ relation with varying $q$ and $\beta=5$.}
    \end{subfigure}
    \caption{\label{fig:s-sigma-relation} The influence of hyperparameters of $\beta$ and $q$ over the $s-\sigma$ relation.}
\end{figure}

From Fig. \ref{fig:dispdisscls3}, it can be concluded that, with the decrease of $\kappa_c$, the optimized coefficients $W_1$ and $W_2$ apply increased dissipation on the resultant numerical scheme. For the compressible flows, the discontinuities and the multiscale flow structures coexist in the flow field. More dissipation is required to suppress the numerical oscillations near discontinuities, and less dissipation is demanded in smooth regions to capture the multiscale structures. Thus, a novel strategy that blends coefficients with different levels of dissipation is proposed in this work. Denoting the coefficients optimized by $\kappa_c = \kappa_0$ as $W_i^{\kappa_0},\,i=1,2$, the ultimate coefficients during simulations are 
\begin{equation}
W_i = s \cdot W_i^{0.6} + (1-s) \cdot W_i^{\kappa_0}  ,\,\,i=1,2.
\label{eq:blendcoeffi}
\end{equation}
In Eq. (\ref{eq:blendcoeffi}), $W_i^{0.6},\, i=1,2,$ are coefficients in discontinuous control volumes with enough dissipation; $W_i^{\kappa_0},\, i=1,2,$ are the coefficients in smooth regions to capture multiscale flow structures. $s$ is a function that transits from $0$ in smooth regions to $1$ in discontinuous regions. $s$ is defined as
\begin{equation}
  s = \tanh(\beta (1-\sigma)^q),
\end{equation}
where $\beta$ and $q$ are hyperparameters. $\sigma$ is a shock detector satisfying
\begin{equation}
\sigma = \left\{
  \begin{array}{lc}
   O(\left(\Delta x\right)^k),\,\, & k \geq 0\,\,\text{if} \,\,\,\, \Omega_i \,\,\,\,\text{locates inside discontinuity}, \\
   1, & \text{otherwise}.
  \end{array}
\right.
\end{equation}
The influence of $\beta$ and $q$ over the $s-\sigma$ relation is shown in Fig. \ref{fig:s-sigma-relation}. With the increase of $\beta$ or the decrease of $q$, $s$ transits to $1$ faster when $\sigma$ moves towards $0$, i.e., the resultant scheme is more dissipative. In this work, $\beta = 5$ and $q = 2$ are utilized in numerical experiments.
\begin{table}
  \centering
\caption{Optimized $W_1$ and $W_2$ under various $\kappa_c$.}\label{tab:w1w2}
\begin{tabular}{lcc}
  \toprule
$\kappa_c$&$W_1$&$W_2$\\
\midrule
0.6 & $4.3062\times 10^{-1}$ & $2.6854\times 10^{-1}$ \\
0.8 & $1.9080\times 10^{-1}$ & $4.2398\times 10^{-2}$ \\
1.0 & $8.2382\times 10^{-2}$ & $1.0720\times 10^{-2}$ \\
1.2 & $3.4334\times 10^{-2}$ & $3.4712\times 10^{-3}$ \\
1.4 & $1.4915\times 10^{-2}$ & $1.3437\times 10^{-3}$ \\
\bottomrule
\end{tabular}
\end{table}

\subsection{Nonlinear Weights}
The nonlinear weights in the WCLS reconstruction are designed as 
\begin{equation}
  W_{i,L} \doteq w_{i,L}^2 = \frac{2\alpha_0}{ \alpha_0 + \alpha_1 }, \,\,
  W_{i,R} \doteq w_{i,R}^2 = \frac{2\alpha_1}{ \alpha_0 + \alpha_1 }, 
\end{equation}
where
\begin{equation}
  \alpha_0 = \frac{1}{\left(I_0+\epsilon_1\right)^p},\,\,
  \alpha_1 = \frac{1}{\left(I_1+\epsilon_1\right)^p},
  \label{eq:nonlinearweight_alpha}
\end{equation}
and
\begin{equation}
 I_{0}   = \left(\frac{2 \lambda_0}{\lambda_{-1} + \lambda_{0}} \left(\overline{u}_{i} - \overline{u}_{i-1}\right)\right)^2,\,
 I_{1}   = \left(\frac{2 \lambda_0}{\lambda_{0} + \lambda_{1}} \left(\overline{u}_{i+1} - \overline{u}_{i}\right)\right)^2,
\end{equation}
\begin{equation}
\lambda_{m}  = \frac{\Delta x_{i+m}}{\Delta x_i},\,\,m=-1,0,1.
\end{equation}
$p$ in Eq. (\ref{eq:nonlinearweight_alpha}) is a free parameter. The value of $p$ does not affect the accuracy of WCLS even near local extrema, which is different from some related work, e.g., the third-order CWENO (CWENO3) \cite{baeza2019central} where $p$ is suggested to be $1$. A larger value of $p$ in Eq. (\ref{eq:nonlinearweight_alpha}) corresponds to a larger dissipation near discontinuities. However, large $p$-values cause ill-conditioning of the WCLS reconstruction matrix.
In our work, $p$ is suggested to be chosen as $0.6$ through numerical experiments. 
$\epsilon_1$ in Eq. (\ref{eq:nonlinearweight_alpha}) is a small number avoiding division by zero and is set to a scale-invariant form as
\begin{equation}
\epsilon_1 = \left[\max\left(1\times 10^{-30} \frac{\sum_{m=-1}^{1}{\left|\overline{u}_{i+m}\right|}}{3}, 1\times 10^{-50}\right)\right]^2.
\label{eq:epsilonren}
\end{equation}

\subsection{Shock Detector\label{sec:shockdetector}}
This work proposes a novel shock detector which is utilized in the linear weights to achieve adaptive dissipation. The shock detector can also be utilized to detect smooth regions, where the nonlinear weighting process can be turned off, which further enhances the high-resolution of the proposed scheme.
The shock detector is derived from the work of Baeza \cite{baeza2020efficient} with some simplification and improvements.
For control volume $\Omega_i$, the shock detector is defined as
\begin{equation}
  \begin{aligned}
  \sigma & = \min\left(1.0, \theta/\theta_c\right),\\
  \theta & = \theta^L\cdot \theta^R,
  \end{aligned}
  \label{eq:owenosd}
\end{equation}
where $\theta_c$ is a critical value chosen as 0.3 in this work and
\begin{equation}
  \theta^L = \frac{J_L + \epsilon_2}{J_L + \tau_L + \epsilon_2},\,\,
  \theta^R = \frac{J_R + \epsilon_2}{J_R + \tau_R + \epsilon_2},
  \label{eq:sigmalr}
\end{equation}
with
\begin{equation}
J_L = \left(I_0 I_1 + I_{-1} I_{-1}\right),\,\,\,\, \tau_L  = d_L(I_0+ I_1),\,\,\,\, d_L  = \left(\zeta_1^L \overline{u}_{i+1} + \zeta_0^L \overline{u}_{i} + \zeta_{-1}^L \overline{u}_{i-1} + \zeta_{-2}^L \overline{u}_{i-2}\right)^2,\label{eq:d3ul}\\
\end{equation}
\begin{equation}
J_R  = \left(I_0 I_1 + I_{2} I_{2}\right),\,\,\,\, \tau_R  = d_R(I_0+ I_1),\,\,\,\, d_R  = \left(\zeta_{-1}^R \overline{u}_{i-1} + \zeta_0^R \overline{u}_{i} + \zeta_{1}^R \overline{u}_{i+1} + \zeta_{2}^R \overline{u}_{i+2}\right)^2\label{eq:d3ur},
\end{equation}
and
\begin{equation}
\begin{array}{c}
  I_{-1}  = \left(\frac{2 \lambda_0}{\lambda_{-2} + \lambda_{-1}} \left(\overline{u}_{i-1} - \overline{u}_{i-2}\right)\right)^2,\,\,\,\, I_{0} = \left(\frac{2 \lambda_0}{\lambda_{-1} + \lambda_{0}} \left(\overline{u}_{i} - \overline{u}_{i-1}\right)\right)^2,\\
  I_{1}   = \left(\frac{2 \lambda_0}{\lambda_{0} + \lambda_{1}} \left(\overline{u}_{i+1} - \overline{u}_{i}\right)\right)^2,\,\,\,\, I_{2}   = \left(\frac{2 \lambda_0}{\lambda_{1} + \lambda_{2}} \left(\overline{u}_{i+2} - \overline{u}_{i+1}\right)\right)^2.
\end{array}\label{eq:I2}
\end{equation}
In Eq. (\ref{eq:sigmalr}), $\epsilon_2$ is a small number avoiding division by zero and set as
\begin{equation}
\epsilon_2 = \left[\max\left(1\times 10^{-6} \frac{\sum_{m=-2}^{2}{\left|\overline{u}_{i+m}\right|}}{5}, 1\times 10^{-50}\right)\right]^4.
\label{eq:epsilonpan}
\end{equation}
In Eq. (\ref{eq:I2}),
\begin{equation}
\lambda_{m}  = \frac{\Delta x_{i+m}}{\Delta x_i},\,\,m=-2,-1,0,1,2.
\end{equation}
The coefficients of $\zeta^L$ and $\zeta^R$ are defined by
\begin{equation}
\zeta_{-2}^L  = -\frac{24}{
  \left(\lambda _{-2}+\lambda _{-1}\right) \left(\lambda _{-2}+\lambda _{-1}+\lambda _0\right) \left(\lambda _{-2}+\lambda _{-1}+\lambda _0+\lambda _1\right)
  },
\end{equation}
\begin{equation}
\zeta_{-1}^L
=\frac{
  24 \left(\lambda _{-2}^2+\left(3 \lambda _{-1}+2 \lambda _0+\lambda _1\right) \lambda _{-2} +3 \lambda _{-1}^2+\lambda _0 \left(\lambda _0+\lambda _1\right)+2 \lambda _{-1} \left(2 \lambda _0+\lambda _1\right)\right)
  }{
    \left(\lambda _{-2}+\lambda _{-1}\right) \left(\lambda _{-1}+\lambda _0\right) \left(\lambda _{-2}+\lambda _{-1}+\lambda _0\right)   \left(\lambda _{-1}+\lambda _0+\lambda _1\right) \left(\lambda _{-2}+\lambda _{-1}+\lambda _0+\lambda _1\right)
  },
\end{equation}
\begin{equation}
\zeta_{0}^L =-\frac{
  24 \left(\lambda _{-1}^2+2 \left(2 \lambda _0+\lambda _1\right) \lambda _{-1}+3 \lambda _0^2  +\lambda _1^2+3 \lambda _0 \lambda _1+\lambda _{-2} \left(\lambda _{-1}+2 \lambda _0+\lambda _1\right)\right)
  }{
    \left(\lambda _{-1}+\lambda _0\right) \left(\lambda _{-2}+\lambda _{-1}+\lambda _0\right) \left(\lambda _0+\lambda _1\right)  \left(\lambda _{-1}+\lambda _0+\lambda _1\right) \left(\lambda _{-2}+\lambda _{-1}+\lambda _0+\lambda _1\right)
  },
\end{equation}
\begin{equation}
\zeta_{1}^L =\frac{24}{
  \left(\lambda _0+\lambda _1\right) \left(\lambda _{-1}+\lambda _0+\lambda _1\right)   \left(\lambda _{-2}+\lambda _{-1}+\lambda _0+\lambda _1\right)
  },
\end{equation}
\begin{equation}
  \zeta_{-1}^R  = -\frac{24}{
    \left(\lambda _{-1}+\lambda _0\right) \left(\lambda _{-1}+\lambda _0+\lambda _1\right) \left(\lambda _{-1}+\lambda _0+\lambda _1+\lambda _2\right)
    },
\end{equation}
\begin{equation}
  \zeta_{0}^R  =\frac{
    24 \left(\lambda _{-1}^2+\left(3 \lambda _0+2 \lambda _1+\lambda _2\right) \lambda _{-1}+3 \lambda _0^2 +\lambda _1 \left(\lambda _1+\lambda _2\right)+2 \lambda _0 \left(2 \lambda _1+\lambda _2\right)\right)
    }{
      \left(\lambda _{-1}+\lambda _0\right) \left(\lambda _0+\lambda _1\right) \left(\lambda _{-1}+\lambda _0+\lambda _1\right)  \left(\lambda _0+\lambda _1+\lambda _2\right) \left(\lambda _{-1}+\lambda _0+\lambda _1+\lambda _2\right)
    },
\end{equation}
\begin{equation}
\zeta_{1}^R  =  -\frac{
  24 \left(\lambda _0^2+2 \left(2 \lambda _1+\lambda _2\right) \lambda _0+3 \lambda _1^2+\lambda _2^2+3 \lambda _1 \lambda _2+\lambda _{-1} \left(\lambda _0+2 \lambda _1+\lambda _2\right)\right)
  }{
    \left(\lambda _0+\lambda _1\right) \left(\lambda _{-1}+\lambda _0+\lambda _1\right) \left(\lambda _1+\lambda _2\right)  \left(\lambda _0+\lambda _1+\lambda _2\right) \left(\lambda _{-1}+\lambda _0+\lambda _1+\lambda _2\right)
  },
\end{equation}
\begin{equation}
\zeta_{2}^R  =  \frac{24}{
  \left(\lambda _1+\lambda _2\right) \left(\lambda _0+\lambda _1+\lambda _2\right)  \left(\lambda _{-1}+\lambda _0+\lambda _1+\lambda _2\right)
  }.
\end{equation}
In smooth regions, $d_L$ in Eq. (\ref{eq:d3ul}) and $d_R$ in Eq. (\ref{eq:d3ur}) are \cite{baeza2020efficient}
\begin{equation}
d_L  = \left(\frac{\partial^3u}{\partial x^3}\right)\left(\Delta x\right)^6 + O(\left(\Delta x\right)^8),\,\,
d_R  = \left(\frac{\partial^3u}{\partial x^3}\right)\left(\Delta x\right)^6 + O(\left(\Delta x\right)^8).
\end{equation}
and in discontinuous regions, 
\begin{equation}
d_L  = O(1),\,\, d_R  = O(1).
\end{equation}

\begin{assumption}
  \label{assump}
  $u(x)$ that is to be reconstructed has a critical point of order $k$, $k \in \left\{0, 1\right\}$; 
  At least one of $I_0$ and $I_1$ are of order $O(\left(\Delta x\right)^{2m}), m \geq 1$;
  If $S_i = \left\{\overline{u}_{i-1}, \overline{u}_i, \overline{u}_{i+1}\right\}$ contains discontinuity, not all of $I_{-1}$ and $I_{2}$ are of order $O(1)$.
\end{assumption}

\begin{theorem}\label{theorem:1}
  Suppose  Assumption \ref{assump} holds,
  then the shock detector $\theta$ in Eq. (\ref{eq:owenosd}) satisfies
  \begin{equation}
  \theta = \left\{
    \begin{array}{ll}
      1+O((\Delta x)^{2m}) + O(\epsilon), m \geq 1, & \,\text{if}\,\,S_i\,\,\text{is smooth}, \\
      O((\Delta x)^{2m}) + O(\epsilon), m \geq 1, & \text{if}\,\,S_i\,\,\text{includes discontinuity}.\\
    \end{array}
  \right.
  \end{equation}
\end{theorem}

Before the proof of Theorem \ref{theorem:1},  let us firstly introduce the other two stencils $S_L$ and $S_R$ as
\begin{align}
S_L = \left\{ \overline{u}_{i-2}, \overline{u}_{i-1}, \overline{u}_{i}, \overline{u}_{i+1} \right\},\\
S_R = \left\{ \overline{u}_{i-1}, \overline{u}_{i}, \overline{u}_{i+1},\overline{u}_{i+2}  \right\}.
\end{align}

\begin{lemma}\label{lemmasrsmooth}\cite{baeza2020efficient}
	Assume that the distribution of $u(x)$ is smooth on $S_R$ and has a $k$-th order critical point, according to Lemma 2 in the work of Baeza et al. \cite{baeza2019efficient}, there holds: If $k=0$, $I_j=O(\left(\Delta x\right)^2)$, $\forall j \in \left\{0,1,2\right\}$. If $k=1$, 
  $\exists j'\in \left\{0,1,2\right\}$ such that $I_{j'} = O(\left(\Delta x\right)^{m_{j'}})$ for some $m_{j'} \in \left\{4,5,6,\cdots\right\}$, and $I_j \in O(\left(\Delta x\right)^4)$, for $j \in \left\{0,1,2\right\}\backslash \left\{j'\right\}$.
\end{lemma}
\begin{lemma}\label{lemmasrdiscontinuous}\cite{baeza2020efficient}
If there is a discontinuity across $S_R$, we have $I_j = O(1)$, for some $j \in \left\{0,1,2\right\}$, and $I_{j'} = O(\left(\Delta x\right)^{2m_{j'}})$ with $m_{j'} \in \left\{1,2,3,\cdots\right\}$, $j' \in \left\{0,1,2\right\}\backslash\left\{j\right\}$.
\end{lemma}
\begin{proposition}\label{propositionsir}
  If the distribution of $u(x)$ has a critical point of order $k$, $k \in \left\{0,1\right\}$, then the magnitude of $\theta^R$ in Eq. (\ref{eq:owenosd}) satisfies
  \begin{equation}
  \theta^R =  \left\{
    \begin{array}{ll}
      1 + O(\left(\Delta x\right)^{4-2k}) + O(\epsilon),\,\,  k = 0,1,&\text{if}\,\,S_R\,\,\text{is smooth}, \\
      1 + O(\left(\Delta x\right)^{2m}) + O(\epsilon),\,\,  m \geq 1,& \text{if}\,\,S_i\,\,\text{is smooth and}\,\,I_2 = O(1), \\
      O(\left(\Delta x\right)^{2\mathfrak{m}}) + O(\epsilon),\,\,  \mathfrak{m} \geq 1, &\text{if}\,\,S_i\,\,\text{includes discontinuity and}\,\,I_2 = O(\left(\Delta x\right)^{2m_2}), m_2 \geq 1\\
      1/(1+O(1)) + O(\epsilon), & \text{if}\,\,S_i\,\,\text{includes discontinuity and} \,\,I_2 = O(1).
    \end{array}
  \right.
  \label{eq:rirorder}
  \end{equation}
\end{proposition}
\begin{proof}

  {\quad}\\$\quad\quad$\textbf{Case I: $S_R$ is smooth.}\\[0.5pt]
  \indent In this condition, from Lemma \ref{lemmasrsmooth}, $I_j = O(\left(\Delta x\right)^{2 m_j})$, for $j \in \left\{0,1,2\right\}$, $m_j \in \left\{1\right\} \cup \left\{\frac{4}{2},\frac{5}{2},\frac{6}{2},\cdots \right\}$. It is easy to verify 
  \begin{equation}
  d_R = O(\left(\Delta x\right)^6).
  \end{equation}
  If the critical point is of order $k = 0$,
  \begin{equation}
  \tau_R  = d_R(I_0 + I_1) = O(\left(\Delta x\right)^6) \times O(\left(\Delta x\right)^2),\,\,\,\, J_R  = I_0 I_1 + I_2 I_2 = O(\left(\Delta x\right)^4).
  \end{equation}
  If the critical point is of order $k = 1$,
  \begin{equation}
  \tau_R  = d_R(I_0 + I_1) = O(\left(\Delta x\right)^6) \times O(\left(\Delta x\right)^4),\,\,\,\, J_R  = I_0 I_1 + I_2 I_2 = O(\left(\Delta x\right)^8).
  \end{equation}
  Then
  \begin{equation}
    \begin{aligned}
    \theta^R  & = \frac{J_R}{J_R+\tau_R+\epsilon}  = \frac{1}{1+\frac{\tau_R}{J_R}} + O(\epsilon) \\
    & =\frac{1}{1 + \frac{O(\left(\Delta x\right)^6) \times O(\left(\Delta x\right)^{2+2k})}{O(\left(\Delta x\right)^{4+4k})}} + O(\epsilon) \\
    & =1 + O(\left(\Delta x\right)^{4-2k}) + O(\epsilon),
    \end{aligned}
  \end{equation}
  where $k\in \left\{0,1\right\}$.\\[1pt]

  \noindent \textbf{Case II: $S_i$ is smooth and $I_2 = O(1)$.}\\[0.5pt]
  \indent From Lemma \ref{lemmasrdiscontinuous}, in this situation, $I_0 =O(\left(\Delta x\right)^{2m_0}) $, $I_1 =O(\left(\Delta x\right)^{2m_1})$, $I_2 = O(1)$, with $m_0, m_1 \in \left\{1,2,3,\cdots\right\}$. In addition, $S_R$ contains a discontinuity and
  \begin{equation}
  d_R = O(1).
  \end{equation}
  Thus, we have
  \begin{equation}
  \begin{aligned}
    \theta^R & = \frac{J_R}{J_R+\tau_R+\epsilon}  = \frac{1}{1+\frac{\tau_R}{J_R}} + O(\epsilon) = \frac{1}{1+\frac{d_R(I_0+I_1)}{I_0I_1+I_2I_2}} + O(\epsilon) \\
    & = \frac{1}{1 + \frac{
      O(1)\times (O(\left(\Delta x\right)^{2m_0}) + O(\left(\Delta x\right)^{2m_1}))
    }{
      O(\left(\Delta x\right)^{2m_0}) \times O(\left(\Delta x\right)^{2m_1}) + O(1) \times O(1)
    }} + O(\epsilon)\\
    & = \frac{1}{1+\frac{O(\left(\Delta x\right)^{2m})}{O(1)}}+O(\epsilon)\\
    & = 1 + O(\left(\Delta x\right)^{2m})+O(\epsilon),
  \end{aligned}
  \end{equation}
  where $m = \min(m_0, m_1)$ and $m \geq 1$.\\[1pt]

  \noindent \textbf{Case III: $S_i$ includes discontinuity and $I_2 = O(\left(\Delta x\right)^{2m_2})$.}\\[0.5pt]
  \indent Denoting $I_j =O(\left(\Delta x\right)^{2m_j}), j=0,1$, one of 
  $m_j, j=0,1$ is of value $0$, and the other is of value $\geq 1$.
  Meanwhile,
  \begin{equation}
  d_R = O(1).
  \end{equation}
  Thus
  \begin{equation}
  \begin{aligned}
    \theta^R & = \frac{J_R}{J_R+\tau_R+\epsilon}  = \frac{1}{1+\frac{\tau_R}{J_R}} + O(\epsilon) = \frac{1}{1+\frac{d_R(I_0+I_1)}{I_0I_1+I_2I_2}} + O(\epsilon) \\
    & = \frac{1}{1 + \frac{ O(1)\times O(1) }{ O(\left(\Delta x\right)^{2M}) +O(\left(\Delta x\right)^{4m_2}) }} + O(\epsilon)\\
    & = \frac{1}{1+\frac{O(1)}{O(\left(\Delta x\right)^{2 \mathfrak{m}})}}+O(\epsilon)  =  O(\left(\Delta x\right)^{2\mathfrak{m}})+O(\epsilon),
  \end{aligned}
  \end{equation}
  where $M = \max(m_0, m_1)$, $\mathfrak{m} = \min(M, 2m_2)$ and $\mathfrak{m} \geq 1$.\\[1pt]

  \noindent \textbf{Case IV: $S_i$ includes discontinuity and $I_2 = O(1)$.}\\[0.5pt]
  \indent Similar with Case III, denoting $I_j =O(\left(\Delta x\right)^{2m_j}), j=0,1$, one of 
  $m_j, j=0,1$ is of value $0$, the other is of value $\geq 1$ and
  \begin{equation}
  d_R = O(1).
  \end{equation}
  Then we have
  \begin{equation}
  \begin{aligned}
    \theta^R & = \frac{J_R}{J_R+\tau_R+\epsilon}  = \frac{1}{1+\frac{\tau_R}{J_R}} + O(\epsilon)  = \frac{1}{1+\frac{d_R(I_0+I_1)}{I_0I_1+I_2I_2}} + O(\epsilon) \\
    & = \frac{1}{1 + \frac{
      O(1)\times (O(1))
    }{
      O(\left(\Delta x\right)^{2M}) 
      +O(1) 
    }} + O(\epsilon)\\
    & = \frac{1}{1+O(1)}+O(\epsilon),\\
  \end{aligned}
  \end{equation}
  where $M = \max(m_0, m_1)$.

  Concluded from Case I to Case IV, Proposition \ref{propositionsir} is proved.
\end{proof}

Symmetrically, we have
  \begin{equation}
  \theta^L =  \left\{
    \begin{array}{ll}
      1 + O(\left(\Delta x\right)^{4-2k}) + O(\epsilon),\,\,  k = 0,1,&\text{if}\,\,S_L\,\,\text{is smooth}, \\
      1 + O(\left(\Delta x\right)^{2m}) + O(\epsilon),\,\,  m \geq 1,& \text{if}\,\,S_i\,\,\text{is smooth and}\,\,I_{-1} = O(1), \\
      O(\left(\Delta x\right)^{2\mathfrak{m}}) + O(\epsilon),\,\,  \mathfrak{m} \geq 1, &\text{if}\,\,S_i\,\,\text{includes discontinuity and}\,\,I_{-1} = O(\left(\Delta x\right)^{2m_{-1}}), m_{-1} \geq 1\\
      1/(1+O(1)) + O(\epsilon), & \text{if}\,\,S_i\,\,\text{includes discontinuity and} \,\,I_{-1} = O(1).
    \end{array}
  \right.
  \label{eq:rilorder}
  \end{equation}

  Combining Eqs. (\ref{eq:rirorder}) and (\ref{eq:rilorder}), and assuming Assumption \ref{assump} holds, Theorem \ref{theorem:1} is proved.

\begin{remark}
In the work of Baeza \cite{baeza2020efficient}, for face $x_{i+1/2}$, the shock detector is constructed as
\begin{equation}
  \theta_{i+1/2}^{\mathrm{OWENO3}} = \frac{J + \epsilon}{J + \tau + \epsilon},
\end{equation}
with
\begin{equation}
J  = I_0 I_1 + 2I_{0}I_{2} + I_{1} I_{2},\,\,\,\, \tau  = d(I_0+ I_1 + I_2), \,\,\,\,d  = \left(\zeta_{-1}^R \overline{u}_{i-1} + \zeta_0^R \overline{u}_{i} + \zeta_{1}^R \overline{u}_{i+1} + \zeta_{2}^R \overline{u}_{i+2}\right)^2.
\end{equation}

In the analysis of Baeza \cite{baeza2020efficient}, Case IV in Proposition \ref{propositionsir} is not considered, and it can be
easily verified that in Case IV,
\[
\theta_{i+1/2}^{\mathrm{OWENO3}} = O(1),
\]
i.e., wth $\theta_{i+1/2}^{\mathrm{OWENO3}}$ alone, it is hard to distinguish whether there is a discontinuity on $S_i$ when $I_2 = O(1)$. Such situation may occur frequently for shock capturing schemes near discontinuities, since a discontinuity may distributed over two or more grid points.
In this paper, by the multiplication of $\theta_{i+1/2}^{\mathrm{OWENO3}}$ and $\theta_{i-1/2}^{\mathrm{OWENO3}}$, together with simplifications on the expressions of $\theta_{i\pm 1/2}^{\mathrm{OWENO3}}$ to $\theta^{R/L}$, when $I_2$ and $I_{-1}$ do not locate inside discontinuities at the same time (which rarely happens in the simulation), $\theta$ approaches $0$ asymptotically when $S_i = \left\{\overline{u}_{i-1}, \overline{u}_i, \overline{u}_{i+1}\right\}$ includes discontinuity.
\end{remark}
\begin{remark}
To better distinguish the smooth regions, a scale sensor proposed in the work of Li \cite{LiYanhui2021} is brought in here. First, $\overline{u}_{i-3}$, $\overline{u}_{i-2}$, $\overline{u}_{i-1}$, $\overline{u}_{i}$, $\overline{u}_{i+1}$ and $\overline{u}_{i+2}$ are incorporated to reconstruct a fifth-order polynomial $p^5_{i+1/2}(x)$, then the scale sensor for interface $x_{i+1/2}$ is constructed as
\begin{equation}
  \omega_{i+1/2} = \sqrt{
    \frac{
    \left|\frac{\mathrm{d}^3 p^5}{\mathrm{d}x^3}h^3\right| + \left|\frac{\mathrm{d}^4 p^5}{\mathrm{d}x^4} h^4\right| 
    }{
    \left|\frac{\mathrm{d} p^5}{\mathrm{d}x} h\right| + \left|\frac{\mathrm{d}^2 p^5}{\mathrm{d}x^2} h^2\right| + \epsilon_3
    }
    },
\end{equation}
where 
\begin{equation}
\epsilon_3 = \max\left(1\times 10^{-3} \frac{\sum_{m=-3}^{2}{\left|\overline{u}_{i+m}\right|}}{6}, 1\times 10^{-50}\right)
\label{eq:epsilon_3}
\end{equation}

As suggested in \cite{LiYanhui2021}, if $\max\left(\omega_{i-1/2}, \omega_{i+1/2}\right) < 1$, $\sigma_i$ in Eq. (\ref{eq:owenosd}) is set as $1$, i.e., the control volume $\Omega_i$ is detected as in smooth regions.
\end{remark}

\section{Extension to Euler Equations \label{sec:euler}}
Taking the following 1D Euler equations as an example, 
\begin{equation}
\frac{\partial \vec{U}}{\partial t} + \frac{\partial \vec{F}}{\partial x} = 0,
\label{eq:euler}
\end{equation}
where
\begin{align}
\vec{U} & = \left(\rho, \rho u, \rho E\right)^T,\\
\vec{F} & = \left(\rho u, \rho u^2 + p, \rho H u\right)^T.
\end{align}
$\rho$, $u$, $p$, $H$ and $E$ are the density, velocity, pressure, total enthalpy and total energy for the fluid, respectively, and
\begin{align}
E  = e + \frac{1}{2}u^2,\,\,\,\, H  = e + \frac{1}{2}u^2 + \frac{p}{\rho},
\end{align}
where $e$ is the specific internal energy. In this paper, the perfect gas is considered and $e = \frac{p}{(\gamma-1)\rho}$, where $\gamma$ is the ratio of specific heat capacity.

When extending the WCLS reconstruction to the Euler equations, the reconstruction system of Eq. (\ref{eq:nonlinearCLSMatrixFormNorm}) should be applied to the characteristic variables.
The Jacobian matrix $\bm{J}$ is $\bm{J} = \frac{\partial \vec{F}}{\partial \vec{U}}$. The left and right eigen matrices of $\bm{J}$ are denoted as $\bm{L}$ and $\bm{R}$. First, the conservation variables are projected to the characteristic space based on the eigen system of $\Omega_i$ as
\begin{equation}
  \vec{\mathcal{W}}_{i+m} = \bm{L}_i\vec{U}_{i+m}, \,\,m=-1,0,1.
\end{equation}
The three characteristic variables in 1D cases are denoted as $\mathcal{W}_1$, $\mathcal{W}_2$ and $\mathcal{W}_3$, respectively.
Second, the reconstruction system for the Euler equations is written as
\begin{equation}
\widetilde{\bm{R}}_i \widetilde{\mathbb{M}}_i^{(-1)} \widetilde{\bm{L}}_i \vec{\widetilde{a}}_{i-1}
+\widetilde{\bm{R}}_i\widetilde{\mathbb{M}}_i^{(0)} \widetilde{\bm{L}} \vec{\widetilde{a}}_{i}
+\widetilde{\bm{R}}_i\widetilde{\mathbb{M}}_i^{(1)} \widetilde{\bm{L}} \vec{\widetilde{a}}_{i+1} = \widetilde{\bm{R}}_i \vec{\widetilde{b}}_i, \label{eq:nonlinearCLSMatrixFormNormEuler}
\end{equation}
where $\widetilde{\bm{L}}$ and $\widetilde{\bm{R}}$ are the expanded left and right eigen matrices. In 1D situation,
\begin{equation}
\widetilde{\bm{L}} = \left(
  \begin{array}{cccccc}
    L_{11} & 0 & L_{12} & 0 & L_{13} & 0 \\ 
    0 & L_{11} & 0 & L_{12} & 0 & L_{13} \\ 
    L_{21} & 0 & L_{22} & 0 & L_{23} & 0 \\ 
    0 & L_{21} & 0 & L_{22} & 0 & L_{23} \\ 
    L_{31} & 0 & L_{32} & 0 & L_{33} & 0 \\ 
    0 & L_{31} & 0 & L_{32} & 0 & L_{33} \\ 
  \end{array}
\right),\,\,\,\,
\widetilde{\bm{R}} = \left(
  \begin{array}{cccccc}
    R_{11} & 0 & R_{12} & 0 & R_{13} & 0 \\ 
    0 & R_{11} & 0 & R_{12} & 0 & R_{13} \\ 
    R_{21} & 0 & R_{22} & 0 & R_{23} & 0 \\ 
    0 & R_{21} & 0 & R_{22} & 0 & R_{23} \\ 
    R_{31} & 0 & R_{32} & 0 & R_{33} & 0 \\ 
    0 & R_{31} & 0 & R_{32} & 0 & R_{33} \\ 
  \end{array}
\right),
\end{equation}
\begin{equation}
\vec{\widetilde{a}}_{i+m} = \left(\vec{a}^{T}_{\rho,i+m},\vec{a}^{T}_{\rho u,i+m},\vec{a}^{T}_{\rho E,i+m}, \right)^T,\,\,\,\,
\vec{\widetilde{b}}_{i} = \left(\vec{b}^{T}_{\mathcal{W}_1,i},\vec{b}^{T}_{\mathcal{W}_2,i},\vec{b}^{T}_{\mathcal{W}_3,i}, \right)^T,
\end{equation}
and 
\begin{equation}
  \widetilde{\mathbb{M}}_i^{(m)} = 
  \begin{pmatrix}
   \widetilde{\bm{M}}_{\mathcal{W}_1, i}^{(m)} & \bm{0} & \bm{0}\\
   \bm{0} & \widetilde{\bm{M}}_{\mathcal{W}_2, i}^{(m)} & \bm{0}\\
   \bm{0} & \bm{0} & \widetilde{\bm{M}}_{\mathcal{W}_3,i}^{(m)}\\
  \end{pmatrix},\,\,m=-1,0,1,
\end{equation}
where $\widetilde{\bm{M}}_{\mathcal{W}_j, i}^{(m)},\,\,m=-1,0,1$ are the nonlinear reconstruction matrices for the $j$-th characteristic variable as in Eq. (\ref{eq:nonlinearCLSMatrixFormNorm}).

The reconstruction matrices of all the conservation variables are coupled together, constituting an enlarged block tridiagonal system. In one-, two- and three-dimensional (1D/2D/3D) cases, the size of each block is $6\times 6$, $8\times 8$ and $10\times 10$, respectively. Taking the non-periodic boundary conditions as an example and denoting the final reconstruction system as
\[
\widetilde{\mathcal{M}} \widetilde{\alpha} = \widetilde{\chi},
\]
where 
\begin{equation}
\widetilde{\mathcal{M}} = 
  \begin{pmatrix}
 \bm{R}_1 \widetilde{\mathbb{M}}_1^{(0)}\bm{L}_1 &  \bm{R}_1\widetilde{\mathbb{M}}_1^{(1)}\bm{L}_1& \bm{0} & \cdots &\cdots& \bm{0}& \bm{0}  \\
 \bm{R}_2\widetilde{\mathbb{M}}_2^{(-1)}\bm{L}_2 & \bm{R}_2\widetilde{\mathbb{M}}_2^{(0)}\bm{L}_2& \bm{R}_2\widetilde{\mathbb{M}}_2^{(1)}\bm{L}_2& \bm{0} &\cdots& \cdots& \bm{0} \\
 \bm{0} & \bm{R}_3\widetilde{\mathbb{M}}_3^{(-1)}\bm{L}_3 &  \bm{R}_3\widetilde{\mathbb{M}}_3^{(0)}\bm{L}_3& \bm{R}_3\widetilde{\mathbb{M}}_3^{(1)} \bm{L}_3& \bm{0} & \cdots&  \bm{0} \\
 \vdots & \vdots & \vdots & \vdots &\vdots & \ddots & \vdots \\
 \bm{0} & \bm{0} &\cdots&\cdots&\bm{0}&\bm{R}_N\widetilde{\mathbb{M}}_{N}^{(-1)}\bm{L}_N & \bm{R}_N\widetilde{\mathbb{M}}_{N}^{(0)}\bm{L}_N
  \end{pmatrix},
\end{equation}
\begin{equation}
  \widetilde{\alpha} = \left(\vec{\widetilde{a}}_{1},\vec{\widetilde{a}}_{2},\vec{\widetilde{a}}_{3},\cdots,\vec{\widetilde{a}}_{N} \right)^T,\,\,\,\,
  \widetilde{\chi} = \left(\vec{\widetilde{\chi}}_{1},\vec{\widetilde{\chi}}_{2},\vec{\widetilde{\chi}}_{3},\cdots,\vec{\widetilde{\chi}}_{N} \right)^T,\,\,\,\,
  \vec{\widetilde{\chi}}_{i} = \bm{R}_i \vec{\widetilde{b}}_{i},\,\,i=1,2,\cdots,N,
\end{equation}
the algorithm (Alg.) of solving such a block tridiagonal system is listed as in Alg. \ref{algorithm:1}.
However, the matrix inversion, matrix multiplication operation for coupled matrices with size $6\times 6$ in 1D, $8\times 8$ in 2D or $10\times 10$ in 3D are much heavier than those for the decoupled matrices with size $2\times 2$, leading the WCLS scheme unfeasible for Euler equations.

In this work, an approximate solution strategy for the coupled block tridiagonal system is proposed as in Alg. \ref{algorithm:2}. First, the shock detector proposed in Sec. \ref{sec:shockdetector} is introduced here. If in smooth regions with $\sigma_i = 1$, the nonlinear weights $W_{i,L}$ and $W_{i,R}$ are both set as $1$. In such situation, $\widetilde{\mathcal{M}}_{i,{i-1}}$, $\widetilde{\mathcal{M}}_{i,{i}}$ and $\widetilde{\mathcal{M}}_{i,{i+1}}$ becomes block diagonal matrices and an integer number $f_1[i]$ is utilized to denote whether the reconstruction system for control volume $\Omega_i$ is decoupled ($f_1[i]=1$) or not ($f_1[i]=0$). Second, it can be observed the $\mathbb{C}'_{i}$ in Alg. \ref{algorithm:1} soon become block diagonal (i.e., decoupled) in smooth regions and its related operation can be carried on in a decoupled manner as in Alg. \ref{algorithm:2}. The discontinuous control volumes occupy only a small fraction of the computational domain. Such approximate solution strategy reduces the simulation time significantly in the meanwhile maintains the numerical accuracy of the WCLS scheme. In Alg. \ref{algorithm:2}, $\mathbb{C}'^{pq}_{i,mn} $ is the $(p,q)$-th element of the $(m,n)$-th block in matrix $\mathbb{C}_i'$. Line 30 in Alg. \ref{algorithm:2} means that if the averaged absolute value of non-diagonal blocks is much smaller than the averaged absolute value of diagonal blocks, $\mathbb{C}_i'$ is approximated by a block diagonal matrix and its block nondiagonal elements are assigned as $\bm{0}$. If the system is decoupled, subtraction, multiplication, and matrix inversion operations are carried on in a decoupled manner, which is denoted as 
  $\overset{*}{-}$, $*$ and $(\cdot)^{-1*}$ in Alg. \ref{algorithm:2}, respectively. 
The approximate solution for a cyclic tridiagonal system can be obtained analogously and is not repeated here.

\begin{algorithm}[!htbp]
  \SetAlgoLined
  \SetKwComment{Comment}{/* }{ */}
  \Comment{In the algorithm, $\mathbb{A}_i = \widetilde{\mathcal{M}}_{i,i-1}$, $\mathbb{B}_i = \widetilde{\mathcal{M}}_{i,i}$ and $\mathbb{C}_i = \widetilde{\mathcal{M}}_{i,i+1}$.}
  \For{$i \gets 1$ \KwTo $N$}{
    \If{$i=1$}{
      $\mathbb{C}'_{1} = \mathbb{B}^{-1}_{1}\cdot \mathbb{C}_{1}$, \quad
      $\widetilde{\chi}_{1} = \mathbb{B}^{-1}_{1} \cdot \widetilde{\chi}_1$
    }\Else{
      $\mathbb{T}_{i} = \mathbb{B}_i - \mathbb{A}_{i}\cdot \mathbb{C}'_{i-1}$,\quad
      $\mathbb{C}'_{i} = \mathbb{T}^{-1}_{i}\cdot \mathbb{C}_{i}$,\quad
      $\widetilde{\chi}_{i} = \mathbb{T}^{-1}_{i}\cdot \left(\widetilde{\chi}_i - \mathbb{A}_i\cdot \widetilde{\chi}_{i-1}\right)$
    }
  }
  $\widetilde{\alpha}_N = \widetilde{\chi}_{N}$\\
  \For{$i \gets N-1$ \KwTo $1$}{
    $\widetilde{\alpha}_{i} = \widetilde{\chi}_{i} - \mathbb{C}'_{i}\cdot \widetilde{\alpha}_{i+1}$
  }
\caption{\label{algorithm:1}Solving a block tridiagonal system $\widetilde{\mathcal{M}} \widetilde{\alpha} = \widetilde{\chi}$.}
\end{algorithm}

\begin{algorithm}[!htbp]
  \SetAlgoLined
  \SetKwComment{Comment}{/* }{ */}
  \Comment{In the algorithm, $\mathbb{A}_i = \widetilde{\mathcal{M}}_{i,i-1}$, $\mathbb{B}_i = \widetilde{\mathcal{M}}_{i,i}$ and $\mathbb{C}_i = \widetilde{\mathcal{M}}_{i,i+1}$. Dim is the physical dimension of the problem.}
  $\text{Count} \gets 0$\\
  \For{$i \gets 1$ \KwTo $N$}{
    \If{$i=1$}{
      \If{$f_1[i] = 1$}{
        $f_2[i] \gets 1$,\quad
        $\mathbb{C}'_{1} = \mathbb{B}^{-1*}_{1} * \mathbb{C}_{1}$,\quad
        $\widetilde{\chi}_{1} = \mathbb{B}^{-1*}_{1} * \widetilde{\chi}_1$
      }\Else{
        $f_2[i] \gets 0$,\quad
        $\mathbb{C}'_{1} = \mathbb{B}^{-1}_{1}\cdot \mathbb{C}_{1}$,\quad
        $\widetilde{\chi}_{1} = \mathbb{B}^{-1}_{1} \cdot \widetilde{\chi}_1$\\
        $\text{Count} \gets \text{Count} +1$
      }
    }\Else{
      \If{$f_1[i]\,\, \& \,\,f_2[i-1]$}{
        $f_2[i] \gets 1$,\quad
        $\mathbb{T}_{i} = \mathbb{B}_i \overset{*}{-} \mathbb{A}_{i} * \mathbb{C}'_{i-1}$,\quad
        $\mathbb{C}'_{i} = \mathbb{T}^{-1*}_{i} * \mathbb{C}_{i}$,\quad
        $\widetilde{\chi}_{i} = \mathbb{T}^{-1*}_{i} * \left(\widetilde{\chi}_i - \mathbb{A}_i * \widetilde{\chi}_{i-1}\right)$\\
        $\text{Count} \gets 0$
      }\ElseIf{$f_1[i] \,\,\& \,\,!f_2[i-1]$}{
        $f_2[i] \gets 0$,\quad
        $\mathbb{T}_{i} = \mathbb{B}_i - \mathbb{A}_{i} * \mathbb{C}'_{i-1}$,\quad
        $\mathbb{C}'_{i} = \mathbb{T}^{-1}_{i} \cdot \mathbb{C}_{i}$,\quad
        $\widetilde{\chi}_{i} = \mathbb{T}^{-1}_{i} \cdot \left(\widetilde{\chi}_i - \mathbb{A}_i * \widetilde{\chi}_{i-1}\right)$\\
        $\text{Count} \gets \text{Count} +1$
      }\ElseIf{ $!f_1[i]\,\, \&\,\,  f_2[i-1]$}{
        $f_2[i] \gets 0$,\quad
        $\mathbb{T}_{i} = \mathbb{B}_i - \mathbb{A}_{i} * \mathbb{C}'_{i-1}$,\quad
        $\mathbb{C}'_{i} = \mathbb{T}^{-1}_{i} \cdot \mathbb{C}_{i}$,\quad
        $\widetilde{\chi}_{i} = \mathbb{T}^{-1}_{i} \cdot \left(\widetilde{\chi}_i - \mathbb{A}_i \cdot \widetilde{\chi}_{i-1}\right)$\\
        $\text{Count} \gets 0$
      }\Else{
        $f_2[i] \gets 0$,\quad
        $\mathbb{T}_{i} = \mathbb{B}_i - \mathbb{A}_{i}\cdot \mathbb{C}'_{i-1}$,\quad
        $\mathbb{C}'_{i} = \mathbb{T}^{-1}_{i}\cdot \mathbb{C}_{i}$,\quad
        $\widetilde{\chi}_{i} = \mathbb{T}^{-1}_{i}\cdot \left(\widetilde{\chi}_i - \mathbb{A}_i\cdot \widetilde{\chi}_{i-1}\right)$\\
        $\text{Count} \gets 0$
      }

      \If{$\text{Count} \geq 1$}{
        \If{ $ \frac{1}{1+\text{Dim}}\frac{ \sum_{1 \leq m,n \leq 2 +\text{Dim},m \neq n}{\sum_{1\leq p, q \leq 2}{\left|\mathbb{C}'\right|^{pq}_{i,mn}}} }{ \sum_{1 \leq m \leq 2 + \text{Dim}}{\sum_{1\leq p, q \leq 2}{\left|\mathbb{C}'\right|^{pq}_{i,mm}} }} < 1\times 10^{-10}$ }{
          $\mathbb{C}'_{i,nm} = \bm{0}, 1 \leq m,n \leq 2+Dim, \,\, m\neq n$,\\
          $f_2[i] \gets 1$
        }
      }
    }
  }
  $\widetilde{\alpha}_N = \widetilde{\chi}_{N}$\\
  \For{$i \gets N-1$ \KwTo $1$}{
    \If{$f_2[i] = 1$}{
    $\widetilde{\alpha}_{i} = \widetilde{\chi}_{i} - \mathbb{C}'_{i} * \widetilde{\alpha}_{i+1}$
    }\Else{
    $\widetilde{\alpha}_{i} = \widetilde{\chi}_{i} - \mathbb{C}'_{i}\cdot \widetilde{\alpha}_{i+1}$
    }
  }
\caption{\label{algorithm:2}Approximate solution of a block tridiagonal system $\widetilde{\mathcal{M}} \widetilde{\alpha} = \widetilde{\chi}$.}
\end{algorithm}

After obtaining the polynomials for each conservation variable on all control volumes, approximate Roe Riemann solver with $h$-type entropy correction \cite{sanders1998multidimensional} is utilized to calculate the numerical inviscid flux.
When extending to multidimensional examples, the WCLS reconstruction is conducted in a dimension-by-dimension manner. And when extending to Navier-Stokes equations, the viscous terms are calculated in a second-order Gauss formula. Interested readers are referred to the work of Wang \cite{wang2015accurate}, Huang \cite{huang2018high,huang2022adaptive} and Pan \cite{jianhua2025Hybrid} for other detailed information of the extension to multidimensional Navier-Stokes equations.

\section{Numerical Experiments\label{sec:numerical_example}}

\subsection{Linear Convection Equation}
\subsubsection{Numerical Accuracy}
Linear convection equation is simulated to validate the accuracy of the WCLS3 scheme. The computational domain is $\Omega=[0,1]$ applied with periodic boundary conditions. The simulation is continued till $t = 1.0$. Third-order SSP-RK scheme is utilized to integrate the linear convection equation in temporal direction with Courant number (CFL) as $0.5$. 
The scheme with nonlinear weights turned on and off in smooth regions are both tested.

\begin{sidewaystable}
  \footnotesize
\centering
\caption{Accuracy test with linear convection equation under $u_0(x) = \sin(2\pi x)$ with CFL = $0.5$ till $t = 1.0$.}
\label{tab:accsin}
\begin{tabular}{cc|cccccc|cccccc}
\hline
\multirow{2}{*}{Schemes} & \multirow{2}{*}{N} &
\multicolumn{6}{c|}{Nonlinear weights turned on in smooth regions} & \multicolumn{6}{c}{Nonlinear weights turned off in smooth regions}\\
\cline{3-14}
& & $\left|\text{Error}\right|_1$ & Order & $\left|\text{Error}\right|_2$ & Order & $\left|\text{Error}\right|_{\infty}$ & Order & $\left|\text{Error}\right|_1$ & Order & $\left|\text{Error}\right|_2$ & Order & $\left|\text{Error}\right|_{\infty}$ & Order  \\ 
\hline
\multirow{6}{*}{CWENO3} 
& $25	$  &$4.38\times 10^{-2}$  &		      &$8.84\times 10^{-2}$&		    &$1.03\times 10^{-1}$	&      &-&-&-&-&-&-\\
& $50	$  &$1.05\times 10^{-2}$	&$2.06$   &$3.14\times 10^{-2}$&$1.50$  &$3.86\times 10^{-2}$ &$1.42$&-&-&-&-&-&-\\
& $100$	 &$2.28\times 10^{-3}$	&$2.20$ 	&$9.70\times 10^{-3}$&$1.69$ 	&$1.34\times 10^{-2}$	&$1.53$&-&-&-&-&-&-\\
& $200$	 &$4.75\times 10^{-4}$	&$2.26$ 	&$2.92\times 10^{-3}$&$1.73$ 	&$4.76\times 10^{-3}$	&$1.49$&-&-&-&-&-&-\\
& $400$	 &$9.63\times 10^{-5}$	&$2.30$ 	&$8.63\times 10^{-4}$&$1.76$ 	&$1.66\times 10^{-3}$	&$1.53$&-&-&-&-&-&-\\
& $800$	 &$1.77\times 10^{-5}$	&$2.44$ 	&$2.52\times 10^{-4}$&$1.78$ 	&$5.38\times 10^{-4}$	&$1.62$&-&-&-&-&-&-\\
\hline
\multirow{6}{*}{\makecell[c]{WCLS \\ $\kappa_0 = 0.8$}} 
&$ 25$&$1.14\times 10^{ -2}$&$    $&$1.49\times 10^{ -2}$&$     $&$3.13\times 10^{ -2}$&$     $&$4.72\times 10^{ -4}$&$     $&$5.25\times 10^{ -4}$&$     $&$7.41\times 10^{ -4}$&$     $\\
&$ 50$&$1.65\times 10^{ -4}$&$6.11$&$2.42\times 10^{ -4}$&$5.95 $&$6.26\times 10^{ -4}$&$5.64 $&$4.63\times 10^{ -5}$&$3.35 $&$5.15\times 10^{ -5}$&$3.35 $&$7.27\times 10^{ -5}$&$3.35 $\\
&$100$&$2.65\times 10^{ -5}$&$2.64$&$4.86\times 10^{ -5}$&$2.32 $&$1.67\times 10^{ -4}$&$1.90 $&$5.33\times 10^{ -6}$&$3.12 $&$5.92\times 10^{ -6}$&$3.12 $&$8.37\times 10^{ -6}$&$3.12 $\\
&$200$&$4.06\times 10^{ -6}$&$2.71$&$9.81\times 10^{ -6}$&$2.31 $&$4.61\times 10^{ -5}$&$1.86 $&$6.51\times 10^{ -7}$&$3.03 $&$7.23\times 10^{ -7}$&$3.03 $&$1.02\times 10^{ -6}$&$3.03 $\\
&$400$&$6.45\times 10^{ -7}$&$2.65$&$2.12\times 10^{ -6}$&$2.21 $&$1.35\times 10^{ -5}$&$1.77 $&$8.09\times 10^{ -8}$&$3.01 $&$8.99\times 10^{ -8}$&$3.01 $&$1.27\times 10^{ -7}$&$3.01 $\\
&$800$&$1.02\times 10^{ -7}$&$2.66$&$4.46\times 10^{ -7}$&$2.25 $&$3.66\times 10^{ -6}$&$1.88 $&$1.01\times 10^{ -8}$&$3.00 $&$1.12\times 10^{ -8}$&$3.00 $&$1.59\times 10^{ -8}$&$3.00 $\\
\hline
\multirow{6}{*}{\makecell[c]{WCLS \\ $\kappa_0 = 1.0$}} 
&$ 25$&$2.88\times 10^{ -3}$&$    $&$3.87\times 10^{ -3}$&$     $&$9.02\times 10^{ -3}$&$     $&$3.68\times 10^{ -4}$&$     $&$4.09\times 10^{ -4}$&$     $&$5.78\times 10^{ -4}$&$     $\\
&$ 50$&$2.25\times 10^{ -4}$&$3.68$&$4.39\times 10^{ -4}$&$3.14 $&$1.69\times 10^{ -3}$&$2.42 $&$4.25\times 10^{ -5}$&$3.11 $&$4.73\times 10^{ -5}$&$3.11 $&$6.68\times 10^{ -5}$&$3.11 $\\
&$100$&$2.39\times 10^{ -5}$&$3.23$&$5.03\times 10^{ -5}$&$3.13 $&$1.99\times 10^{ -4}$&$3.08 $&$5.21\times 10^{ -6}$&$3.03 $&$5.78\times 10^{ -6}$&$3.03 $&$8.18\times 10^{ -6}$&$3.03 $\\
&$200$&$3.55\times 10^{ -6}$&$2.75$&$9.39\times 10^{ -6}$&$2.42 $&$5.08\times 10^{ -5}$&$1.97 $&$6.47\times 10^{ -7}$&$3.01 $&$7.19\times 10^{ -7}$&$3.01 $&$1.02\times 10^{ -6}$&$3.01 $\\
&$400$&$5.89\times 10^{ -7}$&$2.59$&$2.14\times 10^{ -6}$&$2.13 $&$1.53\times 10^{ -5}$&$1.73 $&$8.08\times 10^{ -8}$&$3.00 $&$8.97\times 10^{ -8}$&$3.00 $&$1.27\times 10^{ -7}$&$3.00 $\\
&$800$&$9.49\times 10^{ -8}$&$2.63$&$4.53\times 10^{ -7}$&$2.24 $&$4.07\times 10^{ -6}$&$1.91 $&$1.01\times 10^{ -8}$&$3.00 $&$1.12\times 10^{ -8}$&$3.00 $&$1.59\times 10^{ -8}$&$3.00 $\\
\hline
\multirow{6}{*}{\makecell[c]{WCLS \\ $\kappa_0 = 1.2$}} 
&$ 25$&$1.37\times 10^{ -3}$&$    $&$1.71\times 10^{ -3}$&$     $&$3.33\times 10^{ -3}$&$     $&$3.44\times 10^{ -4}$&$     $&$3.82\times 10^{ -4}$&$     $&$5.40\times 10^{ -4}$&$     $\\
&$ 50$&$3.70\times 10^{ -4}$&$1.89$&$4.98\times 10^{ -4}$&$1.78 $&$1.19\times 10^{ -3}$&$1.48 $&$4.18\times 10^{ -5}$&$3.04 $&$4.64\times 10^{ -5}$&$3.04 $&$6.56\times 10^{ -5}$&$3.04 $\\
&$100$&$3.26\times 10^{ -5}$&$3.51$&$7.31\times 10^{ -5}$&$2.77 $&$3.12\times 10^{ -4}$&$1.93 $&$5.18\times 10^{ -6}$&$3.01 $&$5.75\times 10^{ -6}$&$3.01 $&$8.14\times 10^{ -6}$&$3.01 $\\
&$200$&$4.40\times 10^{ -6}$&$2.89$&$1.28\times 10^{ -5}$&$2.52 $&$7.95\times 10^{ -5}$&$1.97 $&$6.46\times 10^{ -7}$&$3.00 $&$7.18\times 10^{ -7}$&$3.00 $&$1.02\times 10^{ -6}$&$3.00 $\\
&$400$&$6.33\times 10^{ -7}$&$2.80$&$2.37\times 10^{ -6}$&$2.43 $&$1.79\times 10^{ -5}$&$2.15 $&$8.08\times 10^{ -8}$&$3.00 $&$8.97\times 10^{ -8}$&$3.00 $&$1.27\times 10^{ -7}$&$3.00 $\\
&$800$&$9.22\times 10^{ -8}$&$2.78$&$4.54\times 10^{ -7}$&$2.38 $&$4.57\times 10^{ -6}$&$1.97 $&$1.01\times 10^{ -8}$&$3.00 $&$1.12\times 10^{ -8}$&$3.00 $&$1.59\times 10^{ -8}$&$3.00 $\\
\bottomrule
\end{tabular}
\end{sidewaystable}

\begin{sidewaystable}
  \footnotesize
\centering
\caption{Accuracy test with linear convection equation under $u_0(x) = \sin^2(2\pi x)$ with CFL = 0.5 till $t = 1.0$.}\label{tab:accsin2}
\begin{tabular}{cccccccccccccc}
\hline
\multirow{2}{*}{Schemes} & \multirow{2}{*}{N} &
\multicolumn{6}{c|}{Nonlinear weights turned on in smooth regions} & \multicolumn{6}{c}{Nonlinear weights turned off in smooth regions}\\
\cline{3-14}
& & $\left|\text{Error}\right|_1$ & Order & $\left|\text{Error}\right|_2$ & Order & $\left|\text{Error}\right|_{\infty}$ & Order & $\left|\text{Error}\right|_1$ & Order & $\left|\text{Error}\right|_2$ & Order & $\left|\text{Error}\right|_{\infty}$ & Order  \\ 
\hline
\multirow{6}{*}{CWENO3} 
&$25 $&$1.14\times 10^{-1}$&$    $&$1.36\times 10^{-1}$&$    $&$2.22\times 10^{-1}$&$    $&-&-&-&-&-&-\\
&$50 $&$3.34\times 10^{-2}$&$1.77$&$3.95\times 10^{-2}$&$1.78$&$7.59\times 10^{-2}$&$1.55$&-&-&-&-&-&-\\
&$100$&$8.51\times 10^{-3}$&$1.97$&$1.16\times 10^{-2}$&$1.77$&$2.68\times 10^{-2}$&$1.50$&-&-&-&-&-&-\\
&$200$&$1.86\times 10^{-3}$&$2.19$&$3.22\times 10^{-3}$&$1.85$&$9.53\times 10^{-3}$&$1.49$&-&-&-&-&-&-\\
&$400$&$3.85\times 10^{-4}$&$2.27$&$8.59\times 10^{-4}$&$1.90$&$3.32\times 10^{-3}$&$1.52$&-&-&-&-&-&-\\
&$800$&$7.54\times 10^{-5}$&$2.35$&$2.20\times 10^{-4}$&$1.96$&$1.14\times 10^{-3}$&$1.55$&-&-&-&-&-&-\\
\hline
\multirow{6}{*}{\makecell[c]{WCLS \\ $\kappa_0 = 0.8$}} 
&$ 25$&$1.85\times 10^{ -2}$&$     $&$2.39\times 10^{ -2}$&$     $&$6.27\times 10^{ -2}$&$     $&$6.05\times 10^{ -3}$&$     $&$6.72\times 10^{ -3}$&$     $&$9.49\times 10^{ -3}$&$     $\\
&$ 50$&$8.03\times 10^{ -4}$&$4.52 $&$1.17\times 10^{ -3}$&$4.36 $&$3.75\times 10^{ -3}$&$4.06 $&$4.72\times 10^{ -4}$&$3.68 $&$5.24\times 10^{ -4}$&$3.68 $&$7.42\times 10^{ -4}$&$3.68 $\\
&$100$&$1.03\times 10^{ -4}$&$2.96 $&$1.45\times 10^{ -4}$&$3.01 $&$3.84\times 10^{ -4}$&$3.29 $&$4.64\times 10^{ -5}$&$3.35 $&$5.15\times 10^{ -5}$&$3.35 $&$7.28\times 10^{ -5}$&$3.35 $\\
&$200$&$1.63\times 10^{ -5}$&$2.66 $&$2.78\times 10^{ -5}$&$2.38 $&$9.45\times 10^{ -5}$&$2.02 $&$5.33\times 10^{ -6}$&$3.12 $&$5.92\times 10^{ -6}$&$3.12 $&$8.37\times 10^{ -6}$&$3.12 $\\
&$400$&$2.64\times 10^{ -6}$&$2.63 $&$5.89\times 10^{ -6}$&$2.24 $&$2.74\times 10^{ -5}$&$1.78 $&$6.51\times 10^{ -7}$&$3.03 $&$7.23\times 10^{ -7}$&$3.03 $&$1.02\times 10^{ -6}$&$3.03 $\\
&$800$&$4.24\times 10^{ -7}$&$2.64 $&$1.24\times 10^{ -6}$&$2.25 $&$7.41\times 10^{ -6}$&$1.89 $&$8.09\times 10^{ -8}$&$3.01 $&$8.99\times 10^{ -8}$&$3.01 $&$1.27\times 10^{ -7}$&$3.01 $\\
\hline
\multirow{6}{*}{\makecell[c]{WCLS \\ $\kappa_0 = 1.0$}} 
&$ 25$&$6.47\times 10^{ -3}$&$     $&$8.35\times 10^{ -3}$&$     $&$1.99\times 10^{ -2}$&$     $&$3.65\times 10^{ -3}$&$     $&$4.06\times 10^{ -3}$&$     $&$5.74\times 10^{ -3}$&$     $\\
&$ 50$&$5.12\times 10^{ -4}$&$3.66 $&$7.33\times 10^{ -4}$&$3.51 $&$1.76\times 10^{ -3}$&$3.50 $&$3.67\times 10^{ -4}$&$3.31 $&$4.09\times 10^{ -4}$&$3.31 $&$5.77\times 10^{ -4}$&$3.31 $\\
&$100$&$8.86\times 10^{ -5}$&$2.53 $&$1.36\times 10^{ -4}$&$2.43 $&$4.13\times 10^{ -4}$&$2.09 $&$4.26\times 10^{ -5}$&$3.11 $&$4.73\times 10^{ -5}$&$3.11 $&$6.68\times 10^{ -5}$&$3.11 $\\
&$200$&$1.92\times 10^{ -5}$&$2.21 $&$3.71\times 10^{ -5}$&$1.87 $&$1.26\times 10^{ -4}$&$1.71 $&$5.21\times 10^{ -6}$&$3.03 $&$5.78\times 10^{ -6}$&$3.03 $&$8.18\times 10^{ -6}$&$3.03 $\\
&$400$&$2.47\times 10^{ -6}$&$2.96 $&$6.06\times 10^{ -6}$&$2.61 $&$3.14\times 10^{ -5}$&$2.01 $&$6.47\times 10^{ -7}$&$3.01 $&$7.19\times 10^{ -7}$&$3.01 $&$1.02\times 10^{ -6}$&$3.01 $\\
&$800$&$4.05\times 10^{ -7}$&$2.61 $&$1.28\times 10^{ -6}$&$2.25 $&$8.35\times 10^{ -6}$&$1.91 $&$8.08\times 10^{ -8}$&$3.00 $&$8.97\times 10^{ -8}$&$3.00 $&$1.27\times 10^{ -7}$&$3.00 $\\
\hline
\multirow{6}{*}{\makecell[c]{WCLS \\ $\kappa_0 = 1.2$}} 
&$ 25$&$7.42\times 10^{ -3}$&$     $&$9.22\times 10^{ -3}$&$     $&$2.21\times 10^{ -2}$&$     $&$3.02\times 10^{ -3}$&$     $&$3.35\times 10^{ -3}$&$     $&$4.74\times 10^{ -3}$&$     $\\
&$ 50$&$9.04\times 10^{ -4}$&$3.04 $&$1.22\times 10^{ -3}$&$2.92 $&$2.67\times 10^{ -3}$&$3.05 $&$3.44\times 10^{ -4}$&$3.13 $&$3.82\times 10^{ -4}$&$3.13 $&$5.40\times 10^{ -4}$&$3.13 $\\
&$100$&$1.16\times 10^{ -4}$&$2.96 $&$2.00\times 10^{ -4}$&$2.61 $&$6.69\times 10^{ -4}$&$1.99 $&$4.17\times 10^{ -5}$&$3.04 $&$4.64\times 10^{ -5}$&$3.04 $&$6.55\times 10^{ -5}$&$3.04 $\\
&$200$&$1.76\times 10^{ -5}$&$2.73 $&$3.57\times 10^{ -5}$&$2.49 $&$1.58\times 10^{ -4}$&$2.08 $&$5.18\times 10^{ -6}$&$3.01 $&$5.75\times 10^{ -6}$&$3.01 $&$8.14\times 10^{ -6}$&$3.01 $\\
&$400$&$2.62\times 10^{ -6}$&$2.74 $&$6.56\times 10^{ -6}$&$2.44 $&$3.94\times 10^{ -5}$&$2.00 $&$6.46\times 10^{ -7}$&$3.00 $&$7.18\times 10^{ -7}$&$3.00 $&$1.02\times 10^{ -6}$&$3.00 $\\
&$800$&$5.01\times 10^{ -7}$&$2.39 $&$1.60\times 10^{ -6}$&$2.04 $&$1.22\times 10^{ -5}$&$1.69 $&$8.08\times 10^{ -8}$&$3.00 $&$8.97\times 10^{ -8}$&$3.00 $&$1.27\times 10^{ -7}$&$3.00 $\\
 \bottomrule
\end{tabular}
\end{sidewaystable}

\begin{sidewaystable}
  \footnotesize
\centering
\caption{Accuracy test with linear convection equation under $u_0(x) = \sin^3(2\pi x)$ with CFL = 0.5 till $t = 1.0$.\label{tab:accsin3}}
\begin{tabular}{cccccccccccccc}
\hline
\multirow{2}{*}{Schemes} & \multirow{2}{*}{N} &
\multicolumn{6}{c|}{Nonlinear weights turned on in smooth regions} & \multicolumn{6}{c}{Nonlinear weights turned off in smooth regions}\\
\cline{3-14}
& & $\left|\text{Error}\right|_1$ & Order & $\left|\text{Error}\right|_2$ & Order & $\left|\text{Error}\right|_{\infty}$ & Order & $\left|\text{Error}\right|_1$ & Order & $\left|\text{Error}\right|_2$ & Order & $\left|\text{Error}\right|_{\infty}$ & Order  \\ 
\hline
\multirow{6}{*}{CWENO3} 
&$25 $&$1.28\times 10^{-1}$&$    $&$1.46\times 10^{-1}$&$    $&$2.65\times 10^{-1}$&$     $&-&-&-&-&-&-\\
&$50 $&$3.18\times 10^{-2}$&$2.01$&$4.24\times 10^{-2}$&$1.78$&$1.12\times 10^{-1}$&$1.24 $&-&-&-&-&-&-\\
&$100$&$6.95\times 10^{-3}$&$2.19$&$1.21\times 10^{-2}$&$1.81$&$4.00\times 10^{-2}$&$1.48 $&-&-&-&-&-&-\\
&$200$&$1.53\times 10^{-3}$&$2.18$&$3.38\times 10^{-3}$&$1.84$&$1.43\times 10^{-2}$&$1.49 $&-&-&-&-&-&-\\
&$400$&$3.10\times 10^{-4}$&$2.30$&$9.08\times 10^{-4}$&$1.90$&$4.98\times 10^{-3}$&$1.52 $&-&-&-&-&-&-\\
&$800$&$5.95\times 10^{-5}$&$2.38$&$2.33\times 10^{-4}$&$1.96$&$1.69\times 10^{-3}$&$1.56 $&-&-&-&-&-&-\\
\hline
\multirow{6}{*}{\makecell[c]{WCLS \\ $\kappa_0 = 0.8$}} 
&$ 25$&$2.63\times 10^{ -2}$&$    $&$2.96\times 10^{ -2}$&$    $&$4.81\times 10^{ -2}$&$    $&$2.13\times 10^{ -2}$&$    $&$2.38\times 10^{ -2}$&$    $&$3.61\times 10^{ -2}$&$    $\\
&$ 50$&$2.85\times 10^{ -3}$&$3.21$&$3.75\times 10^{ -3}$&$2.98$&$9.51\times 10^{ -3}$&$2.34$&$1.53\times 10^{ -3}$&$3.80$&$1.91\times 10^{ -3}$&$3.64$&$4.40\times 10^{ -3}$&$3.04$\\
&$100$&$3.39\times 10^{ -4}$&$3.07$&$5.59\times 10^{ -4}$&$2.75$&$1.84\times 10^{ -3}$&$2.37$&$1.56\times 10^{ -4}$&$3.30$&$1.95\times 10^{ -4}$&$3.30$&$6.26\times 10^{ -4}$&$2.81$\\
&$200$&$4.47\times 10^{ -5}$&$2.92$&$8.45\times 10^{ -5}$&$2.73$&$3.49\times 10^{ -4}$&$2.40$&$1.79\times 10^{ -5}$&$3.12$&$2.33\times 10^{ -5}$&$3.07$&$9.31\times 10^{ -5}$&$2.75$\\
&$400$&$5.79\times 10^{ -6}$&$2.95$&$1.29\times 10^{ -5}$&$2.72$&$6.77\times 10^{ -5}$&$2.37$&$2.21\times 10^{ -6}$&$3.02$&$3.22\times 10^{ -6}$&$2.85$&$1.66\times 10^{ -5}$&$2.49$\\
&$800$&$7.71\times 10^{ -7}$&$2.91$&$2.11\times 10^{ -6}$&$2.61$&$1.32\times 10^{ -5}$&$2.36$&$2.70\times 10^{ -7}$&$3.03$&$4.43\times 10^{ -7}$&$2.86$&$3.07\times 10^{ -6}$&$2.43$\\
\hline
\multirow{6}{*}{\makecell[c]{WCLS \\ $\kappa_0 = 1.0$}} 
&$ 25$&$1.85\times 10^{ -2}$&$    $&$2.20\times 10^{ -2}$&$    $&$4.25\times 10^{ -2}$&$    $&$1.40\times 10^{ -2}$&$    $&$1.59\times 10^{ -2}$&$    $&$3.10\times 10^{ -2}$&$    $\\
&$ 50$&$1.93\times 10^{ -3}$&$3.26$&$2.54\times 10^{ -3}$&$3.12$&$6.70\times 10^{ -3}$&$2.67$&$1.35\times 10^{ -3}$&$3.38$&$1.67\times 10^{ -3}$&$3.25$&$4.44\times 10^{ -3}$&$2.80$\\
&$100$&$2.58\times 10^{ -4}$&$2.90$&$3.95\times 10^{ -4}$&$2.68$&$1.35\times 10^{ -3}$&$2.31$&$1.61\times 10^{ -4}$&$3.07$&$2.11\times 10^{ -4}$&$2.99$&$7.03\times 10^{ -4}$&$2.66$\\
&$200$&$3.55\times 10^{ -5}$&$2.86$&$6.30\times 10^{ -5}$&$2.65$&$2.66\times 10^{ -4}$&$2.34$&$1.87\times 10^{ -5}$&$3.10$&$2.59\times 10^{ -5}$&$3.02$&$1.08\times 10^{ -4}$&$2.71$\\
&$400$&$4.81\times 10^{ -6}$&$2.88$&$1.03\times 10^{ -5}$&$2.61$&$5.19\times 10^{ -5}$&$2.36$&$2.20\times 10^{ -6}$&$3.09$&$3.24\times 10^{ -6}$&$3.00$&$1.78\times 10^{ -5}$&$2.60$\\
&$800$&$6.55\times 10^{ -7}$&$2.88$&$1.80\times 10^{ -6}$&$2.52$&$1.28\times 10^{ -5}$&$2.02$&$2.66\times 10^{ -7}$&$3.05$&$4.43\times 10^{ -7}$&$2.87$&$3.27\times 10^{ -6}$&$2.44$\\
\hline
\multirow{6}{*}{\makecell[c]{WCLS \\ $\kappa_0 = 1.2$}} 
&$ 25$&$2.47\times 10^{ -2}$&$    $&$2.96\times 10^{ -2}$&$    $&$6.25\times 10^{ -2}$&$    $&$2.04\times 10^{ -2}$&$    $&$2.49\times 10^{ -2}$&$    $&$5.41\times 10^{ -2}$&$    $\\
&$ 50$&$3.71\times 10^{ -3}$&$2.74$&$5.03\times 10^{ -3}$&$2.56$&$1.43\times 10^{ -2}$&$2.13$&$3.17\times 10^{ -3}$&$2.69$&$4.71\times 10^{ -3}$&$2.41$&$1.46\times 10^{ -2}$&$1.89$\\
&$100$&$4.45\times 10^{ -4}$&$3.06$&$7.54\times 10^{ -4}$&$2.74$&$2.42\times 10^{ -3}$&$2.56$&$3.97\times 10^{ -4}$&$3.00$&$7.48\times 10^{ -4}$&$2.65$&$2.87\times 10^{ -3}$&$2.35$\\
&$200$&$5.42\times 10^{ -5}$&$3.04$&$1.09\times 10^{ -4}$&$2.79$&$4.72\times 10^{ -4}$&$2.36$&$4.36\times 10^{ -5}$&$3.18$&$9.60\times 10^{ -5}$&$2.96$&$3.92\times 10^{ -4}$&$2.87$\\
&$400$&$5.63\times 10^{ -6}$&$3.27$&$1.25\times 10^{ -5}$&$3.12$&$6.58\times 10^{ -5}$&$2.84$&$4.68\times 10^{ -6}$&$3.22$&$1.32\times 10^{ -5}$&$2.86$&$8.36\times 10^{ -5}$&$2.23$\\
&$800$&$6.64\times 10^{ -7}$&$3.08$&$1.83\times 10^{ -6}$&$2.77$&$1.39\times 10^{ -5}$&$2.25$&$3.54\times 10^{ -7}$&$3.72$&$8.23\times 10^{ -7}$&$4.01$&$6.47\times 10^{ -6}$&$3.69$\\
 \bottomrule
\end{tabular}
\end{sidewaystable}

Three different initial conditions are considered, i.e., $u_0(x) = \sin(2\pi x)$, $u_0(x) = \sin^2(2\pi x)$ and $u_0(x) = \sin^3(2\pi x)$. 
The first two initial conditions of $u_0(x) = \sin(2\pi x)$ and $u_0(x) = \sin^2(2\pi x)$ contain first-order extrema, i.e., $u_0'(x) = 0, u_0''(x) \neq 0$. The last initial condition of $u_0(x) = \sin^3(2\pi x)$ contains second-order extrema, i.e., $u_0'(x) = 0, u_0''(x) = 0$ and $u_0'''(x) = 0$.

The errors for the problem with $u_0(x) = \sin(2\pi x)$, $u_0(x) = \sin^2(2\pi x)$ and $u_0(x) = \sin^3(2\pi x)$ are shown in Tabs. \ref{tab:accsin}-\ref{tab:accsin3}, respectively. In Tabs. \ref{tab:accsin}-\ref{tab:accsin3}, we study the accuracy of the WCLS scheme under different coefficients in smooth regions, i.e.,  coefficients with different $\kappa_0$ in Eq. (\ref{eq:blendcoeffi}). 
As a comparison, the errors of the CWENO3 scheme \cite{baeza2019central} are also presented. Due to the existence of extrema, the CWENO3 scheme exhibits order reduction to the second order of accuracy in terms of $L_1$ and $L_2$ errors. The posterior order of the CWENO3 scheme is even smaller than $2$ in terms of the $L_\infty$ errors. 

When the nonlinear weights are turned off in smooth regions, with the increase of $\kappa_0$, the overall accuracy of the WCLS scheme is improved and
the prior third order of accuracy of the WCLS scheme is achieved under three different initial conditions considering the $L_1$ and $L_2$ errors.
In terms of the $L_\infty$ errors, the prior third order of accuracy is achieved for $u_0(x) = \sin(2\pi x)$ and $u_0(x) = \sin^2(2\pi x)$ and 
slight order degradation is observed when $u_0(x) = \sin^3(2\pi x)$.

When the nonlinear weights are turned on in smooth regions, order-degradation is observed due to the inclusion of dissipation terms in Eq. \ref{eq:nonlinearCLS}. However, the errors of the proposed WCLS3 scheme are still orders of magnitude smaller than the CWENO3 scheme. In addition, the posterior order of accuracy for the WCLS3 scheme with nonlinear weights turned on in smooth regions is higher than the CWENO3 scheme.

\subsubsection{Problem with Discontinuity\label{sec:cw}}
The following initial condition with a discontinuity at the middle of the computational domain $\Omega = [-1,1]$ is considered in this section,
\begin{equation}
  u_0(x) = \left\{
    \begin{array}{lr}
      -\sin(\pi x) -\frac{1}{2}x^3,& -1 \leq x \leq 0,\\
      -\sin(\pi x) -\frac{1}{2}x^3 + 1,& 0 < x \leq 1.\\
    \end{array}
  \right.
\end{equation}
The computational domain is discretized by $200$ uniform cells and periodic boundary conditions are applied at the two ends. Third-order SSP-RK method is utilized to advance the problem to $t = 6.0$ with CFL = $0.5$.

\begin{figure}[!htbp]
  \centering
    \begin{subfigure}[b]{0.3\textwidth}
    \includegraphics[width=\textwidth]{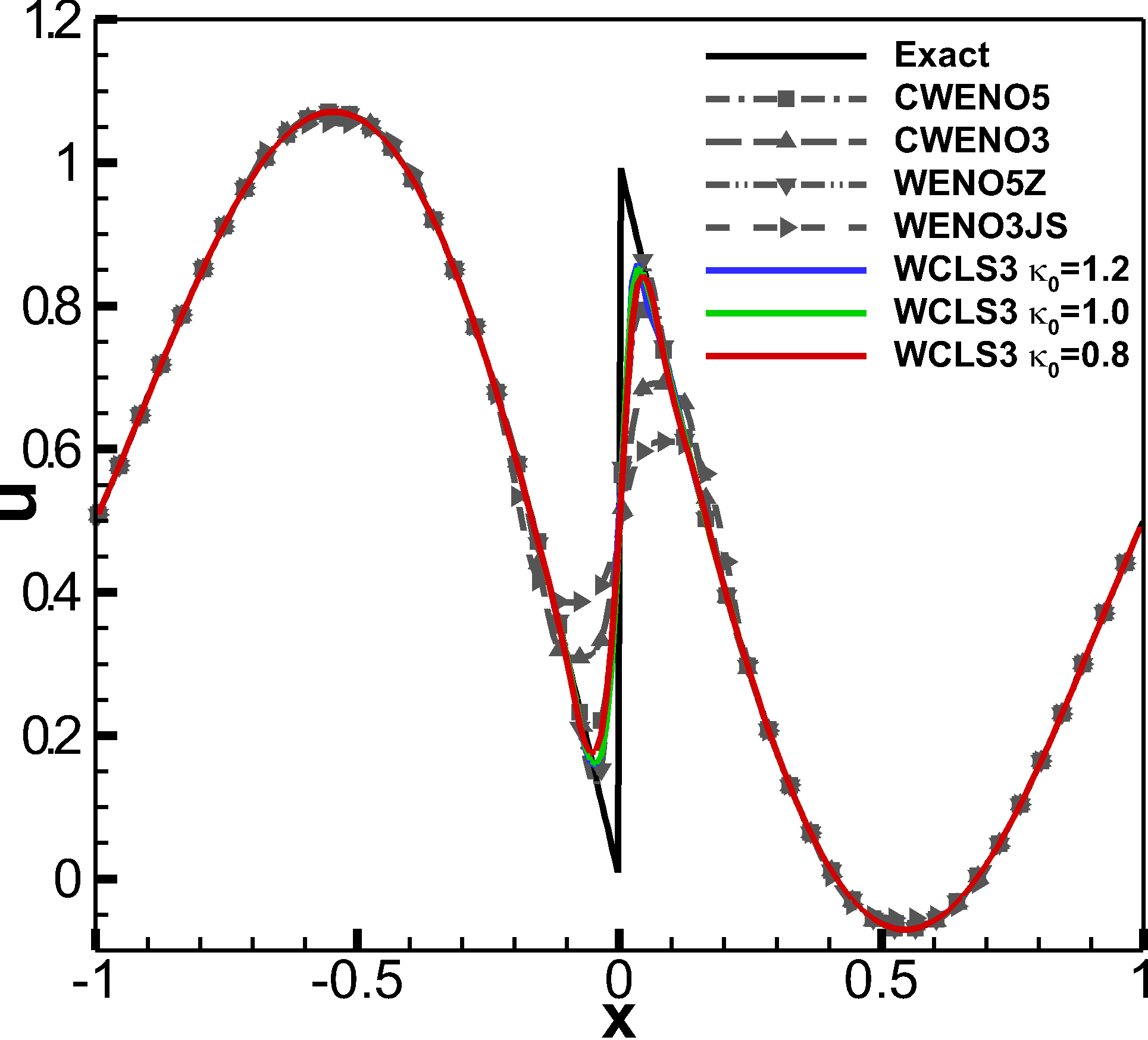}
    \caption{Overview.}
    \end{subfigure}
    \quad 
    \begin{subfigure}[b]{0.3\textwidth}
    \includegraphics[width=\textwidth]{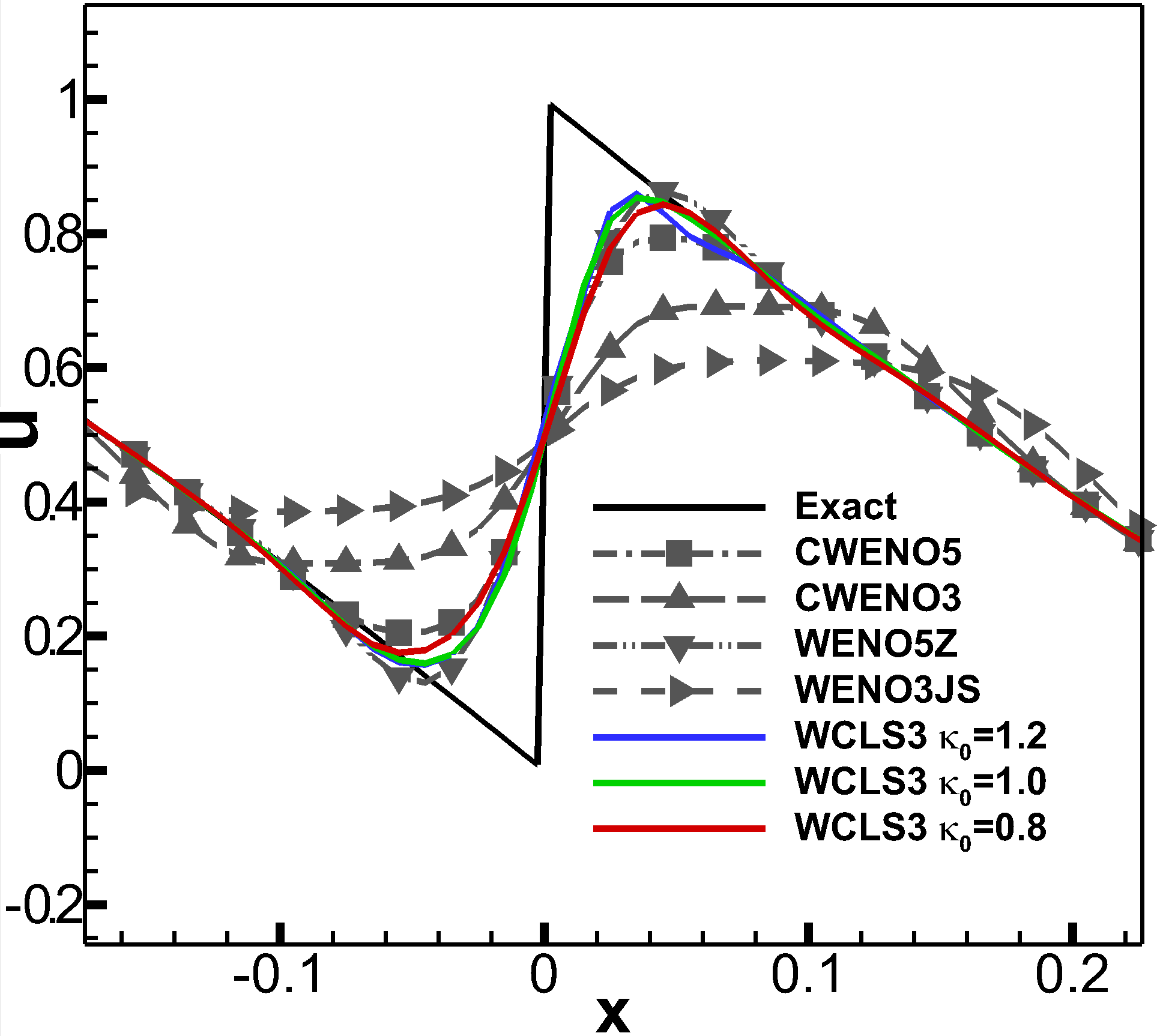}
      \caption{Close view.}
    \end{subfigure}
    \caption{\label{fig:cw} Results of the WCLS3 scheme for linear convection problem with discontinuity. $N = 200$, $t = 6.0$ and $CFL = 0.5$.}
\end{figure}

Figure \ref{fig:cw} shows the results of the WCLS3 scheme with different coefficients. Nonlinear weights are turned on across the whole computational domain. As a comparison, the result of third-order WENO-JS \cite{jiang1996efficient} (WENO3-JS), fifth-order WENO-Z scheme \cite{jacobs2009high} (WENO5-Z), third- and fifth-order CWENO schemes \cite{baeza2019central} (CWENO3 and CWENO5) are also presented.
At extrema of $x = \pm 0.5$, no degradation is observed for the WCLS3 scheme. Near discontinuity of $x=0$, the WCLS scheme captures the steep gradient even comparative with the WENO5-Z scheme and better than the CWENO5 schemes without oscillations.
In this problem, with the increase of $\kappa_0$, the WCLS scheme gives better resolution of the discontinuity.

\subsubsection{Problem with Gaussian-Square-Triangle-Ellipse Waves\label{sec:gste}}

The initial condition is
\begin{equation}
  u_0(x)  =  \left\{
  \begin{array}{lr}
    \frac{1}{6}\left( G(x,\beta,z-\delta) +G(x,\beta, z+\delta) +4G(x,\beta,z) \right), & 0.2 \leq x \leq 0.4,\\
    1, & 0.6 \leq x \leq 0.8,\\
    1-\left|10(x-1.1)\right|,& 1.0 \leq x \leq 1.2,\\
    \frac{1}{6}\left( F(x,\alpha,a-\delta) +F(x,\alpha, a+\delta)+4 F(x,\alpha,a) \right), & 1.4 \leq x \leq 1.6,\\
    0,& \text{elsewise},
  \end{array}
  \right.
  \label{eq:gste}
\end{equation}
where 
\begin{equation}
G(x,\beta,z)  = e^{\beta(x-1-z)^2},\quad F(x,\alpha,a)  = \sqrt{\max(1-\alpha^2(x-1-a^2), 0)},
\end{equation}
$z = -0.7$, $\delta = 0.005$, $\beta = \log_{10} 2/\left(36\delta^2\right)$, $\alpha = 10$ and $a = 0.5$.

The computational domain is $\Omega = [0,2]$ with periodic boundary condition and discretized by $N=200$ uniform cells. Third-order SSP-RK method is utilized to advance the simulation to $t = 10.0$ with CFL = $0.5$.

\begin{figure}[!htbp]
  \centering
    \begin{subfigure}[b]{0.3\textwidth}
    \includegraphics[width=\textwidth]{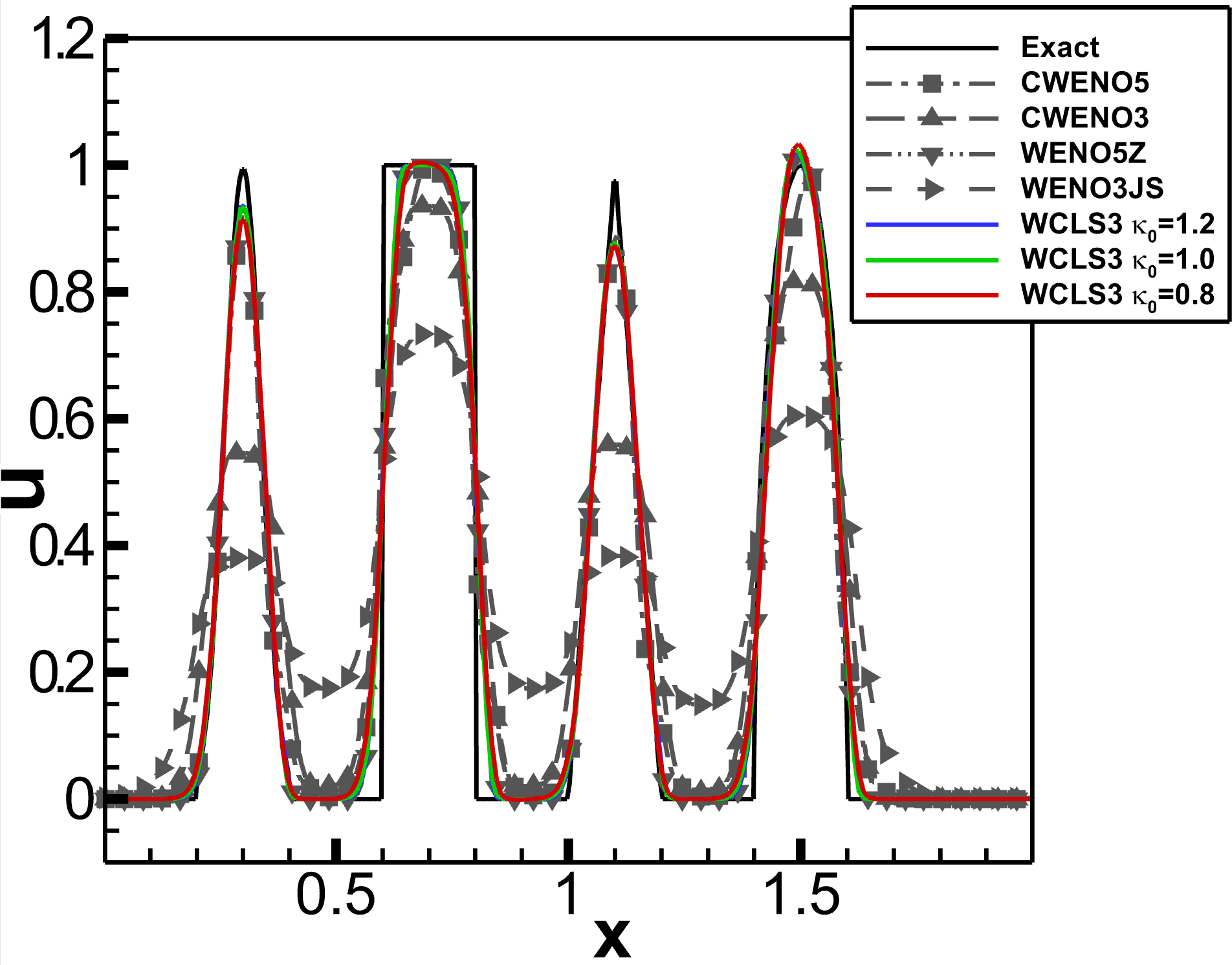}
    \caption{Overview.}
    \end{subfigure}
    \quad
    \begin{subfigure}[b]{0.3\textwidth}
    \includegraphics[width=\textwidth]{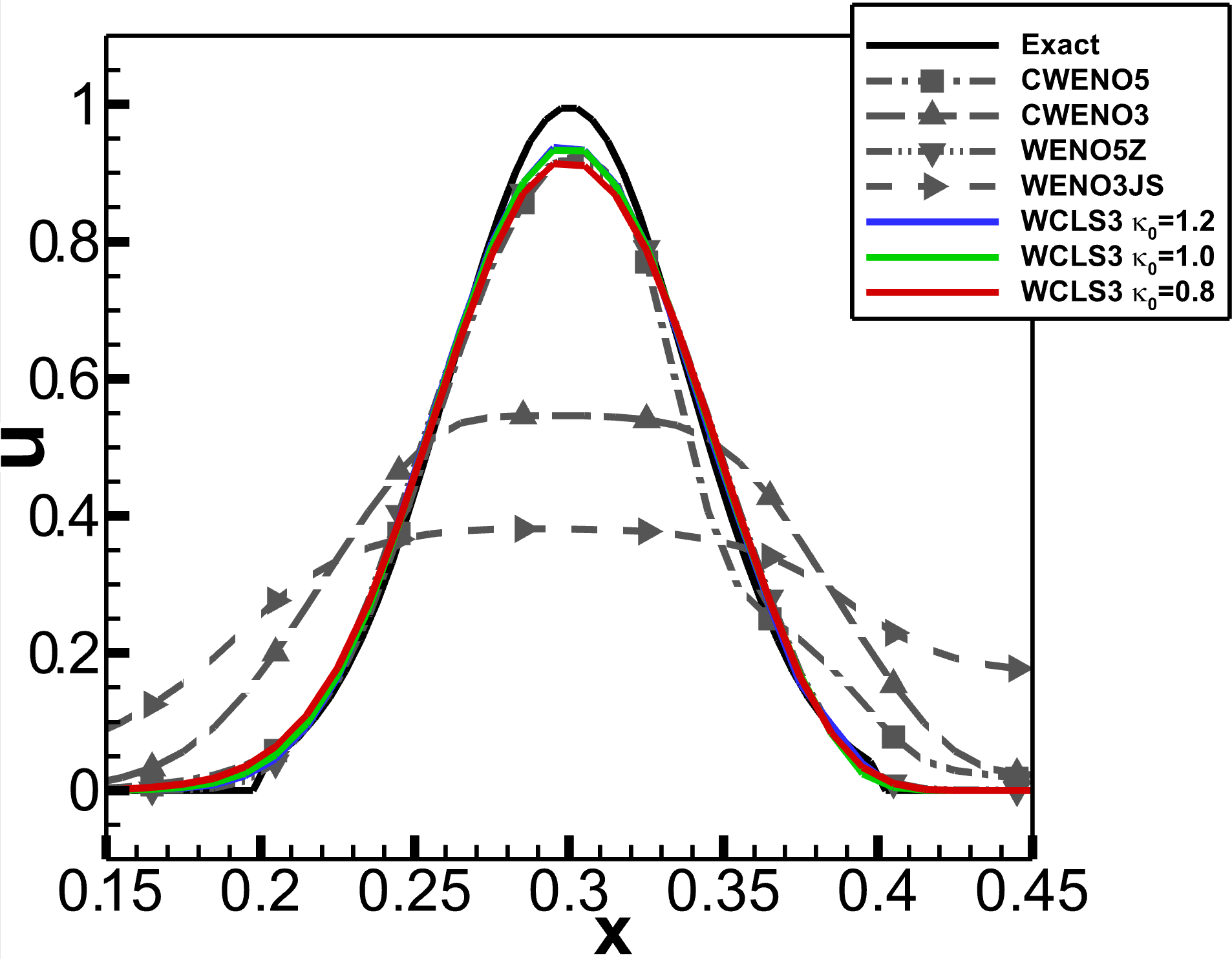}
    \caption{Gaussian wave.}
    \end{subfigure}
    \quad 
    \begin{subfigure}[b]{0.3\textwidth}
    \includegraphics[width=\textwidth]{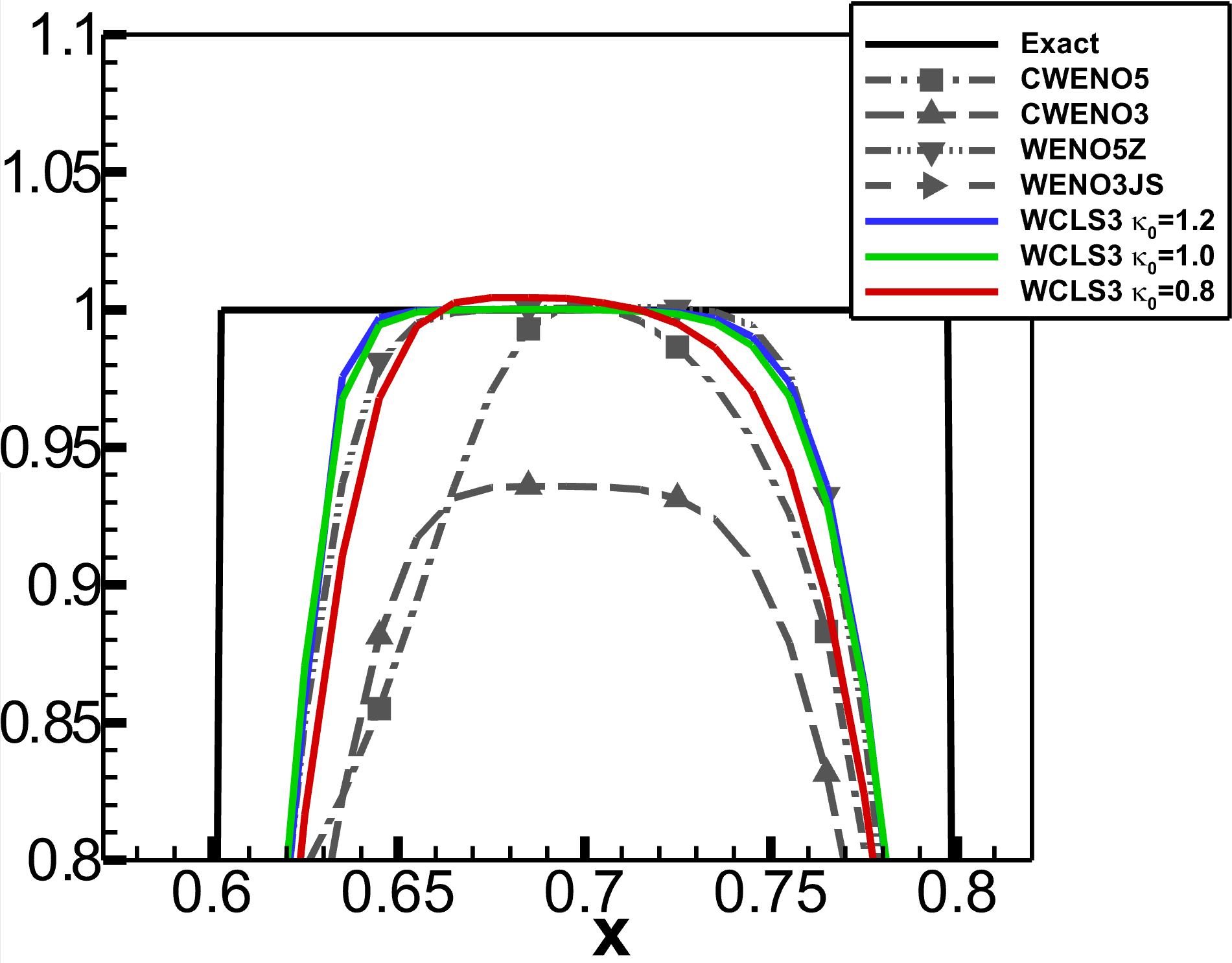}
    \caption{Square wave.}
    \end{subfigure}\\
    \begin{subfigure}[b]{0.3\textwidth}
    \includegraphics[width=\textwidth]{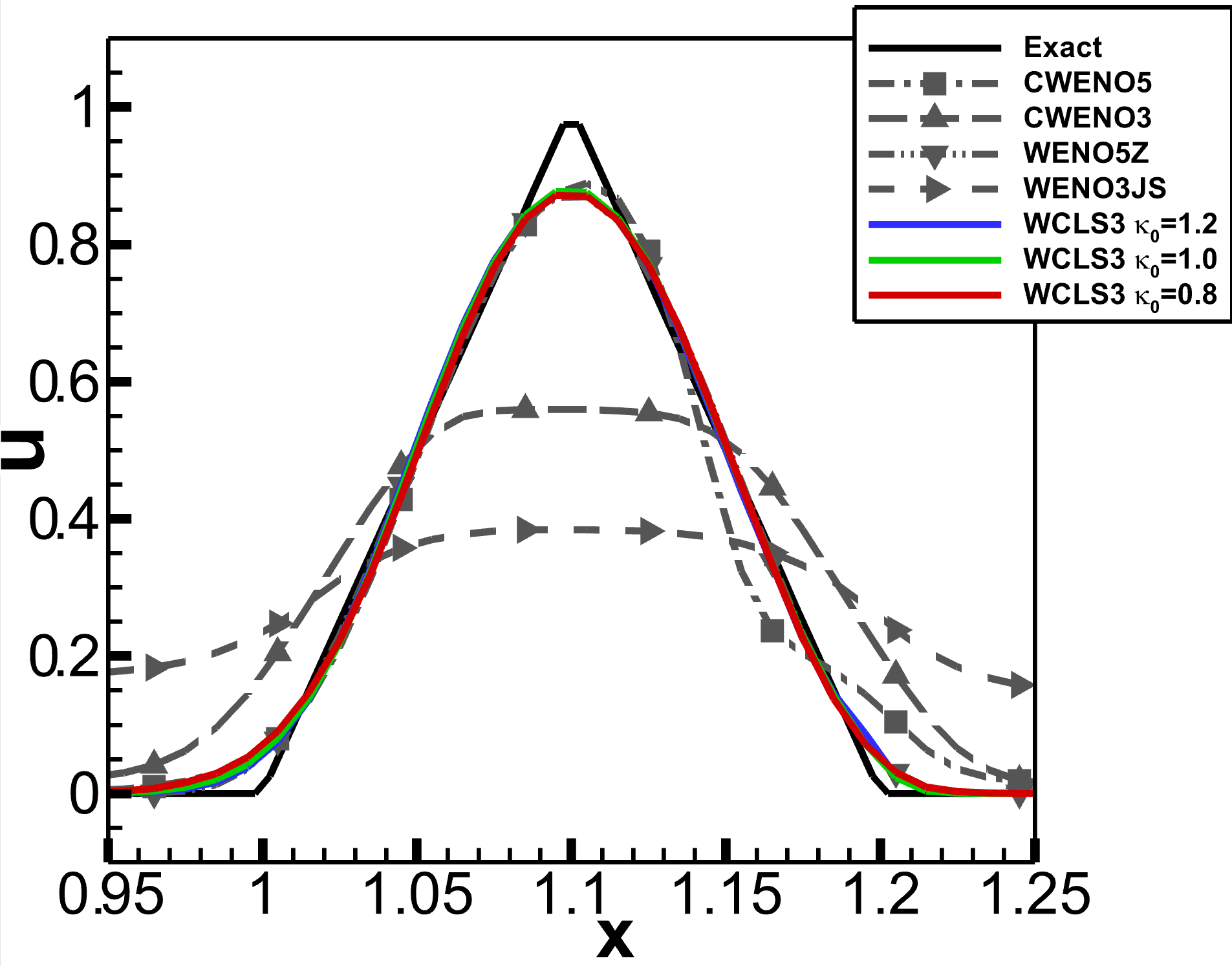}
    \caption{Triangle wave.}
    \end{subfigure}
    \quad 
    \begin{subfigure}[b]{0.3\textwidth}
    \includegraphics[width=\textwidth]{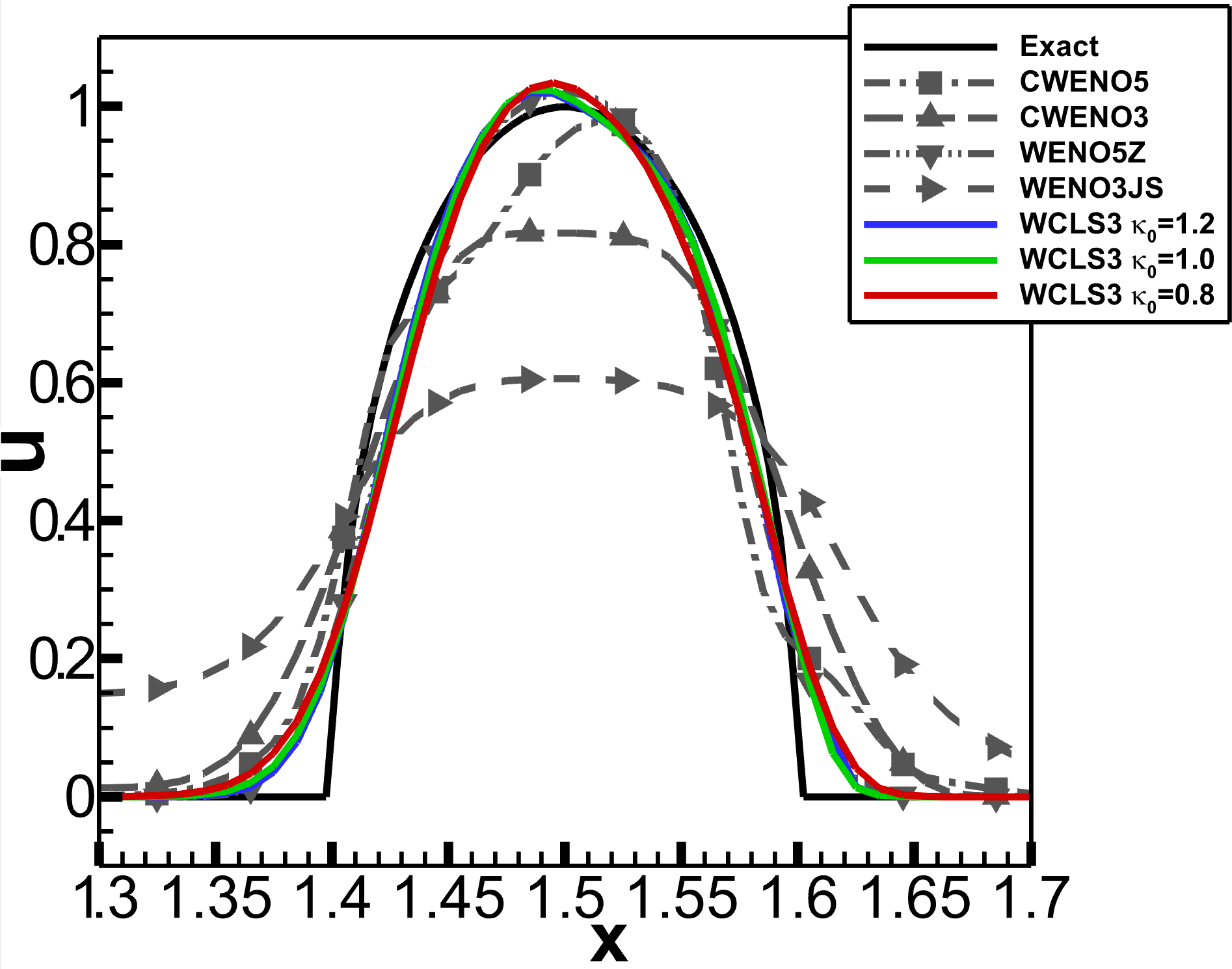}
    \caption{Ellipse wave.}
    \end{subfigure}
    \caption{\label{fig:GSTE} Results of the WCLS scheme for linear convection problem with Gaussian-square-triangle-ellipse waves. $N = 200$, $t = 10.0$ and $CFL = 0.5$.}
\end{figure}

The results of the WCLS3 scheme with different coefficients are shown as in Fig. \ref{fig:GSTE}. Nonlinear weights are turned on across the computational domain.
Even though the vanilla weights of WENO3-JS are utilized in the WCLS3 scheme, 
no order degradation is observed at smooth extrema of Gaussian, triangle or ellipse waves. 
The dissipation terms in Eq. \ref{eq:nonlinearCLS} only take in effect when the nonlinear weights of $W_L$ or $W_R$ are extremely small, does not poisoning the resolution of the WCLS scheme at smooth extrema.
It should be mentioned that the resolution of proposed WCLS3 scheme is even higher than the WENO3-NN scheme by Bezgin et al. \cite{bezgin_weno3-nn_2022} aided by neural-networks, and is comparative with or better than the WENO5-Z and CWENO5 schemes.
For the WENO3-JS scheme and CWENO3 scheme, the smooth extrema are dissipated due to the inability to identify first-order extrema in smooth regions. 

\subsection{Euler Equations}
In the simulation of Euler equations and Navier-Stokes equations in Sec. \ref{sec:ns}, the nonlinear weights are turned off in smooth regions with the help of shock detector.
\subsubsection{Propagation of Broadband Sound Waves}
\begin{figure}[!htbp]
  \centering
    \begin{subfigure}[b]{0.3\textwidth}
    \includegraphics[width=\textwidth]{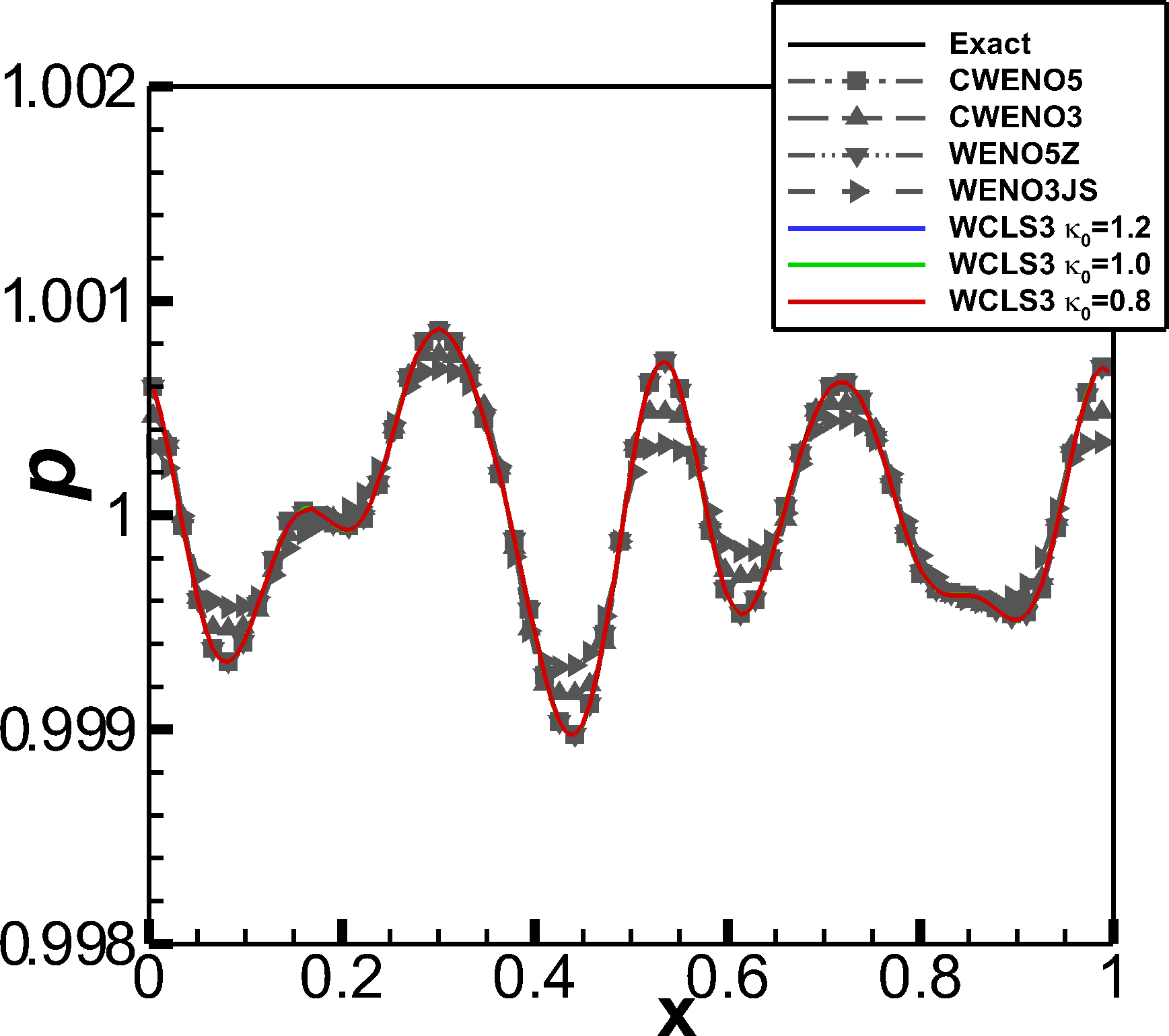}
    \caption{$k_0=4$.\label{fig:bwp4}}
    \end{subfigure}
    \quad 
    \begin{subfigure}[b]{0.3\textwidth}
    \includegraphics[width=\textwidth]{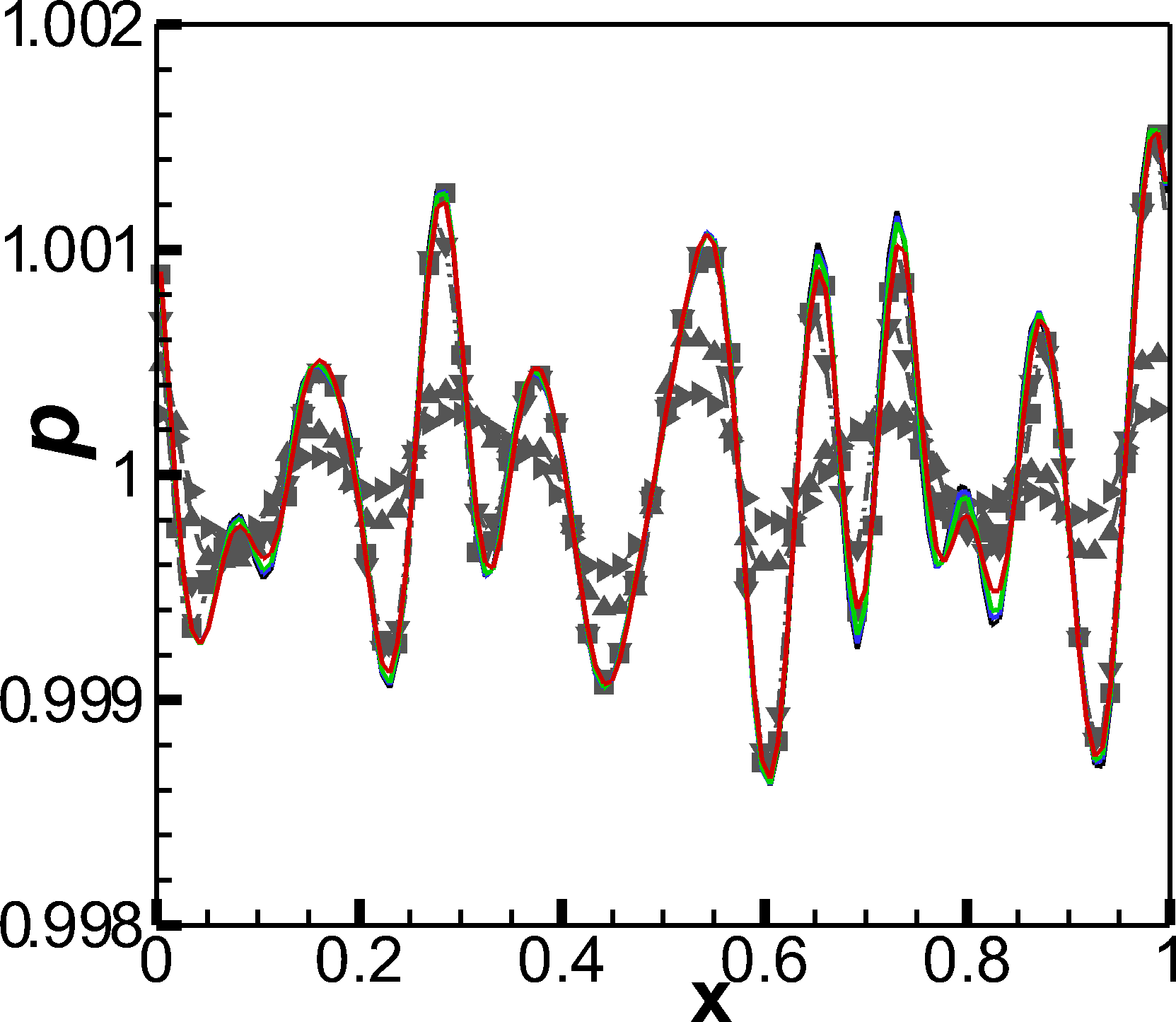}
    \caption{$k_0=8$.\label{fig:bwp8}}
    \end{subfigure}
    \begin{subfigure}[b]{0.3\textwidth}
    \includegraphics[width=\textwidth]{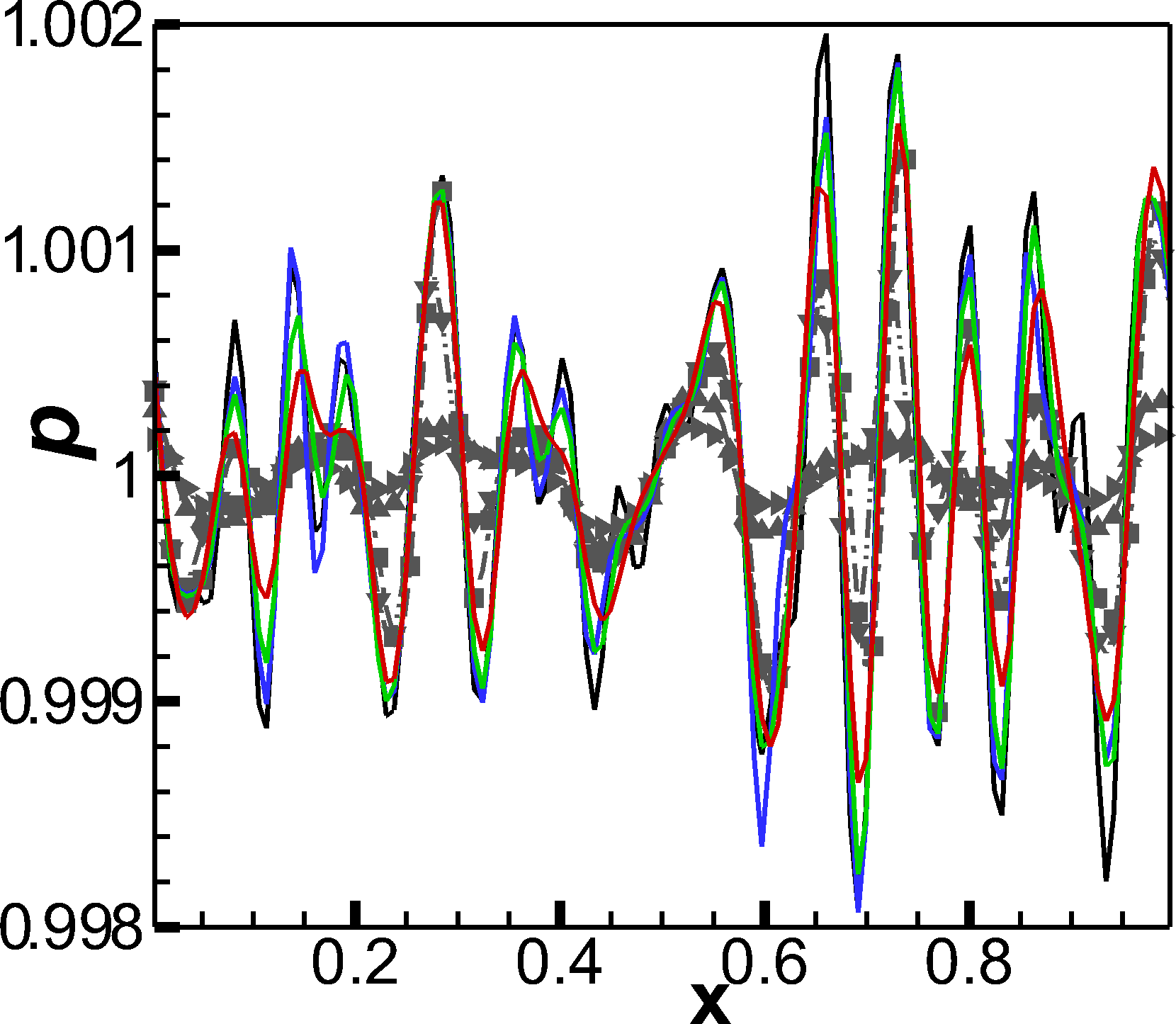}
    \caption{$k_0=12$.\label{fig:bwp12}}
    \end{subfigure}
    \caption{\label{fig:bwp} Propagation of sound waves. $N = 128$, $t = 1.0/(1+\sqrt{\gamma})$ and $CFL = 0.5$.}
\end{figure}

Euler equations with small disturbance, i.e., acoustic waves are simulated. The initial condition is
\begin{equation}
\begin{aligned}
  p(x, t=0) & = p_0\left( 1+ \epsilon \sum_{k=1}^{N/2}{\left[E_p(k)\right]^{0.5} \sin(2\pi k(x+\psi_k))}\right),\\
  \rho(x,t=0) & = \rho_0 \left(\frac{p(x,0)}{p_0}\right)^{1/\gamma},\\
  u(x,t=0) &= u_0 + \frac{2}{\gamma-1}\left(c(x,t=0)-c_0\right),
\end{aligned}
\end{equation}
where 
$
  E_p(k) = \left(k/k_0\right)^4e^{-2(k/k_0)^2}
$
is the energy spectrum. $p_0 = 1$, $u_0 = 1$ and $\rho_0 = 1$. 
The periodic computational domain is $\Omega = [0,1]$ discretized by $N = 128$ uniform cells. $\psi_k,\,k=1,2,3,\cdots,N/2$ are random numbers ranging from 0 to 1. $\epsilon = 0.001$ stands for the intensity of the acoustic turbulence. $c = \sqrt{\gamma p/\rho}$ is the speed of sound and $\gamma=1.4$. The solution at $t = 1.0/(1+\sqrt{\gamma})$ are shown from Fig. \ref{fig:bwp4} to Fig \ref{fig:bwp12} with $k_0 = 4, 8, 12$, respectively.

When $k_0 = 4$, the WCLS3 scheme with different coefficients obtain almost identical results as shown in Fig. \ref{fig:bwp4}. When $k_0$ increases to $8$ and $12$, coefficients with increasing $\kappa_0$ capture a broader range of bandwidths, as shown in Figs. \ref{fig:bwp8} and \ref{fig:bwp12}. In smooth regions, the WCLS3 scheme performs even better than the fifth-order WENO5-Z and CWENO5 schemes.

\begin{figure}[!htbp]
  \centering
    \begin{subfigure}[b]{0.4\textwidth}
    \includegraphics[width=\textwidth]{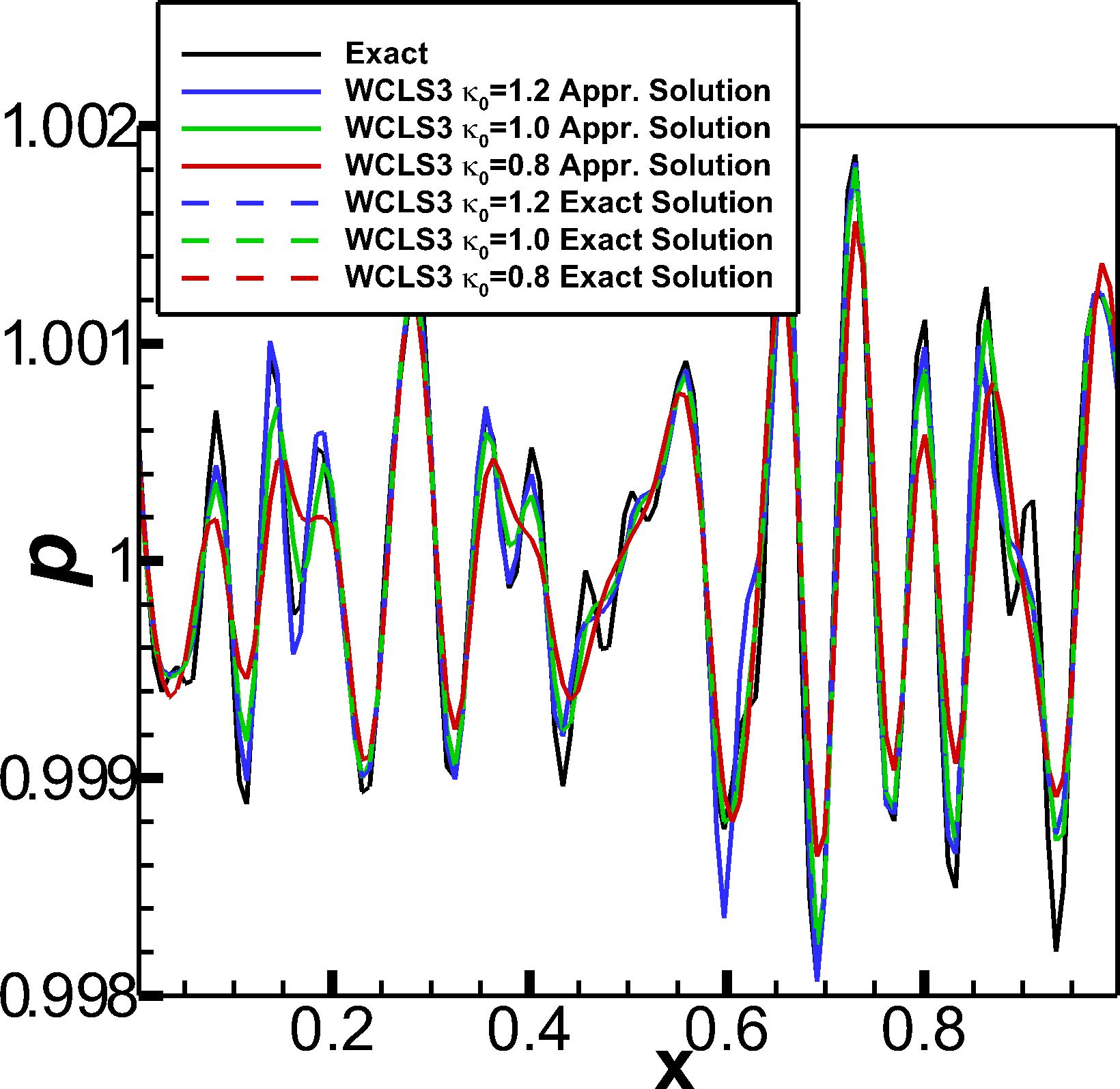}
    \caption{Overview for $k_0=12$.\label{fig:bwp12_comp}}
    \end{subfigure}
    \quad 
    \begin{subfigure}[b]{0.4\textwidth}
    \includegraphics[width=\textwidth]{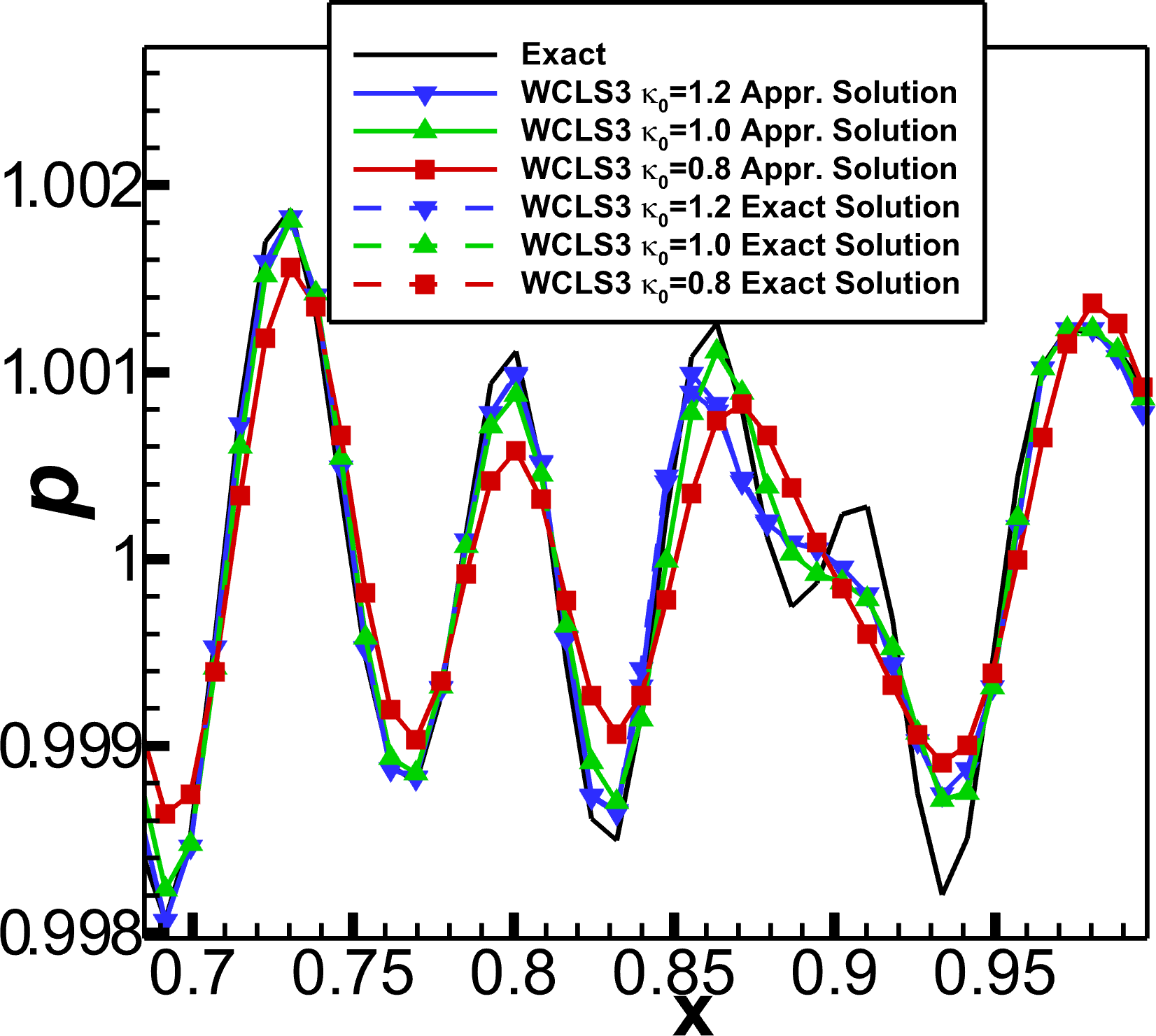}
    \caption{Close view for $k_0=12$.\label{fig:bwp12_comp_cv}}
    \end{subfigure}
    \caption{\label{fig:bwp_cmp} Comparison between exact cyclic block tridiagonal solver and approximate cyclic block tridiagonal solver for the propagation of sound waves.}
\end{figure}
Figure \ref{fig:bwp_cmp} also compares the results with exact cyclic block tridiagonal solver and approximate cyclic block tridiagonal solver. Only slight differences are observed for the WCLS3 scheme with coefficients of $\kappa_0 = 1.2$.

\subsubsection{Sod Problem \cite{sod1978survey}}
\begin{figure}[!htbp]
  \centering
    \begin{subfigure}[b]{0.4\textwidth}
    \includegraphics[width=\textwidth]{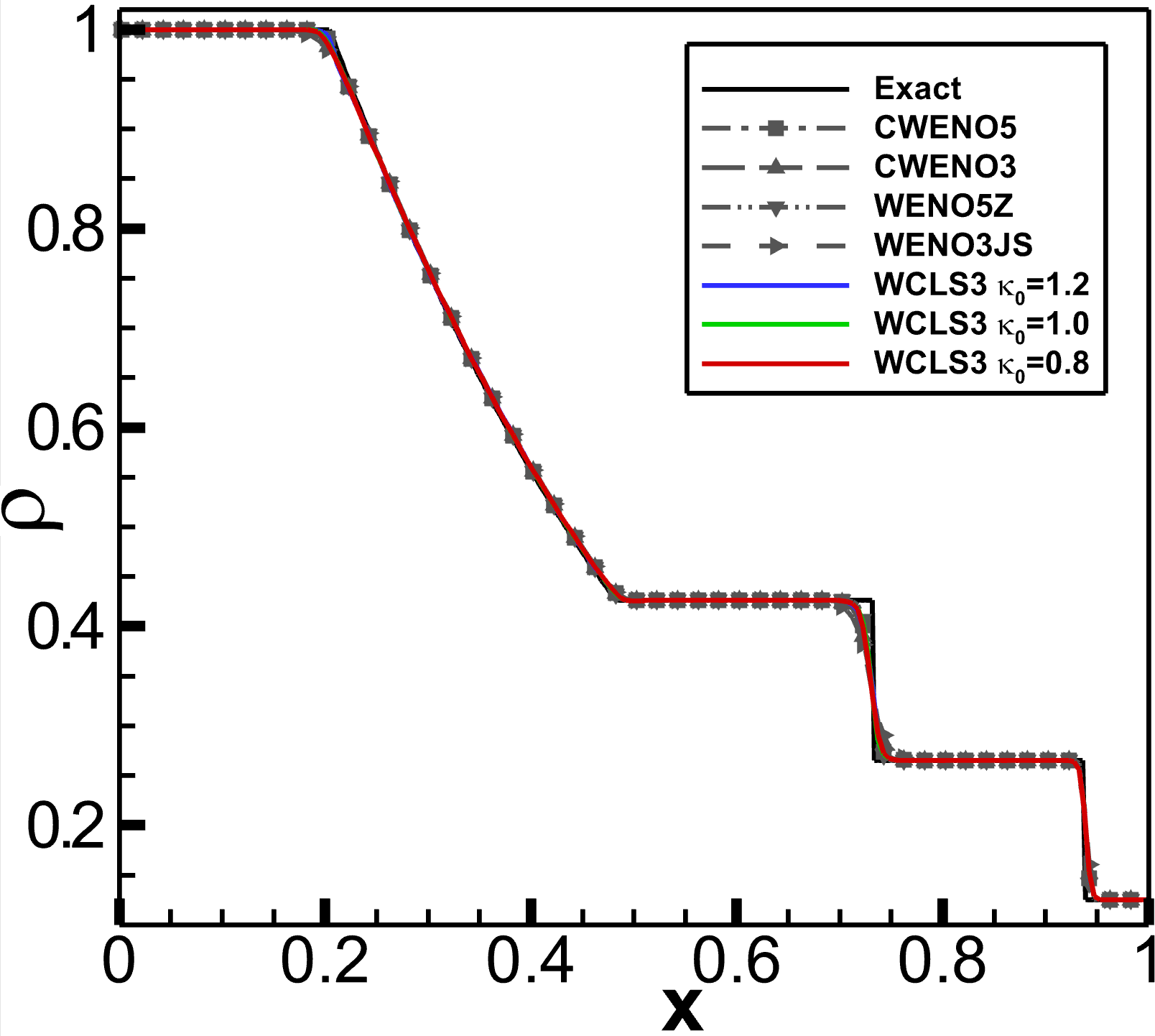}
    \caption{Density.}
    \end{subfigure}
    \quad 
    \begin{subfigure}[b]{0.4\textwidth}
    \includegraphics[width=\textwidth]{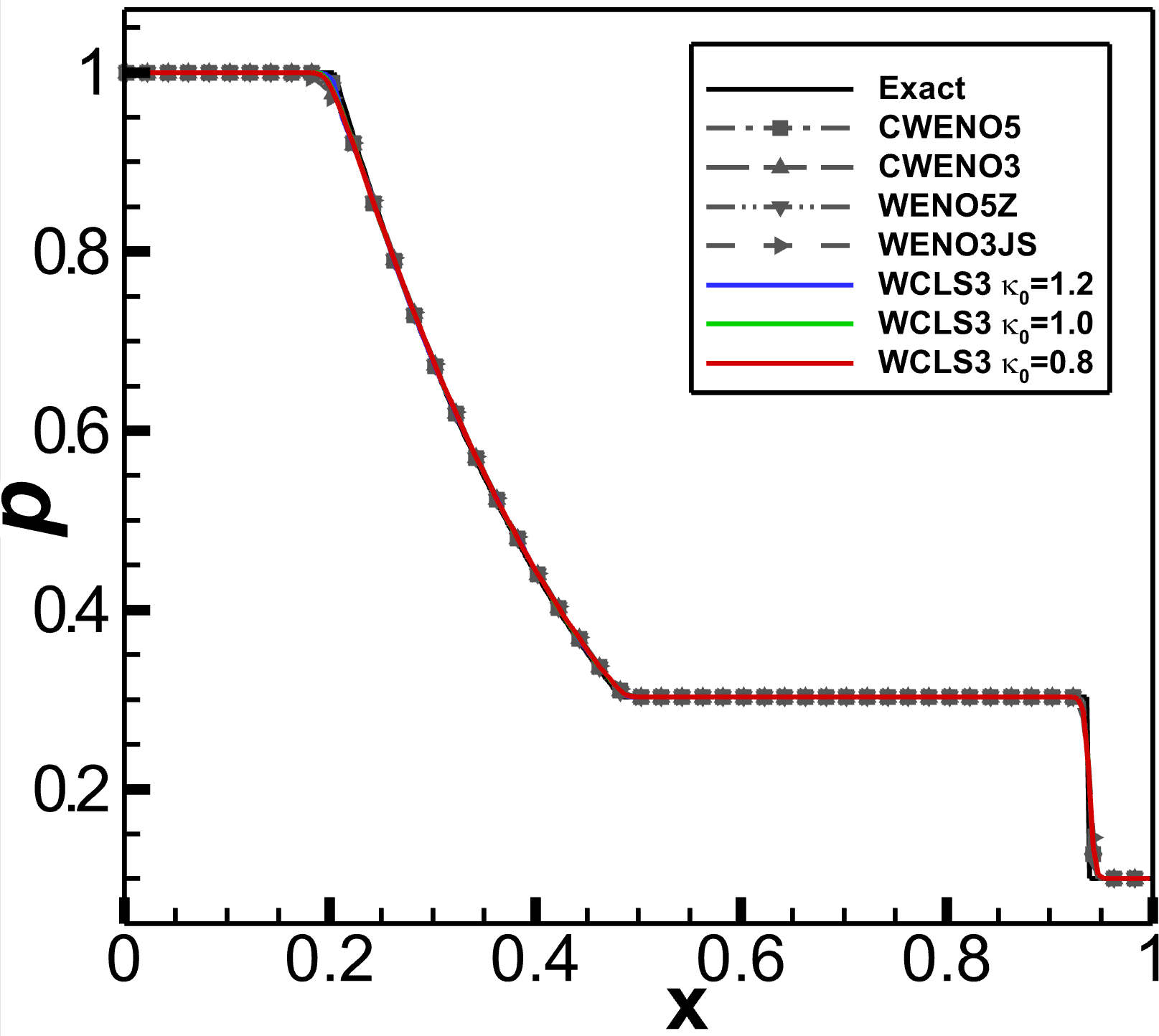}
    \caption{Pressure.}
    \end{subfigure}\\
    \begin{subfigure}[b]{0.4\textwidth}
    \includegraphics[width=\textwidth]{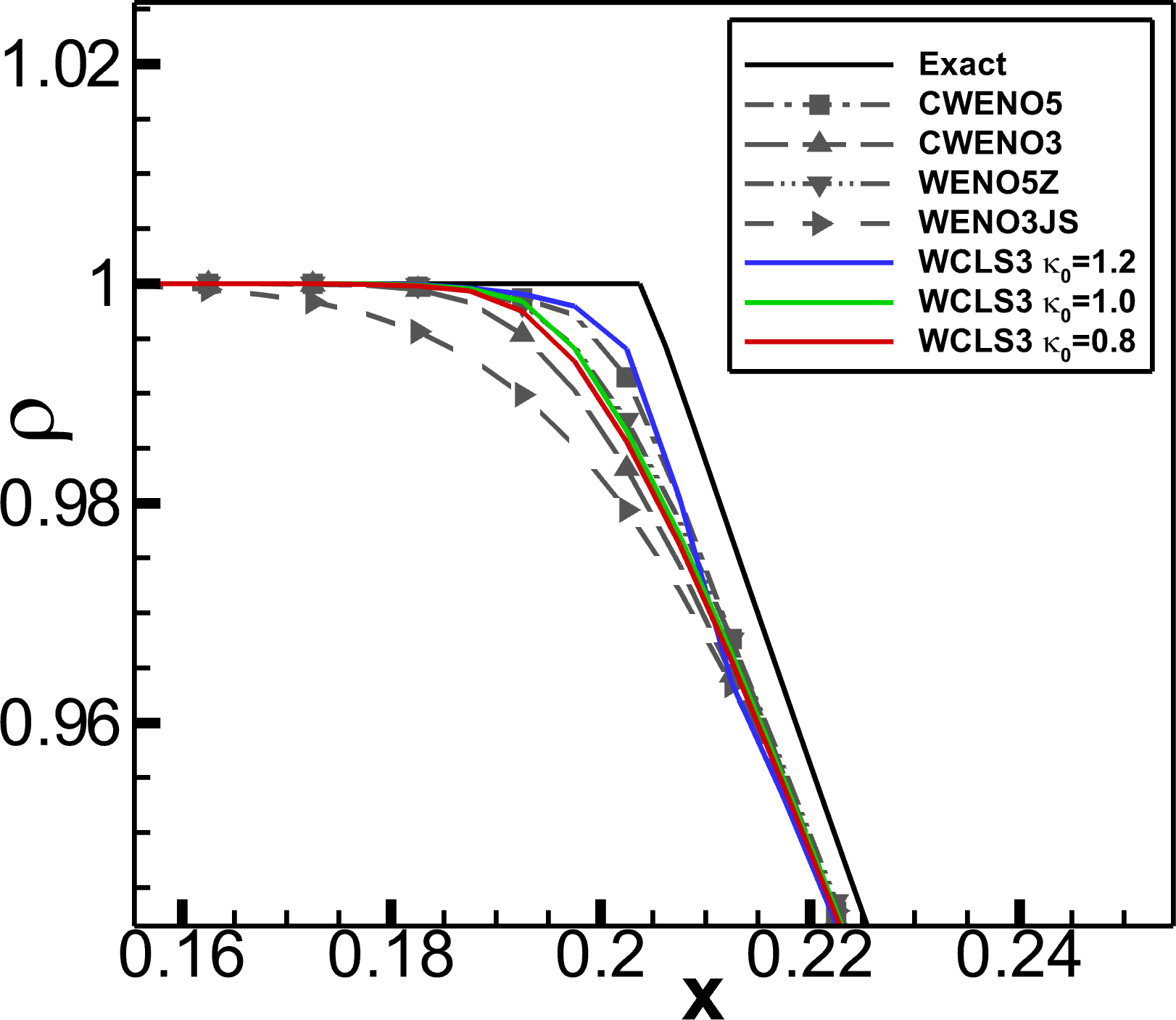}
    \caption{Close view at the head of rarefaction wave.\label{fig:sod_rw_closeview}}
    \end{subfigure}
    \quad 
    \begin{subfigure}[b]{0.4\textwidth}
    \includegraphics[width=\textwidth]{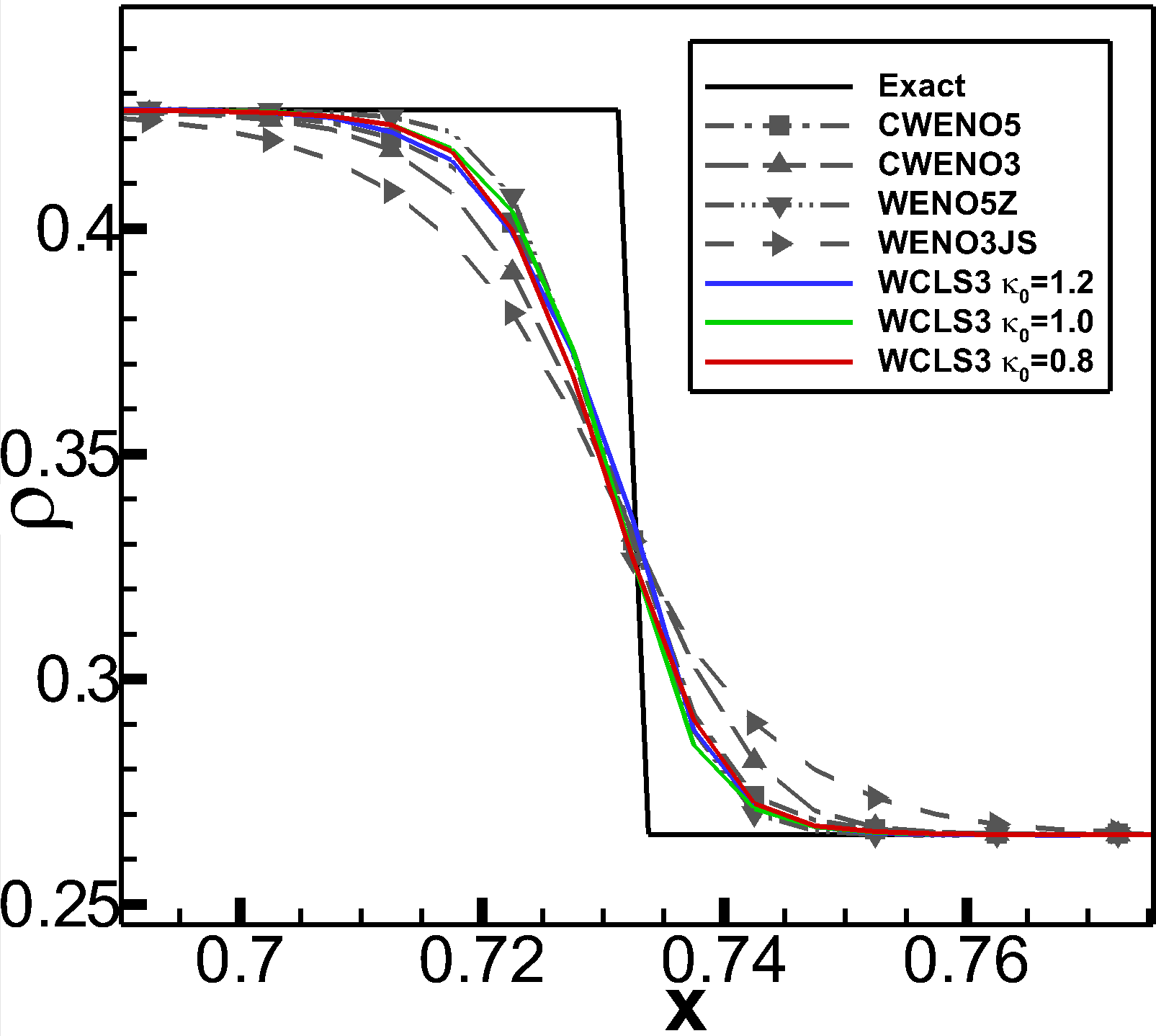}
    \caption{Close view at the contact discontinuity.\label{fig:sod_cd_closeview}}
    \end{subfigure}
    \caption{\label{fig:sod} Sod shock tube problem. $N = 200$, $t = 0.25$ and $CFL = 0.5$.}
\end{figure}
Sod shock tube problem is a popular benchmark in testing the performance of shock-capturing schemes. The initial condition is given by
\begin{equation}
  \left(\rho, u, p\right) = \left\{
  \begin{array}{ll}
    1, 0, 1, & \text{if}\,\,0.0\leq x < 0.5,\\
    0.125, 0, 0.1, & \text{if}\,\,0.5 \leq x\leq 1.0.\\
  \end{array}
  \right.
\end{equation}
The computational domain $\Omega = [0,1]$ is discretized by $200$ uniform cells. The final simulation time is $t = 0.25$ and the CFL number is $0.5$ with third-order SSP-RK integrator in temporal direction.

Figure \ref{fig:sod} shows the density and pressure distribution obtained by the WCLS3 scheme. As Fig. \ref{fig:sod} illustrates, the results by the WCLS3 scheme are oscillation-free near discontinuities and high-resolution in smooth regions.
Figures \ref{fig:sod_rw_closeview} and \ref{fig:sod_cd_closeview} present the close view of density profile at the head of rarefaction wave and around the contact discontinuity. The resolution of the WCLS3 scheme are higher than the WENO3-JS and CWENO3 schemes and even comparative with or better than the CWENO5 and WENO5-Z schemes with the increase of $\kappa_0$.

\subsubsection{Lax Problem \cite{lax}}
The Lax shock tube problem is more severe than the Sod problem and is utilized in this section to further validate the robustness and high-resolution of the proposed WCLS3 scheme. The initial condition for the Lax problem is
\begin{equation}
  \left(\rho, u, p\right) = \left\{
  \begin{array}{ll}
    0.445, 0.698, 3.528, & \text{if}\,\,0.0\leq x < 0.5,\\
    0.5, 0, 0.571, & \text{if}\,\,0.5 \leq x\leq 1.0.\\
  \end{array}
  \right.
\end{equation}
The computational domain is $\Omega = [0,1]$ discretized by $200$ uniform cells. The final simulation time is $t = 0.1$ and the CFL number is 0.5 with third-order SSP-RK time integrator.

\begin{figure}[!htbp]
  \centering
    \begin{subfigure}[b]{0.4\textwidth}
    \includegraphics[width=\textwidth]{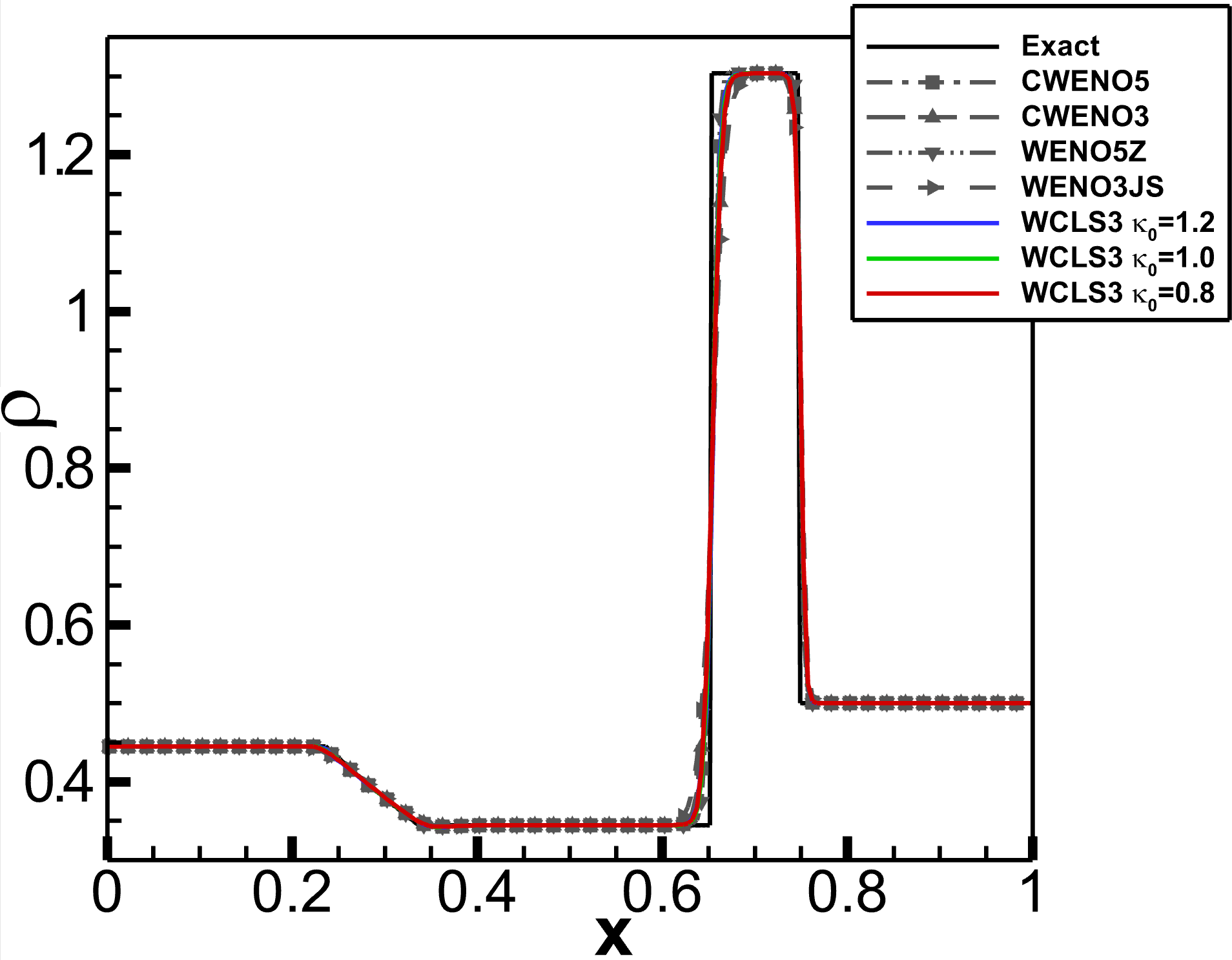}
    \caption{Density.}
    \end{subfigure}
    \quad 
    \begin{subfigure}[b]{0.4\textwidth}
    \includegraphics[width=\textwidth]{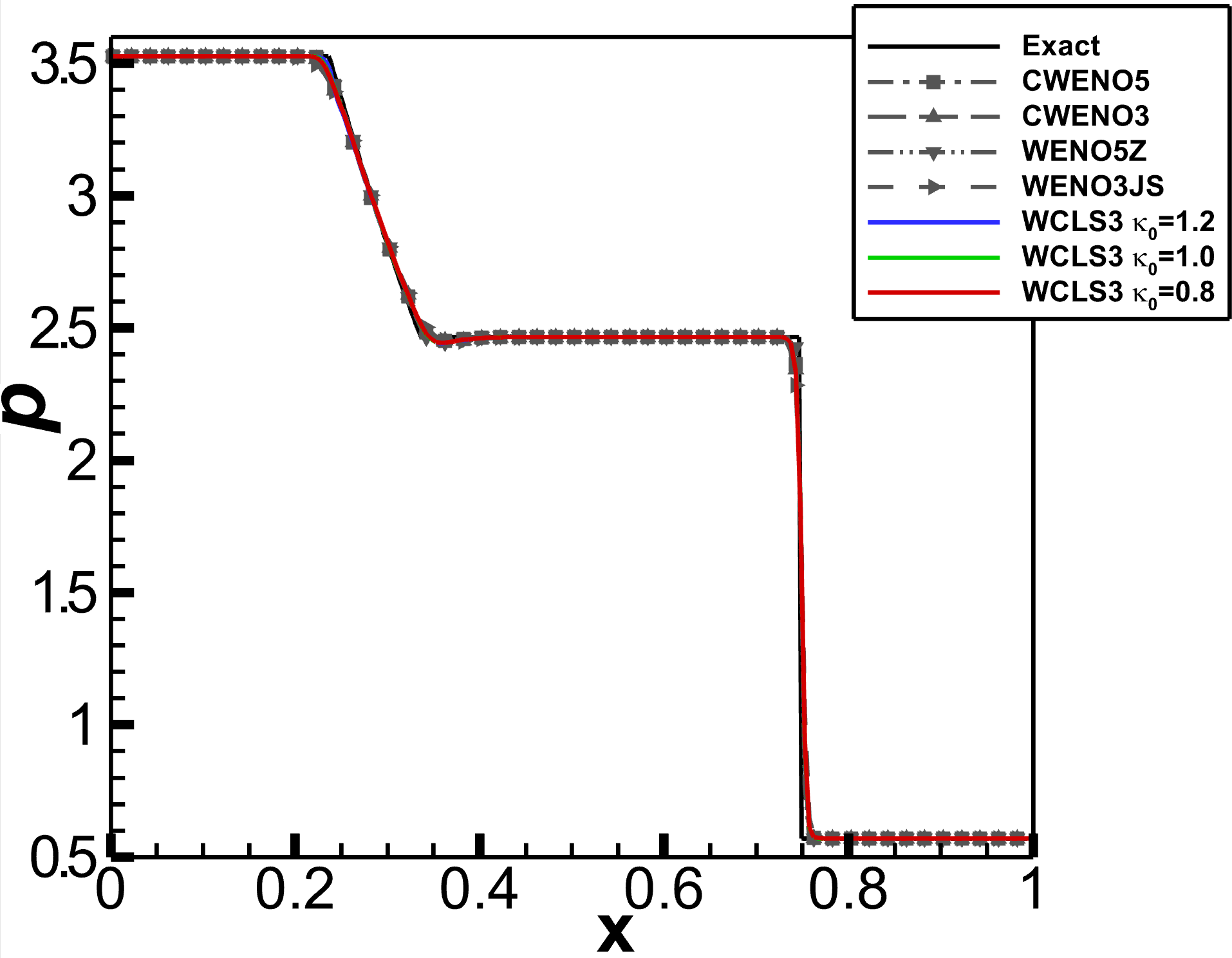}
    \caption{Pressure.}
    \end{subfigure}\\
    \begin{subfigure}[b]{0.4\textwidth}
    \includegraphics[width=\textwidth]{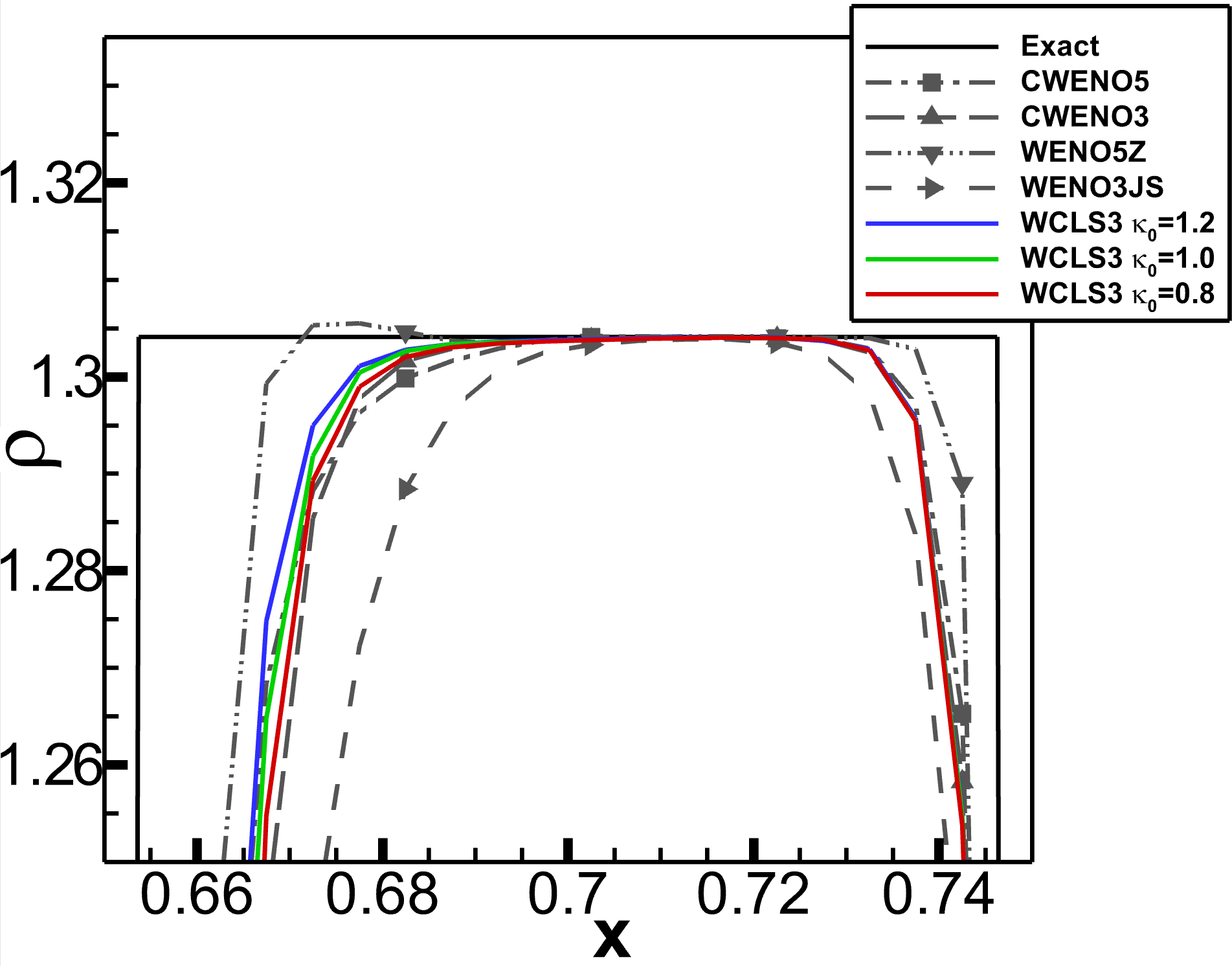}
    \caption{Density close view.}
    \end{subfigure}
    \quad 
    \begin{subfigure}[b]{0.4\textwidth}
    \includegraphics[width=\textwidth]{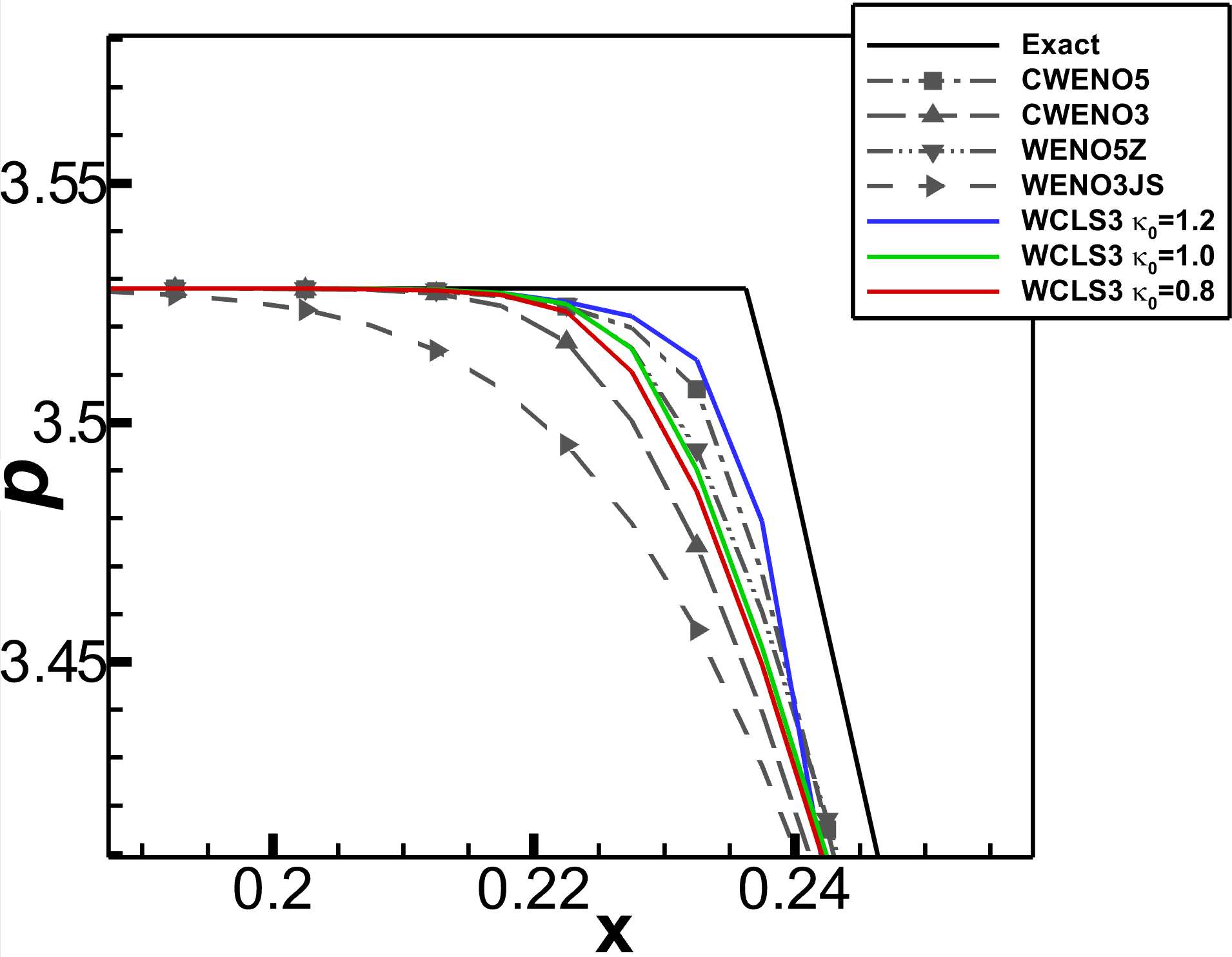}
    \caption{Pressure close view.\label{fig:lax_p_closeview}}
    \end{subfigure}
    \caption{\label{fig:lax} Lax shock tube problem. $N = 200$, $t = 0.1$ and $CFL = 0.5$.}
\end{figure}

The results of the WCLS3 scheme are shown as in Fig. \ref{fig:lax}. As shown in the close views of density and pressure profiles, the WCLS3 scheme captures the shock and contact discontinuities without oscillations. The resolution at discontinuities of the WCLS3 is higher than the WENO3-JS and CWENO3 schemes and is comparative with CWENO5 scheme in this case. At the front of rarefaction wave, as shown in Fig. \ref{fig:lax_p_closeview}, the resolution of the WCLS3 scheme with coefficients of $\kappa_0 = 1.2$ is even higher than the WENO5-Z scheme.

\begin{figure}[!htbp]
  \centering
    \includegraphics[width=0.4\textwidth]{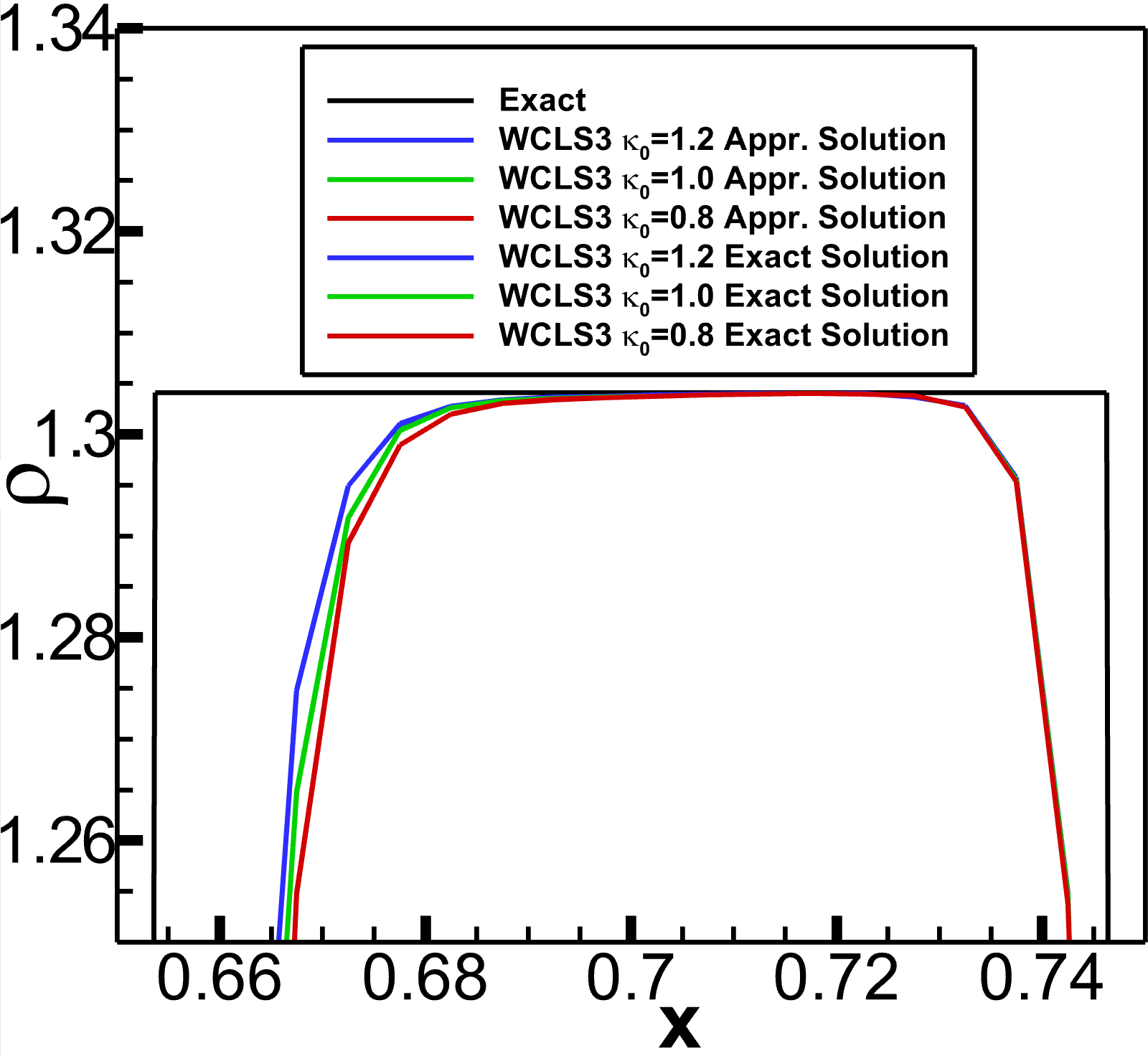}
    \caption{\label{fig:lax_compare} Comparison between Alg. \ref{algorithm:1} and Alg. \ref{algorithm:2} for the lax shock tube problem.}
\end{figure}

Figure \ref{fig:lax_compare} compares the results of the WCLS3 scheme with exact block tridiagonal solver as in Alg. \ref{algorithm:1} and approximate solver as in Alg. \ref{algorithm:2}. In this case, the two algorithms give almost identical results.

\subsubsection{Interacting Blast Waves \cite{woodward_numerical_1984}}
In this problem, two shocks interact with each other in a computational domain with reflective boundary conditions. The initial condition is
\begin{equation}
  \left(\rho, u, p\right) = \left\{
  \begin{array}{ll}
    1, 0, 1000, & \text{if}\,\,0.0\leq x < 0.1,\\
    1, 0, 0.01, & \text{if}\,\,0.1\leq x < 0.9,\\
    1, 0, 100, & \text{if}\,\,0.9\leq x < 1.0.\\
  \end{array}
  \right.
\end{equation}
The computational domain is $[0,1]$ discretized by $400$ uniform control volumes. The final simulation time is $t = 0.038 $ advanced by the third-order SSP-RK method with CFL number of $0.5$.

\begin{figure}[!htbp]
  \centering
    \begin{subfigure}[b]{0.3\textwidth}
    \includegraphics[width=\textwidth]{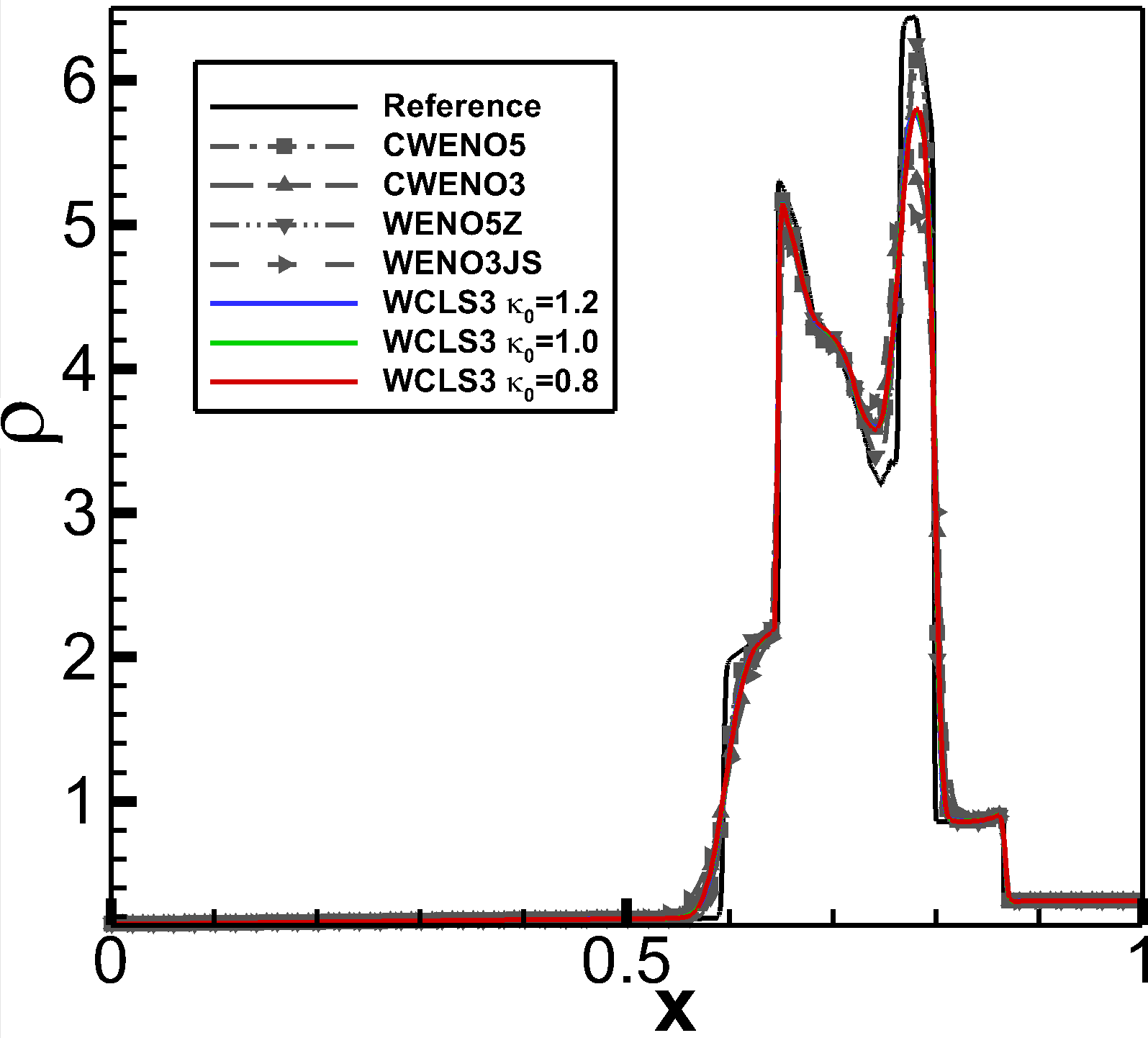}
    \caption{Density.}
    \end{subfigure}
    \quad
    \begin{subfigure}[b]{0.3\textwidth}
    \includegraphics[width=\textwidth]{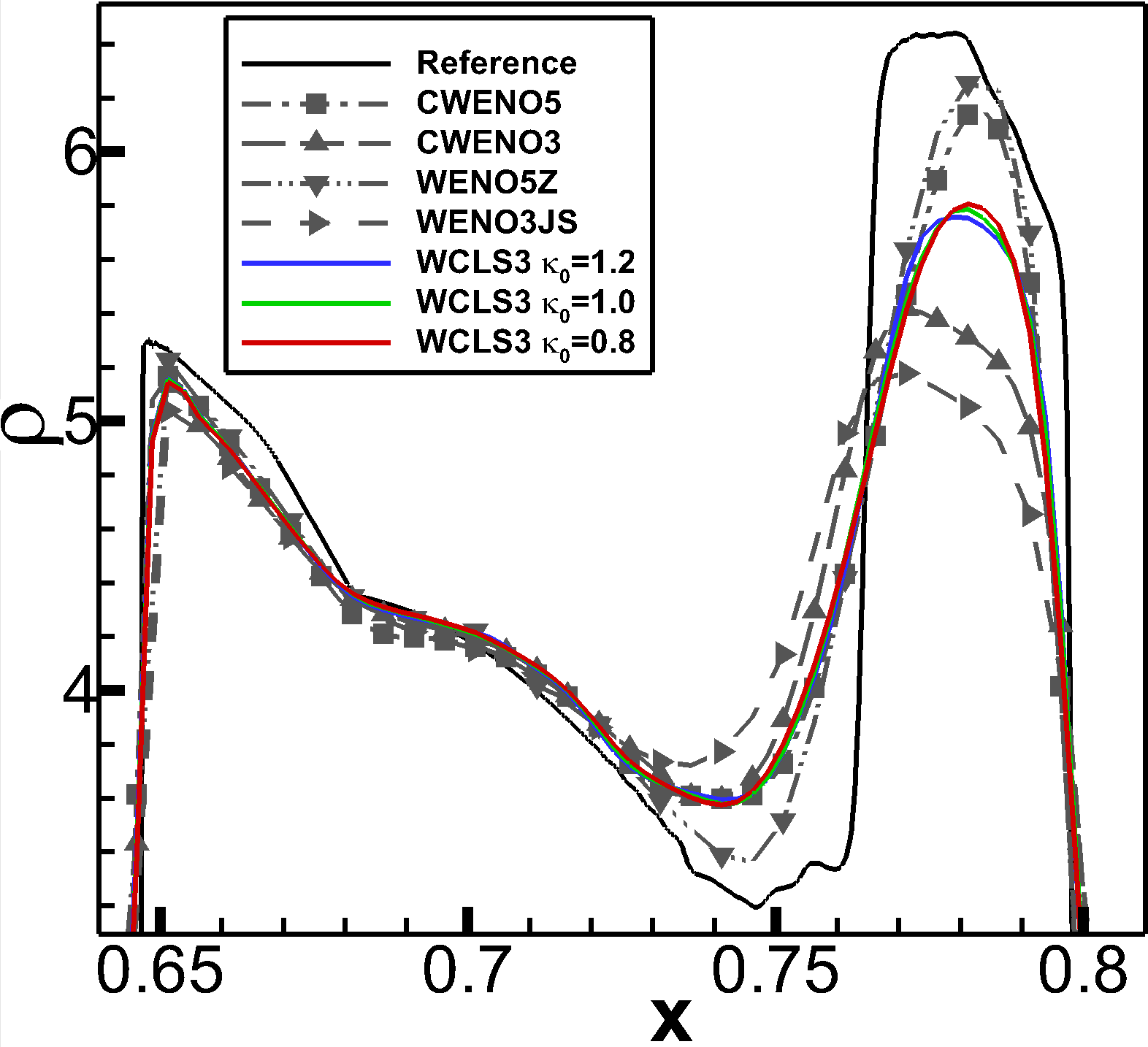}
    \caption{Close view.\label{fig:ibw_closeview}}
    \end{subfigure}
    \quad
    \begin{subfigure}[b]{0.3\textwidth}
    \includegraphics[width=\textwidth]{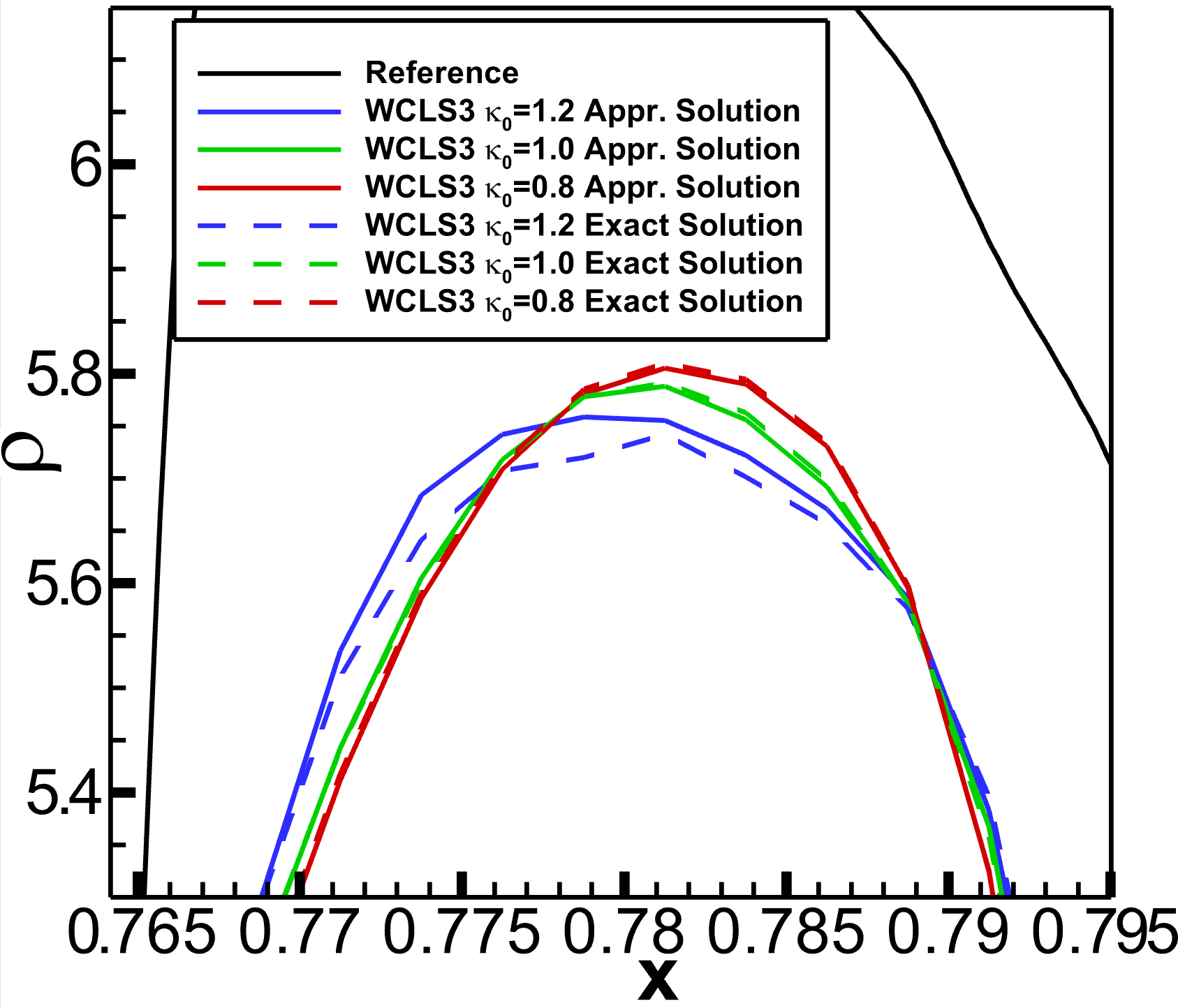}
    \caption{Comparison between Alg. \ref{algorithm:1} and Alg. \ref{algorithm:2}.\label{fig:ibw_cv_comp}}
    \end{subfigure}
    \caption{\label{fig:ibw} Interacting blast waves. $N = 400$, $t = 0.038$ and $CFL = 0.5$.}
\end{figure}

Figure \ref{fig:ibw} depicts the density profile of the WCLS3 scheme. The reference line is the result obtained by WENO5-Z scheme with $5000$ uniform control volumes. Again, the WCLS3 scheme demonstrates higher resolution than the WENO3-JS and the CWENO3 schemes. Figure \ref{fig:ibw} also confirms the oscillation-free property of the proposed WCLS3 scheme in complex shock interacting problems. Figure \ref{fig:ibw_cv_comp} compares the density profile of the WCLS3 scheme with block tridiagonal solver of Alg. \ref{algorithm:1} and Alg. \ref{algorithm:2}. Only slight differences are observed in the close view of Fig. \ref{fig:ibw_cv_comp}, which validates the effectiveness of the proposed approximate block tridiagonal solver.

\subsubsection{Shu-Osher Problem}
In Shu-Osher problem \cite{shu1989efficient}, the shock interacts with the disturbance of density, generating multiscale flow structures. The initial condition for the Shu-Osher problem is
\begin{equation}
  \left(\rho, u, p\right) = \left\{
  \begin{array}{ll}
    3.857143, 2.629369, 10.333333, & \text{if}\,\,0.0\leq x < 1.0,\\
    1+0.2\sin(5x), 0, 1, & \text{if}\,\,1.0 \leq x\leq 10.\\
  \end{array}
  \right.
\end{equation}

\begin{figure}[!htbp]
  \centering
    \begin{subfigure}[b]{0.33\textwidth}
    \includegraphics[width=\textwidth]{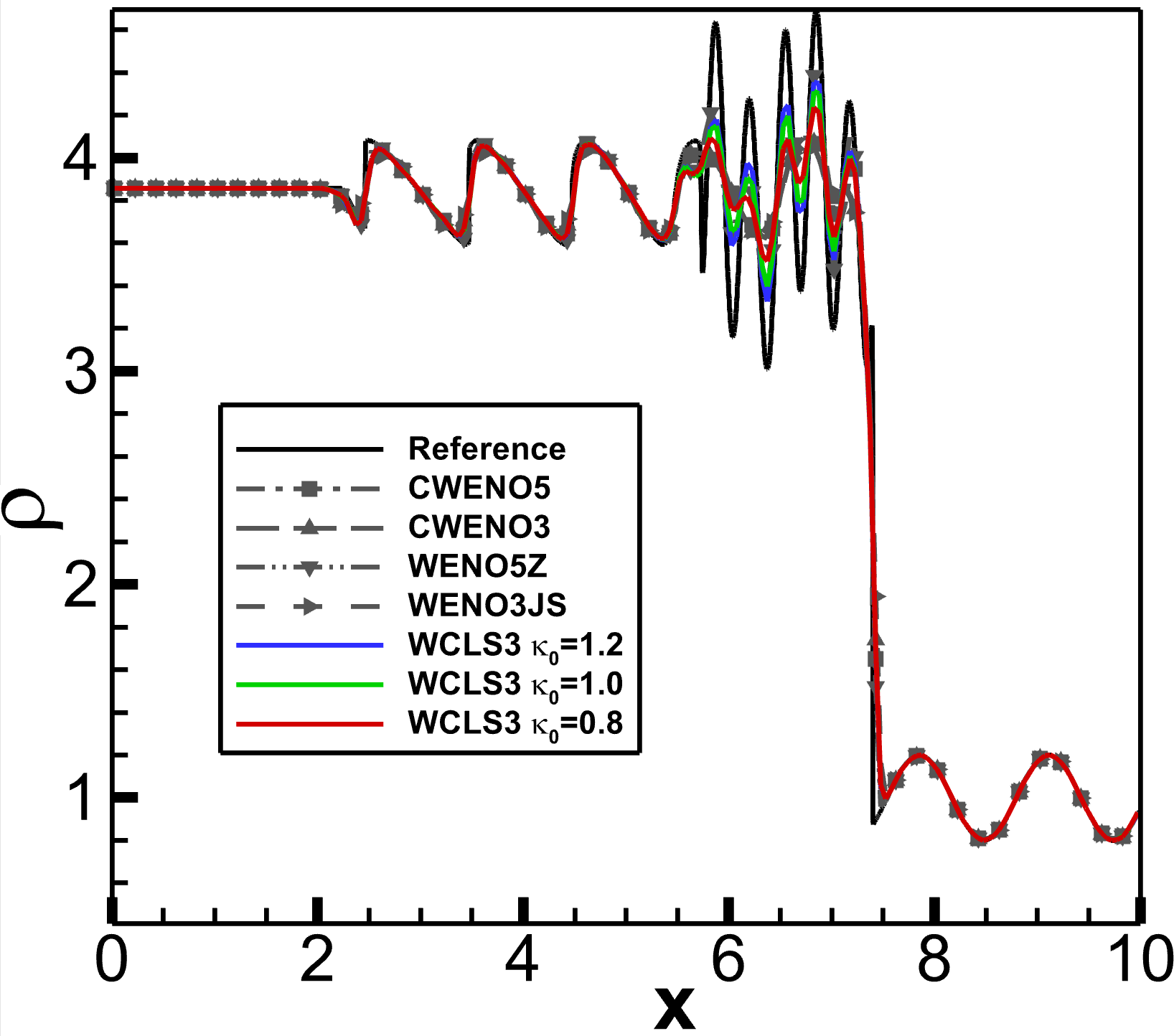}
    \caption{Density.}
    \end{subfigure}
    \begin{subfigure}[b]{0.33\textwidth}
    \includegraphics[width=\textwidth]{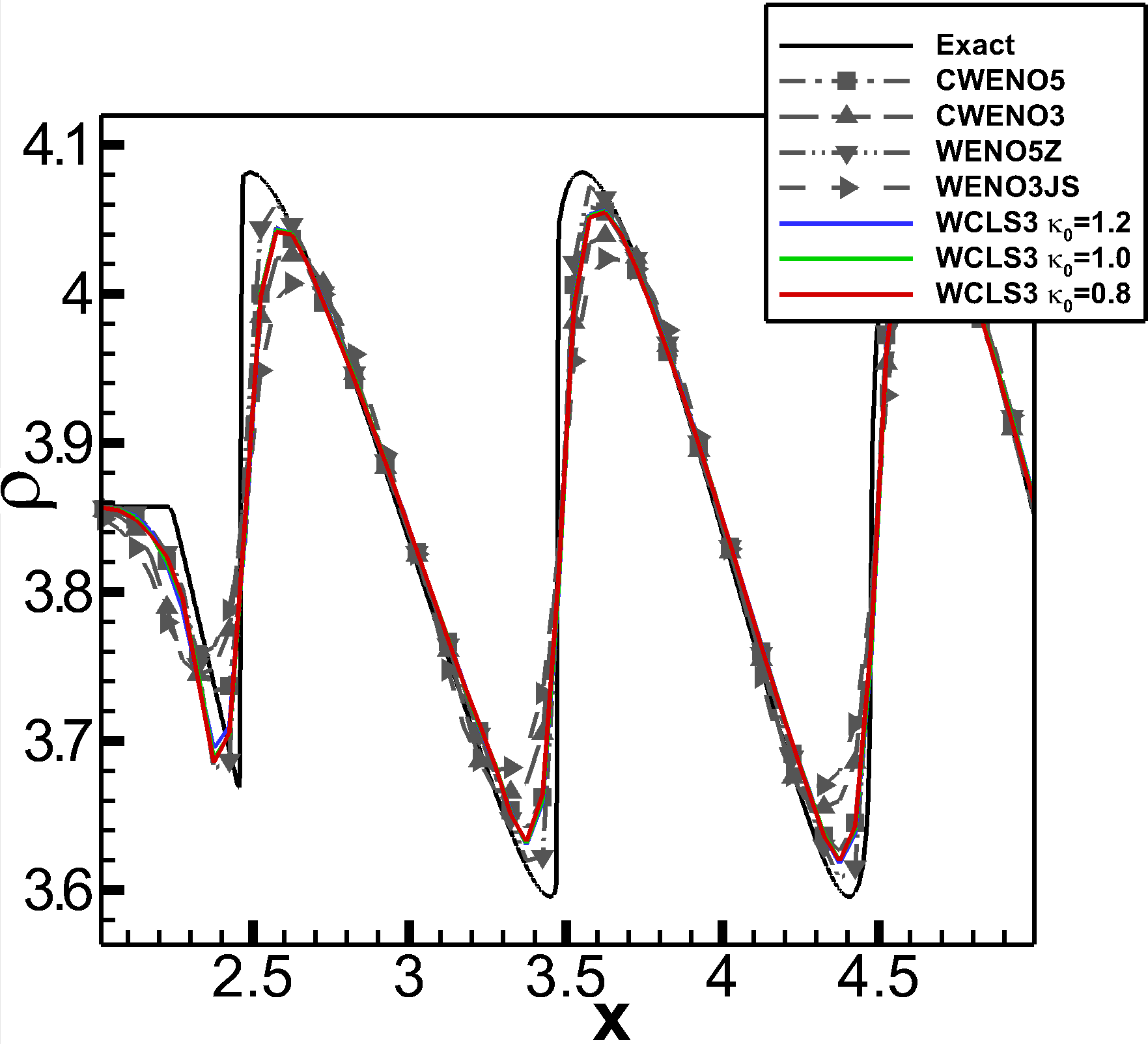}
    \caption{Close view of shocks.\label{fig:SO_closeview1}}
    \end{subfigure}
    \begin{subfigure}[b]{0.33\textwidth}
    \includegraphics[width=\textwidth]{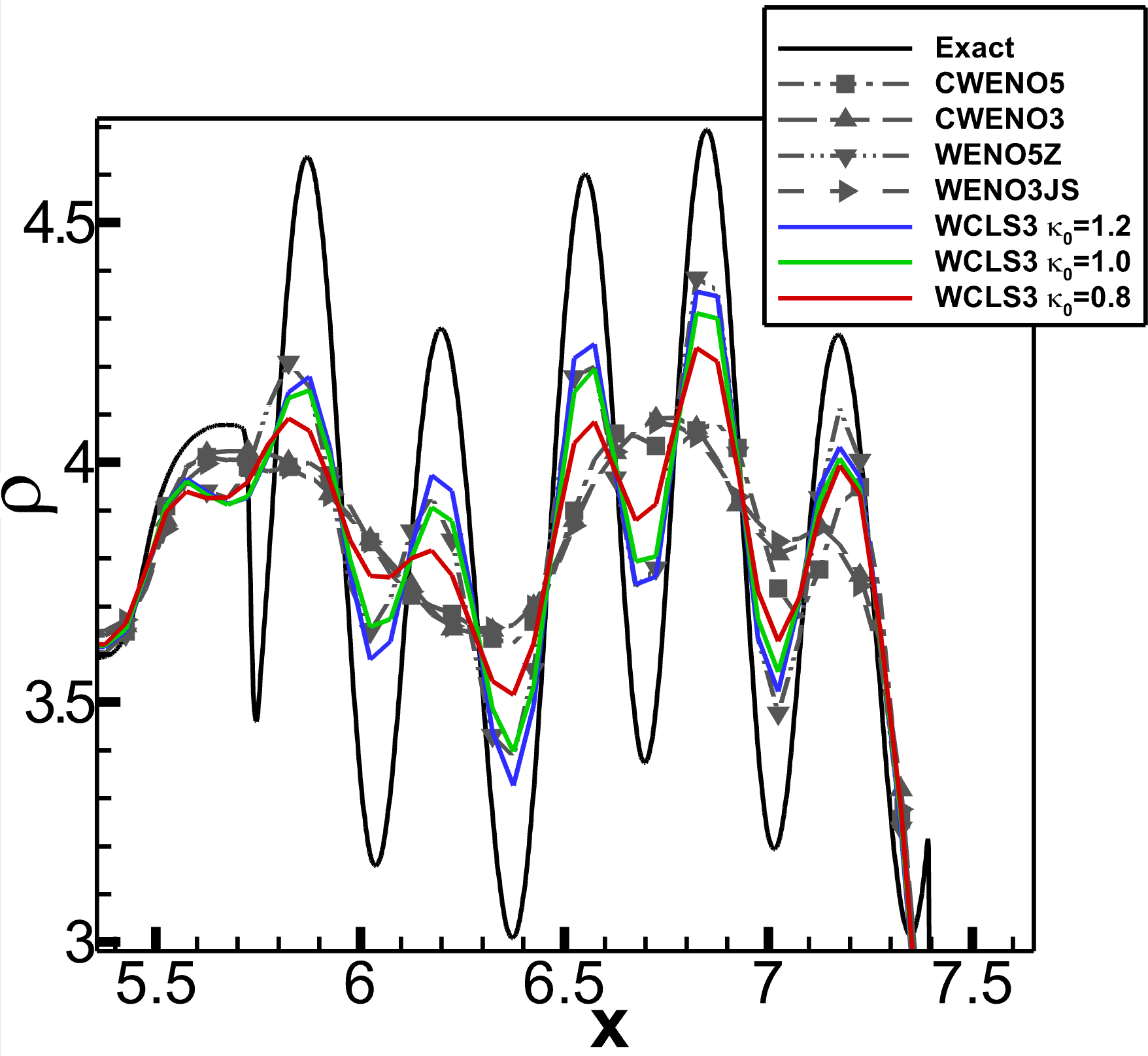}
    \caption{Close view of density disturbance.\label{fig:SO_closeview2}}
    \end{subfigure}
    \caption{\label{fig:so} Shu-Osher problem. $N = 200$, $t = 1.8$ and $CFL = 0.5$.}
\end{figure}

The computational domain is $\Omega = [0,10]$ discretized by $200$ uniform control volumes. The simulation end time is $t = 1.8$ advanced by the third-order SSP-RK method with CFL = $0.5$.
The problem serves as a benchmark to check both the resolution and the robustness of shock-capturing schemes for compressible flows. Figure \ref{fig:so} shows the density profile obtained by the WCLS3 scheme. 
The reference result is obtained by the WENO5-Z scheme with $4000$ uniform control volumes. Close views around the shocks and the density disturbance are given as in Figs. \ref{fig:SO_closeview1} and \ref{fig:SO_closeview2}, respectively.
In Fig. \ref{fig:SO_closeview1}, the shocks are captured by the WCLS3 scheme without oscillations; and in Fig. \ref{fig:SO_closeview2}, the waves of density disturbance are well-resolved by the WCLS3 scheme. The resolution of the WCLS3 scheme is higher than the WENO3-JS, CWENO3 and CWENO5 schemes, and comparative with the WENO5-Z scheme in this case. As expected, with the increase of $\kappa_0$, the resolution of the WCLS3 scheme becomes higher.

\subsubsection{Titarev-Toro Problem \cite{titarev2004finite}}
The Titarev-Toro problem is an upgraded version of the Shu-Osher problem. The initial condition is given by
\begin{equation}
  \left(\rho, u, p\right) = \left\{
  \begin{array}{ll}
    1.515695, 0.523346, 1.805000, & \text{if}\,\,0.0\leq x < 0.5,\\
    1+0.1\sin(20 \pi x), 0, 1, & \text{if}\,\,0.5 \leq x\leq 10.\\
  \end{array}
  \right.
\end{equation}

\begin{figure}[!htbp]
  \centering
    \begin{subfigure}[b]{0.3\textwidth}
    \includegraphics[width=\textwidth]{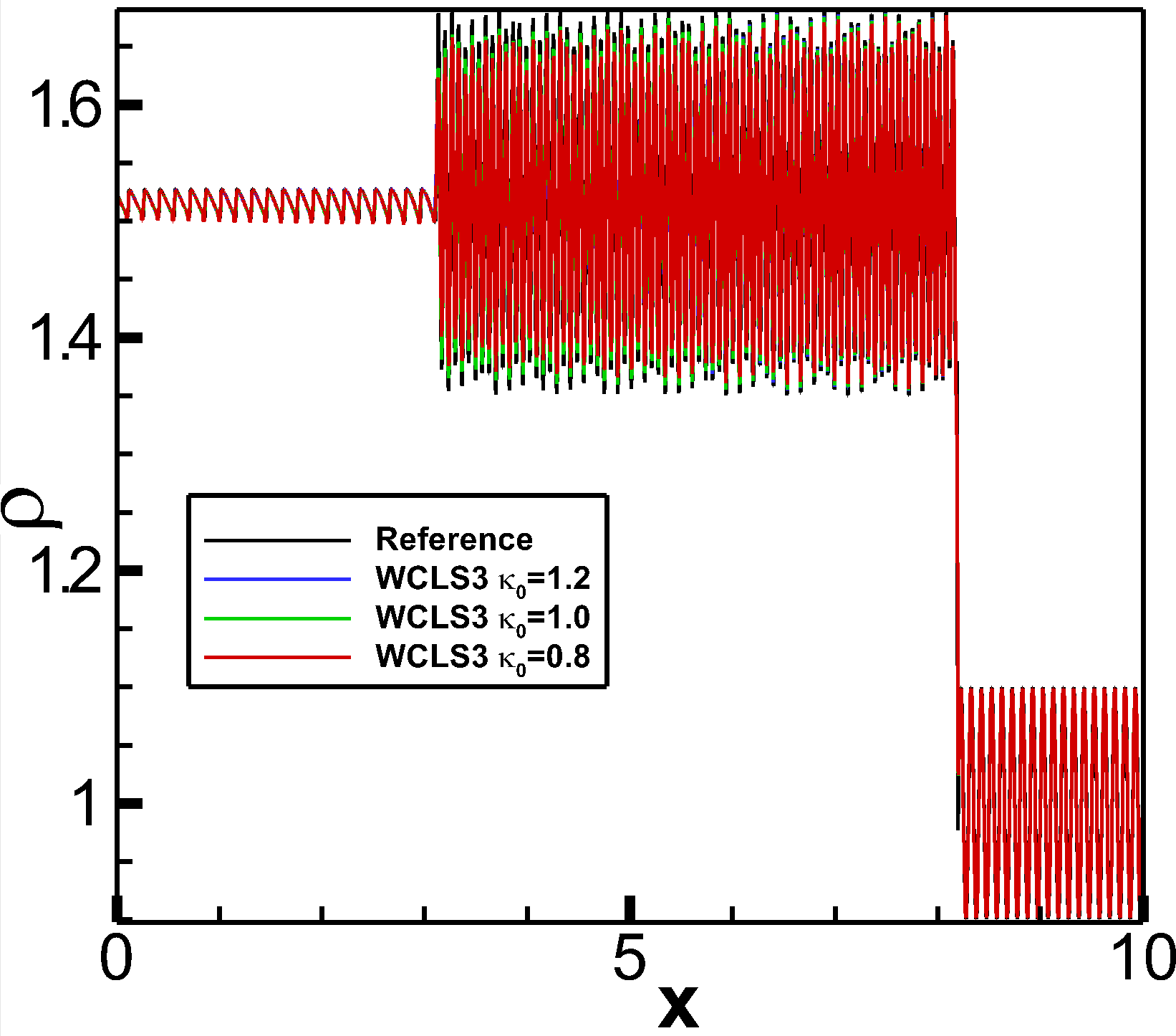}
    \caption{Density.}
    \end{subfigure}
    \quad
    \begin{subfigure}[b]{0.3\textwidth}
    \includegraphics[width=\textwidth]{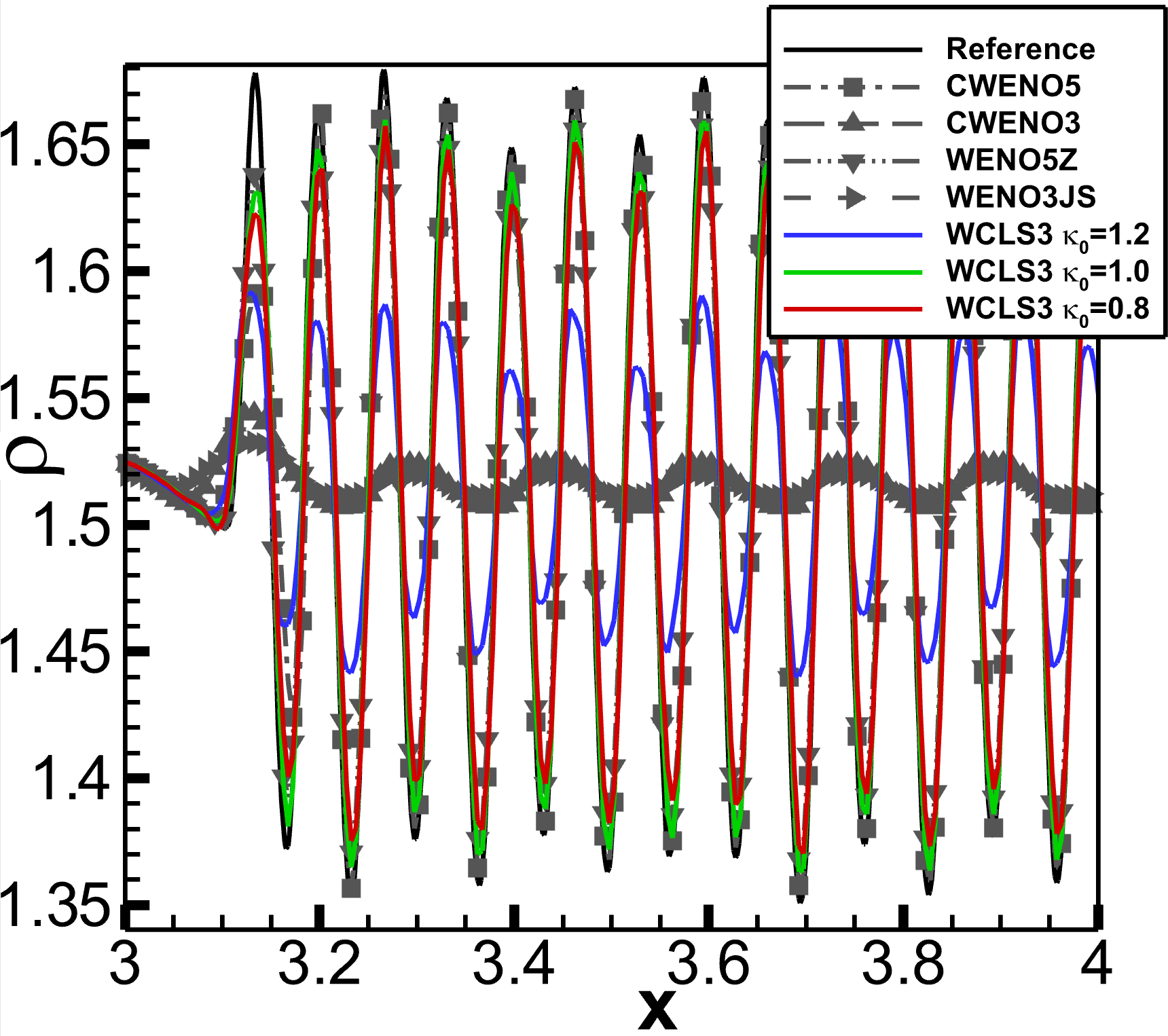}
    \caption{Close view 1.\label{fig:TT_closeview1}}
    \end{subfigure}\\
    \begin{subfigure}[b]{0.3\textwidth}
    \includegraphics[width=\textwidth]{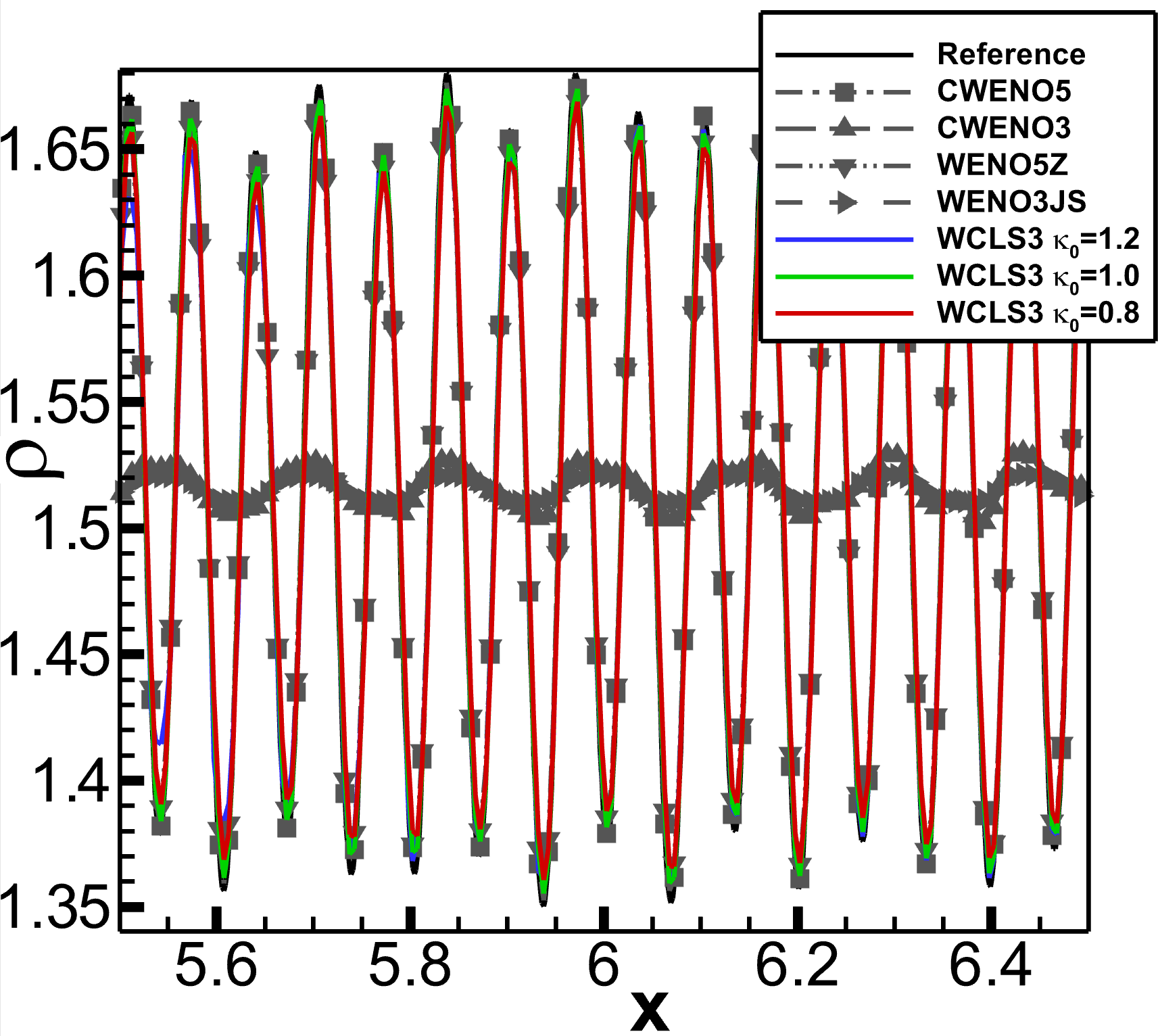}
    \caption{Close view 2.\label{fig:TT_closeview2}}
    \end{subfigure}
    \quad
    \begin{subfigure}[b]{0.3\textwidth}
    \includegraphics[width=\textwidth]{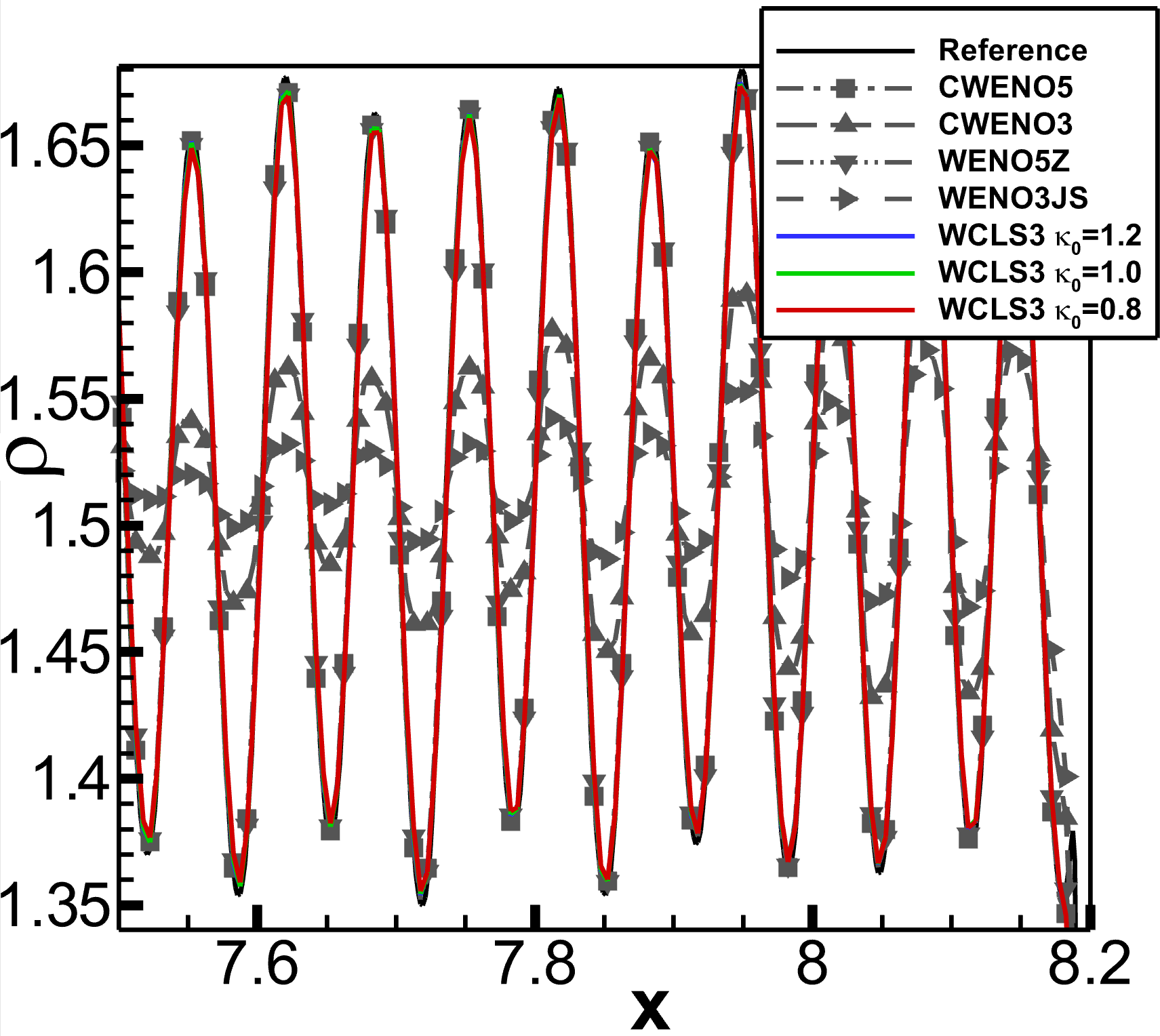}
    \caption{Close view 3.\label{fig:TT_closeview3}}
    \end{subfigure}
    \caption{\label{fig:tt} Titarev-Toro problem. $N = 2000$, $t = 5.0$ and $CFL = 0.5$.}
\end{figure}

Figure \ref{fig:tt} depicts the results obtained by the WCLS3 scheme with $2000$ uniform control volumes at $t = 5.0$. The time integrator is the third-order SSP-RK method with CFL number as 0.5. The reference line is the result obtained by WENO5-Z scheme using $6000$ uniform control volumes. As shown in the close views from Fig. \ref{fig:TT_closeview1} to Fig. \ref{fig:TT_closeview3}, the WCLS3 scheme captures the fine structures in a resolution comparative with the fifth-order CWENO5 and WENO5-Z schemes. An interesting phenomenon is observed, that is, in regions of $x\in [3,4]$ the resolution of the WCLS3 scheme with $\kappa_0 = 1.2$ is however lower than the WCLS3 with $\kappa_0 = 0.8$ or $\kappa_0 = 1.0$. A reasonable explanation is that, due to the smaller dissipation of the WCLS3 scheme with $\kappa_0 = 1.2$, numerical oscillations tend to be triggered more easily near discontinuities, then the mechanism of adaptive coefficients apply more dissipations to suppress the numerical oscillations to ensure the robustness of the WCLS3 scheme. In smooth regions, e.g., Fig. \ref{fig:TT_closeview2}, with the increase of $\kappa_0$, the resolution becomes higher due to smaller dissipation error and a broader range of resolved wavenumber.

\subsubsection{Double Shear Layer Problem\label{sec:dsl} \cite{daru2000evaluation}}
\begin{figure}[!htbp]
  \centering
    \begin{subfigure}[b]{0.3\textwidth}
    \includegraphics[width=\textwidth]{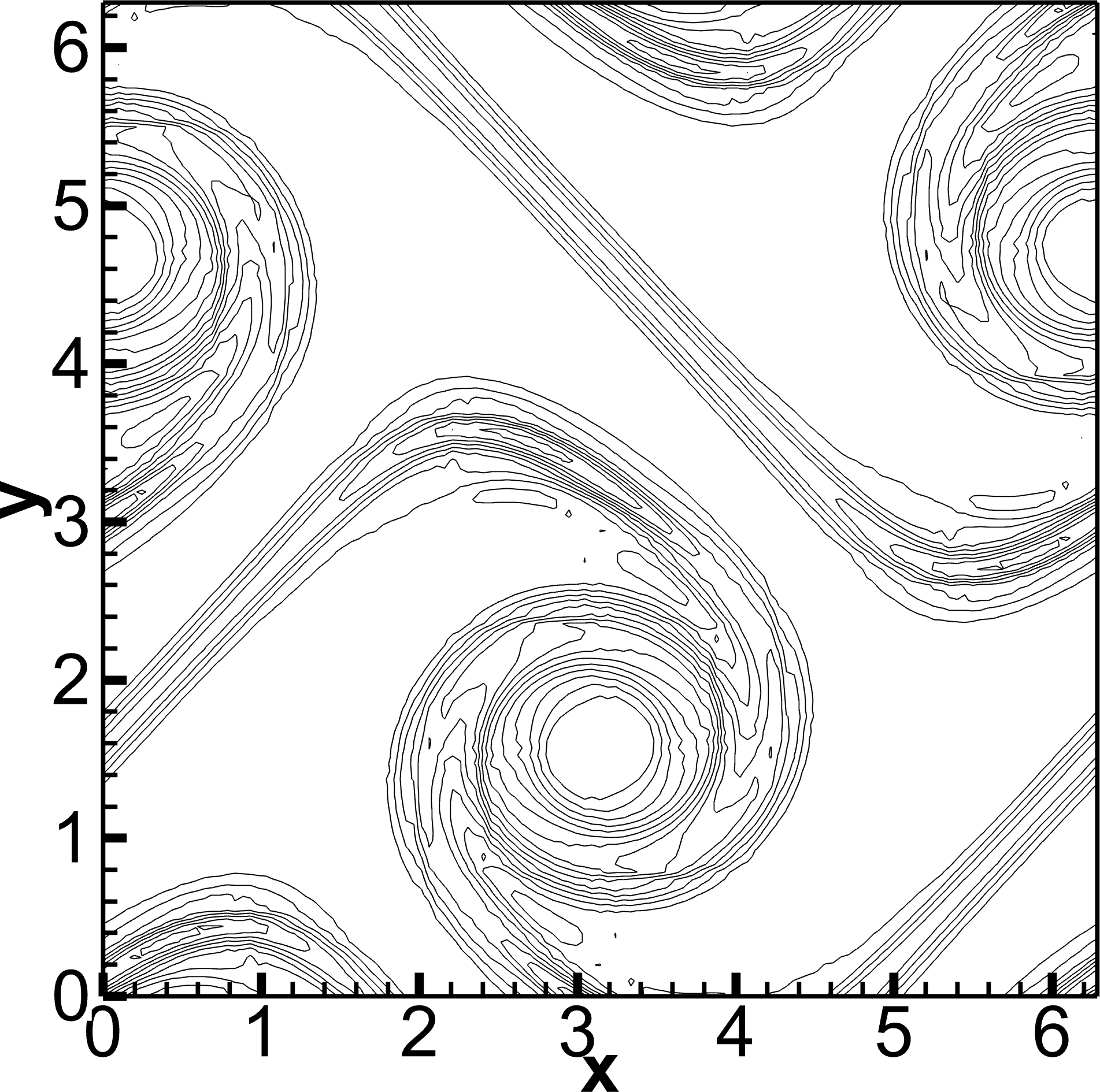}
    \caption{CWENO3.\label{fig:dsl_cweno_3rd}}
    \end{subfigure}
    \quad
    \begin{subfigure}[b]{0.3\textwidth}
    \includegraphics[width=\textwidth]{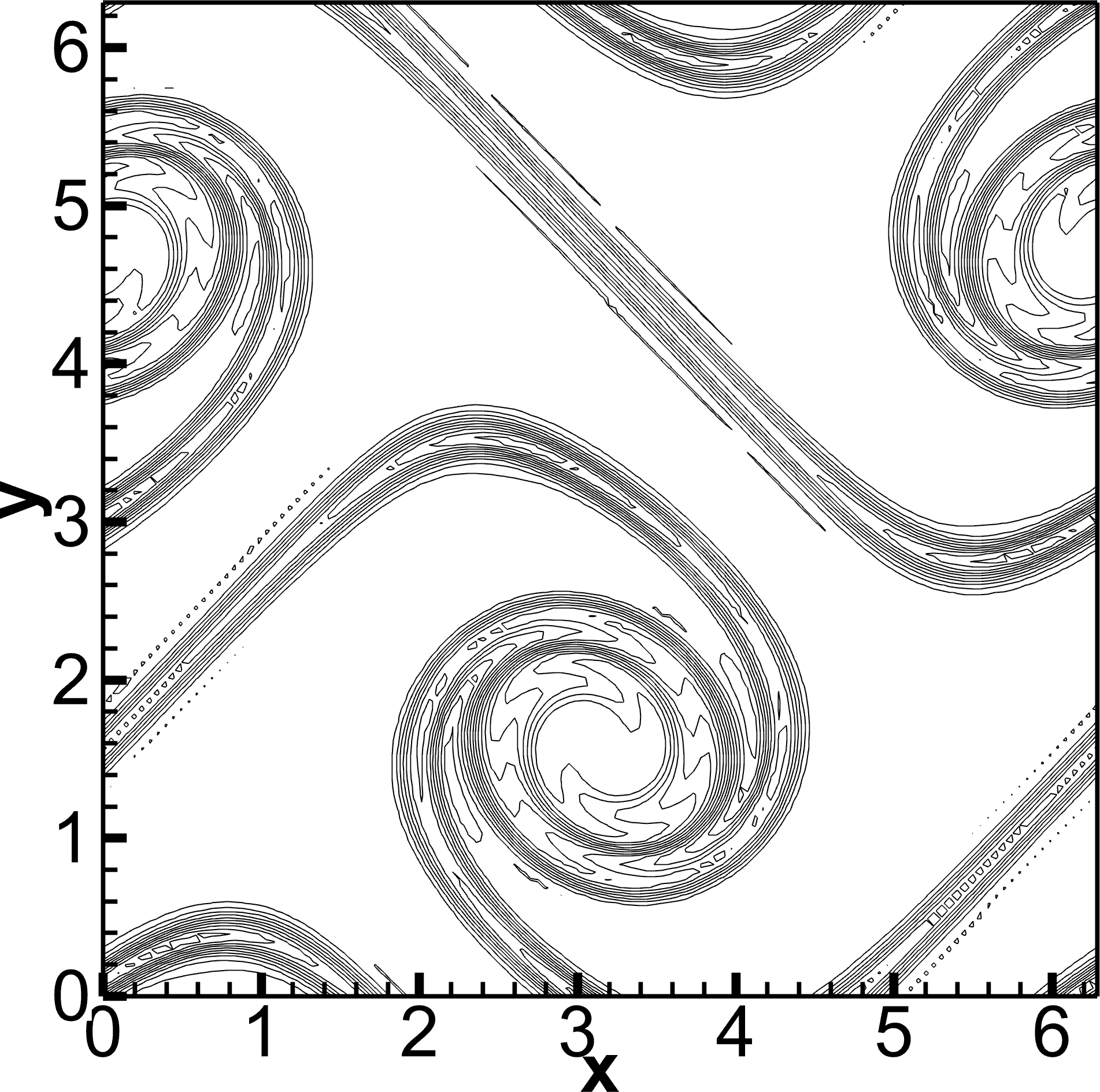}
    \caption{CWENO5.\label{fig:dsl_cweno_5th}}
    \end{subfigure}\\
    \begin{subfigure}[b]{0.3\textwidth}
    \includegraphics[width=\textwidth]{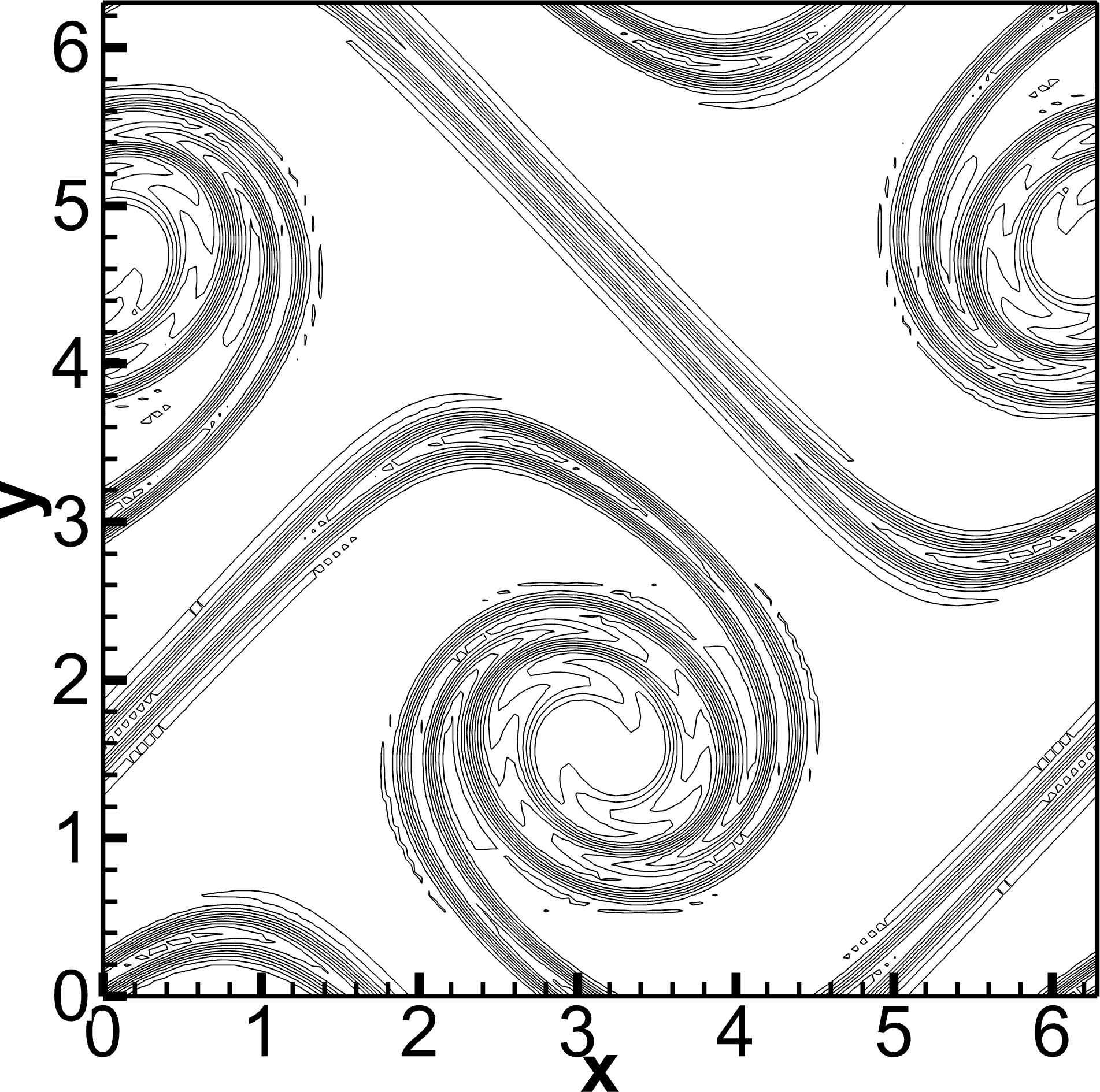}
    \caption{WCLS3 with $\kappa_0 = 0.8$.\label{fig:dsl_k0.8}}
    \end{subfigure}
    \begin{subfigure}[b]{0.3\textwidth}
    \includegraphics[width=\textwidth]{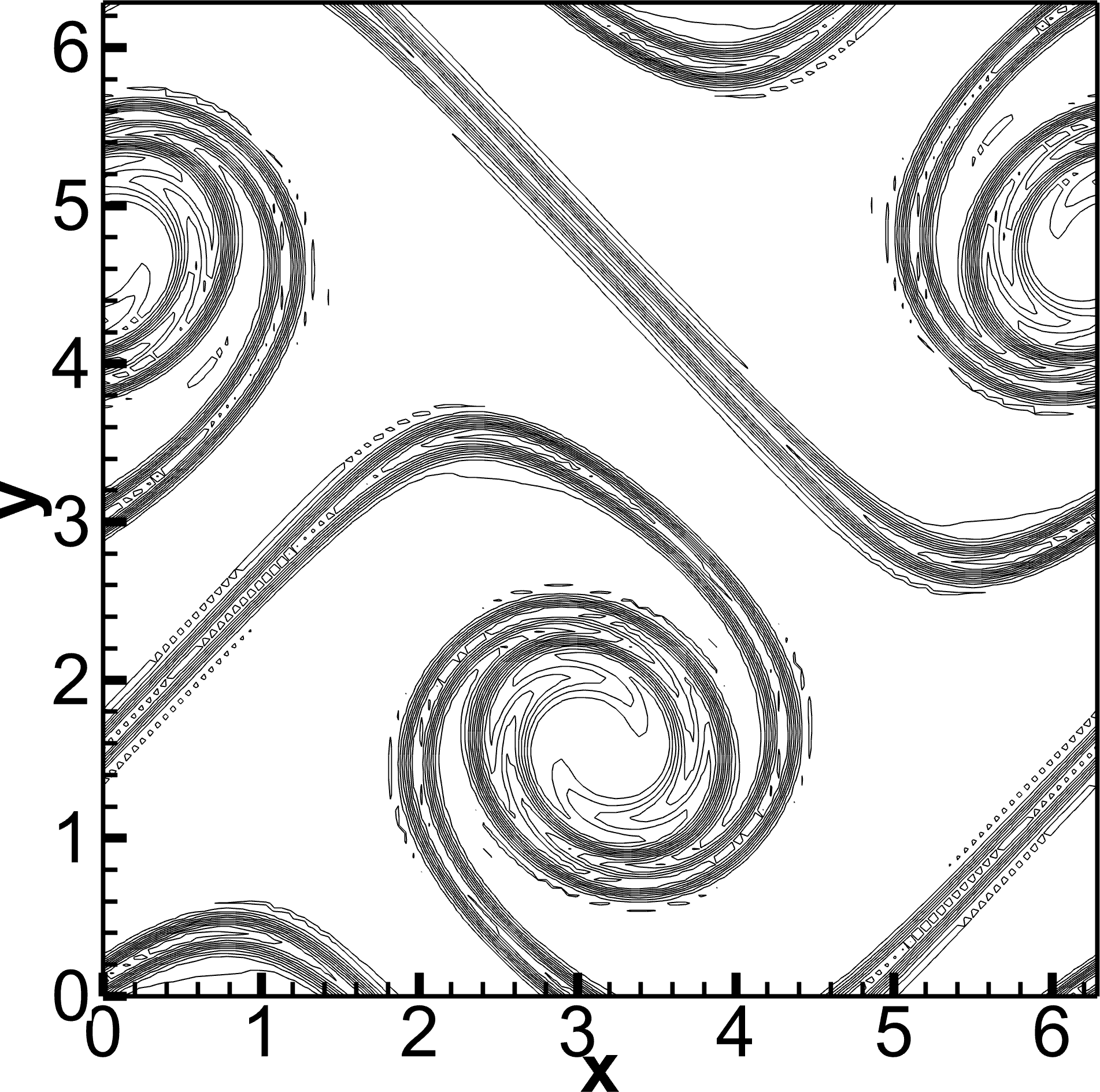}
    \caption{WCLS3 with $\kappa_0 = 1.0$.\label{fig:dsl_k1.0}}
    \end{subfigure}
    \begin{subfigure}[b]{0.3\textwidth}
    \includegraphics[width=\textwidth]{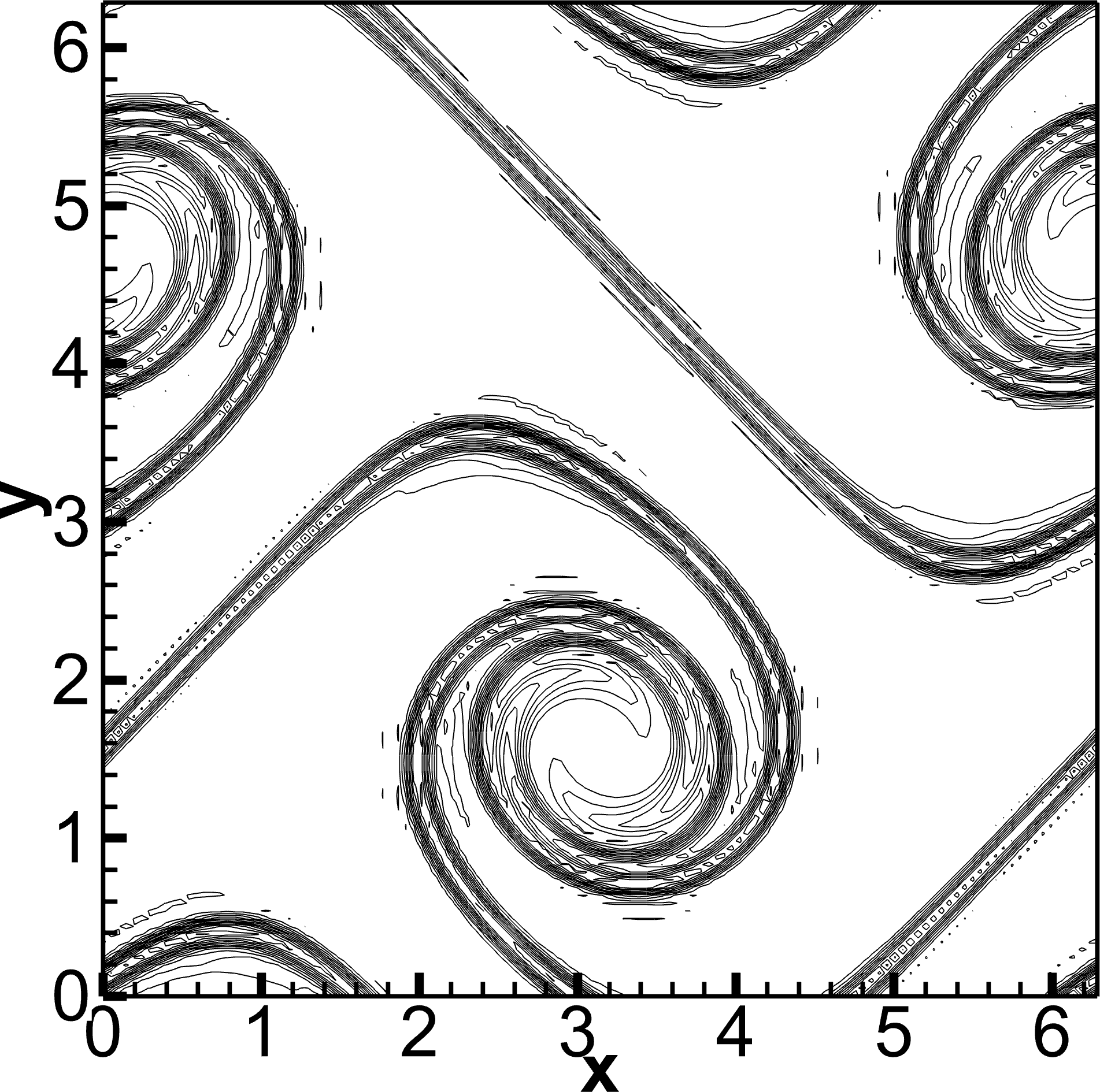}
    \caption{WCLS3 with $\kappa_0 = 1.2$.\label{fig:dsl_k1.2}}
    \end{subfigure}
    \caption{Z-vorticity contours for the double shear layer problem with 30 lines ranging from -4.3 to 4.3. $N_x \times N_y = 128 \times 128$, $t = 8.0$ and CFL = $0.8$.
    \label{fig:dsl}}
\end{figure}

In this section, Euler equations in 2D periodic computational domain $\Omega = [0, 2\pi]\times [0,2\pi]$ are solved to validate the performance of the WCLS3 scheme.
The initial condition is given as
\begin{align}
  u(x,y) & = \left\{
  \begin{array}{ll}
    \tanh\left(\frac{15(y-0.5\pi)}{\pi}\right), & y\leq \pi, \\
    \tanh\left(\frac{15(1.5\pi-y)}{\pi}\right), & y >  \pi,
  \end{array}
  \right. \\
  v(x,y) & = 0.05 \sin(x),\\
  \rho(x,y) &= 1.0.
\end{align}
The pressure is initially uniform. The Mach number  based on the mean velocity $\overline{\sqrt{u^2+v^2}}$ is 0.1, where $\overline{\left(\cdot \right)}$ is an average operator over the whole computational domain. The computational domain is discretized by $128\times 128$ uniform control volumes. The simulation is advanced by the third-order SSP-RK method with CFL number as 0.8 till $t=8.0$.

Figures \ref{fig:dsl} presents the contours of z-vorticity for the WCLS3 schemes. As a comparison, the results of the CWENO3 and CWENO5 schemes are also presented in Figs. \ref{fig:dsl_cweno_3rd} and \ref{fig:dsl_cweno_5th}, respectively. Thirty contour lines ranging from -4.3 to 4.3 are plotted. The WCLS3 scheme resolves finer vortex structures than the CWENO3 scheme. The resolution of the proposed WCLS3 scheme is comparative with or even higher than the CWENO5 scheme with the increase of $\kappa_0$.

\subsubsection{Shock Vortex Interaction \cite{jiang1996efficient}}
A vortex interacts with a stationary shock of Mach number 1.1 in this problem. The shock is positioned at $x = 0.5$ and is perpendicular to the $x$-axis. The gas state at the upstream of the shock is $\left(\rho,u,v,p\right) = \left(1, 1.1\sqrt{\gamma}, 0, 1\right)$. An isentropic perturbation initially centered at $(x_c, y_c)$ is added to the mean flow. The perturbation is in the form as
\begin{equation}
  \left(\delta u, \delta v\right) = \frac{\varepsilon}{r_c} e^{\vartheta (1-\tau^2)}\left(-(y-y_c), x-x_c\right), \,\,\,\, \delta T = -\frac{(\gamma-1)\varepsilon^2}{4\vartheta \gamma}e^{2\vartheta (1-\tau^2)},\,\,\,\,\delta S = 0,
\end{equation}
where $r = \sqrt{(x-x_c)^2+(y-y_c)^2}$, $\tau = r/r_c$, $x_c = 0.25$, $y_c = 0.5$, $r_c = 0.05$, $\varepsilon = 0.3$ and $\vartheta = 0.204$.

\begin{figure}[!htbp]
  \centering
    \begin{subfigure}[b]{0.45\textwidth}
    \includegraphics[width=\textwidth]{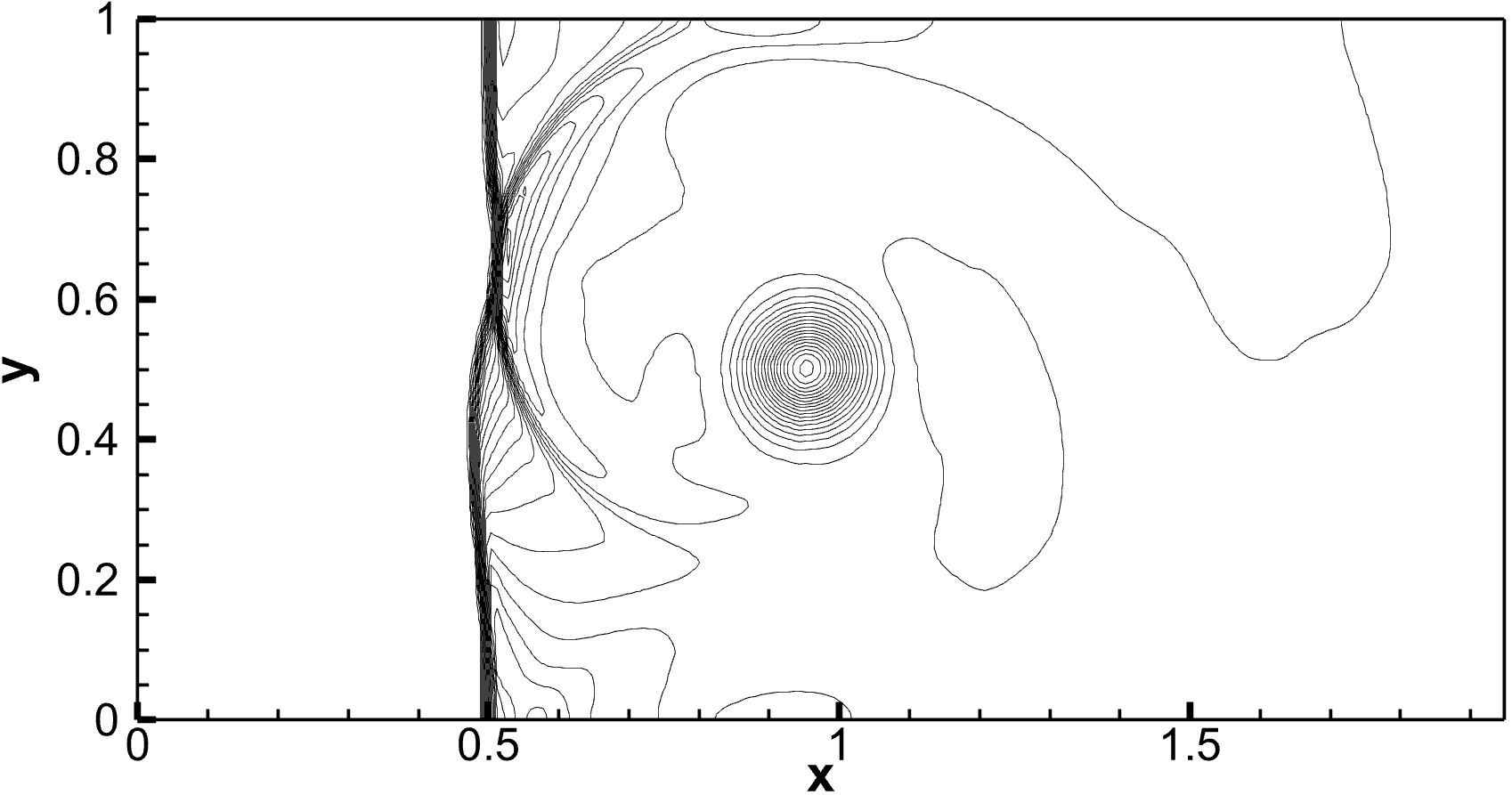}
    \caption{CWENO3.\label{fig:svi_cweno3_125}}
    \end{subfigure}
    \quad
    \begin{subfigure}[b]{0.45\textwidth}
    \includegraphics[width=\textwidth]{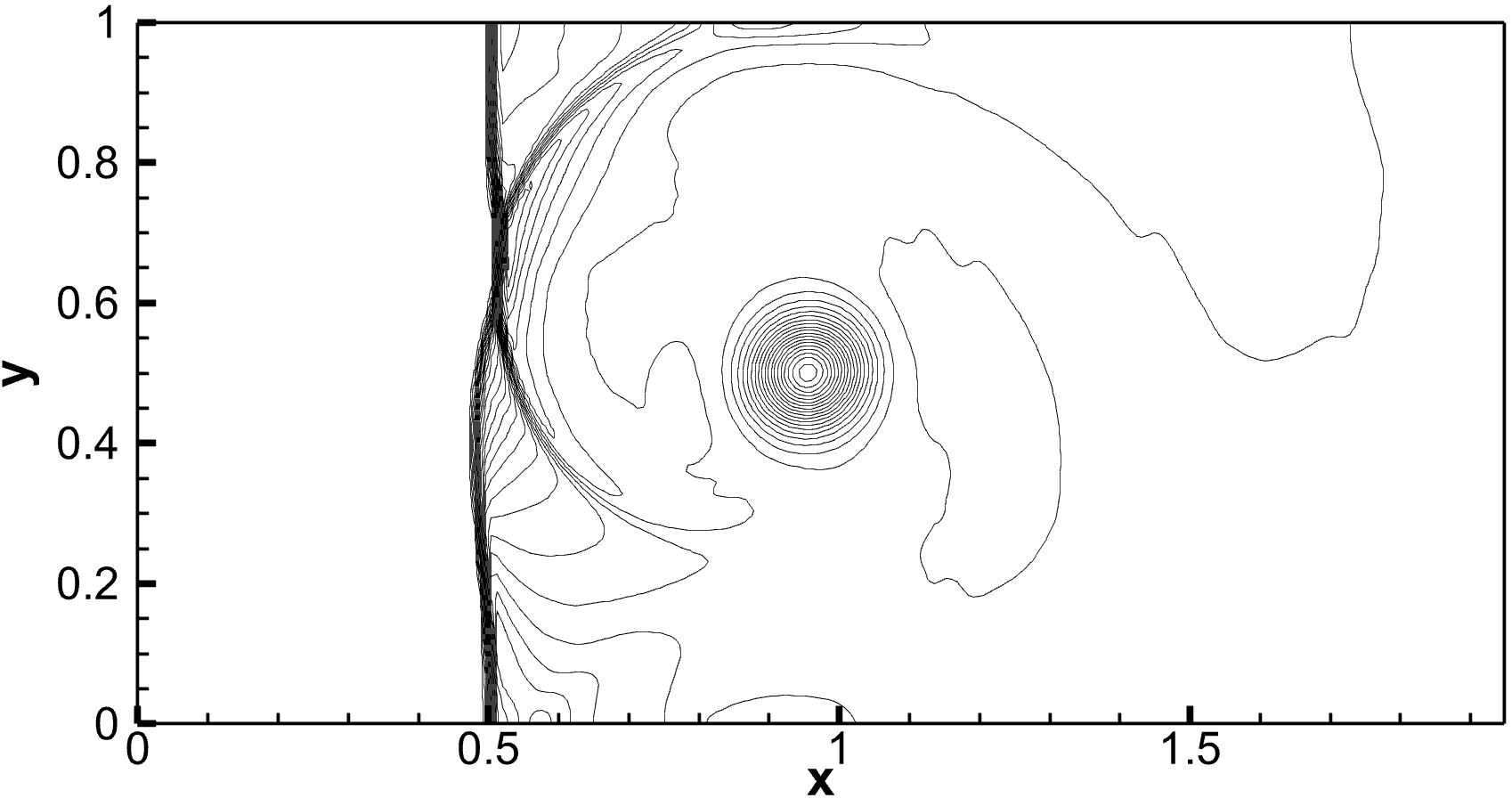}
    \caption{CWENO5.\label{fig:dsl_cweno5_125}}
    \end{subfigure}\\
    \begin{subfigure}[b]{0.45\textwidth}
    \includegraphics[width=\textwidth]{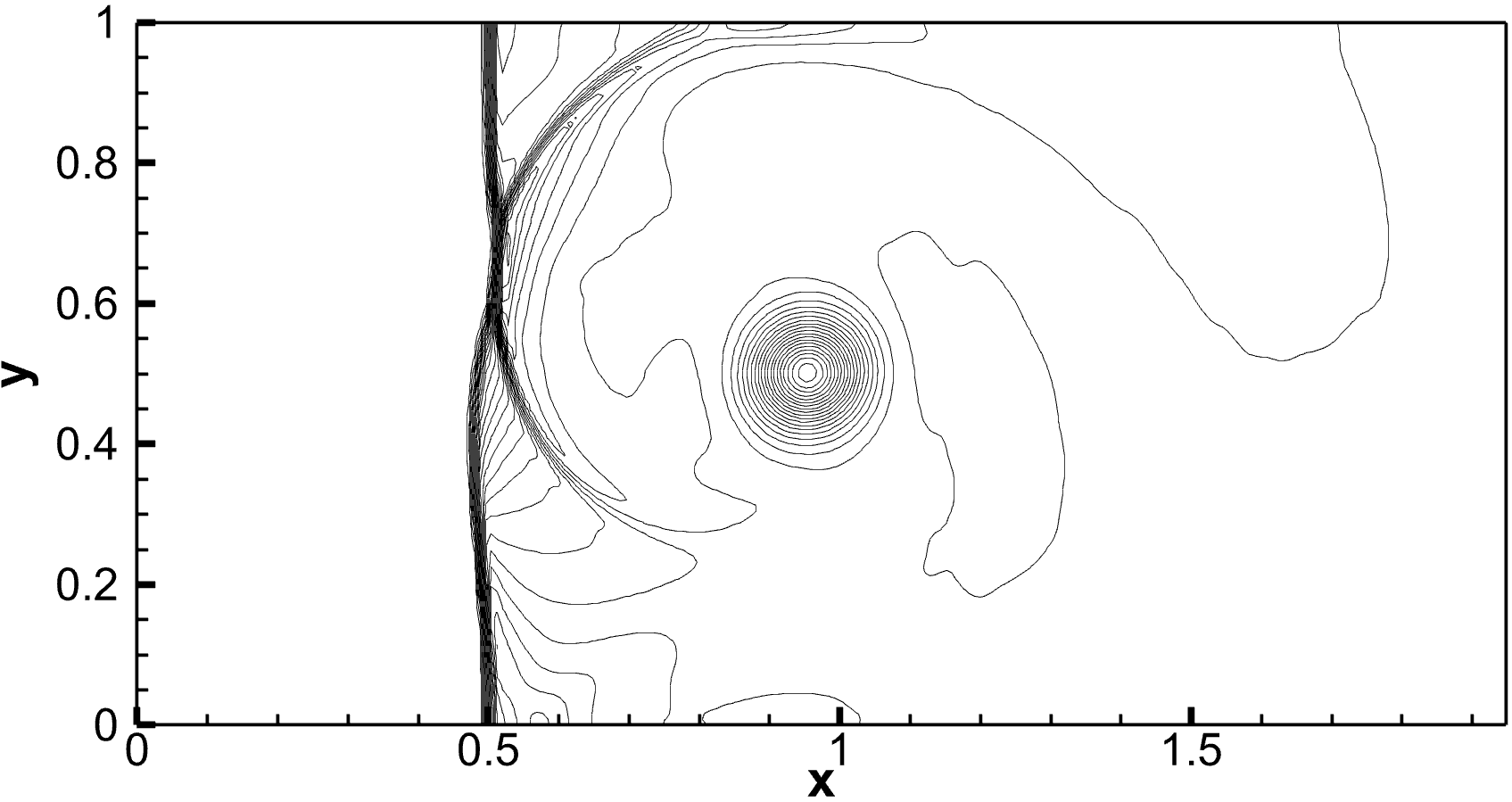}
    \caption{WCLS3 with $\kappa_0 = 0.8$.\label{fig:svi_k0.8_125}}
    \end{subfigure}
    \begin{subfigure}[b]{0.45\textwidth}
    \includegraphics[width=\textwidth]{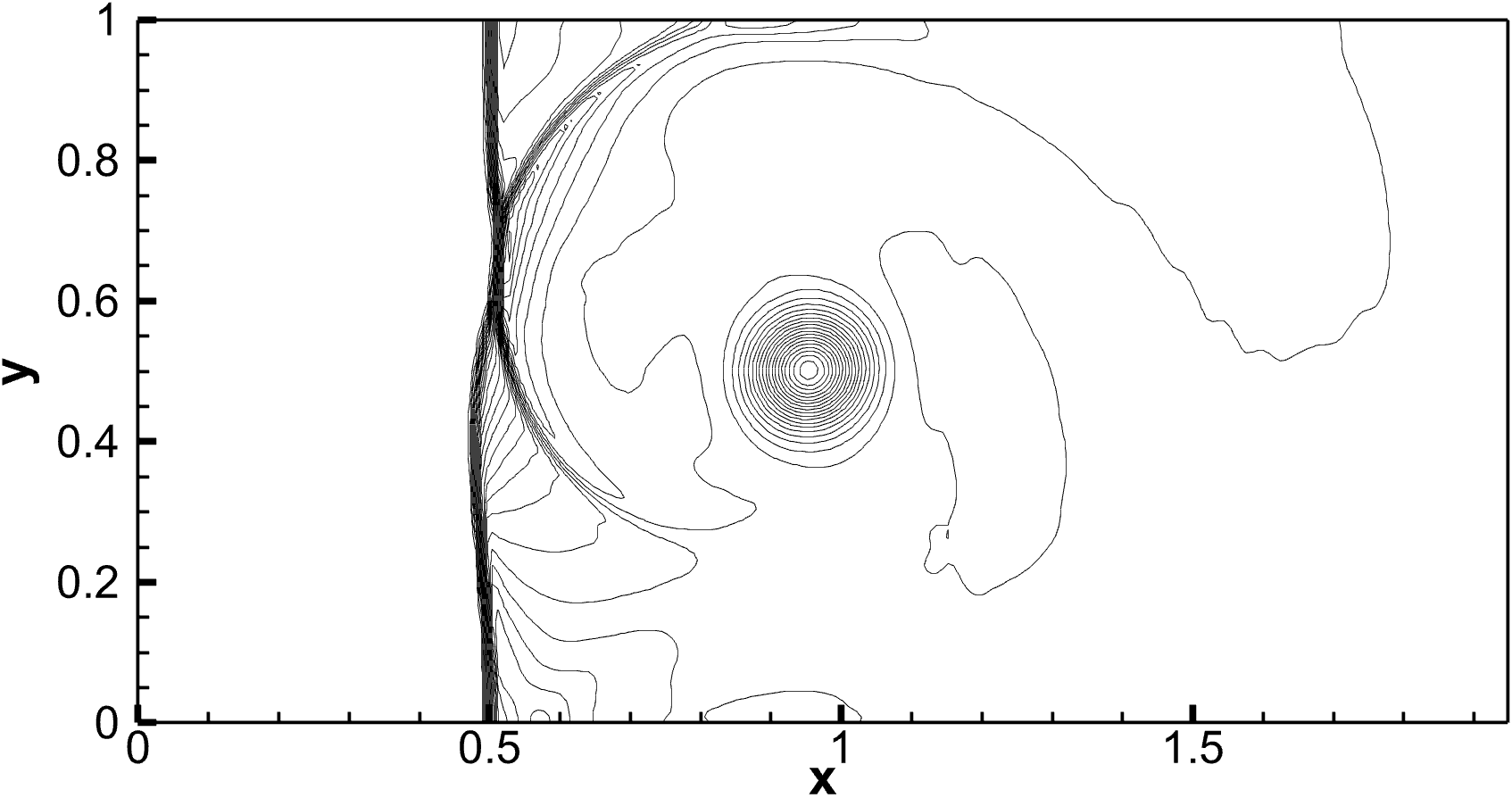}
    \caption{WCLS3 with $\kappa_0 = 1.0$.\label{fig:svi_k1.0_125}}
    \end{subfigure}\\
    \begin{subfigure}[b]{0.45\textwidth}
    \includegraphics[width=\textwidth]{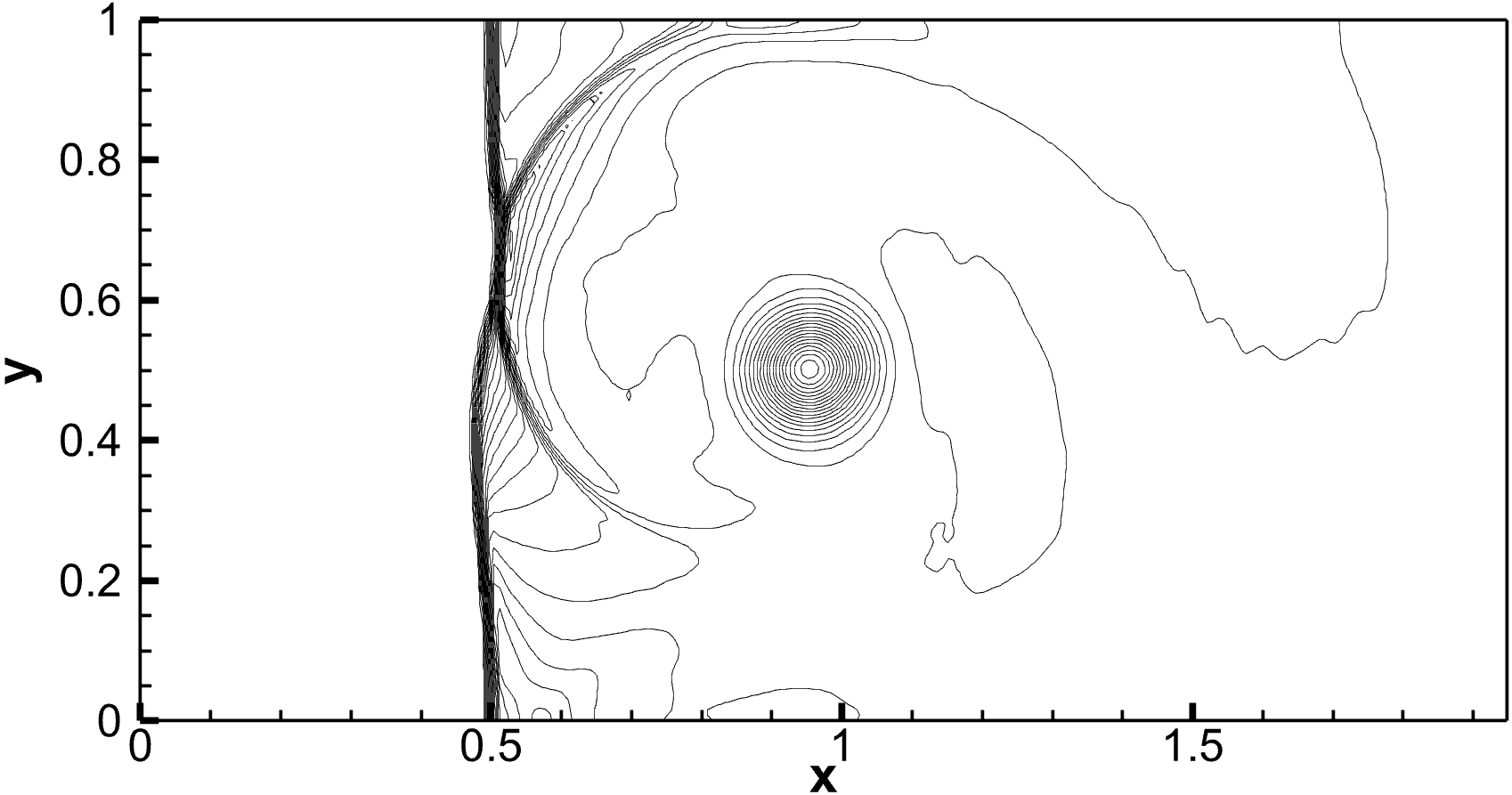}
    \caption{WCLS3 with $\kappa_0 = 1.2$.\label{fig:svi_k1.2_125}}
    \end{subfigure}
    \caption{Density contours for the shock vortex interaction with 30 lines ranging from 1.01 to 1.22. $N_x \times N_y = 250 \times 125$, $t = 0.6$ and CFL = $0.6$.
    \label{fig:svi_125}}
\end{figure}

\begin{figure}[!htbp]
  \centering
    \begin{subfigure}[b]{0.45\textwidth}
    \includegraphics[width=\textwidth]{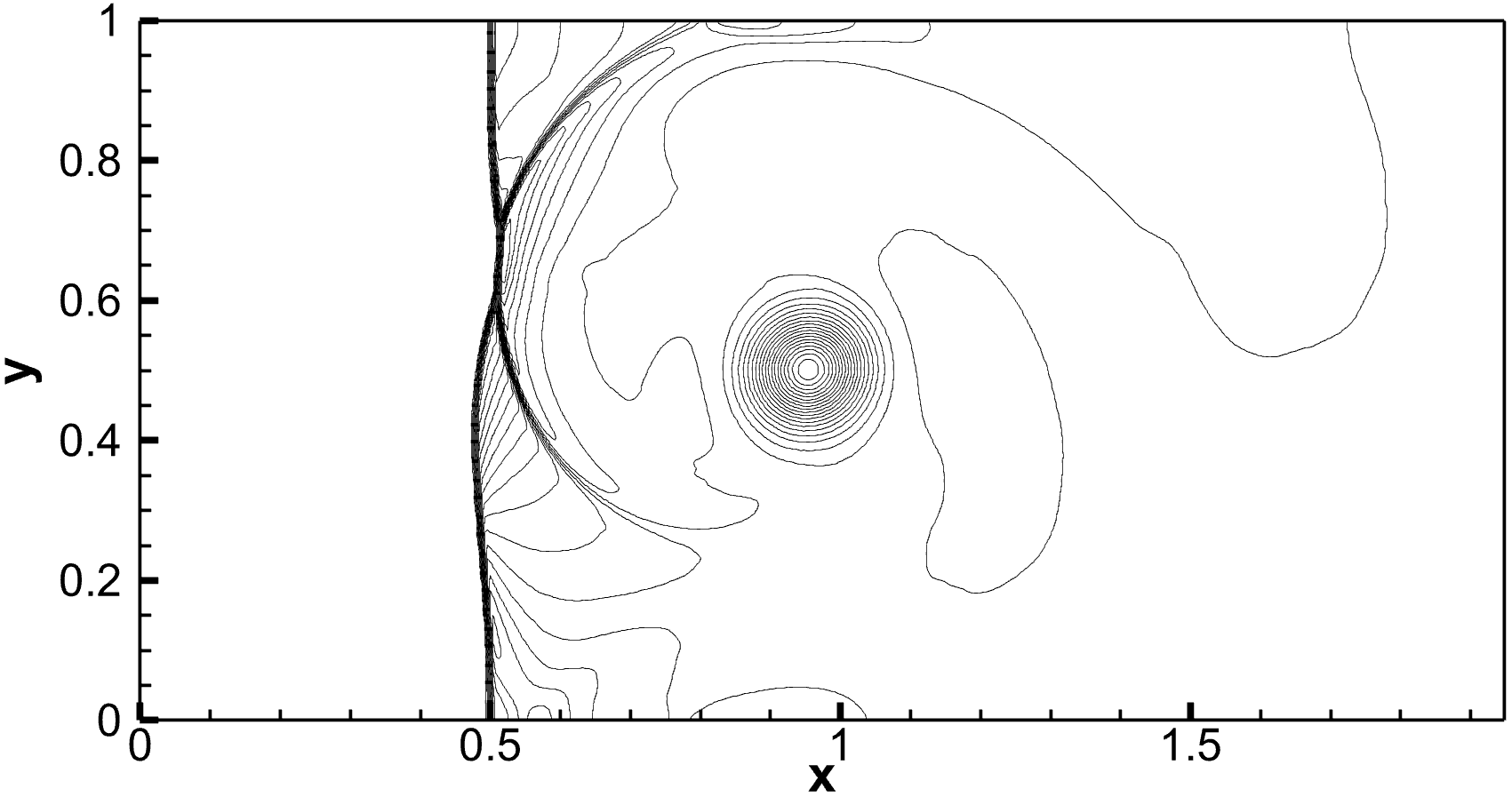}
    \caption{CWENO3.\label{fig:svi_cweno3_250}}
    \end{subfigure}
    \quad
    \begin{subfigure}[b]{0.45\textwidth}
    \includegraphics[width=\textwidth]{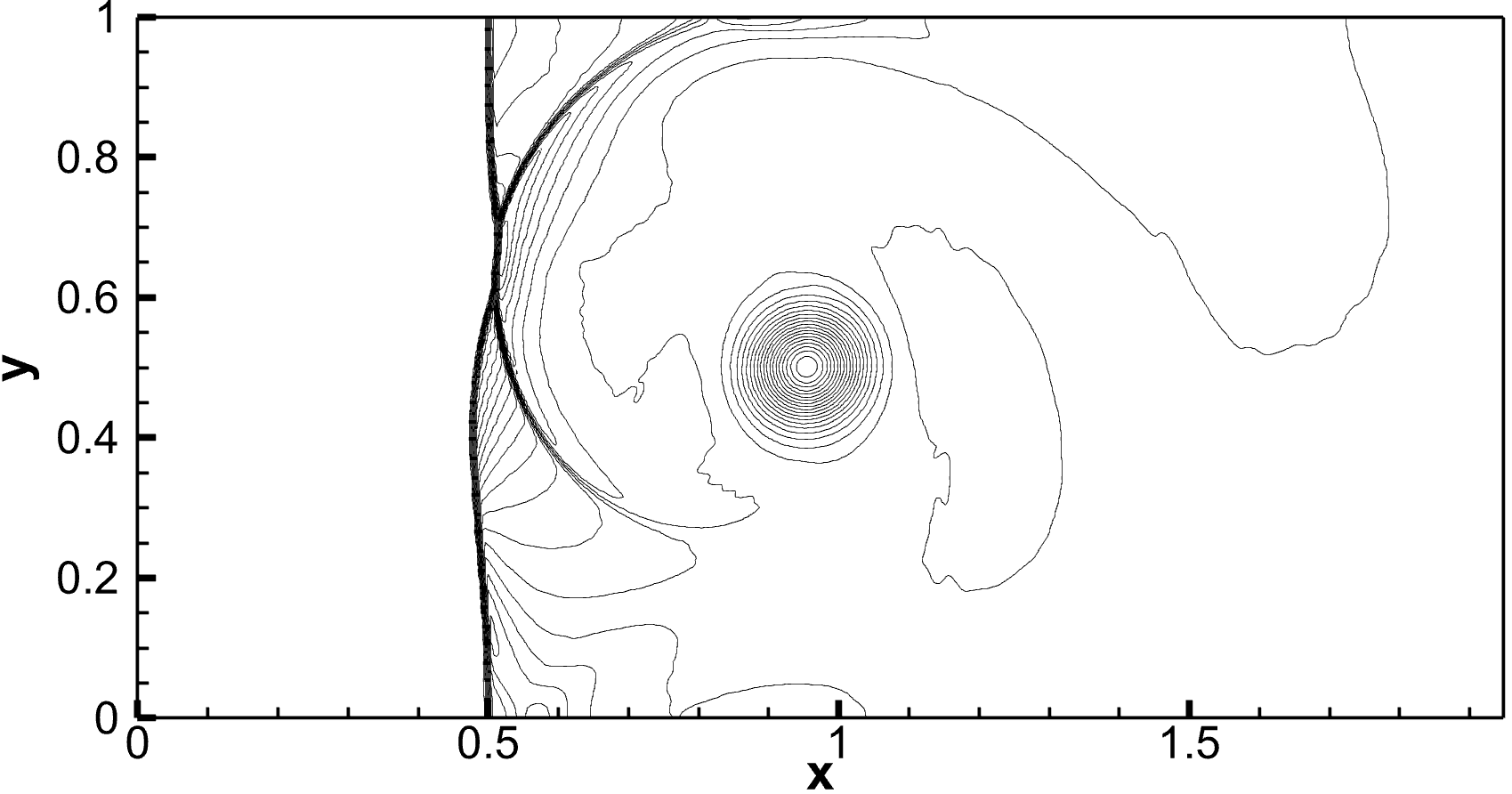}
    \caption{CWENO5.\label{fig:dsl_cweno5_250}}
    \end{subfigure}\\
    \begin{subfigure}[b]{0.45\textwidth}
    \includegraphics[width=\textwidth]{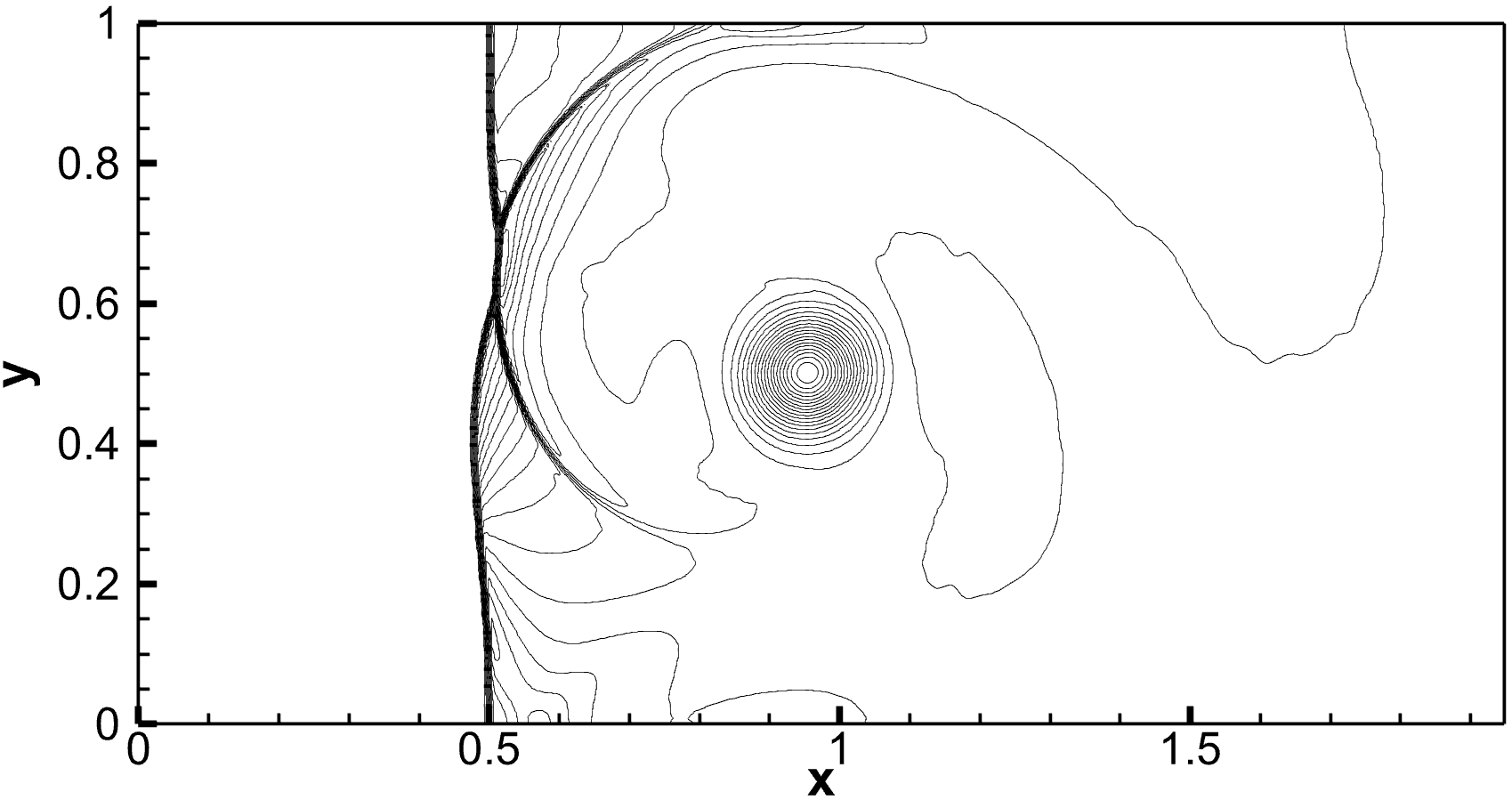}
    \caption{WCLS3 with $\kappa_0 = 0.8$.\label{fig:svi_k0.8_250}}
    \end{subfigure}
    \begin{subfigure}[b]{0.45\textwidth}
    \includegraphics[width=\textwidth]{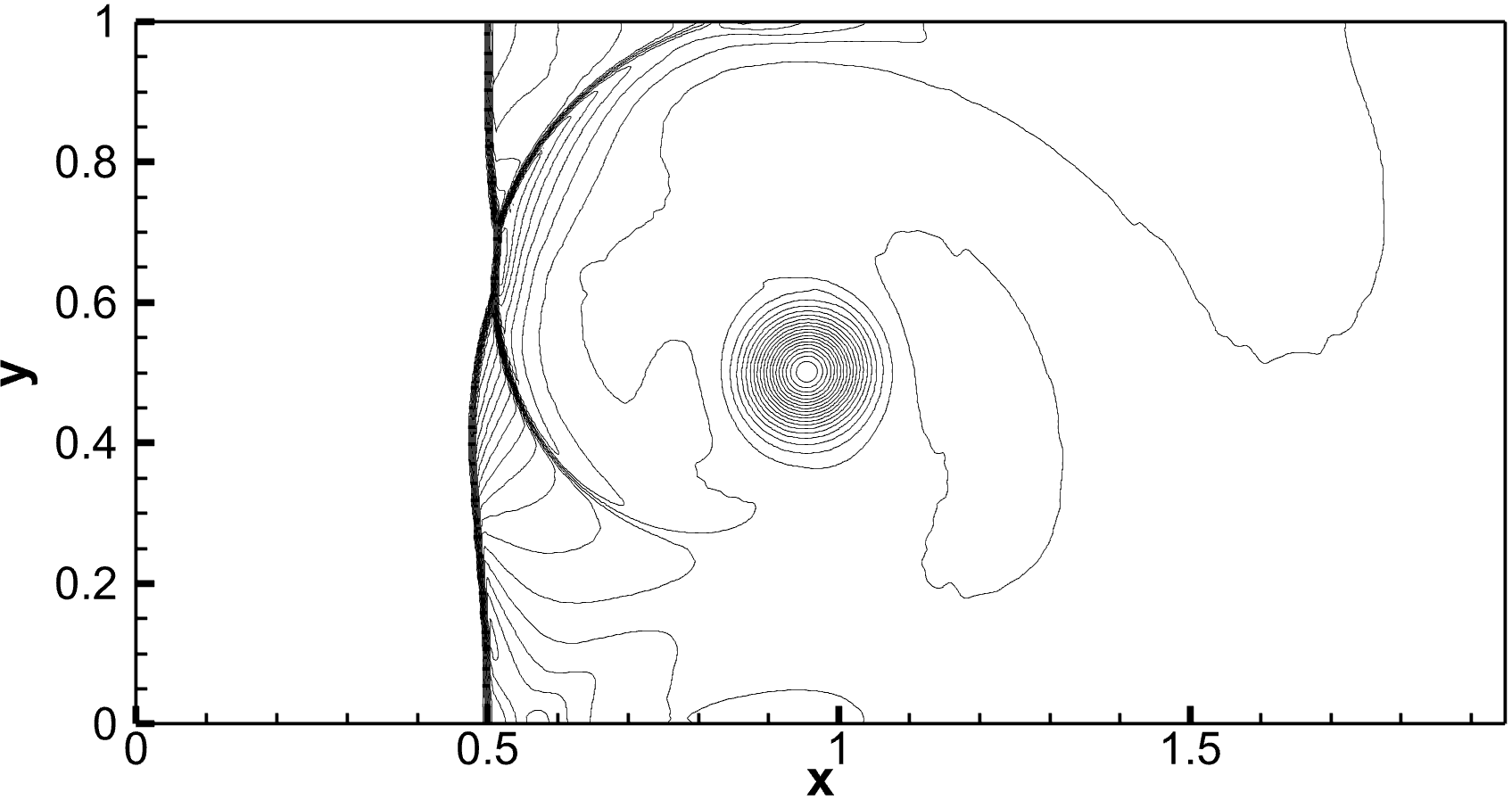}
    \caption{WCLS3 with $\kappa_0 = 1.0$.\label{fig:svi_k1.0_250}}
    \end{subfigure}\\
    \begin{subfigure}[b]{0.45\textwidth}
    \includegraphics[width=\textwidth]{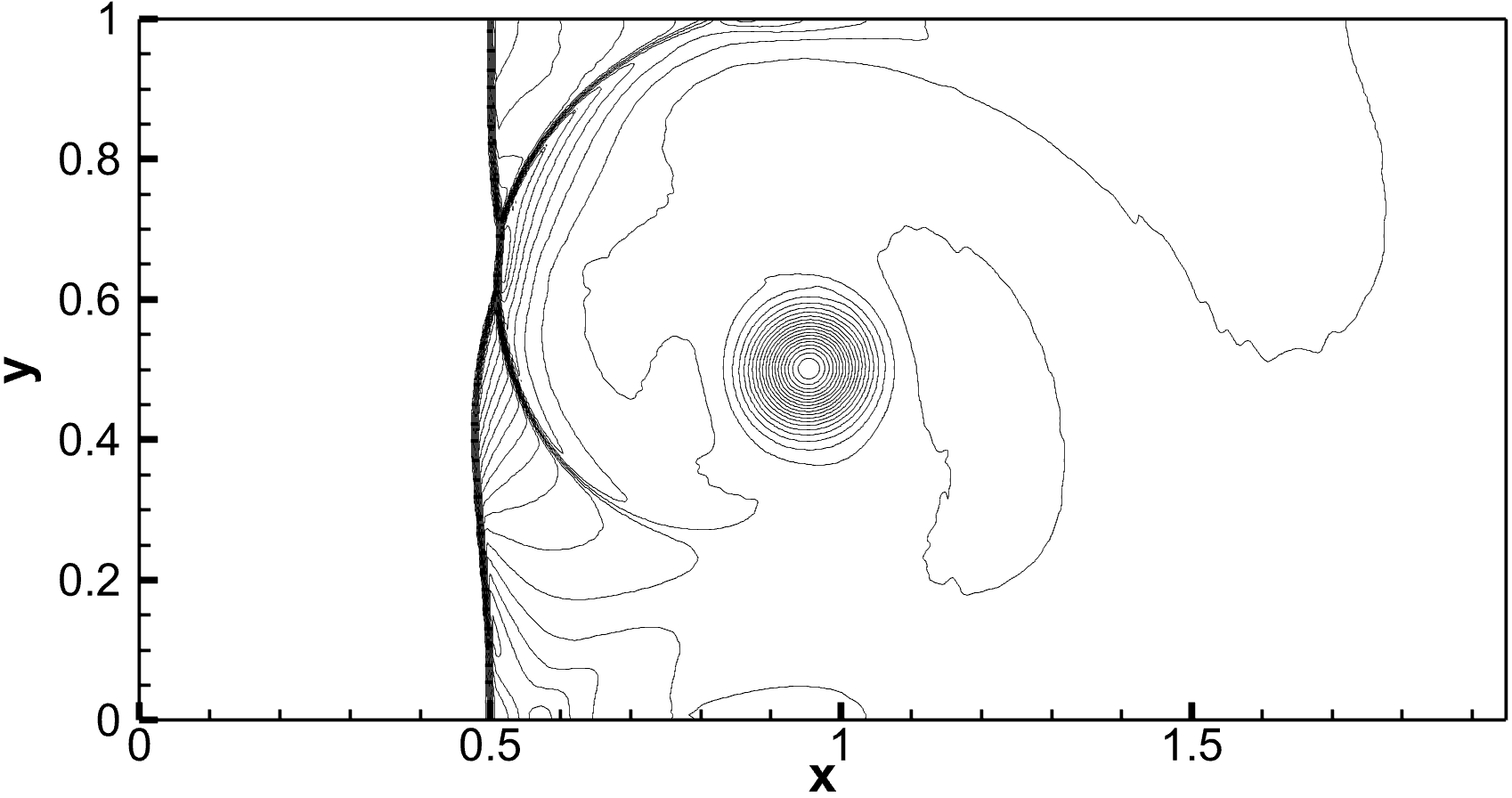}
    \caption{WCLS3 with $\kappa_0 = 1.2$.\label{fig:svi_k1.2_250}}
    \end{subfigure}
    \caption{Density contours for the shock vortex interaction with 30 lines ranging from 1.01 to 1.22. $N_x \times N_y = 500 \times 250$, $t = 0.6$ and CFL = $0.6$.
    \label{fig:svi_250}}
\end{figure}

The computational domain is $\Omega = [0,2]\times[0,1]$. The left and right boundaries are applied with inflow and outflow boundary conditions, respectively. Inviscid walls are positioned at the top and bottom sides. The simulation is advanced till $t = 0.6$ by the third-order SSP RK method with CFL = $0.6$. 
Figures \ref{fig:svi_125} and \ref{fig:svi_250} present the density contours at $t = 0.6$ with grid number being $250 \times 125$ and $500 \times 250$, respectively. As demonstrated, the WCLS3 scheme captures the discontinuities robustly and oscillation-freely.

\subsubsection{Double Mach Reflection \cite{weinan1994numerical} \label{sec:dmr}}

A shock of Mach number 10 moves right-downward with an inclined angle of $60$ degrees with $x$-axis. When the shock hits the inviscid wall which is located along $x \geq 1/6$, a bow shock is formed with Mach stem on the right side. 
The initial condition is
\begin{equation}
  \left(\rho, u, v, p\right) = \left\{
  \begin{array}{ll}
    8.0, 7.1447, -4.125, 116.5, & x \leq 1/6 + \sqrt{3}y/3,\\
    1.4, 0, 0, 1, & x > 1/6 + \sqrt{3}y/3,
  \end{array}
  \right.
\end{equation}

The simulation domain is $[0,4]\times [0,1]$. The simulation end time is $t= 0.2$. The time discretization method is the third-order SSP-RK method with CFL number being $0.6$.
\begin{figure}[!htbp]
  \centering
    \begin{subfigure}[b]{0.45\textwidth}
    \includegraphics[width=\textwidth]{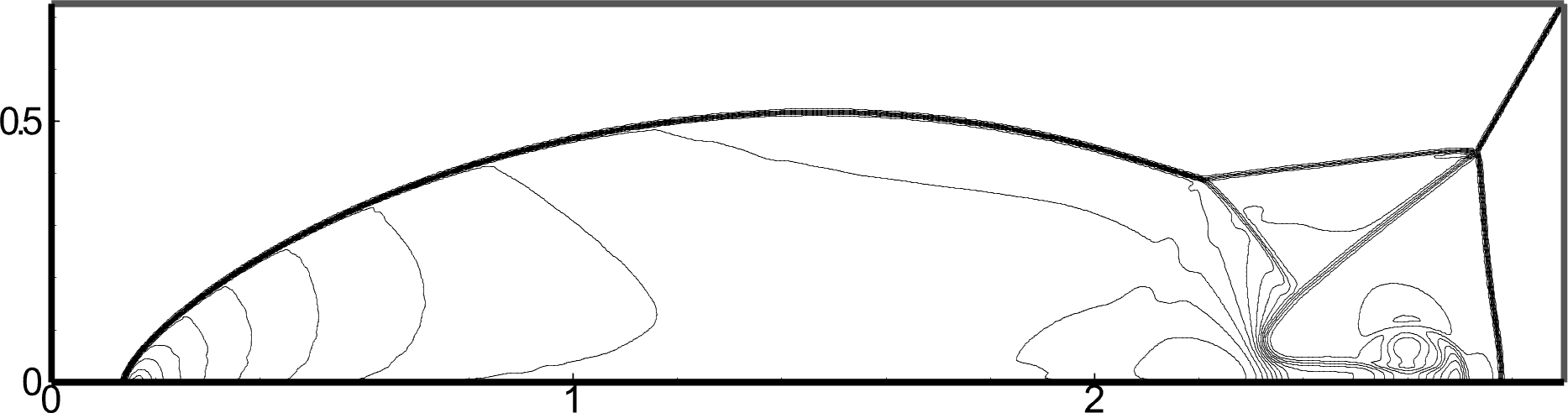}
    \caption{CWENO3.\label{fig:dmr_cweno3_240}}
    \end{subfigure}
    \quad
    \begin{subfigure}[b]{0.45\textwidth}
    \includegraphics[width=\textwidth]{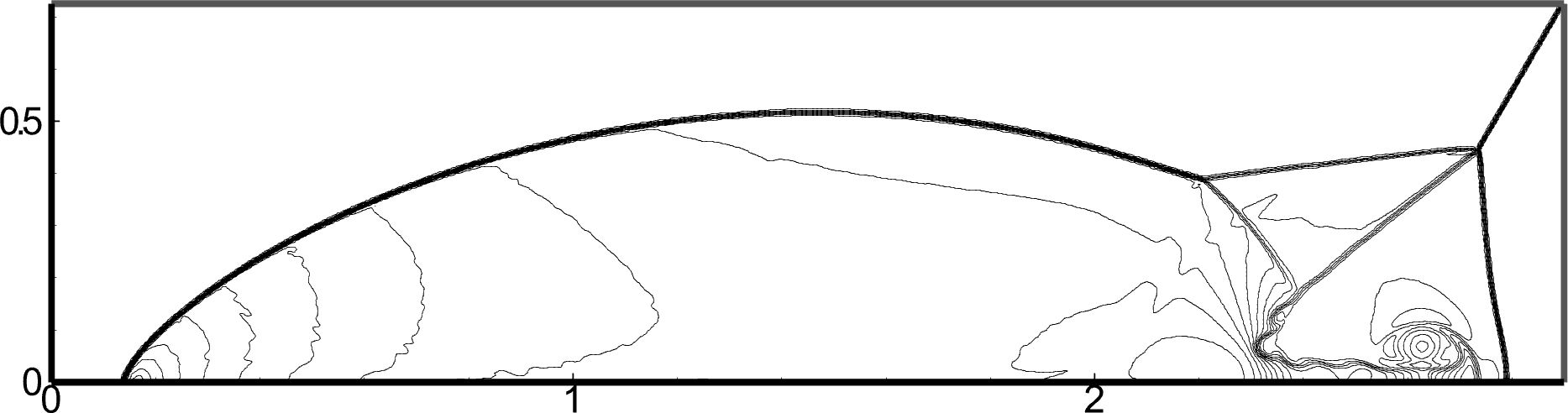}
    \caption{CWENO5.\label{fig:dmr_cweno5_240}}
    \end{subfigure}\\
    \begin{subfigure}[b]{0.45\textwidth}
    \includegraphics[width=\textwidth]{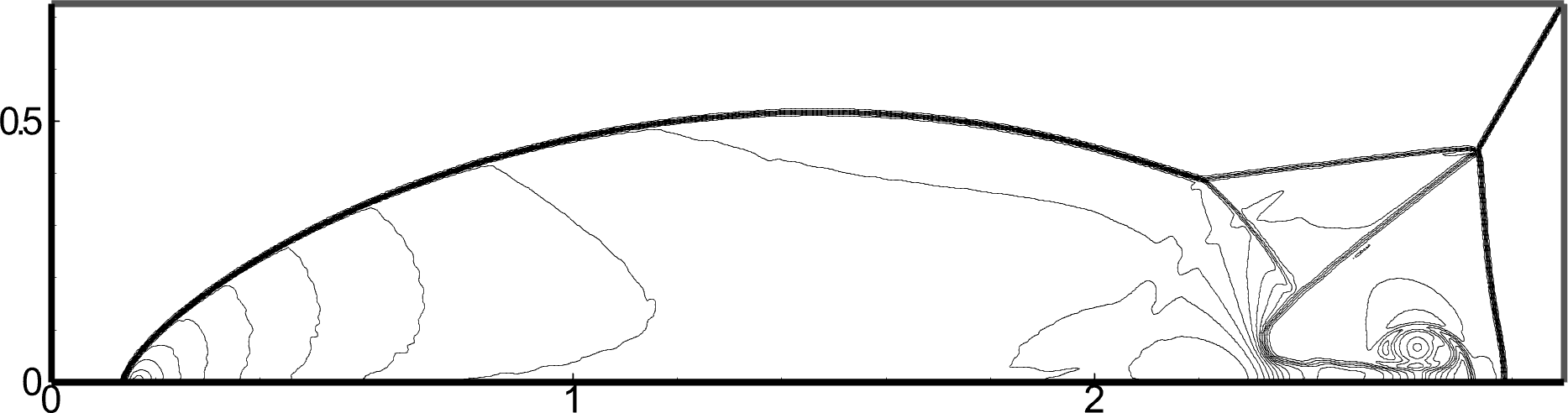}
    \caption{WCLS3 with $\kappa_0 = 0.8$.\label{fig:dmr_k0.8_240}}
    \end{subfigure}
    \begin{subfigure}[b]{0.45\textwidth}
    \includegraphics[width=\textwidth]{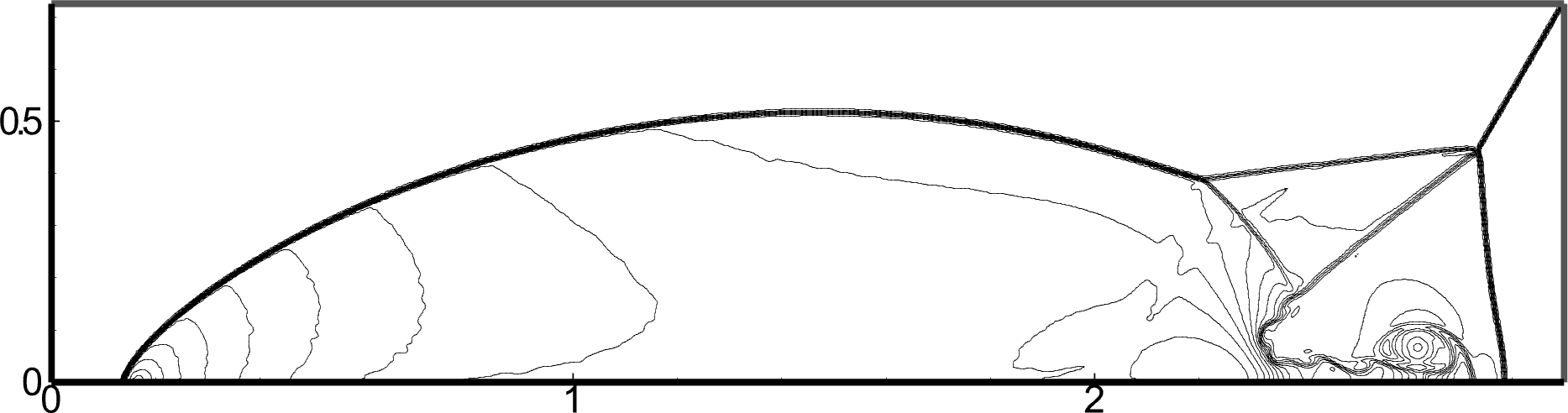}
    \caption{WCLS3 with $\kappa_0 = 1.0$.\label{fig:dmr_k1.0_240}}
    \end{subfigure}\\
    \begin{subfigure}[b]{0.45\textwidth}
    \includegraphics[width=\textwidth]{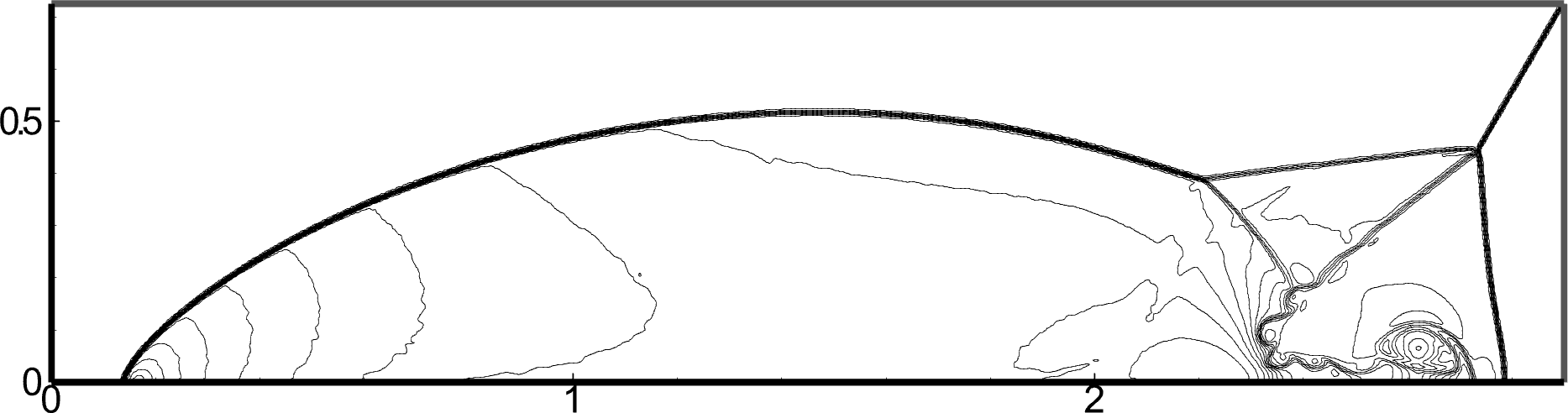}
    \caption{WCLS3 with $\kappa_0 = 1.2$.\label{fig:dmr_k1.2_240}}
    \end{subfigure}
    \caption{Density contours for the double Mach reflection with 30 lines ranging from 2.1 to 22. $N_x \times N_y = 960 \times 240$, $t = 0.2$ and CFL = $0.6$.
    \label{fig:dmr_240}}
\end{figure}

\begin{figure}[!htbp]
  \centering
    \begin{subfigure}[b]{0.3\textwidth}
    \includegraphics[width=\textwidth]{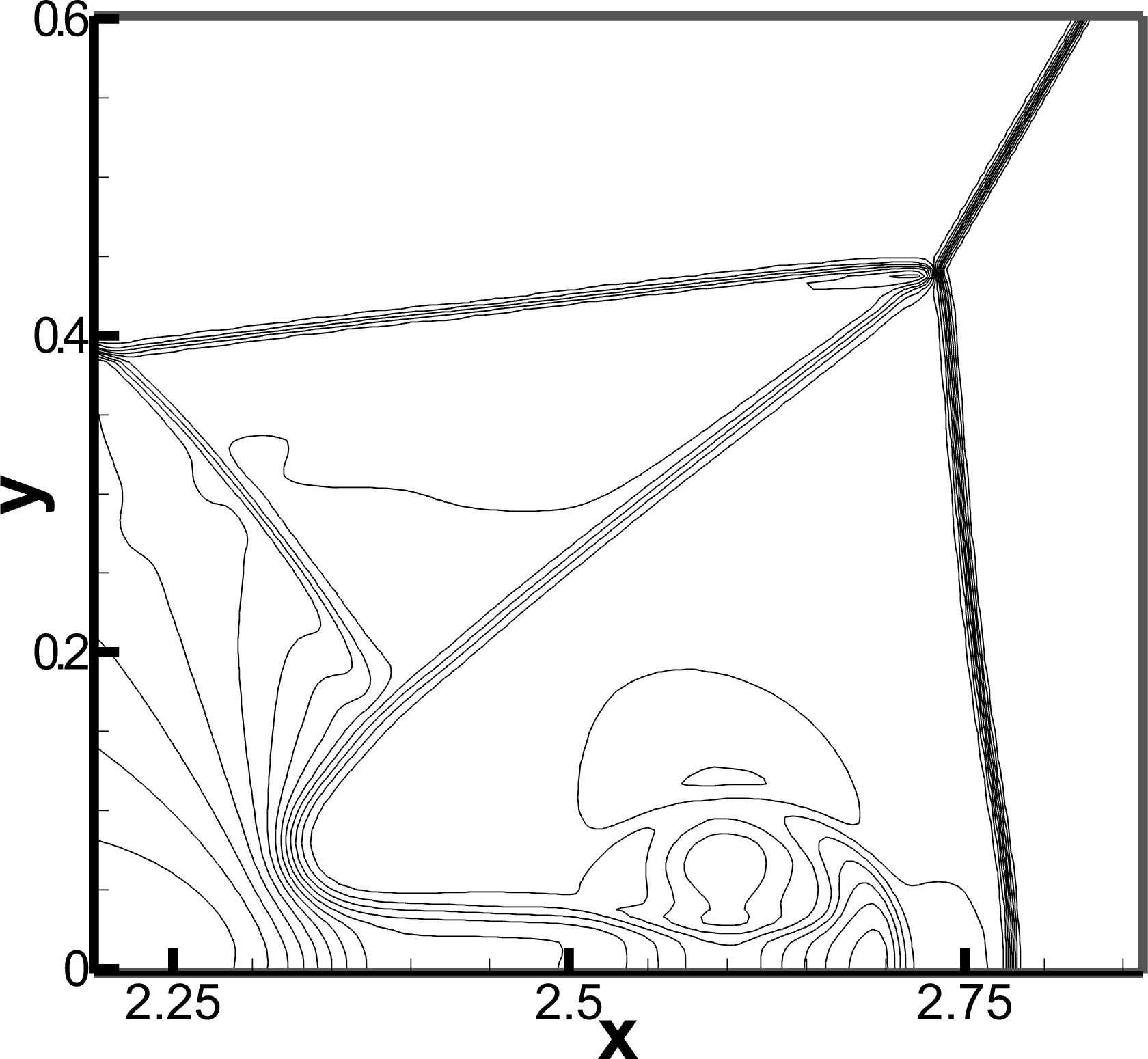}
    \caption{CWENO3.\label{fig:dmr_cweno3_240_cv}}
    \end{subfigure}
    \quad
    \begin{subfigure}[b]{0.3\textwidth}
    \includegraphics[width=\textwidth]{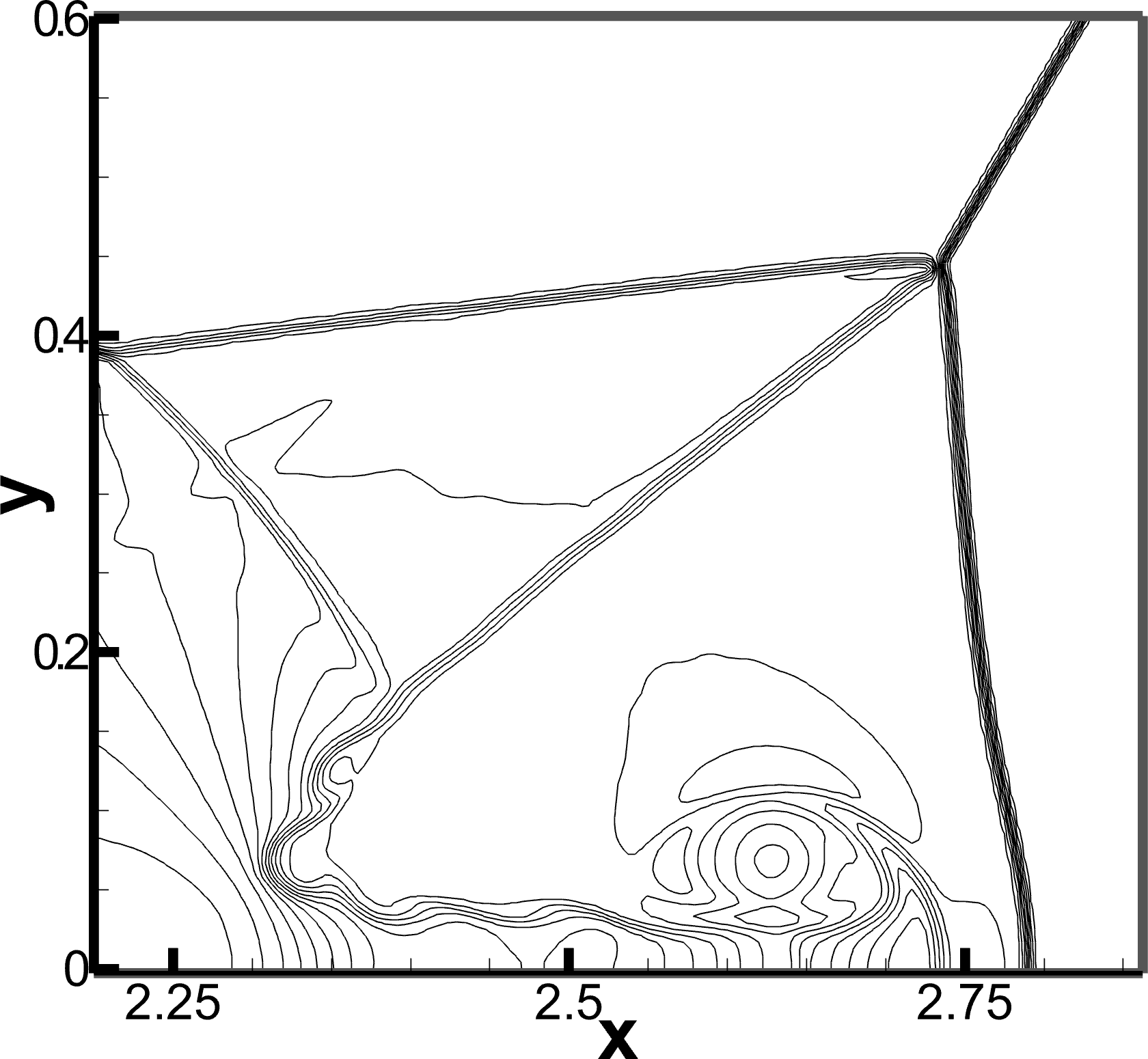}
    \caption{CWENO5.\label{fig:dmr_cweno5_240_cv}}
    \end{subfigure}\\
    \begin{subfigure}[b]{0.3\textwidth}
    \includegraphics[width=\textwidth]{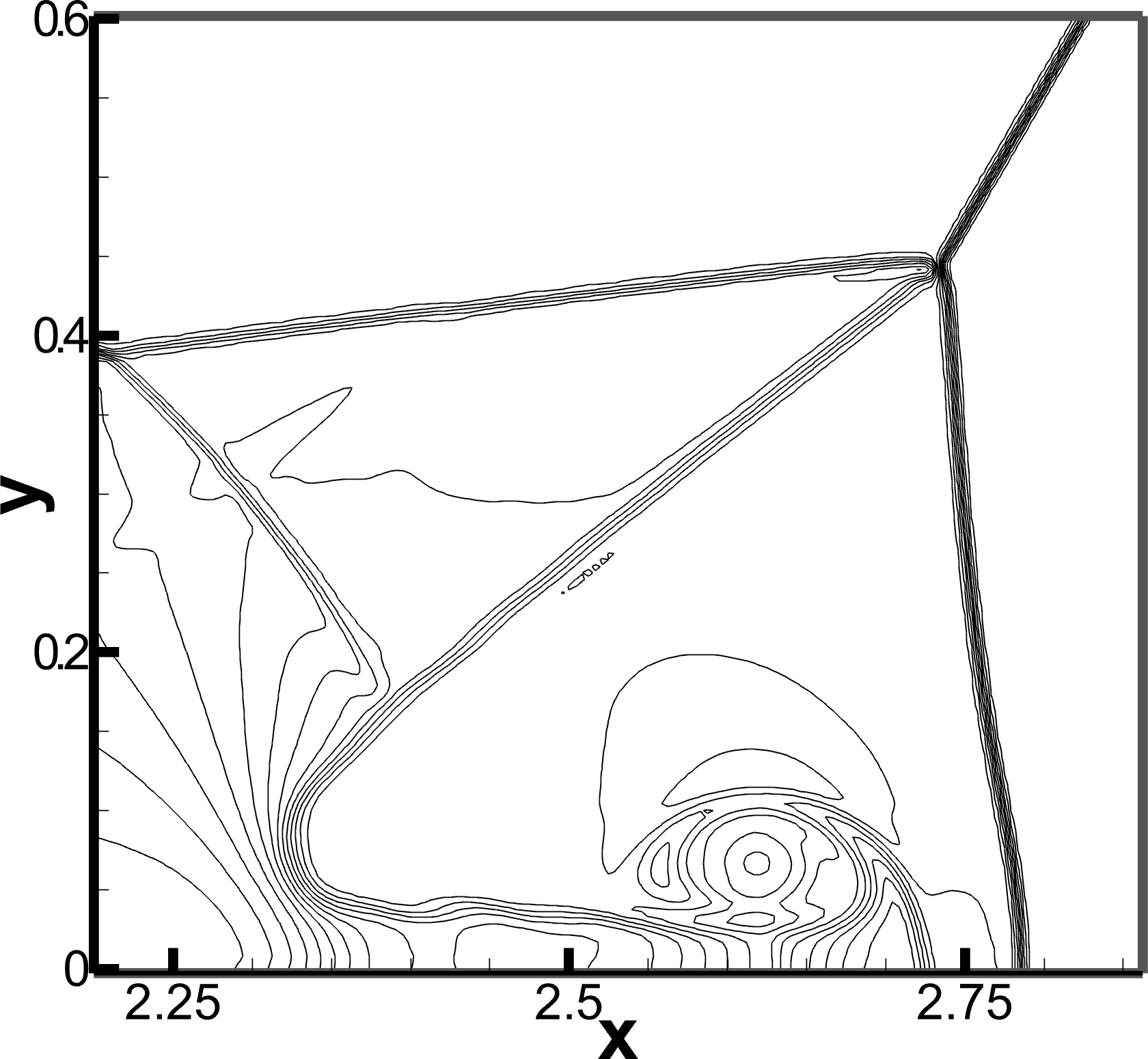}
    \caption{WCLS3 with $\kappa_0 = 0.8$.\label{fig:dmr_k0.8_240_cv}}
    \end{subfigure}
    \begin{subfigure}[b]{0.3\textwidth}
    \includegraphics[width=\textwidth]{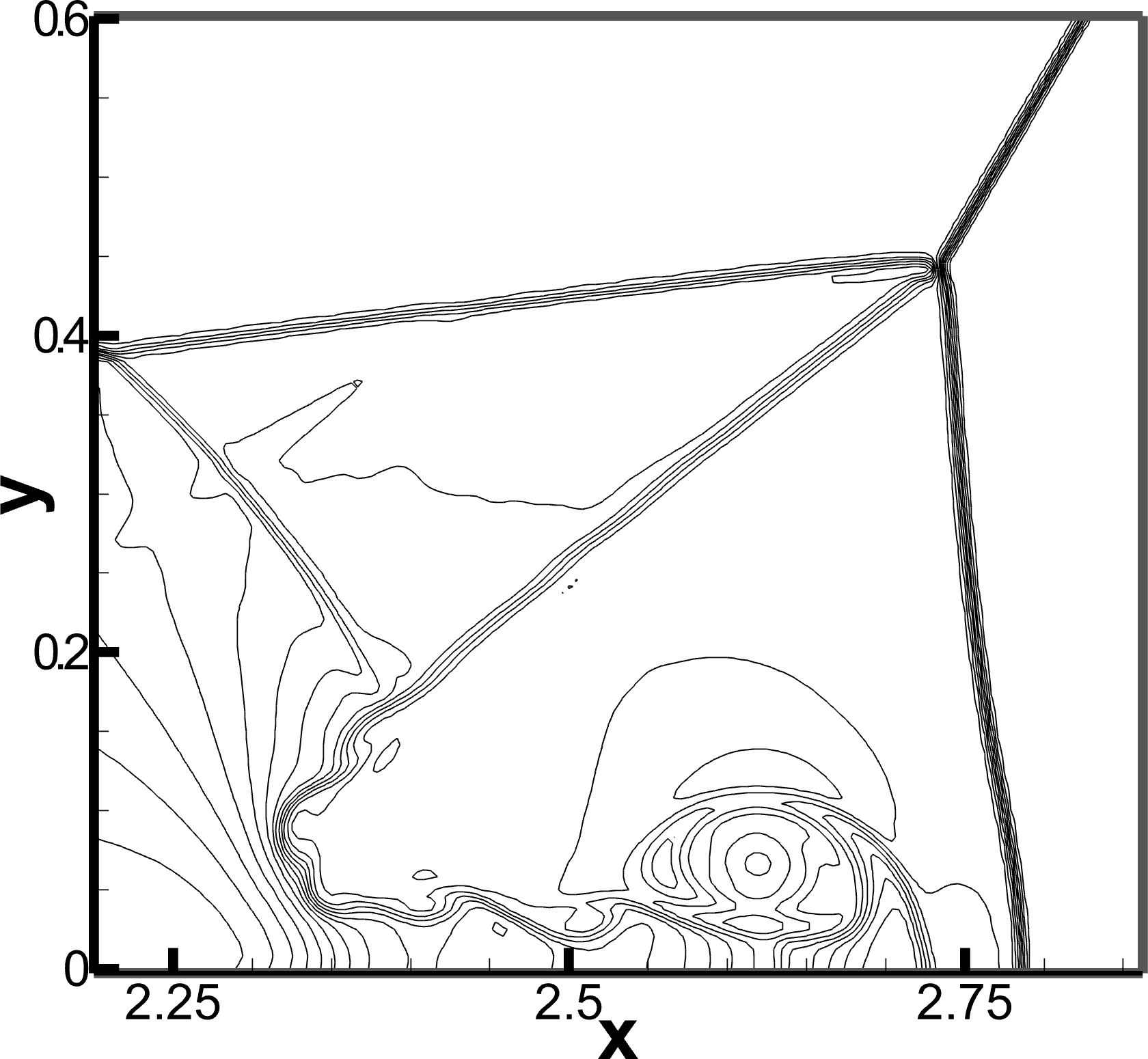}
    \caption{WCLS3 with $\kappa_0 = 1.0$.\label{fig:dmr_k1.0_240_cv}}
    \end{subfigure}
    \begin{subfigure}[b]{0.3\textwidth}
    \includegraphics[width=\textwidth]{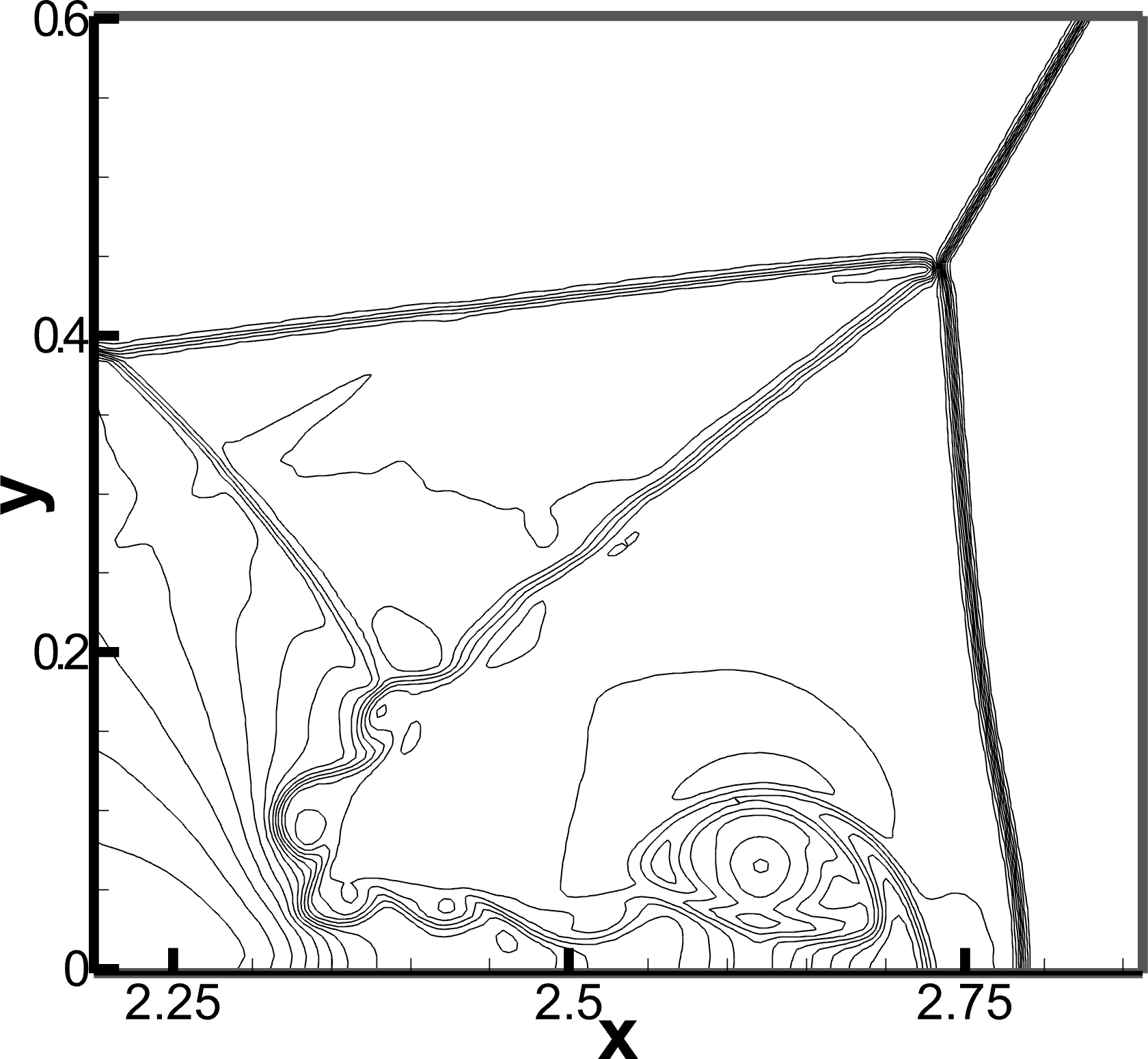}
    \caption{WCLS3 with $\kappa_0 = 1.2$.\label{fig:dmr_k1.2_240_cv}}
    \end{subfigure}
    \caption{Close view of density contours for the double Mach reflection with 30 lines ranging from 2.1 to 22. $N_x \times N_y = 960 \times 240$, $t = 0.2$ and CFL = $0.6$.
    \label{fig:dmr_240_cv}}
\end{figure}

\begin{figure}[!htbp]
  \centering
    \begin{subfigure}[b]{0.45\textwidth}
    \includegraphics[width=\textwidth]{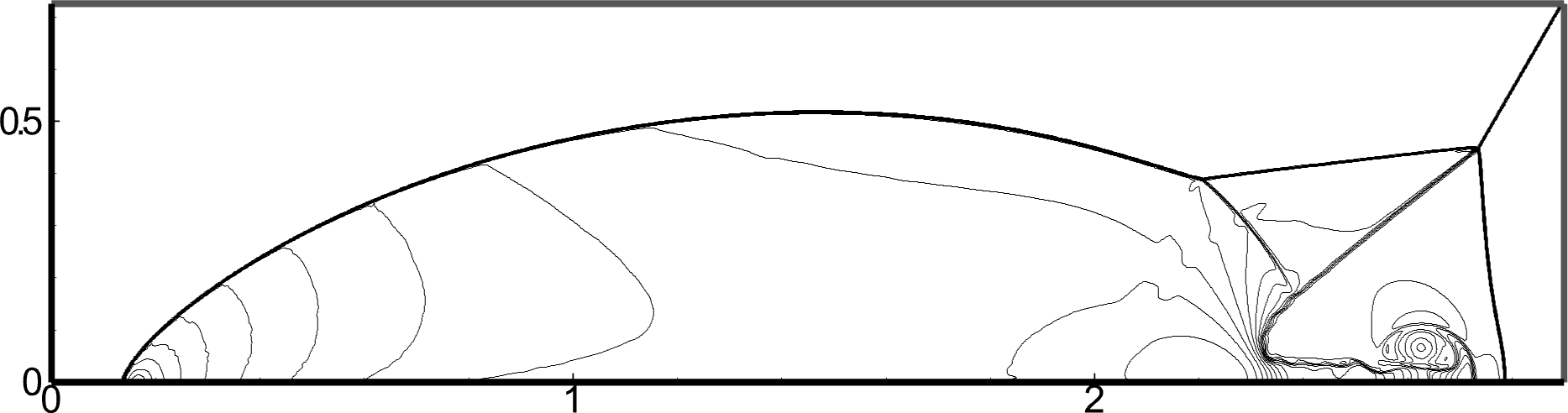}
    \caption{CWENO3.\label{fig:dmr_cweno3_480}}
    \end{subfigure}
    \quad
    \begin{subfigure}[b]{0.45\textwidth}
    \includegraphics[width=\textwidth]{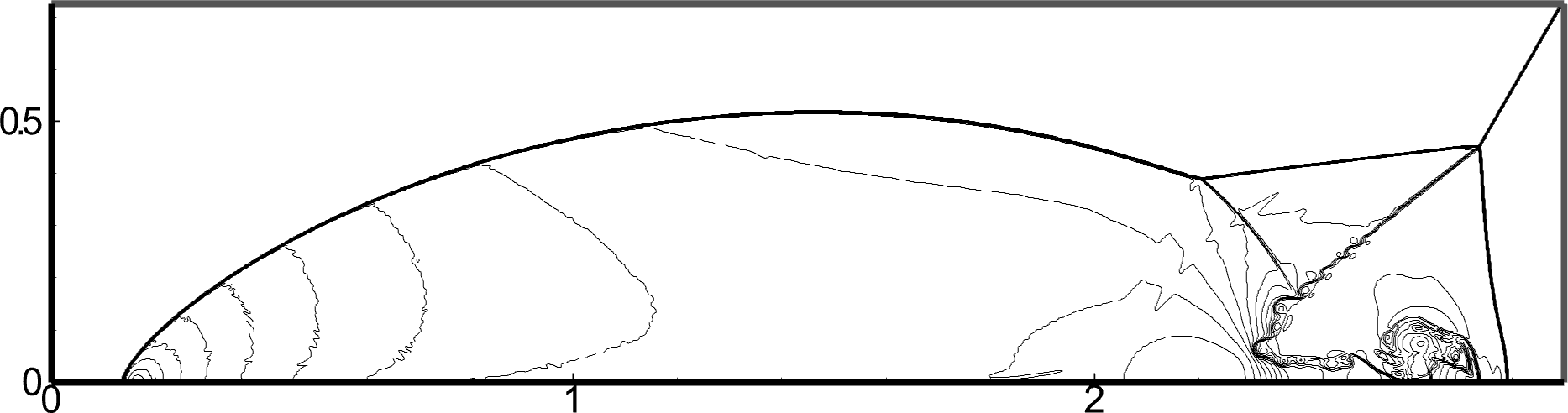}
    \caption{CWENO5.\label{fig:dmr_cweno5_480}}
    \end{subfigure}\\
    \begin{subfigure}[b]{0.45\textwidth}
    \includegraphics[width=\textwidth]{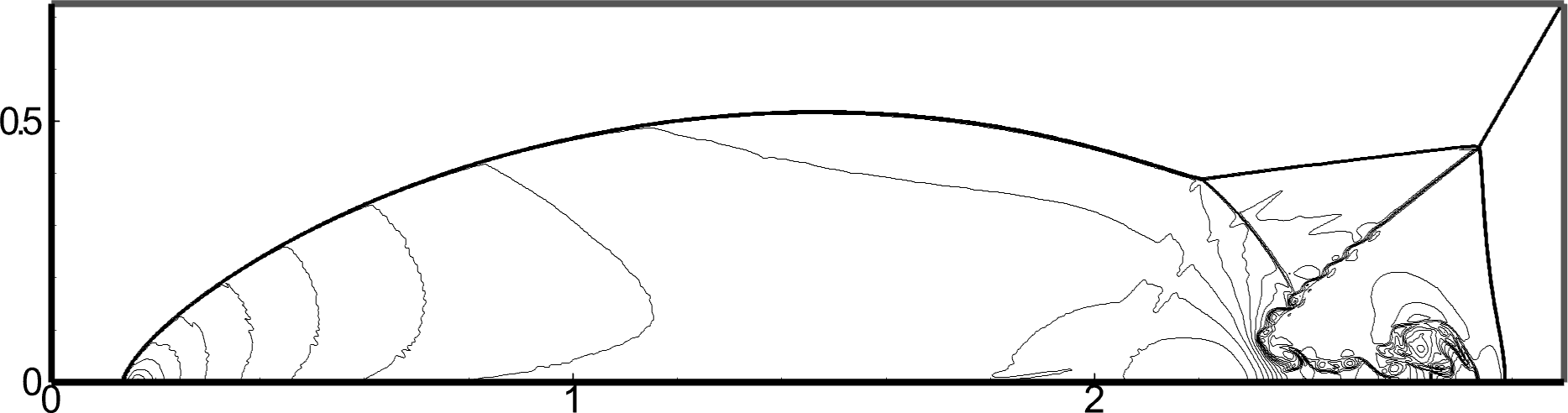}
    \caption{WCLS3 with $\kappa_0 = 0.8$.\label{fig:dmr_k0.8_480}}
    \end{subfigure}
    \begin{subfigure}[b]{0.45\textwidth}
    \includegraphics[width=\textwidth]{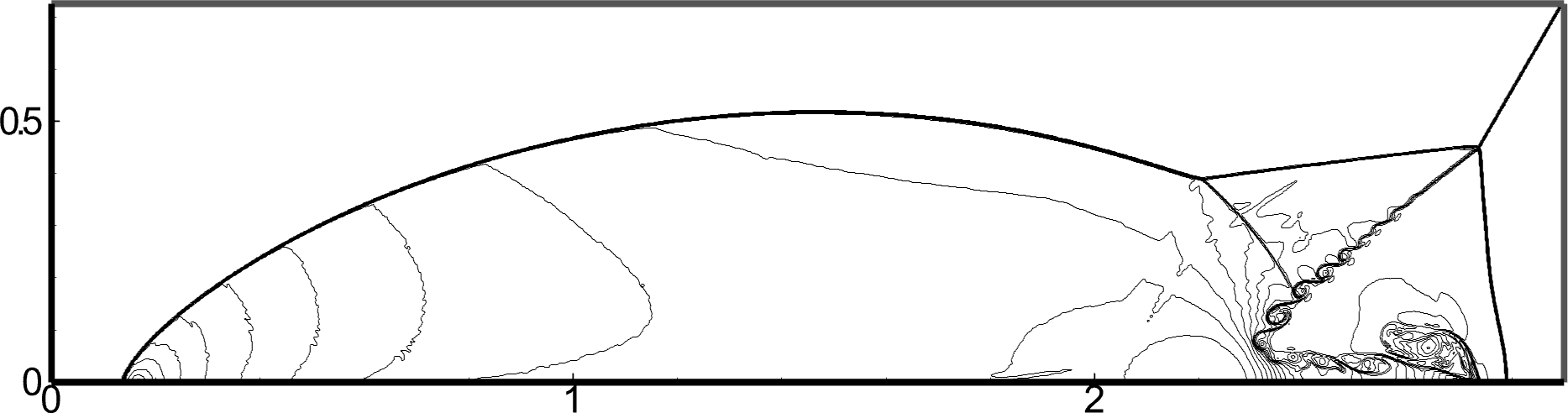}
    \caption{WCLS3 with $\kappa_0 = 1.0$.\label{fig:dmr_k1.0_480}}
    \end{subfigure}\\
    \begin{subfigure}[b]{0.45\textwidth}
    \includegraphics[width=\textwidth]{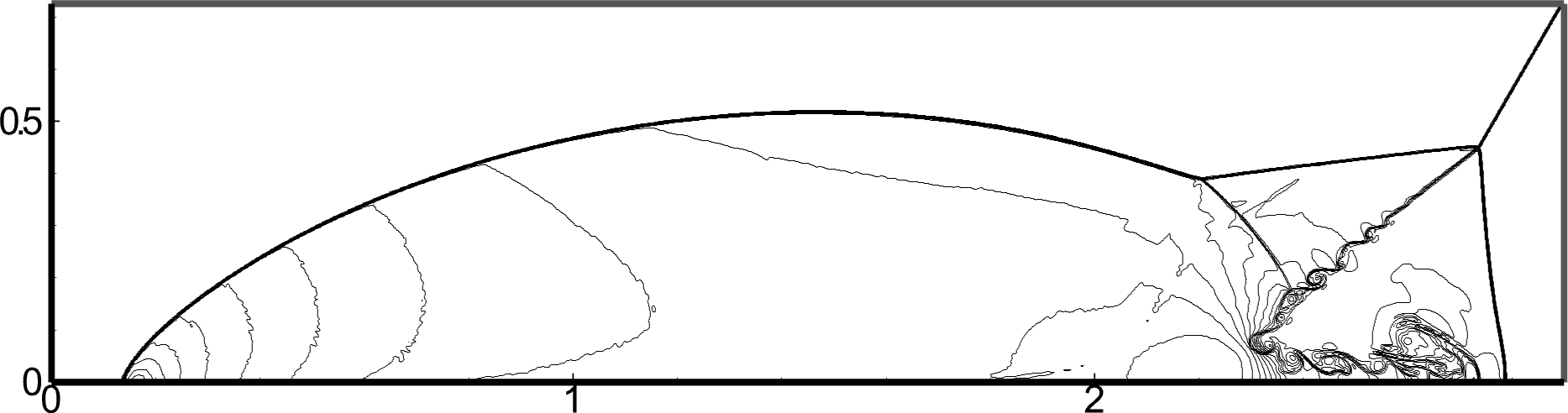}
    \caption{WCLS3 with $\kappa_0 = 1.2$.\label{fig:dmr_k1.2_480}}
    \end{subfigure}
    \caption{Density contours for the double Mach reflection with 30 lines ranging from 2.1 to 22. $N_x \times N_y = 960 \times 480$, $t = 0.2$ and CFL = $0.6$.
    \label{fig:dmr_480}}
\end{figure}

\begin{figure}[!htbp]
  \centering
    \begin{subfigure}[b]{0.3\textwidth}
    \includegraphics[width=\textwidth]{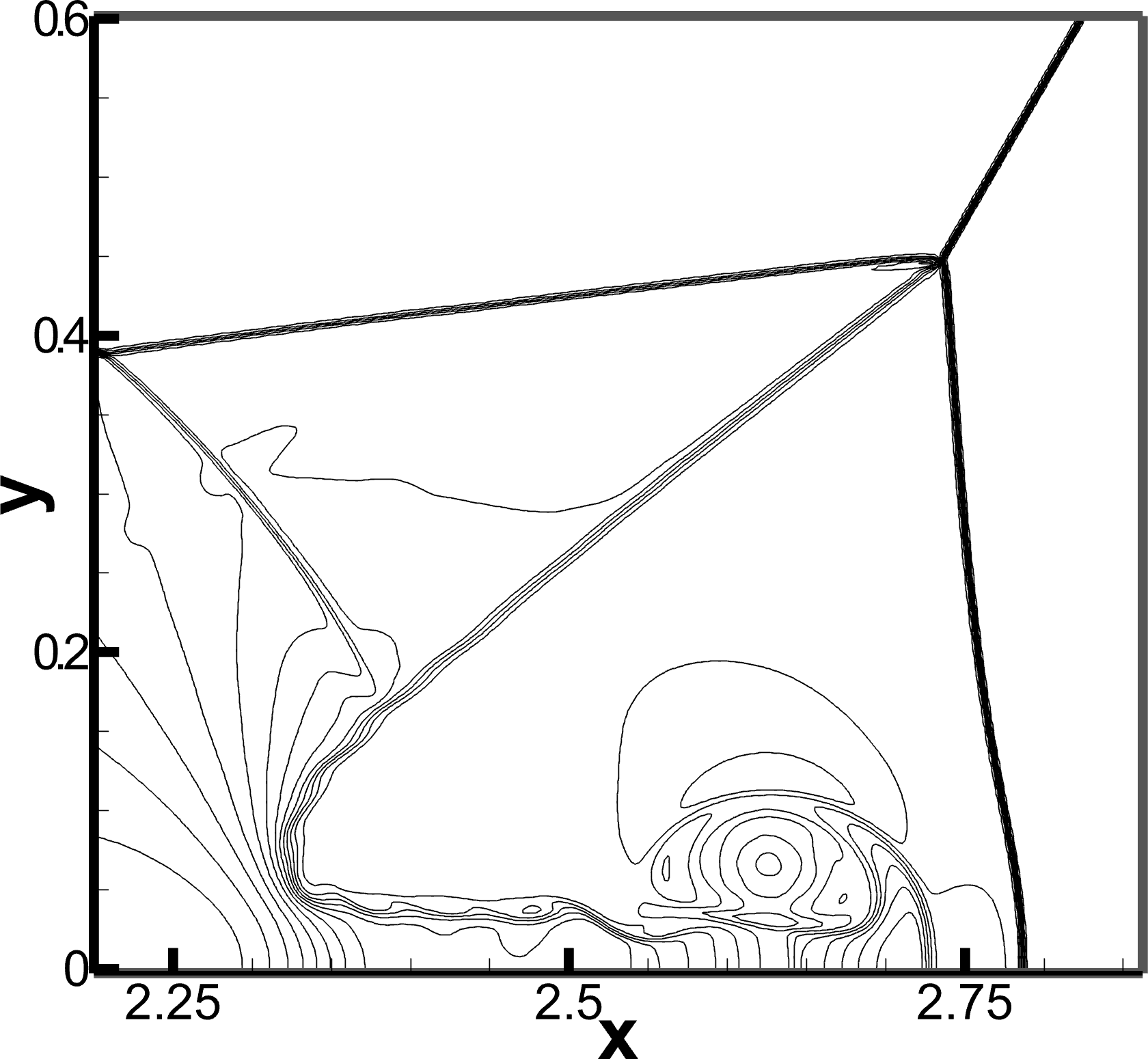}
    \caption{CWENO3.\label{fig:dmr_cweno3_480_cv}}
    \end{subfigure}
    \quad
    \begin{subfigure}[b]{0.3\textwidth}
    \includegraphics[width=\textwidth]{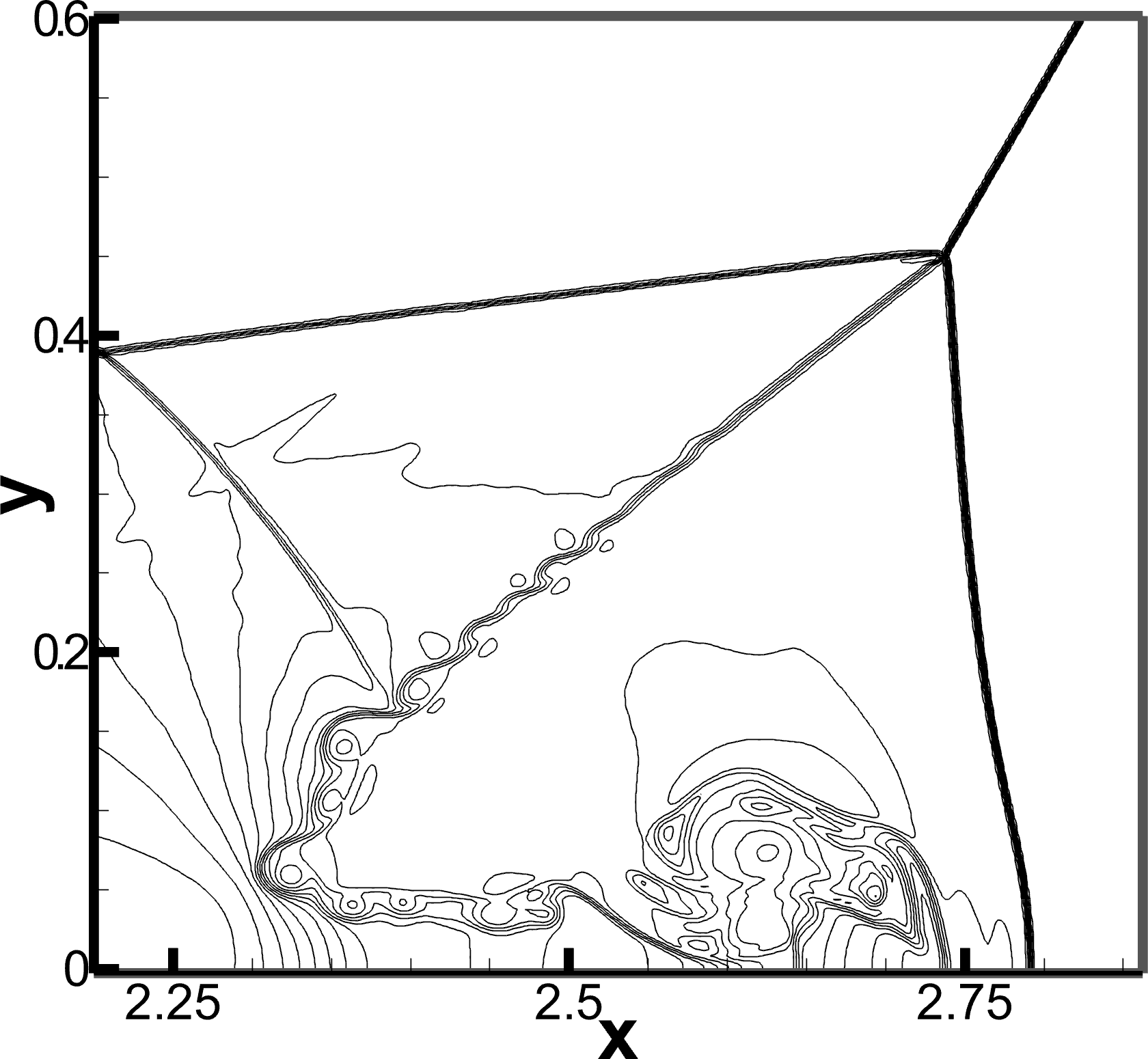}
    \caption{CWENO5.\label{fig:dmr_cweno5_480_cv}}
    \end{subfigure}\\
    \begin{subfigure}[b]{0.3\textwidth}
    \includegraphics[width=\textwidth]{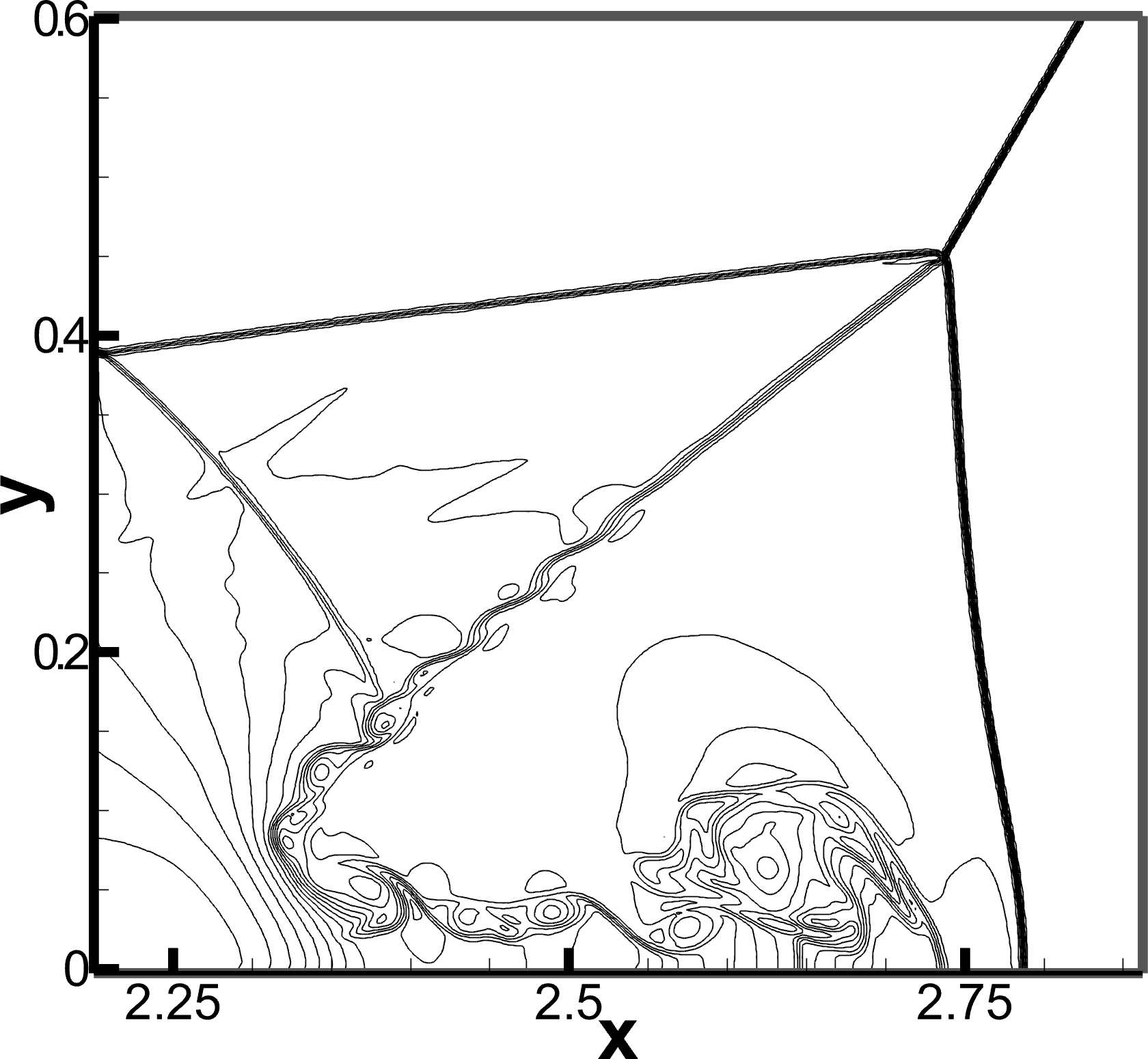}
    \caption{WCLS3 with $\kappa_0 = 0.8$.\label{fig:dmr_k0.8_480_cv}}
    \end{subfigure}
    \begin{subfigure}[b]{0.3\textwidth}
    \includegraphics[width=\textwidth]{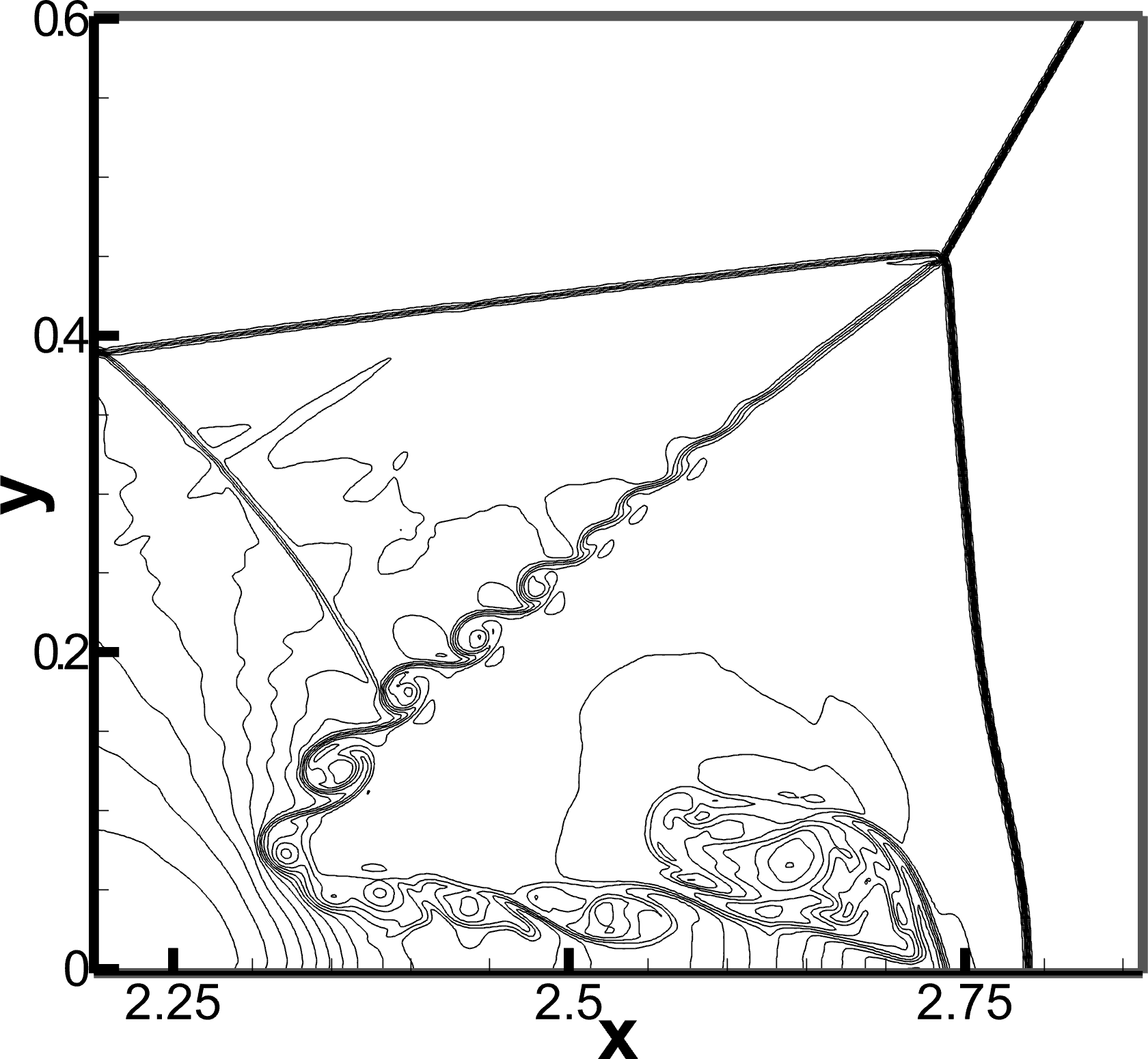}
    \caption{WCLS3 with $\kappa_0 = 1.0$.\label{fig:dmr_k1.0_480_cv}}
    \end{subfigure}
    \begin{subfigure}[b]{0.3\textwidth}
    \includegraphics[width=\textwidth]{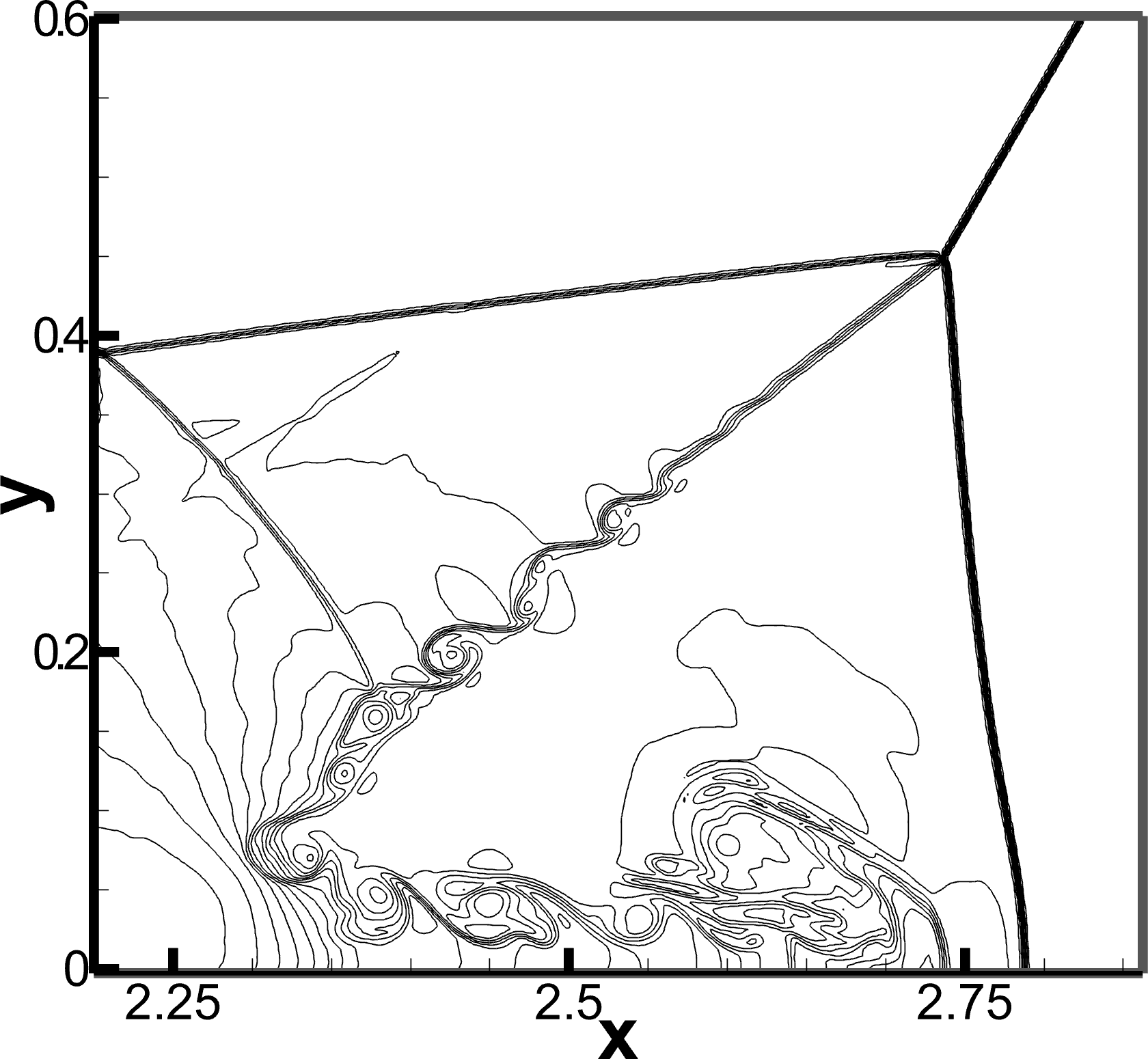}
    \caption{WCLS3 with $\kappa_0 = 1.2$.\label{fig:dmr_k1.2_480_cv}}
    \end{subfigure}
    \caption{Close view of density contours for the double Mach reflection with 30 lines ranging from 2.1 to 22. $N_x \times N_y = 1920 \times 480$, $t = 0.2$ and CFL = $0.6$.
    \label{fig:dmr_480_cv}}
\end{figure}

\begin{figure}[!htbp]
  \centering
    \begin{subfigure}[b]{0.6\textwidth}
    \includegraphics[width=\textwidth]{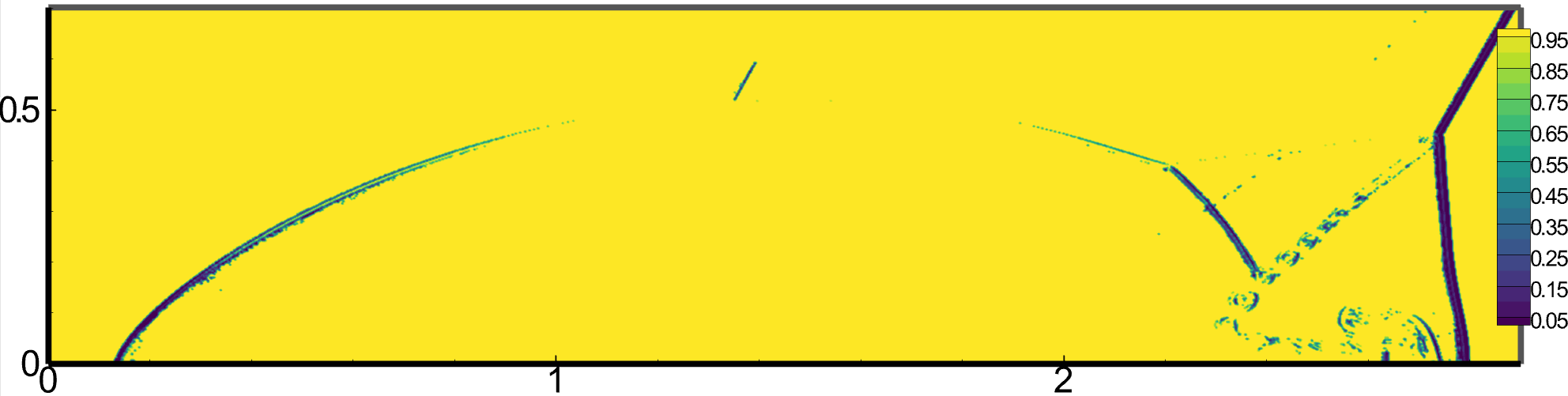}
    \caption{$\sigma$ in $x$-direction.}
    \end{subfigure}
    \quad
    \begin{subfigure}[b]{0.6\textwidth}
    \includegraphics[width=\textwidth]{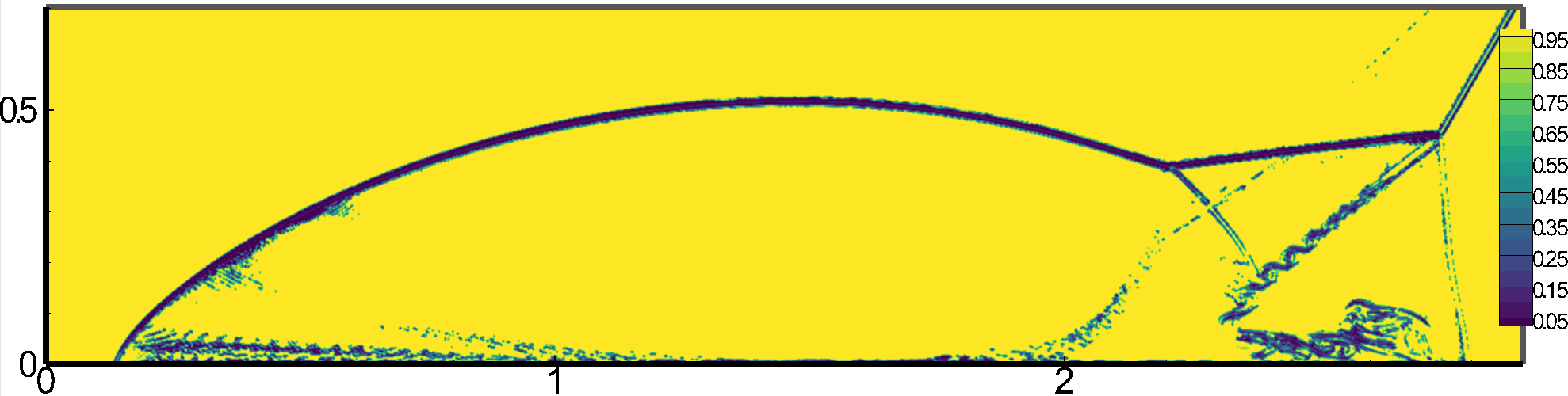}
    \caption{$\sigma$ in $y$-direction.}
    \end{subfigure}
    \caption{Shock detector for the double Mach reflection problem. $N_x \times N_y = 1920 \times 480$, $t = 0.2$ and CFL = $0.6$.
    \label{fig:dmr_shock}}
\end{figure}

The density contours on meshes with grid number being $960 \times 240$ and $1920 \times 480$ are shown as in Figs. \ref{fig:dmr_240} and \ref{fig:dmr_480}, respectively. The proposed WCLS3 scheme captures the discontinuities robustly and resolves the Kelvin-Helmholtz instabilities along the slip lines in high resolution. In this example, the resolution of the WCLS3 scheme is even higher than or comparative with the CWENO5 scheme as shown in Figs. \ref{fig:dmr_240_cv} and \ref{fig:dmr_480_cv}.
Figure \ref{fig:dmr_shock} also presents the shock detector distribution for the WCLS3 scheme with $\kappa_0 = 1.0$. As demonstrated, the proposed shock detector identifies the troubled cells accurately.

\subsubsection{Rayleigh-Taylor Instability}
Problem involves Rayleigh-Taylor instabilities is utilized to validate the robustness and high resolution of the proposed WCLS3 scheme. The initial condition is 
\begin{equation}
  \left(\rho, u,v, p\right) = \left\{
    \begin{array}{lr}
      2, 0, -0.025\sqrt{\gamma p/\rho} \cos(8\pi x), 2y+1, & 0.0\leq y < 0.5,\\
      1, 0, -0.025\sqrt{\gamma p/\rho} \cos(8\pi x), y+1.5,& 0.5\leq y < 1.0,
    \end{array}
  \right.
\end{equation}
where the ratio of specific heat capability is $\gamma = 5/3$.
Gravity is added with the strength of body force as $(0, 1)^T$. 
The computational domain is $[0, 0.25]\times [0,1]$.
The left and right boundaries employ inviscid walls, while the top and bottom boundaries maintain fixed values as
\begin{equation}
  \left(\rho, u,v, p\right) = \left\{
    \begin{array}{lr}
    1,0,0,2.5, &  y = 1.0,\\
    2,0,0,1.0, &  y = 0.0.\\
    \end{array}
  \right.
\end{equation}

Initially, there is a contact discontinuity at the interface $y = 0.5$ with a small disturbance between two fluids of different density. Under gravity, Rayleigh-Taylor instability is triggered.  The heavier fluid forms finger-like structures penetrating upward through the lighter fluid, while the lighter fluid develops bubble-like structures moving downward into the heavier fluid.

\begin{figure}[!htbp]
  \centering
    \begin{subfigure}[b]{0.16\textwidth}
    \includegraphics[width=\textwidth]{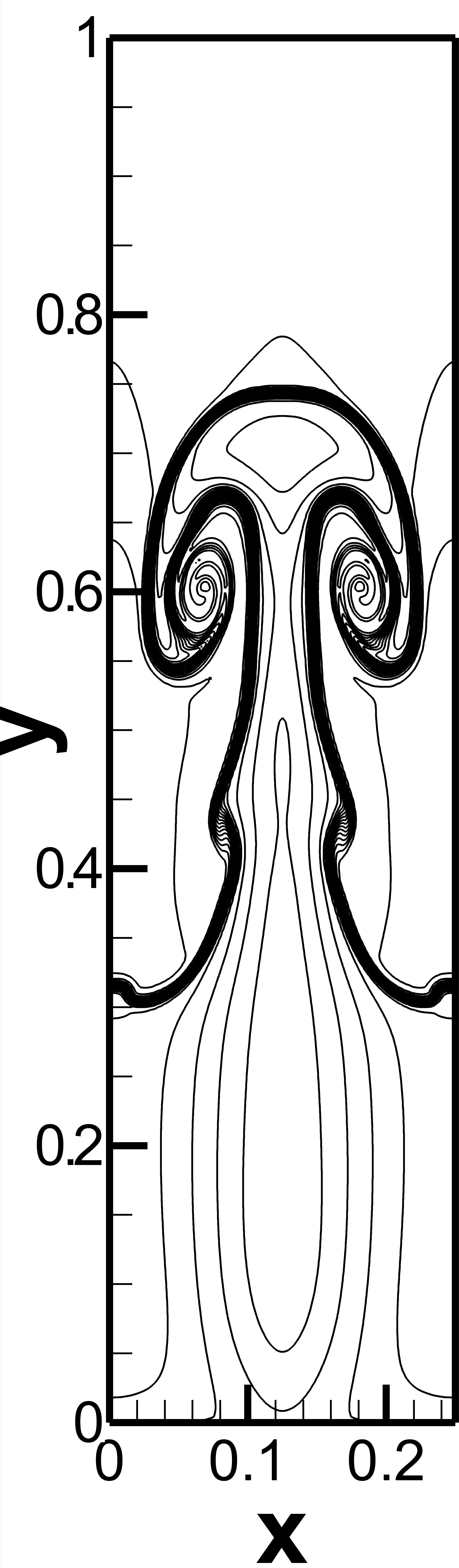}
    \caption{CWENO3.\label{fig:rt_cweno3_480}}
    \end{subfigure}
    \begin{subfigure}[b]{0.16\textwidth}
    \includegraphics[width=\textwidth]{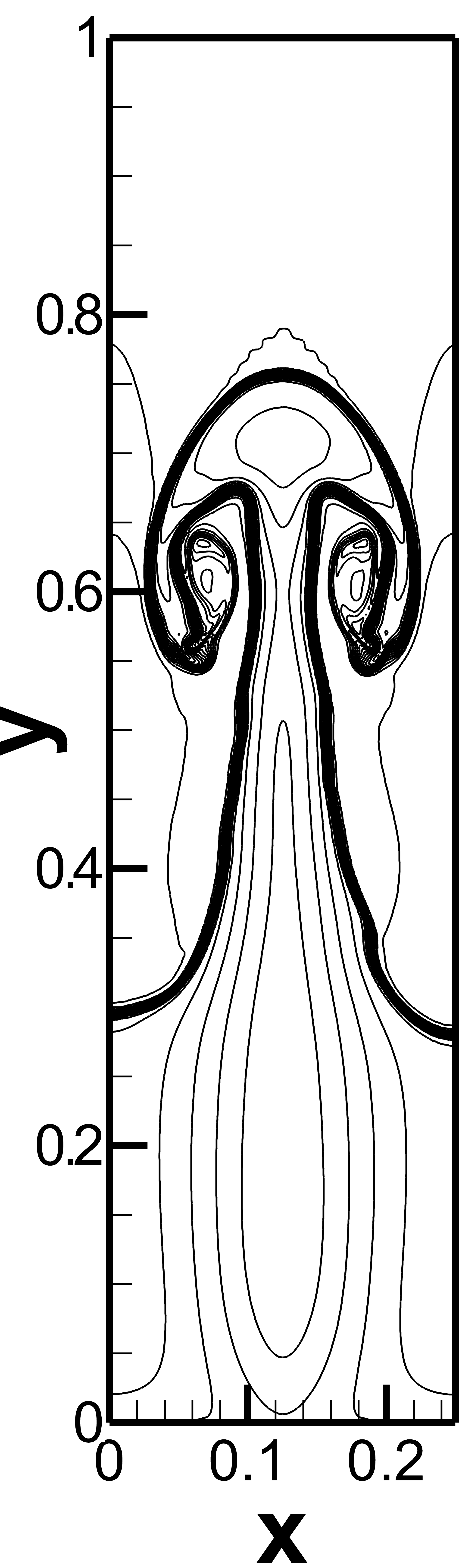}
    \caption{CWENO5.\label{fig:rt_cweno5_480}}
    \end{subfigure}
    \begin{subfigure}[b]{0.16\textwidth}
    \includegraphics[width=\textwidth]{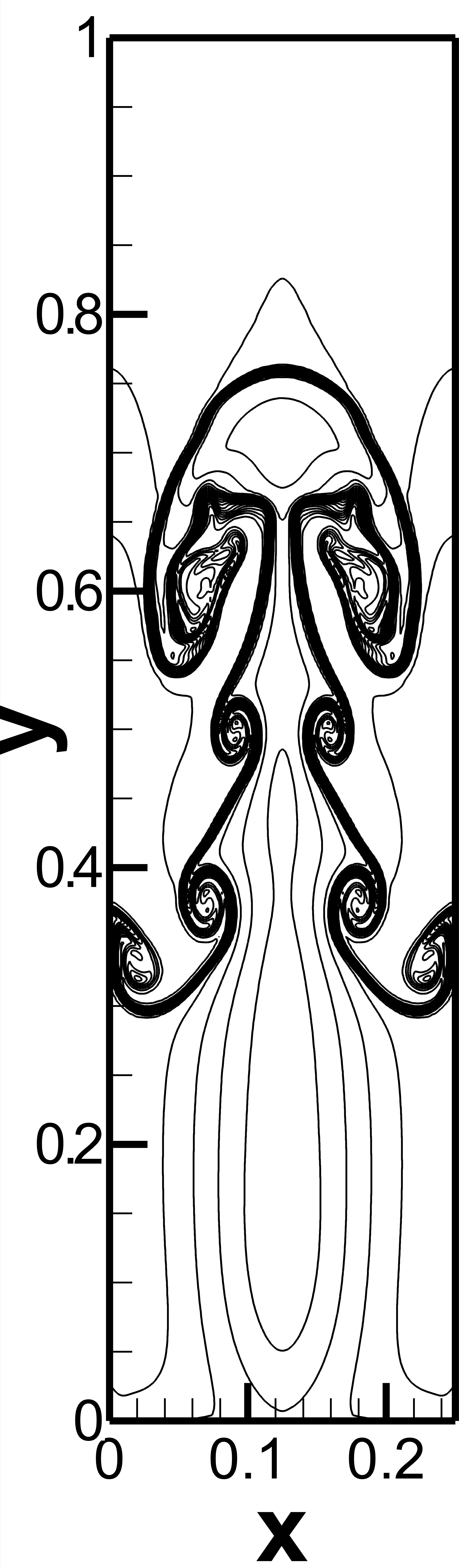}
    \caption{WCLS3 with $\kappa_0 = 0.8$.\label{fig:RT_k0.8_480}}
    \end{subfigure}
    \begin{subfigure}[b]{0.16\textwidth}
    \includegraphics[width=\textwidth]{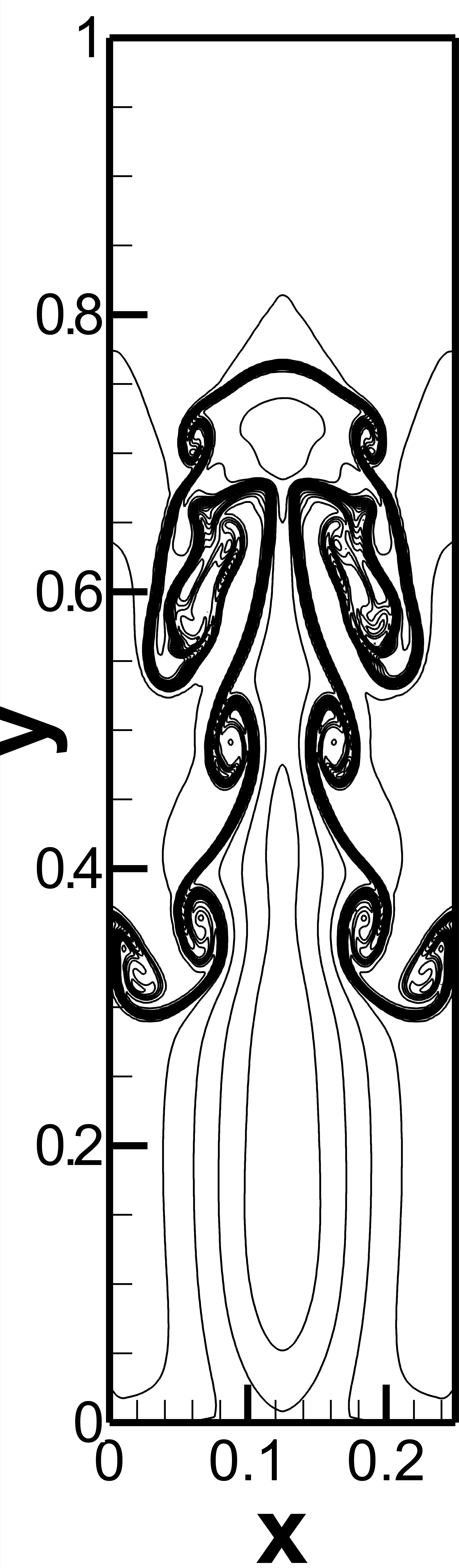}
    \caption{WCLS3 with $\kappa_0 = 1.0$.\label{fig:RT_k1.0_480}}
    \end{subfigure}
    \begin{subfigure}[b]{0.16\textwidth}
    \includegraphics[width=\textwidth]{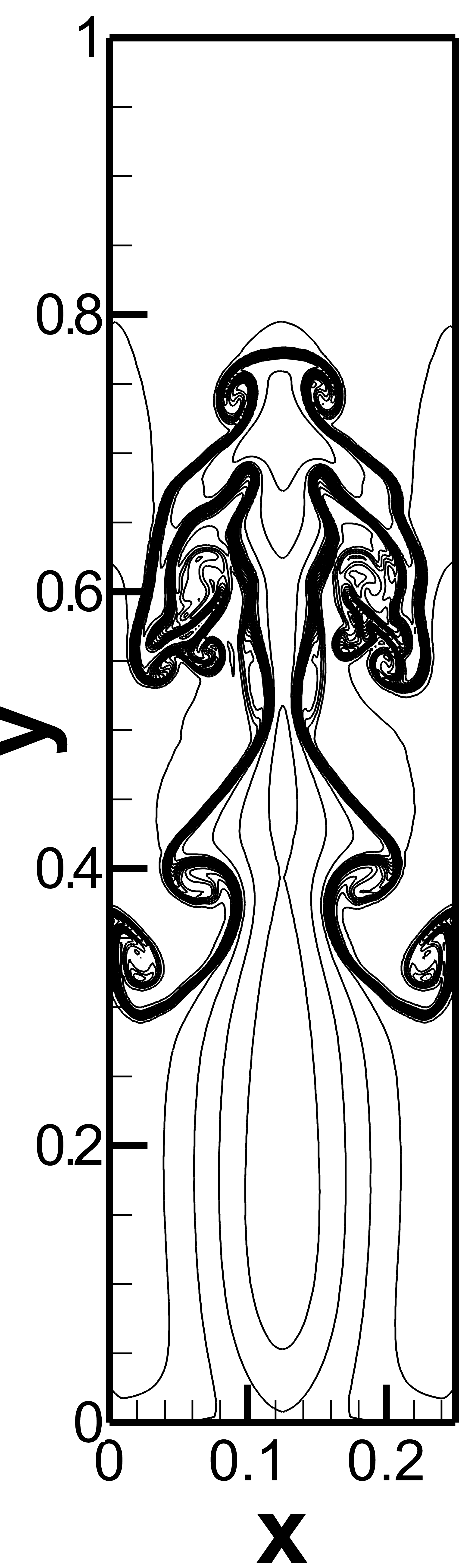}
    \caption{WCLS3 with $\kappa_0 = 1.2$.\label{fig:RT_k1.2_480}}
    \end{subfigure}
    \caption{Density contours for the Rayleigh-Taylor instabilities with 20 lines ranging from 0.9 to 2.2. $N_x \times N_y = 120 \times 480$, $t = 1.95$ and CFL = $0.5$.
    \label{fig:RT_480}}
\end{figure}

\begin{figure}[!htbp]
  \centering
    \begin{subfigure}[b]{0.16\textwidth}
    \includegraphics[width=\textwidth]{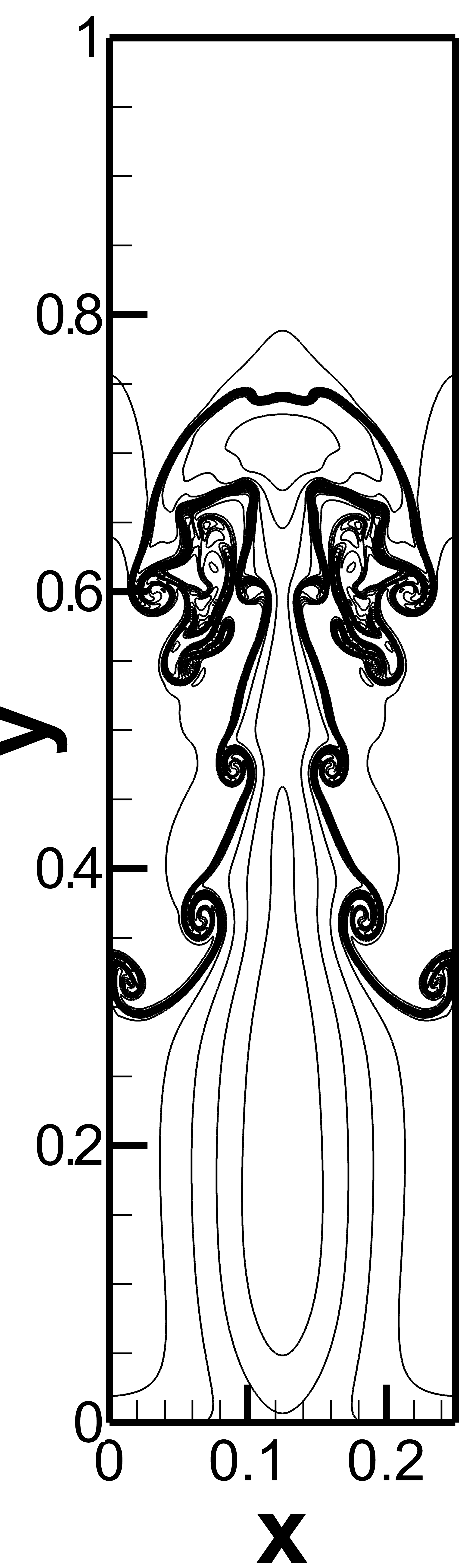}
    \caption{CWENO3.\label{fig:rt_cweno3_960}}
    \end{subfigure}
    \begin{subfigure}[b]{0.16\textwidth}
    \includegraphics[width=\textwidth]{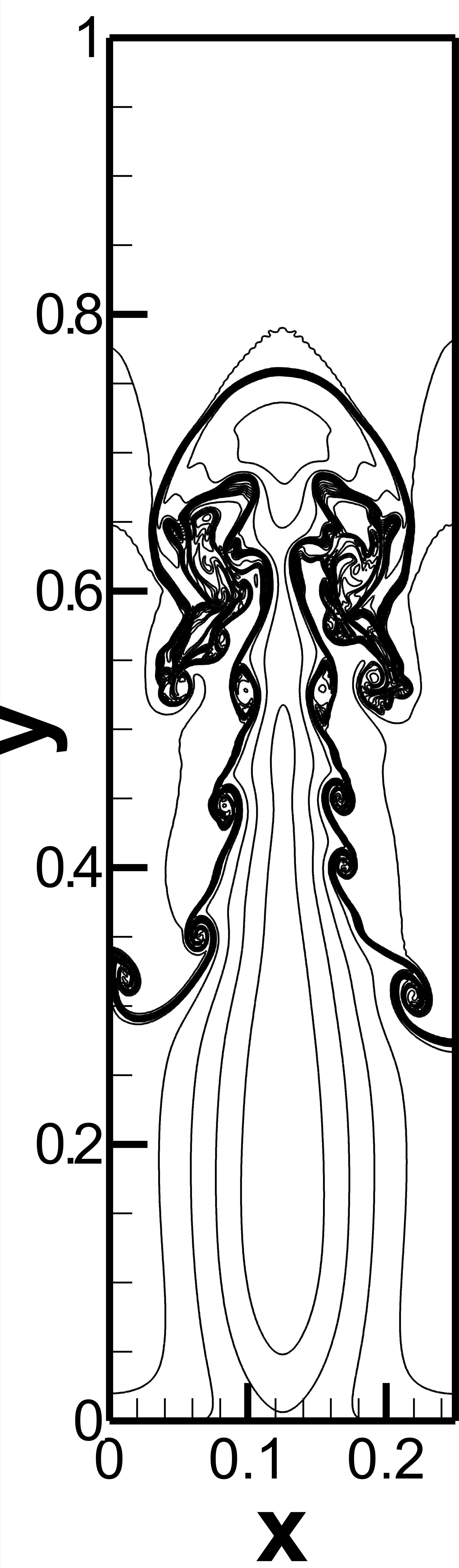}
    \caption{CWENO5.\label{fig:rt_cweno5_960}}
    \end{subfigure}
    \begin{subfigure}[b]{0.16\textwidth}
    \includegraphics[width=\textwidth]{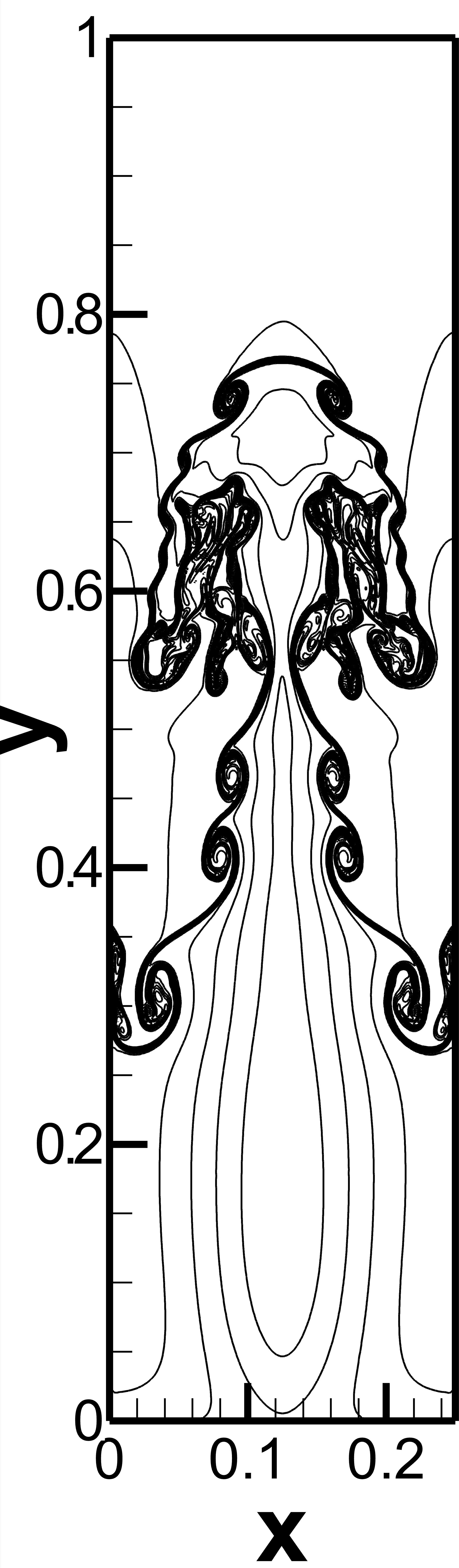}
    \caption{WCLS3 with $\kappa_0 = 0.8$.\label{fig:RT_k0.8_960}}
    \end{subfigure}
    \begin{subfigure}[b]{0.16\textwidth}
    \includegraphics[width=\textwidth]{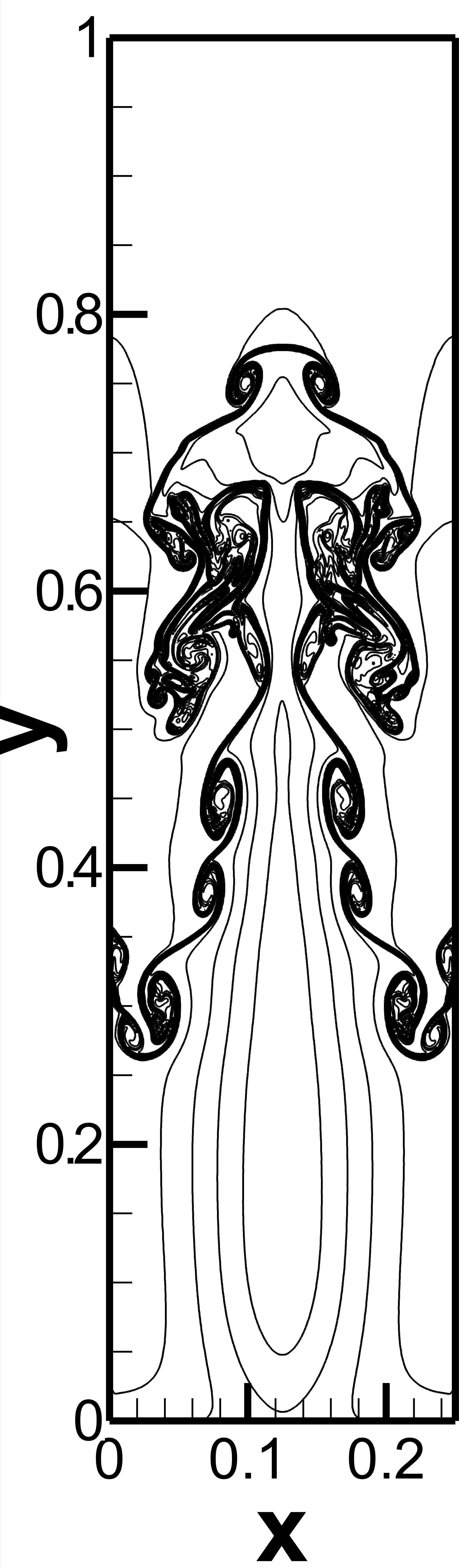}
    \caption{WCLS3 with $\kappa_0 = 1.0$.\label{fig:RT_k1.0_960}}
    \end{subfigure}
    \begin{subfigure}[b]{0.16\textwidth}
    \includegraphics[width=\textwidth]{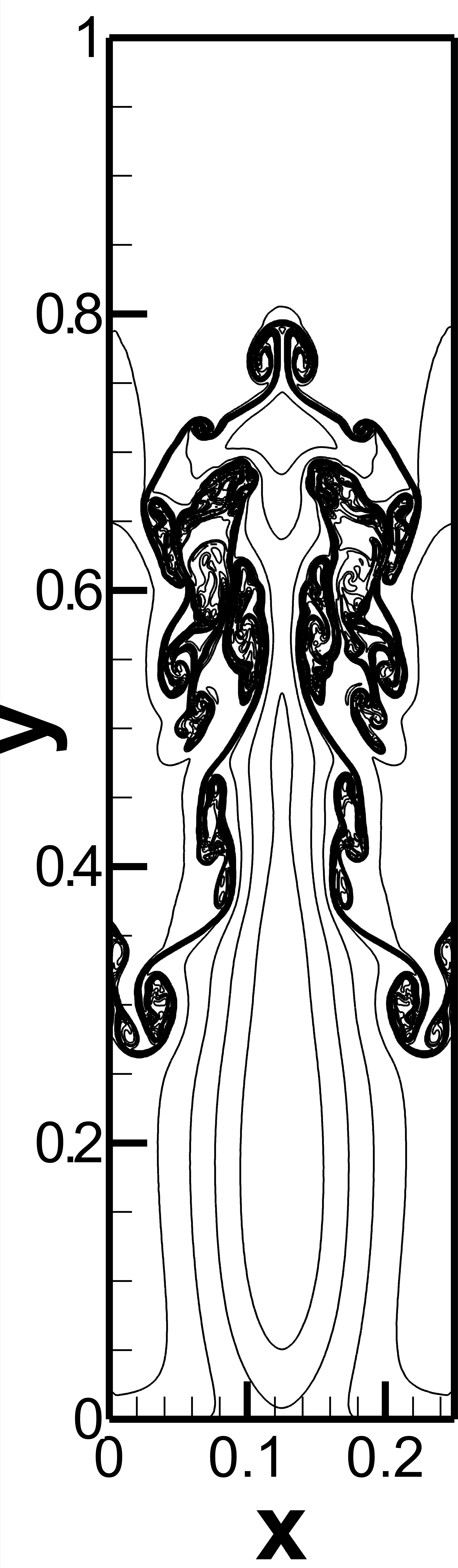}
    \caption{WCLS3 with $\kappa_0 = 1.2$.\label{fig:RT_k1.2_960}}
    \end{subfigure}
    \caption{Density contours for the Rayleigh-Taylor instabilities with 20 lines ranging from 0.9 to 2.2. $N_x \times N_y = 240 \times 960$, $t = 1.95$ and CFL = $0.5$.
    \label{fig:RT_960}}
\end{figure}

The results at $t = 1.95$ advanced by the third-order SSP-RK method with CFL = $0.5$ are presented in Figs. \ref{fig:RT_480} and \ref{fig:RT_960} with grid number of $120\times 480$ and $240\times 960$, respectively. The resolution of the proposed WCLS3 scheme is much higher than the CWENO3 scheme, even much better than the CWENO5 scheme in this experiment. With the increase of $\kappa_0$, the WCLS3 scheme resolves finer flow structures. Although no special treatment is taken to preserve the symmetry of the problem as in \cite{wakimura_symmetry-preserving_2022}, the symmetry of the large-scale structures is preserved quite well in the WCLS3 results with finer mesh as in Fig. \ref{fig:RT_960}.

\subsubsection{2D Riemann Problem\label{sec:2drp}}
The initial condition of the 2D Riemann problem is given as
\begin{equation}
  \left(\rho, u, v, p\right)  =  \left\{
  \begin{array}{llllr}
    1.5,& 0,& 0,& 1.5, & 0.8 \leq x \leq 1, 0.8 \leq y \leq 1,\\
    0.5323, & 1.206, & 0, & 0.3, &0 \leq x < 0.8, 0.8 \leq y \leq 1,\\
    0.138, & 1.206, & 1.206, & 0.029, &0 \leq x < 0.8, 0 \leq y < 0.8,\\
    0.5323, & 0, & 1.206, & 0.3, & 0.8 \leq x \leq 1, 0 \leq y < 0.8.\\
  \end{array}
  \right.
\end{equation}
Shock waves propagate and interact with each other, forming complex shock diffraction patterns. 
Along the slip lines, Kelvin-Helmholtz instabilities are observed with rich multiscale structures.

\begin{figure}[!htbp]
  \centering
    \begin{subfigure}[b]{0.3\textwidth}
    \includegraphics[width=\textwidth]{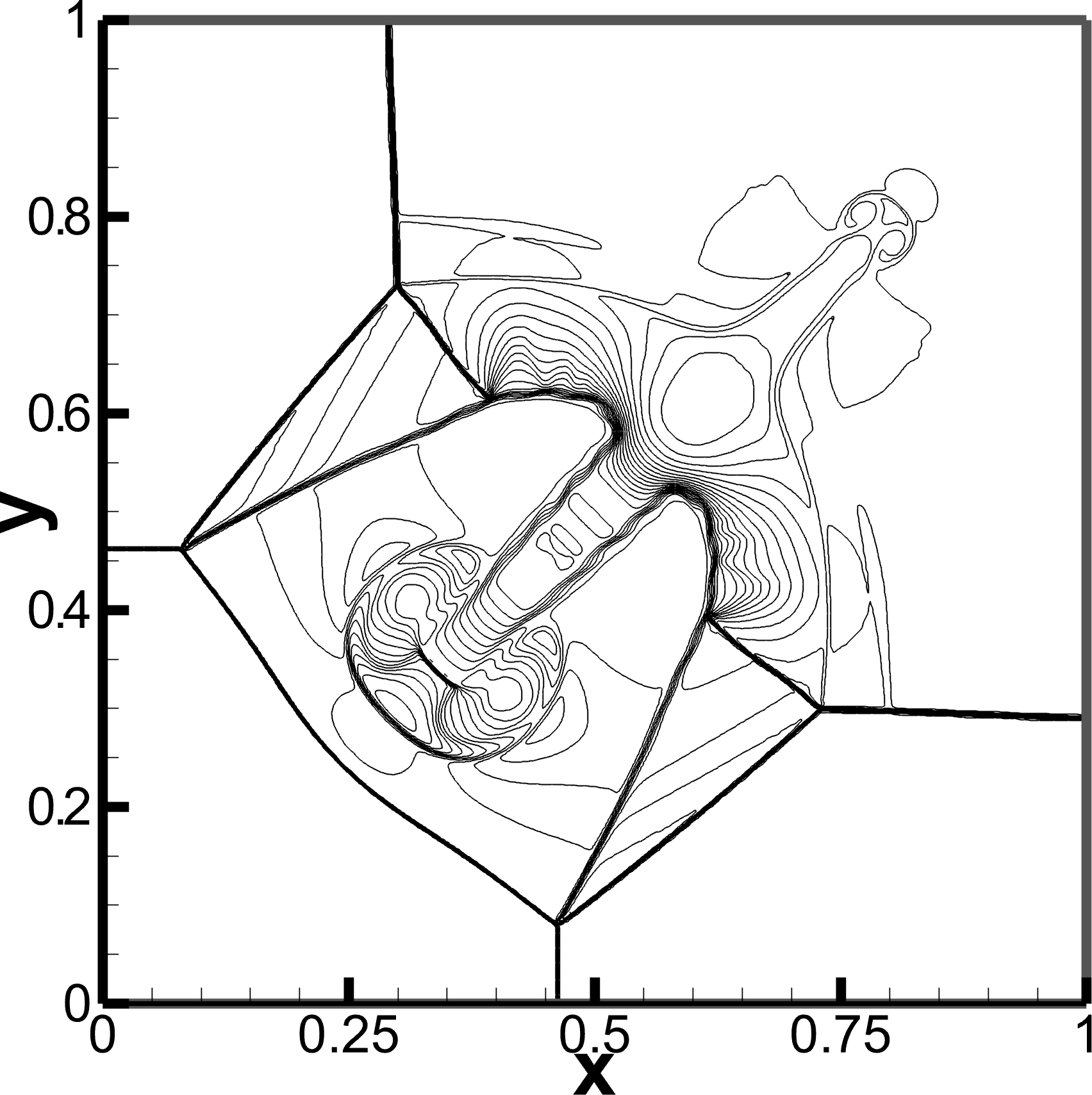}
    \caption{CWENO3.\label{fig:2dr_cweno3_400}}
    \end{subfigure}
    \quad
    \begin{subfigure}[b]{0.3\textwidth}
    \includegraphics[width=\textwidth]{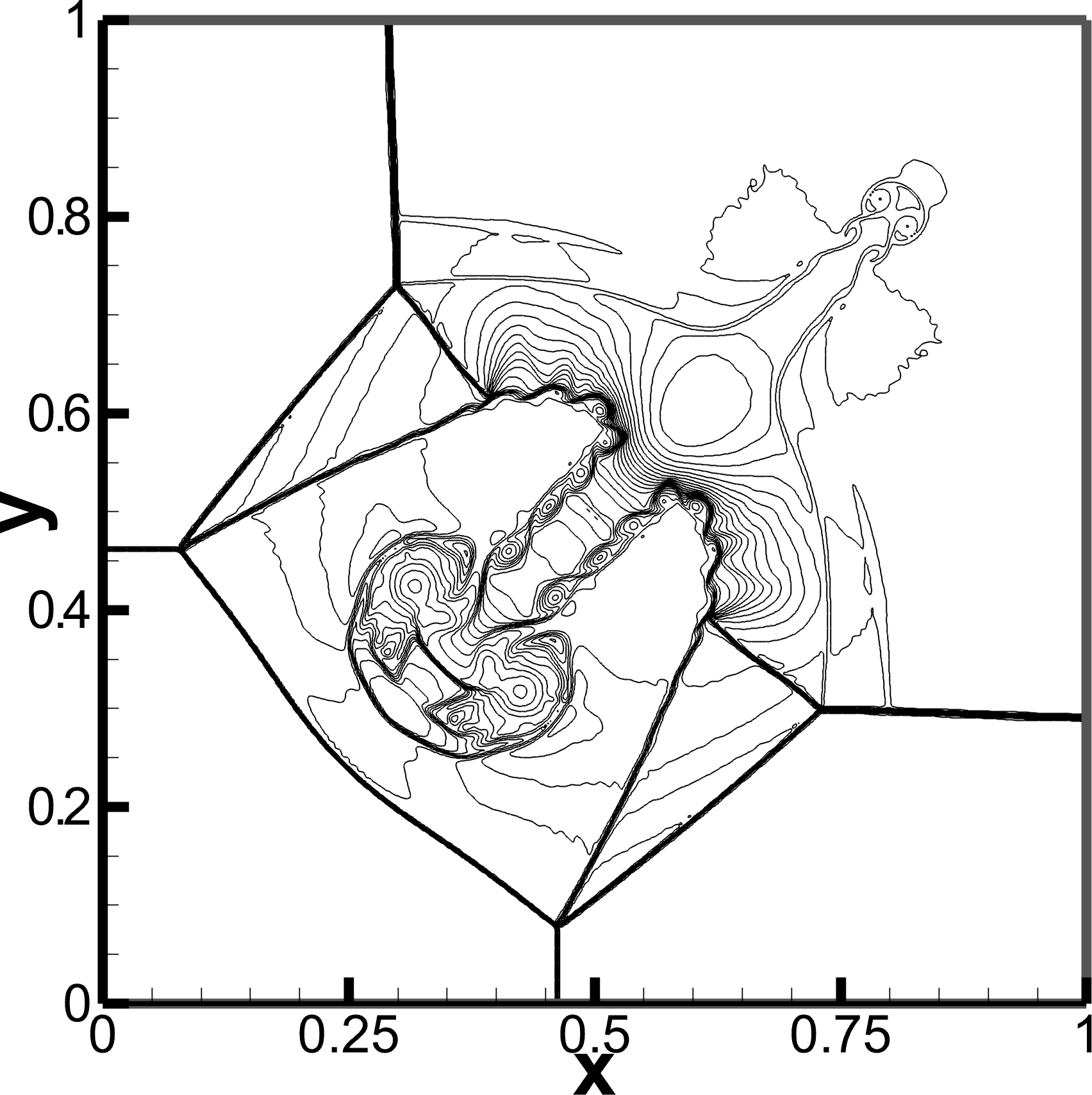}
    \caption{CWENO5.\label{fig:2dr_cweno5_400}}
    \end{subfigure}\\
    \begin{subfigure}[b]{0.3\textwidth}
    \includegraphics[width=\textwidth]{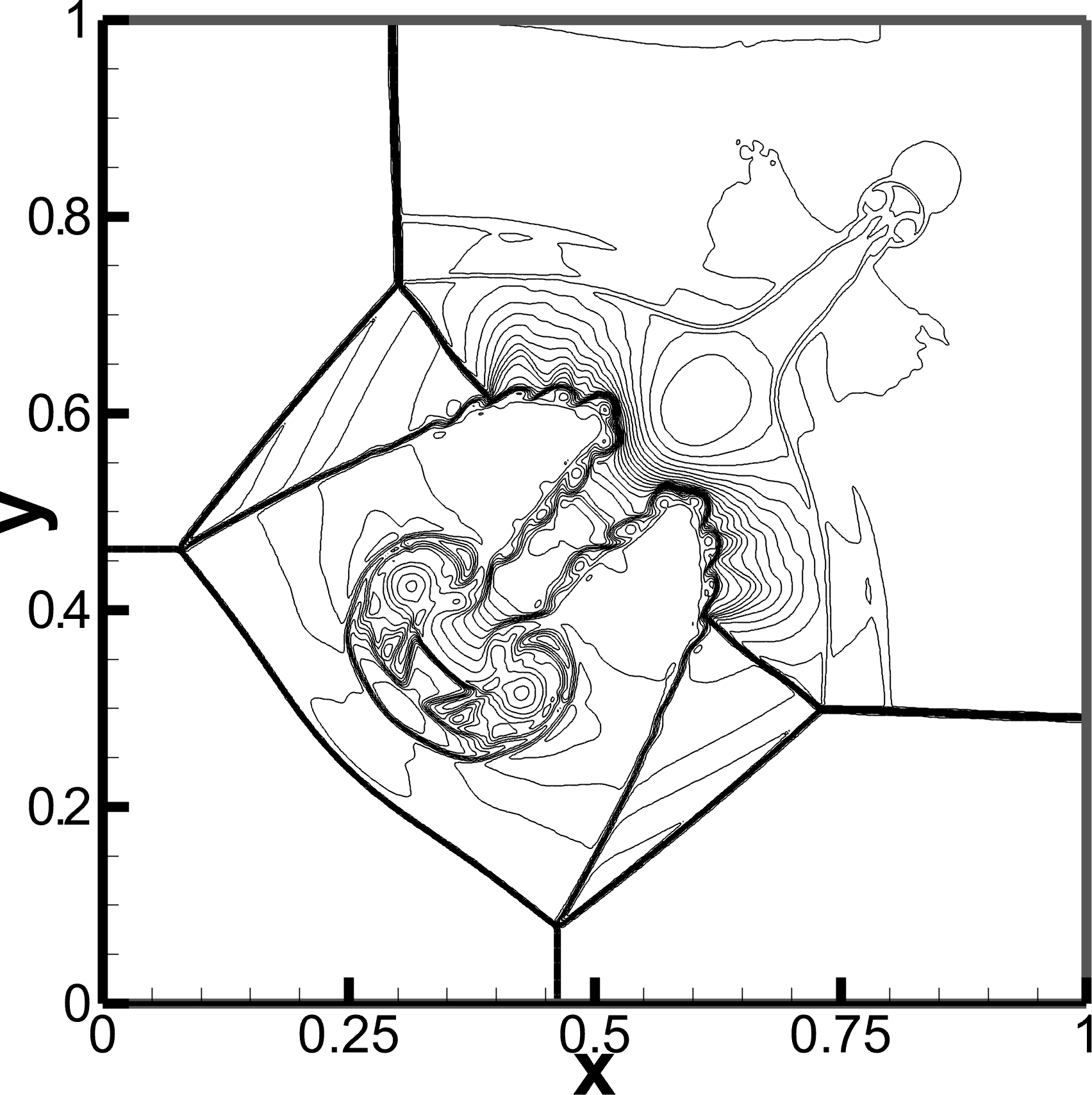}
    \caption{WCLS3 with $\kappa_0 = 0.8$.\label{fig:2dr_k0.8_400}}
    \end{subfigure}
    \begin{subfigure}[b]{0.3\textwidth}
    \includegraphics[width=\textwidth]{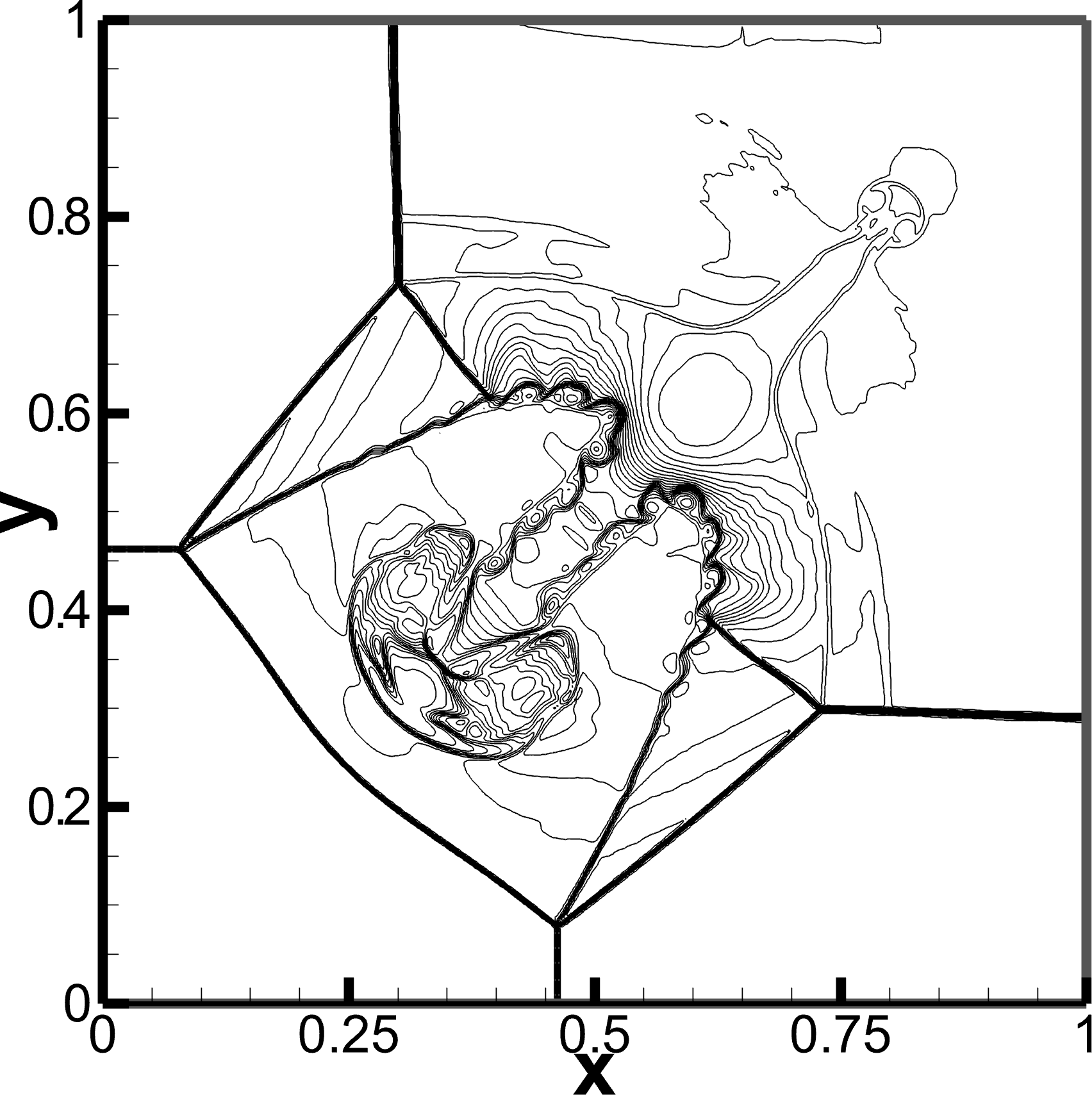}
    \caption{WCLS3 with $\kappa_0 = 1.0$.\label{fig:2dr_k1.0_400}}
    \end{subfigure}
    \begin{subfigure}[b]{0.3\textwidth}
    \includegraphics[width=\textwidth]{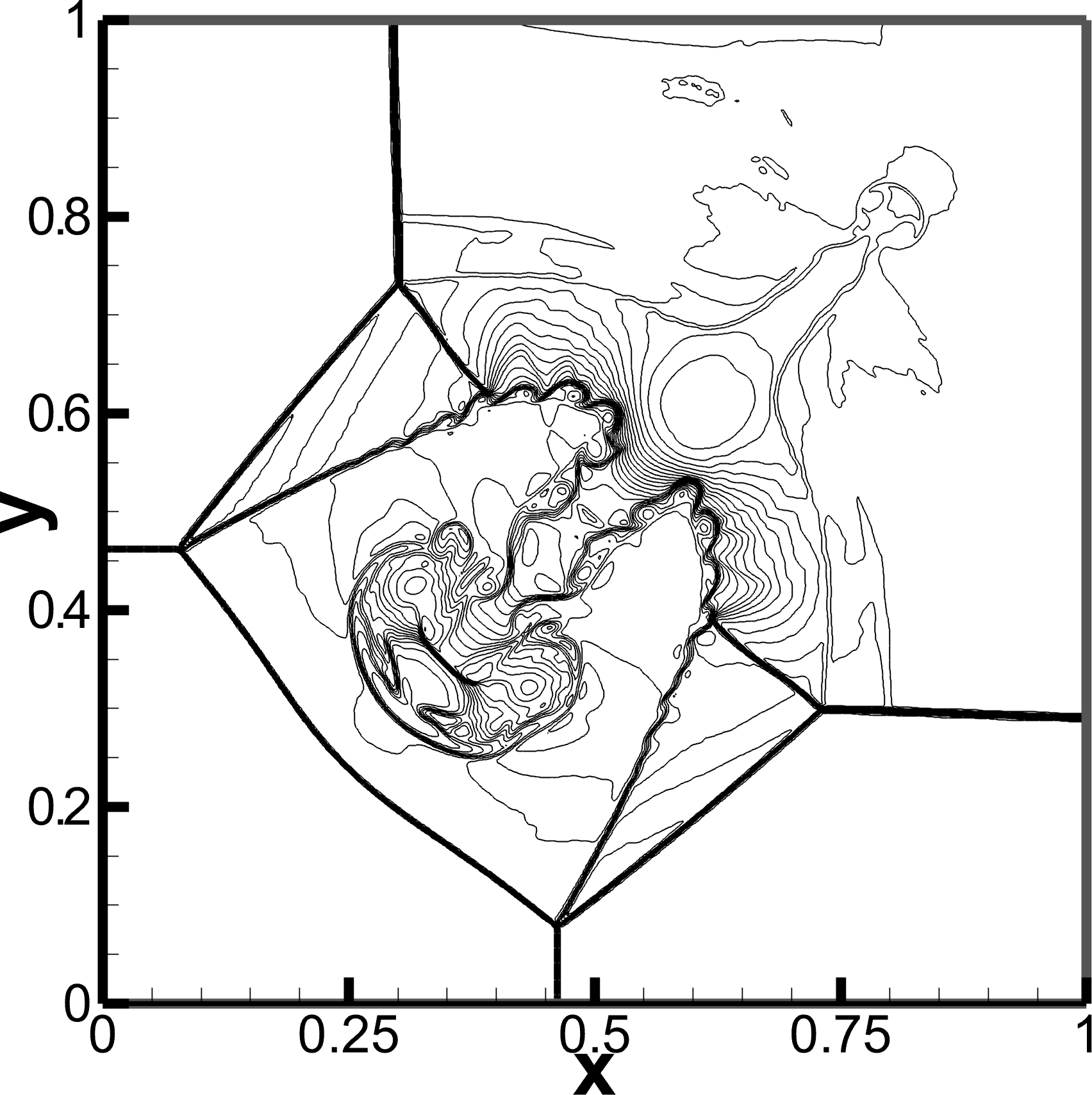}
    \caption{WCLS3 with $\kappa_0 = 1.2$.\label{fig:2dr_k1.2_400}}
    \end{subfigure}
    \caption{Density contours for the 2D Riemann problem with 30 lines ranging from 0.18 to 1.7. $N_x \times N_y = 400 \times 400$, $t = 0.8$ and CFL = $0.6$.
    \label{fig:2dr_400}}
\end{figure}

\begin{figure}[!htbp]
  \centering
    \begin{subfigure}[b]{0.3\textwidth}
    \includegraphics[width=\textwidth]{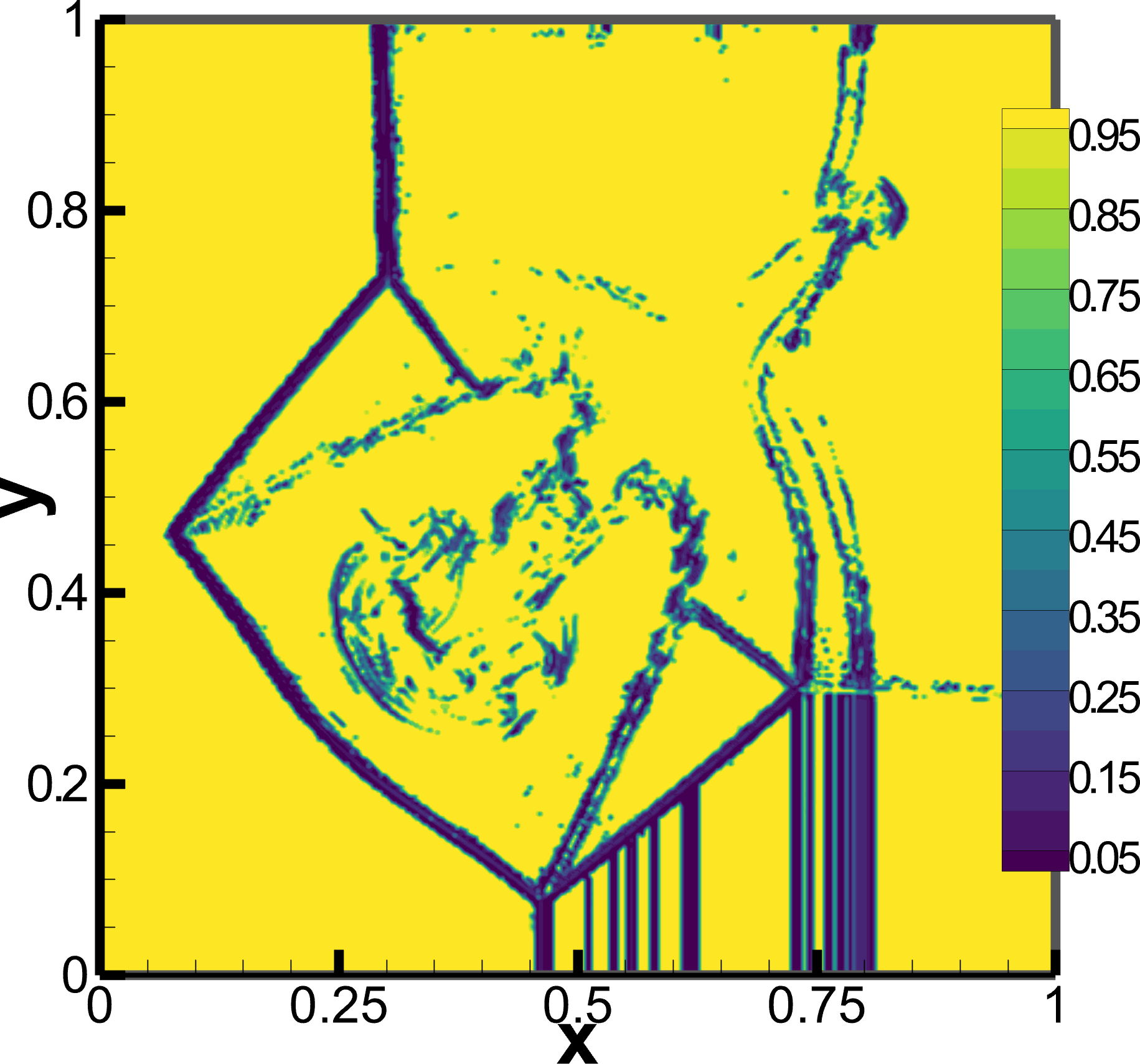}
    \caption{$\sigma$ in $x$-direction.}
    \end{subfigure}
    \quad
    \begin{subfigure}[b]{0.3\textwidth}
    \includegraphics[width=\textwidth]{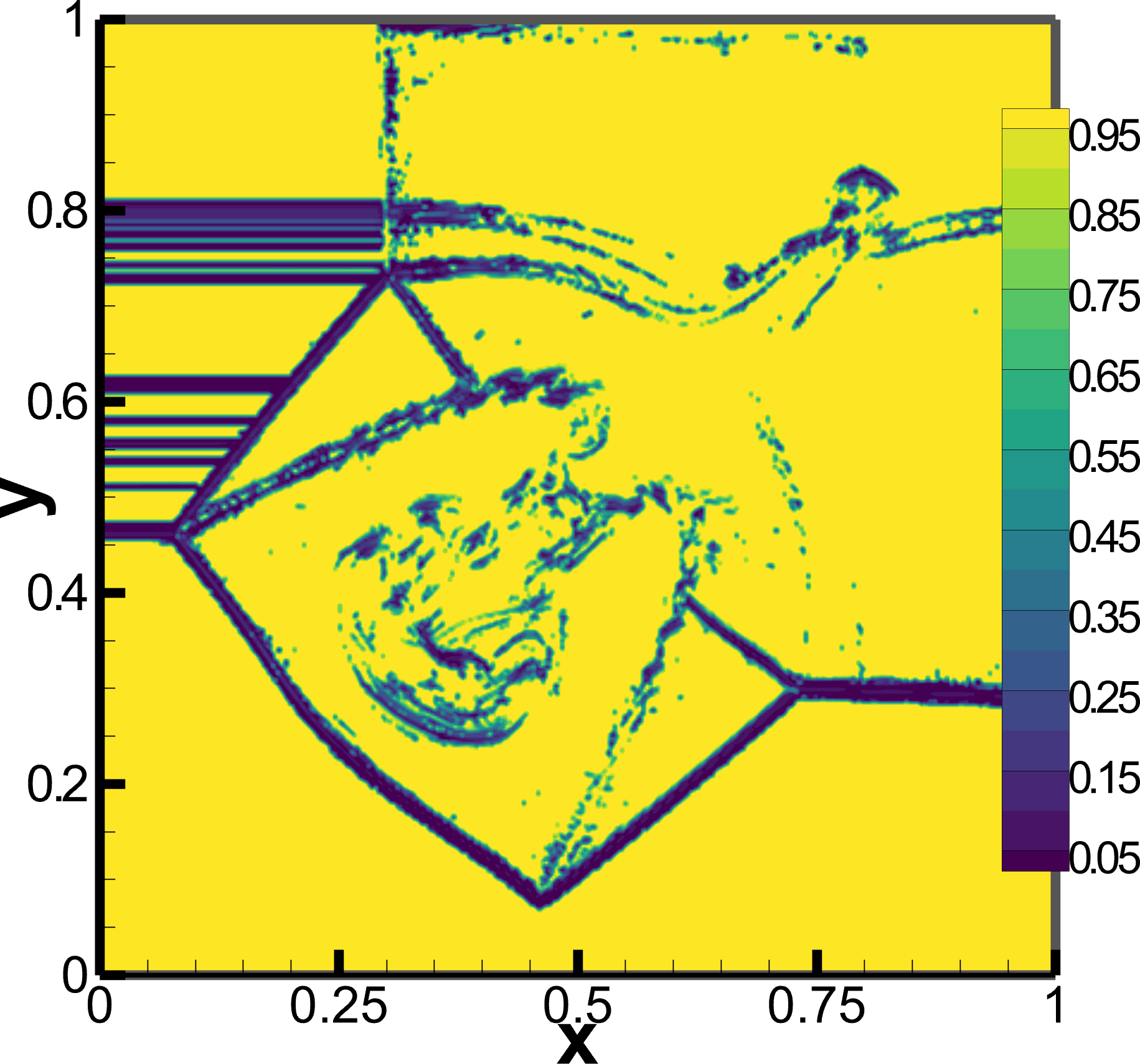}
    \caption{$\sigma$ in $y$-direction.}
    \end{subfigure}
    \caption{Shock detector for the 2D Riemann problem. $N_x \times N_y = 400 \times 400$, $t = 0.8$ and CFL = $0.6$.
    \label{fig:2dr_shock}}
\end{figure}

The computational domain is $\Omega = [0,1]\times[0,1]$ discretized by $400 \times 400$ uniform control volumes.
The problem is advanced to $t = 0.8$ utilizing third-order SSP-RK method with CFL number 0.6.
Figure \ref{fig:2dr_400} depicts the density contours for the WCLS3 scheme. Once again, the WCLS3 scheme performs higher resolution than the CWENO3 scheme and comparative resolution with the CWENO5 scheme near the Kelvin-Helmholtz instabilities.
Figure \ref{fig:2dr_shock} presents the shock detector distribution, which demonstrates that the proposed shock detector identifies the troubled cells accurately.

\subsection{Navier-Stokes Equations\label{sec:ns}}
\subsubsection{Viscous Shock Tube\label{sec:vst}}
A viscous shock tube problem with Reynolds number $200$ \cite{daru_numerical_2009} is taken to finalize the numerical experiments.
Strong shock, boundary layer and vortex structures interact with each in this problem. The initial condition is given as
\begin{equation}
  \left(\rho, u, v, p\right)  = \left\{
    \begin{array}{ll}
      120,0,0,\frac{120}{\gamma},& x < 0.5,\\
      1.2,0,0,\frac{1.2}{\gamma}, & x\geq 0.5,
    \end{array}
  \right.
\end{equation}
where $\gamma=1.4$.
The computational domain is $\Omega = [0,1]\times[0,1]$ with four adiabatic viscous walls.
Initially, the perfect gas is separated into left and right sides with different density and pressure.  Upon start, a shock wave propagates rightward, followed by a contact discontinuity. The moving shock and contact discontinuity interact with the viscous walls, forming a thin boundary layer. When the shock reflects from the right wall, it interacts with both the contact discontinuity and the boundary layer, forming a $\lambda$-shaped shock structure. This structure further induces boundary layer separation bubbles, unstable slip lines, and vortex structures downstream of the bubbles.

The computational domain is discretized by $300\times 300$ and $600\times 600$ uniform control volumes. The simulation is advanced by the third-order SSP-RK method to $t = 1.0$ with CFL number as $0.6$. 
When the number of grids is $300 \times 300$, the height of the main vortex of the WCLS3 scheme is approximately 0.133. When the grid number is increased to $600 \times 600$, a secondary vortex is generated near the main vortex, and the height of the main vortex is $0.163$ which is closer to the reference result \cite{daru_numerical_2009}.

\begin{figure}[!htbp]
  \centering
    \begin{subfigure}[b]{0.3\textwidth}
    \includegraphics[width=\textwidth]{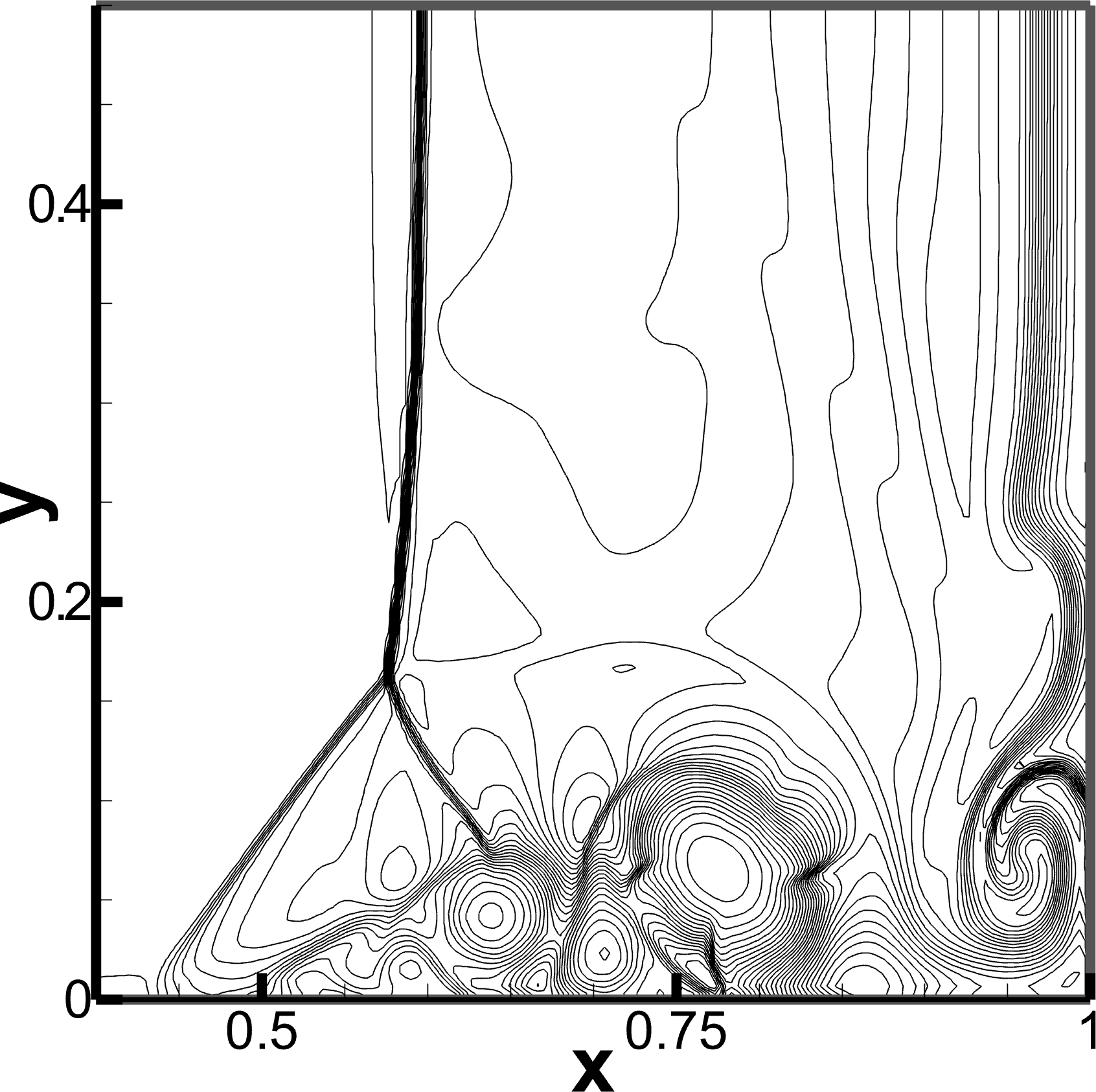}
    \caption{CWENO3.\label{fig:vst_cweno3_300}}
    \end{subfigure}
    \quad
    \begin{subfigure}[b]{0.3\textwidth}
    \includegraphics[width=\textwidth]{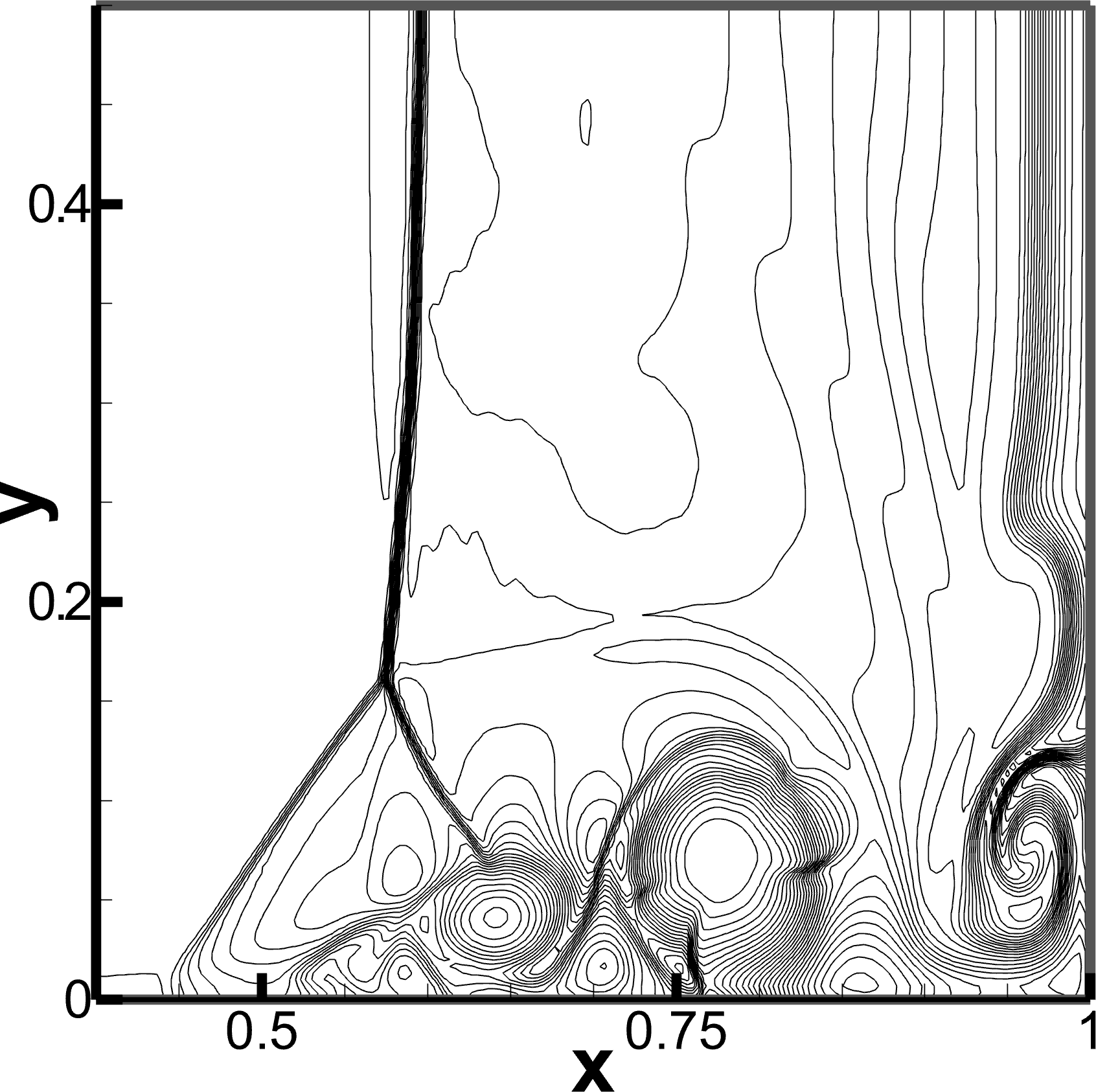}
    \caption{CWENO5.\label{fig:vst_cweno5_300}}
    \end{subfigure}\\
    \begin{subfigure}[b]{0.3\textwidth}
    \includegraphics[width=\textwidth]{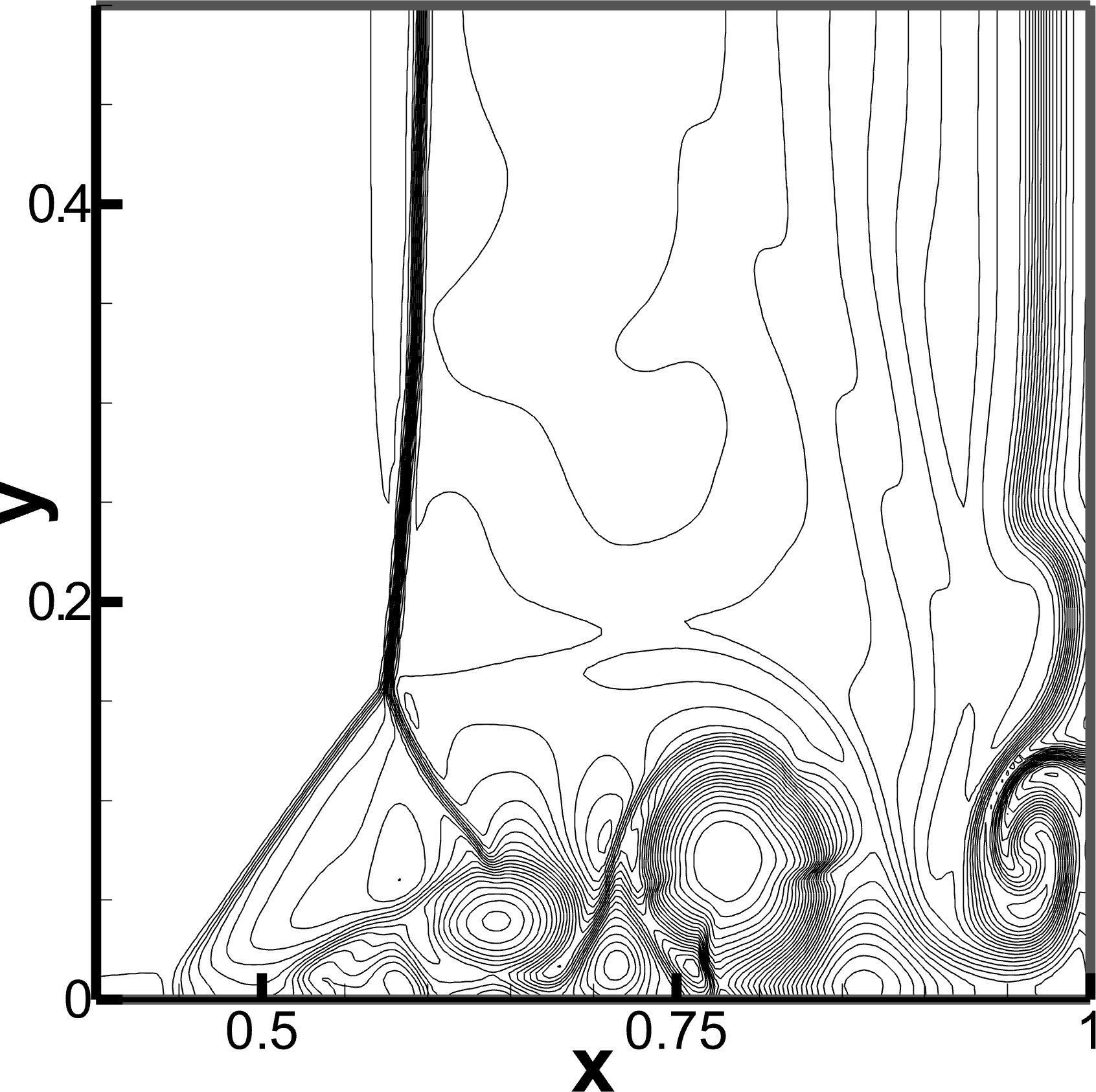}
    \caption{WCLS3 with $\kappa_0 = 0.8$.\label{fig:vst_k0.8_300}}
    \end{subfigure}
    \begin{subfigure}[b]{0.3\textwidth}
    \includegraphics[width=\textwidth]{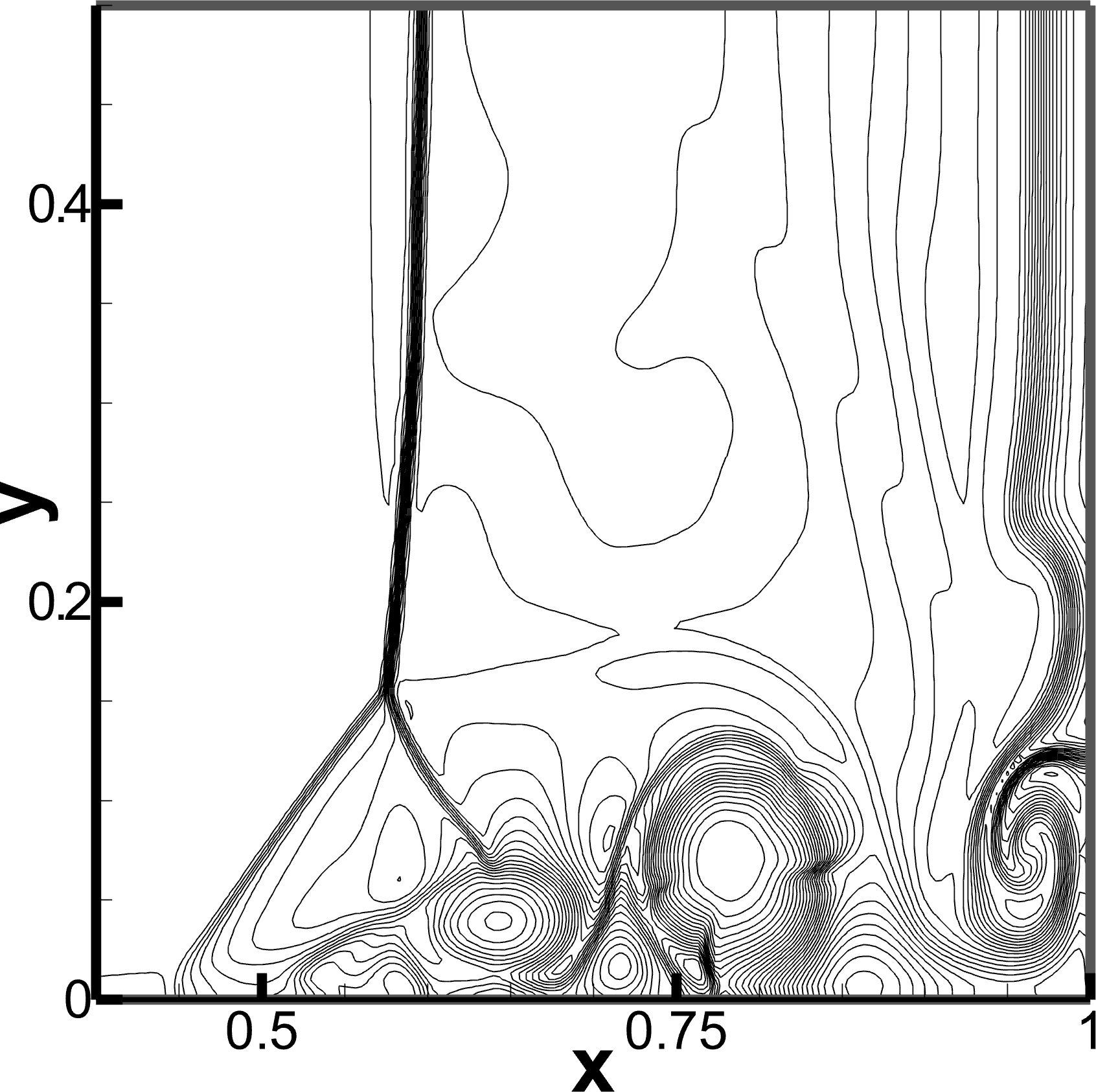}
    \caption{WCLS3 with $\kappa_0 = 1.0$.\label{fig:vst_k1.0_300}}
    \end{subfigure}
    \begin{subfigure}[b]{0.3\textwidth}
    \includegraphics[width=\textwidth]{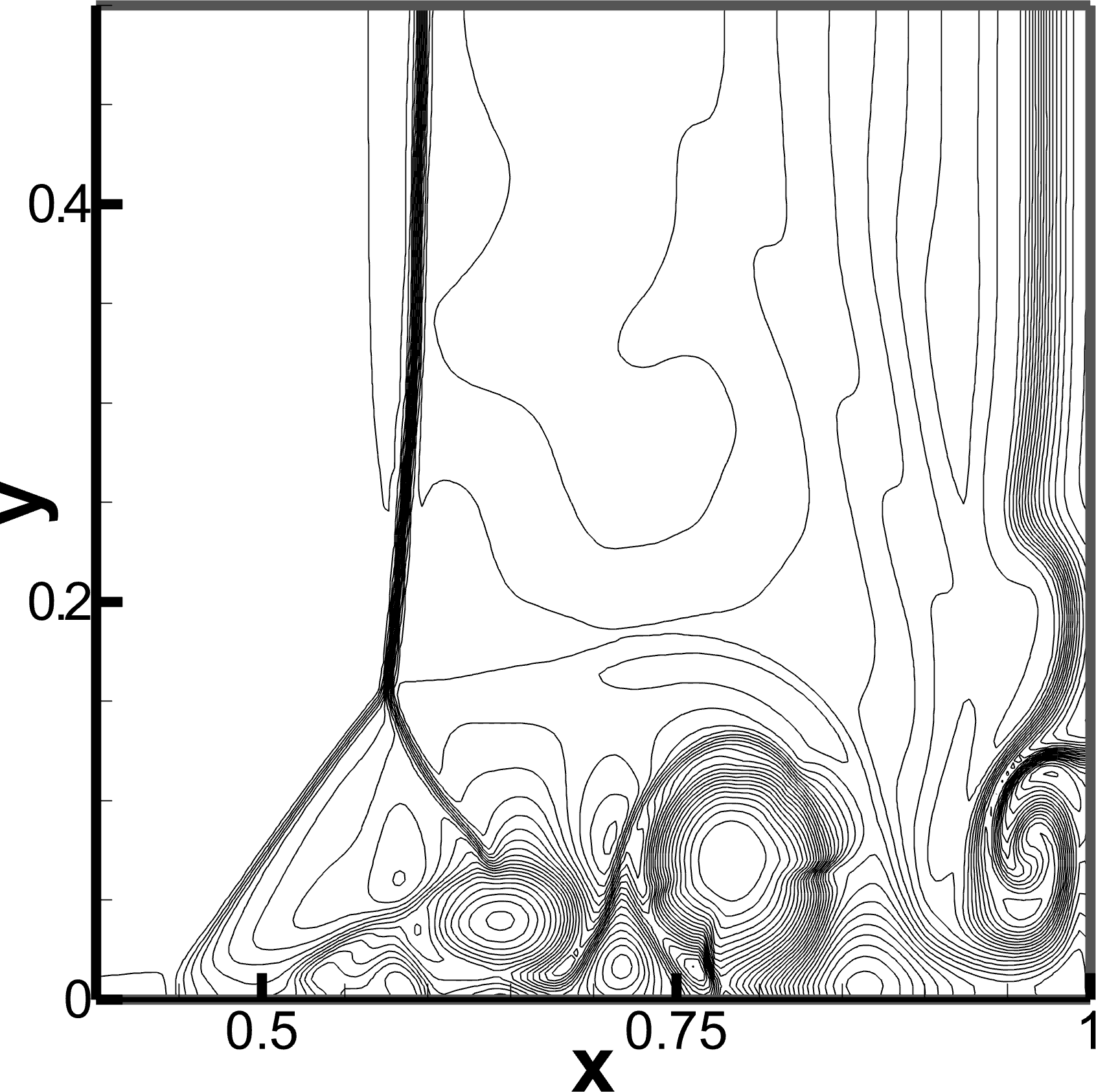}
    \caption{WCLS3 with $\kappa_0 = 1.2$.\label{fig:vst_k1.2_300}}
    \end{subfigure}
    \caption{Density contours for the viscous shock tube with 30 lines ranging from 21 to 131. $N_x \times N_y = 300 \times 300$, $t = 1.0$ and CFL = $0.6$.
    \label{fig:vst_300}}
\end{figure}

\begin{figure}[!htbp]
  \centering
    \begin{subfigure}[b]{0.3\textwidth}
    \includegraphics[width=\textwidth]{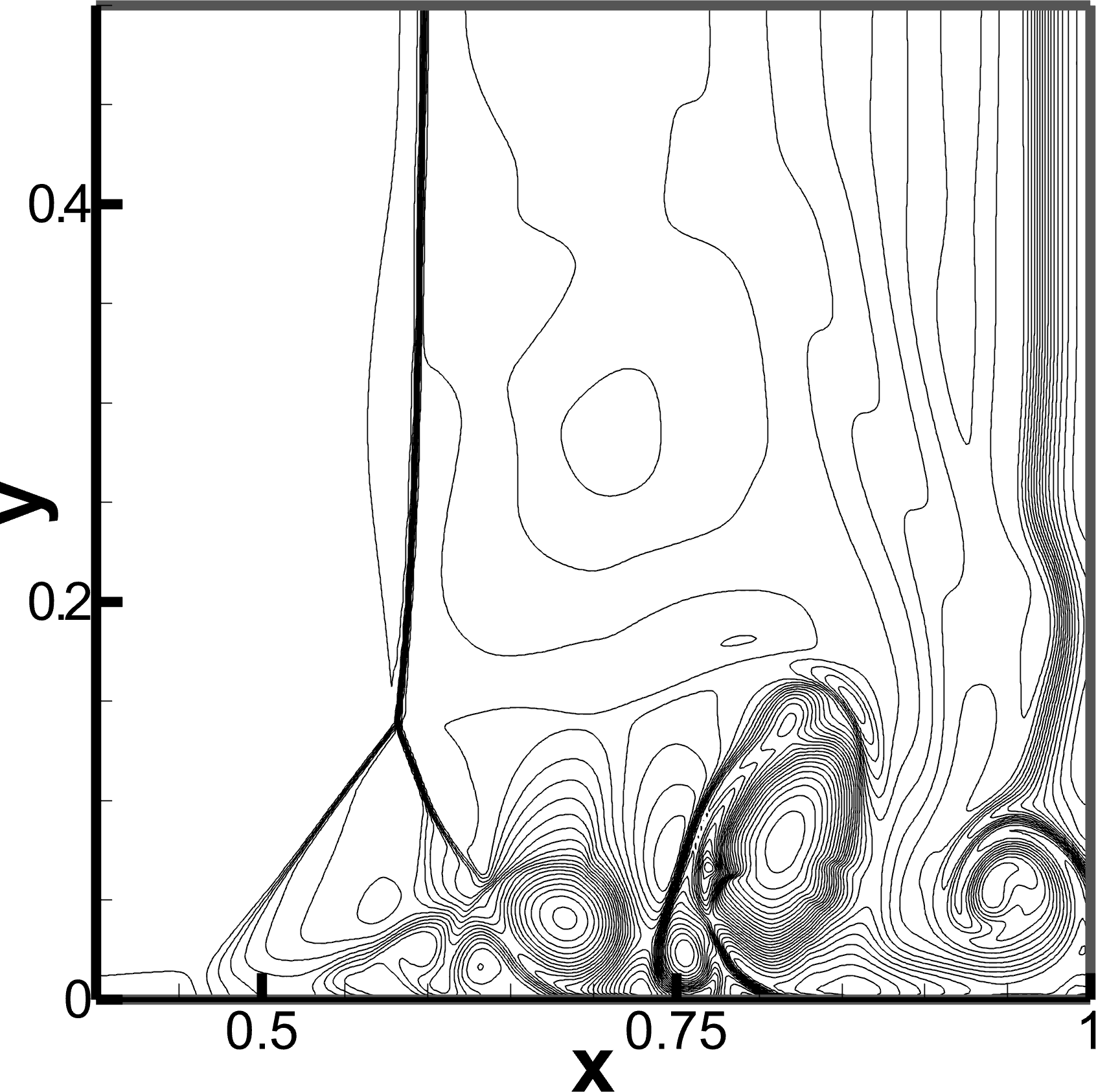}
    \caption{CWENO3.\label{fig:vst_cweno3_600}}
    \end{subfigure}
    \quad
    \begin{subfigure}[b]{0.3\textwidth}
    \includegraphics[width=\textwidth]{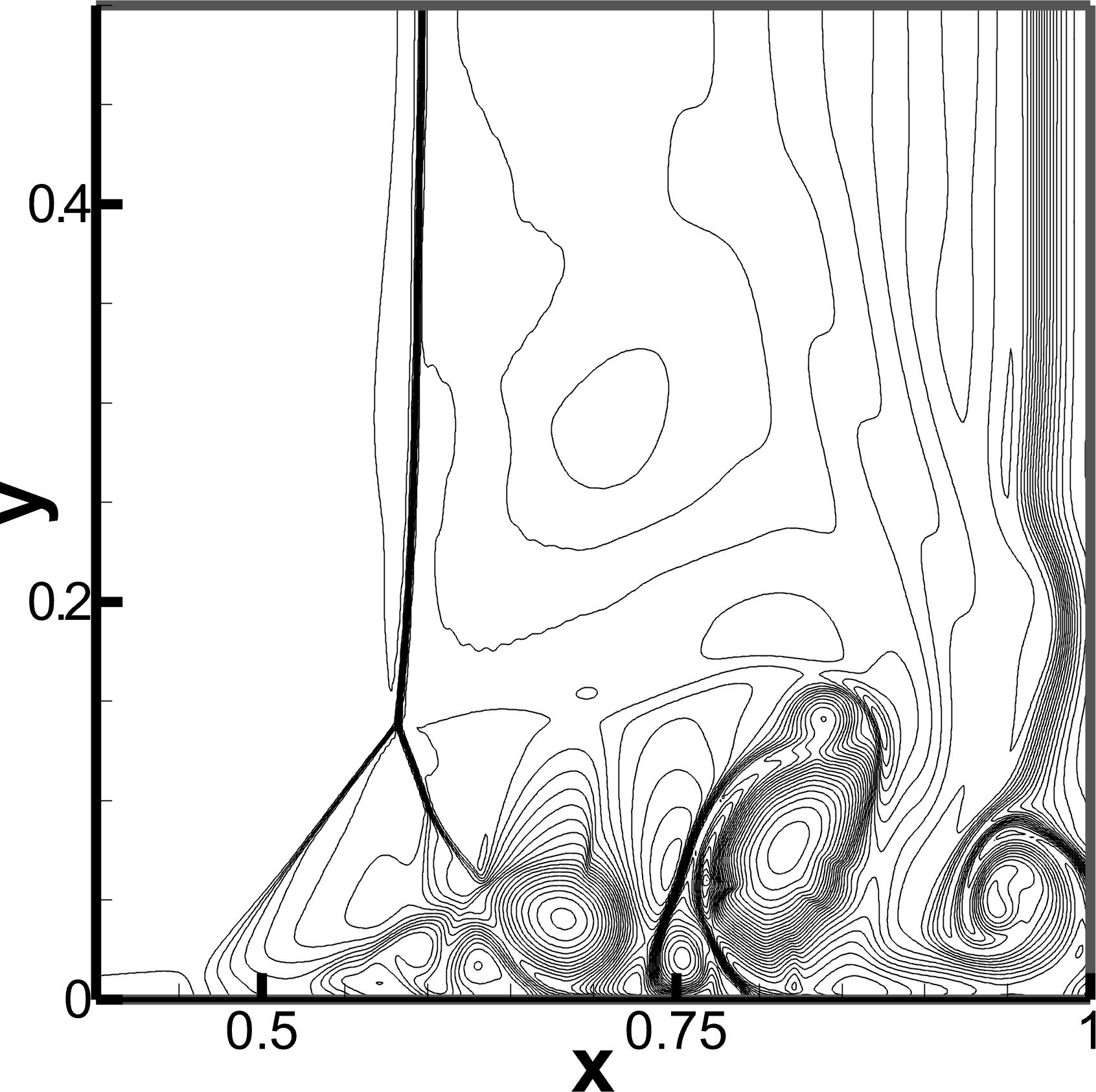}
    \caption{CWENO5.\label{fig:vst_cweno5_600}}
    \end{subfigure}\\
    \begin{subfigure}[b]{0.3\textwidth}
    \includegraphics[width=\textwidth]{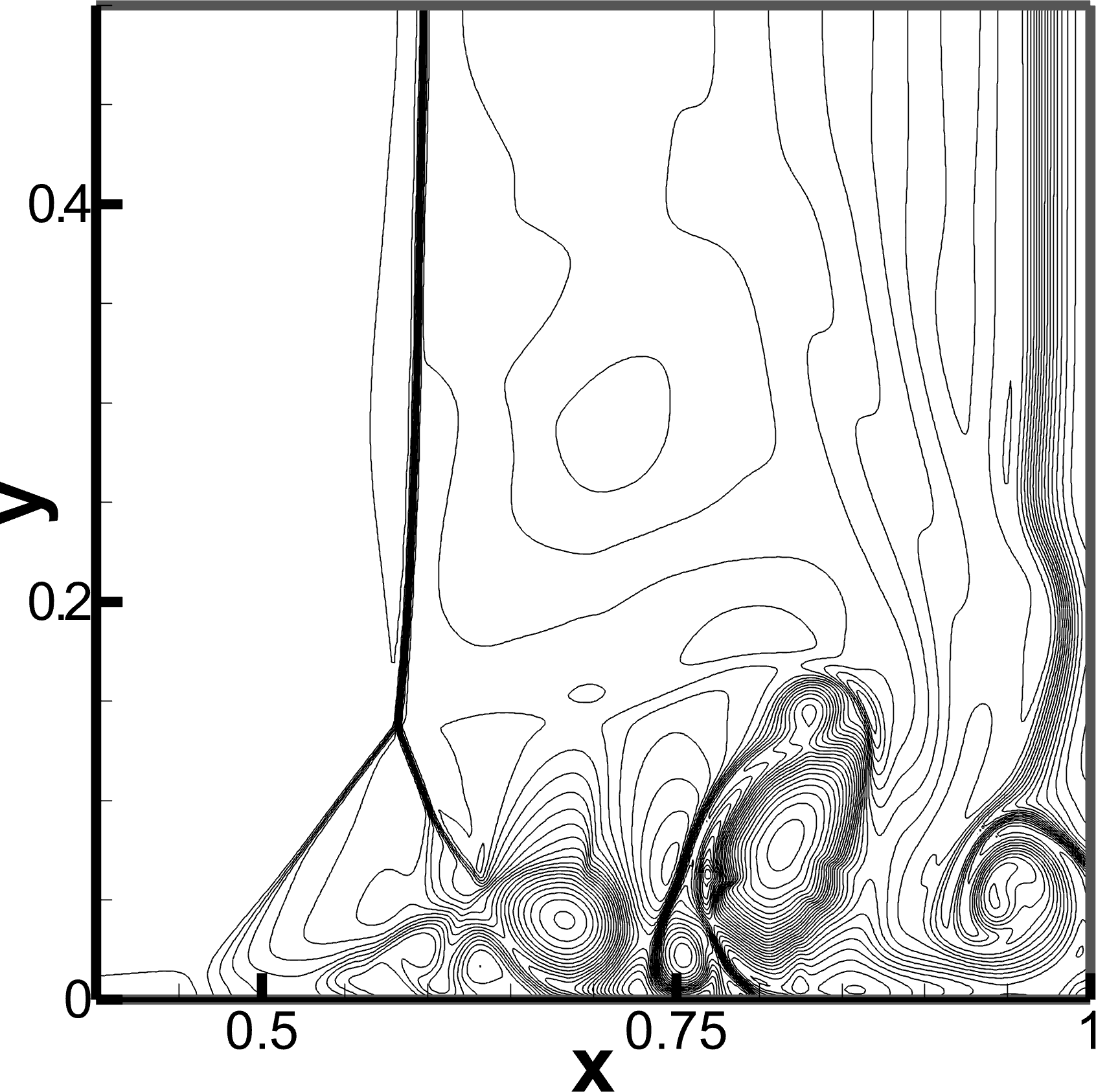}
    \caption{WCLS3 with $\kappa_0 = 0.8$.\label{fig:vst_k0.8_600}}
    \end{subfigure}
    \begin{subfigure}[b]{0.3\textwidth}
    \includegraphics[width=\textwidth]{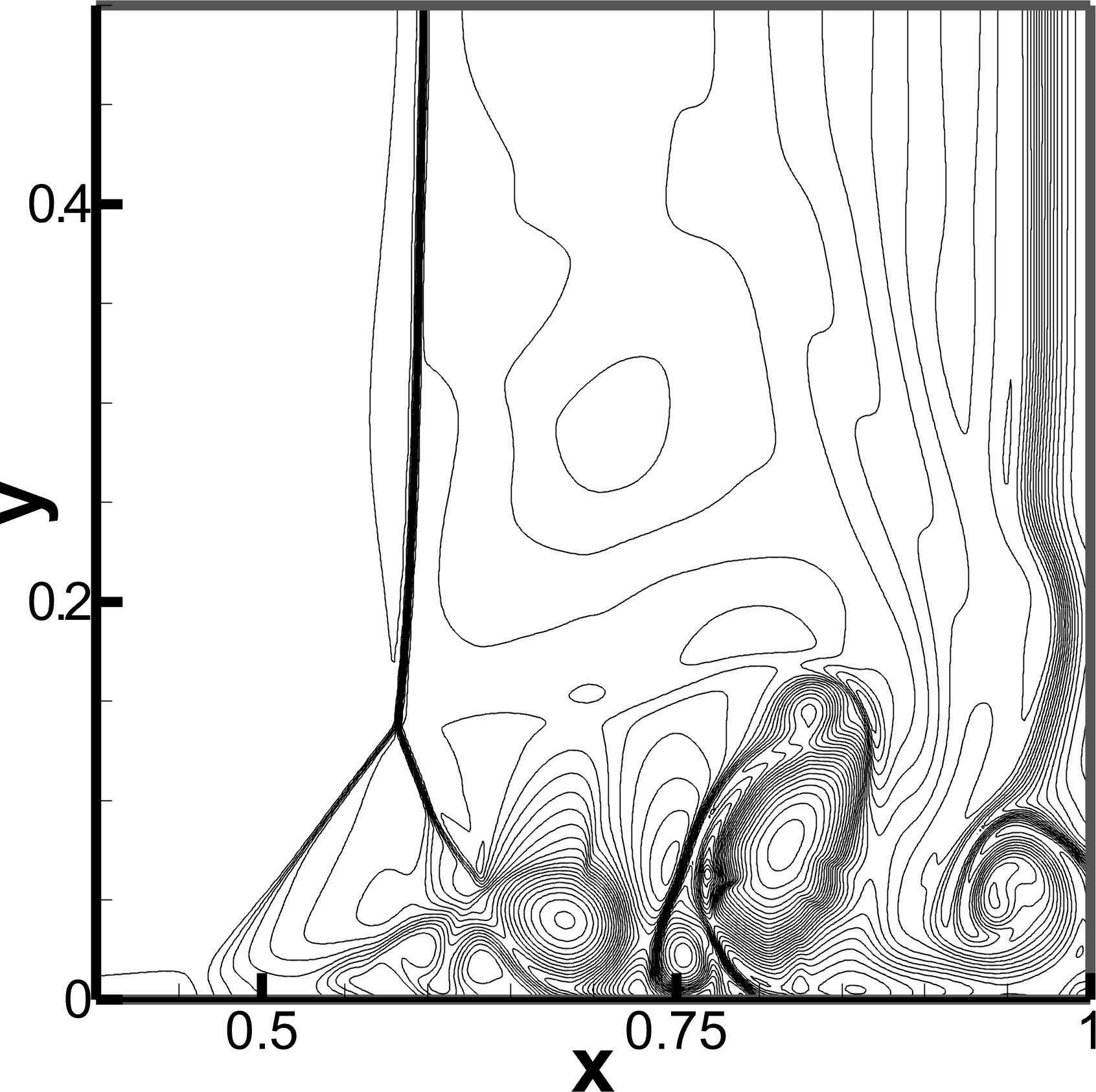}
    \caption{WCLS3 with $\kappa_0 = 1.0$.\label{fig:vst_k1.0_600}}
    \end{subfigure}
    \begin{subfigure}[b]{0.3\textwidth}
    \includegraphics[width=\textwidth]{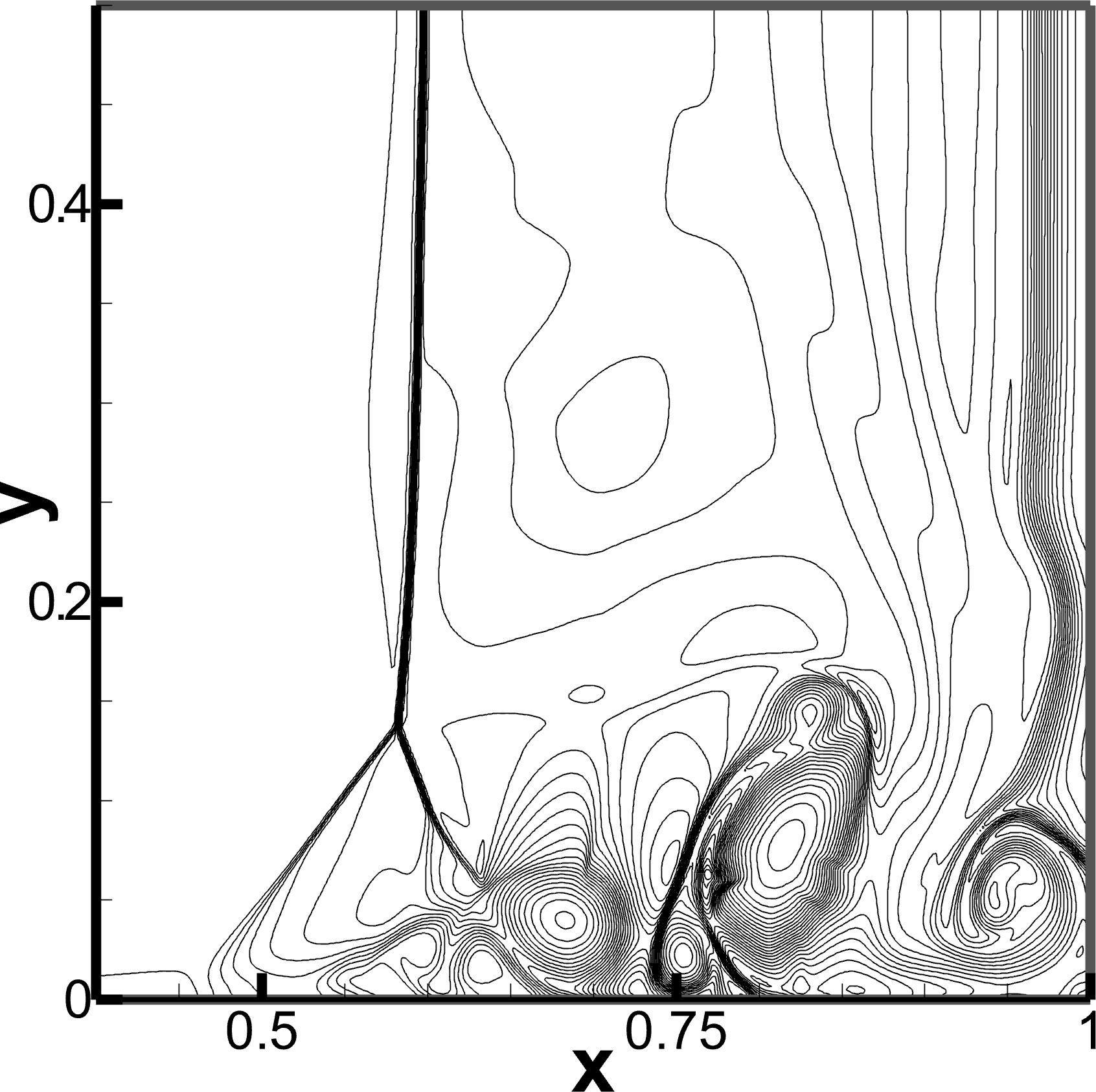}
    \caption{WCLS3 with $\kappa_0 = 1.2$.\label{fig:vst_k1.2_600}}
    \end{subfigure}
    \caption{Density contours for the viscous shock tube with 30 lines ranging from 21 to 131. $N_x \times N_y = 600 \times 600$, $t = 1.0$ and CFL = $0.6$.
    \label{fig:vst_600}}
\end{figure}

\begin{remark}
Table \ref{tab:simtime} lists the simulation time for several 2D numerical examples. The reconstruction process for the CWENO schemes is conducted in the characteristic space for all control volumes. The reconstruction process for the WCLS3 scheme includes the shock detector computation, the characteristic decomposition for troubled cells and the solution of block tridiagonal systems.
The authors admit that the solution of block tridiagonal systems is computational heavy and is the most critical limitation of the proposed WCLS3 scheme. However, the resolution gain is also significant with the WCLS3 scheme. And when extending to implicit method \cite{wang2016compact2,wang2017compact,jianhua2018high}, the authors believe the WCLS3 scheme can achieve the high accuracy and high resolution efficiently.
\begin{table}
  \centering
\caption{Simulation time of 2D cases. Unit: s.\label{tab:simtime}}
\begin{tabular}{lccc}
  \toprule
Case & CWENO3  & CWENO5 & WCLS3 with $\kappa_0 = 1.0$\\
\hline
Double shear layer  & 178 & 253 & 525 \\
Shock vortex interaction $250 \times 125$ & 50  & 68 & 99   \\
Shock vortex interaction $500 \times 250$ & 426 & 575 & 740 \\
Double Mach reflection $1920 \times 480$  & 18120&24580 & 22300\\
2D Riemann problem & 1293 & 1776 & 3499 \\
Viscous shock tube $300 \times 300$ &1184  &	1530  & 2319 \\
Viscous shock tube $600 \times 600$ &18900 &	19320 & 23460\\
\bottomrule
\end{tabular}
\end{table}
\end{remark}

\section{Conclusions \label{sec:concl}}
This paper proposes a third-order weighted essentially non-oscillatory compact least-squares scheme designed for curvilinear non-uniform grids. The scheme includes an adaptive strategy for the linear coefficients, which ensures robustness in discontinuous regions and high resolution in smooth regions. Through a nonlinear weighting process, the scheme achieves enhanced shock-capturing capabilities. Coupled with an improved shock detector developed in this work, the method avoids order degradation in both linear convection and nonlinear Euler equations. The third-order weighted essentially non-oscillatory compact least-squares scheme is promising with a resolution that is higher than the third-order weighted essentially non-oscillatory schemes and comparative with the fifth-order weighted essentially non-oscillatory schemes. A series of numerical experiments confirm the robustness and high resolution of the proposed scheme. Further work includes the extension to higher order and to implicit framework which could further improve the efficiency of the proposed method.

\section*{Acknowledgements}
This work was supported by the National Natural Science Foundation of China (No. 12102211), the Science and Technology Innovation 2025 Major Project of Ningbo, China (No. 2022Z213), the Major Program of the National Natural Science Foundation of China (Nos. 12292980,12292982) and Project in Northeast Normal University (No. GFPY202505).

\bibliographystyle{elsarticle-num}
\bibliography{WCLS}
\end{document}